\documentclass[12pt,oneside,openany]{book}
\begin{document}
\mainmatter

\begin{titlepage}
\begin{center}
{\large\bf Group-Variation Equations for the Coefficients in the\\ $1/N$
Expansions of Physical Quantities in\\ $\textrm{SU}(N)$ Gauge Theories in $D=3+1$\\}
{Chris Austin (chrisaustin@ukonline.co.uk)\\
33 Collins Terrace, Maryport, Cumbria CA15 8DL, England\\
}
\end{center}
\begin{center}
{\bf Abstract}
\end{center}
\noindent The coefficients in the $1/N$ expansions of the vacuum
expectation values and correlation functions of Wilson loops, in continuum
$\textrm{SU}(N)$ gauge theories in 3+1 dimensions, are shown to be 
determined by a closed
and complete set of equations, called the Group-Variation Equations,
that exhibit a simple and robust mechanism for the emergence of massive
glueballs and the Wilson area law.
The equations predict that the cylinder-topology 
minimal-area spanning surface term in the two-glueball correlation 
function, when it exists, must be multiplied by a pre-exponential 
factor, which for large area $A$ of the minimal-area cylinder-topology 
surface, decreases with increasing $A$ at least as fast as $\frac{1}{\ln(\sigma A)}$.  
If this factor decreases faster than $\frac{1}{\ln(\sigma A)}$, then the
mass $m_{0^{++}}$ of the lightest glueball, and the coefficient $\sigma$ of the 
area in the Wilson area law, are determined in a precisely parallel 
manner, and the equations give a zeroth-order estimate of
$m_{0^{++}}/\surd\sigma$ of 2.38,
about 33\% less than the best lattice value, without the need for a full calculation of any of the terms in the right-hand sides.  The large
distance behaviour of the vacuum expectation values and correlation
functions is completely determined by terms called island diagrams,
the dominant contributions to which come from islands of \emph{fixed}
size of about $\frac{1}{\sqrt{\sigma}}$.  The value of $\sigma$ is 
determined by the point at which $\left|\frac{\beta(g)}{g}\right|$
reaches a critical value, and since the large distance behaviour of all
physical quantities is determined by islands of the \emph{fixed} size
$\frac{1}{\sqrt{\sigma}}$, the running coupling $g^2$ never increases
beyond the value at which $\left|\frac{\beta(g)}{g}\right|$ reaches the
critical value.  Evidence is given, based on 't Hooft's demonstrations of
the geometric convergence of sums of planar diagrams, and the fact that
the coefficients in $\beta(g)$ all have the same sign, in natural
renormalization schemes, that the sums in the right-hand sides of the
Group-Variation Equations will converge geometrically, and that the
critical value of $g^2$ will be \emph{strictly smaller} than the radius of
convergence, which would imply that the Group-Variation Equations provide a basis for the calculation of all physical quantities within the
framework of the $1/N$ expansion, if a systematic method exists for
solving the equations, for example, the iterative substitution of the
left-hand sides into the right-hand sides.

\end{titlepage}

\setcounter{page}{2}
\pagestyle{plain}

\tableofcontents

\newpage
\section*{Introduction}
\label{intro}

The continuing desirability of a reliable analytic method of calculating 
physical quantities, at large distances,
in QCD, has recently been emphasized by Gross, 
\cite{25 Years}, \cite{Strings 2000}, and Witten, 
\cite{Witten New Perspectives}.  In this paper I give a new, complete, set 
of equations, called the Group-Variation Equations, for the coefficients 
in the $\frac{1}{N}$ expansions of vacuum expectation values and 
correlation functions of Wilson loops in $\textrm{SU}(N)$ Yang-Mills 
theory in $3+1$ dimensions, which exhibit a simple and robust mechanism
for the emergence of massive glueballs and the Wilson area
law.\footnote{By a Wilson loop, I mean the trace 
of a general closed-loop path-ordered phase factor,
 not just rectangular loops.  We perform a standard Wick
rotation to 4 Euclidean dimensions.}  There is a separate 
Group-Variation Equation for each non-vanishing coefficient in the 
$\frac{1}{N}$ expansion of a vacuum expectation value or correlation 
function, and the Group-Variation Equation for a coefficient expresses 
the derivative of that coefficient, with respect to the coupling 
constant $g^2$, in a natural way, in terms of that coefficient itself, 
and the other coefficients, of the same, or lower, order in the 
$\frac{1}{N}$ expansions.

If we consider a set of one or more Wilson loops, and define a 
one-parameter family of sets of loops, by multiplying the sizes and 
separations of the loops in the original set by a scale factor $L$, then 
by the renormalization group, the Group-Variation Equation for the 
vacuum expectation value, or the correlation function, of any set of 
loops in the family, can be expressed as an equation for
the derivative with 
respect to $L$, of that vacuum expectation value or correlation 
function.  The Group-Variation Equation, for that vacuum expectation 
value or correlation function, can then be integrated with respect to 
$L$, starting from boundary conditions at small $L$, as given by 
renormalization-group-improved perturbation theory, and continuing to 
\emph{arbitrarily large} $L$, since the structure of the Group-Variation 
Equations guarantees that the Wilson area law, and the massive glueball 
saturation of the correlation functions, solve the equations 
self-consistently, at long distances.

The Group-Variation Equations for the lowest non-vanishing coefficients 
in the $\frac{1}{N}$ expansions of the vacuum expectation value of one 
Wilson loop, and the correlation functions of two or more Wilson loops, 
close among themselves, and their right-hand sides have a simple 
representation in terms of planar diagrams, where a simply connected 
window in a diagram represents the vacuum expectation value of the 
Wilson loop, defined on the closed loop of paths that forms the border 
of the window, a multiply connected window represents the correlation 
function of the Wilson loops, defined on the closed loops of paths that 
form the border of the window, and the propagators in the diagrams 
represent sums over paths, that are to be taken with kinematic weights 
given explicitly in the paper.  \footnote{Window-weighted path integrals 
were studied by Migdal and Makeenko \cite{MigMak}, in connection with 
the analysis of the Migdal-Makeenko equations, but the Group-Variation 
Equations have no connection with the Migdal-Makeenko equations.}

Each diagram is multiplied by an integer coefficient, 
$\left.\frac{\mathrm{d}}{\mathrm{d}M}\mathbf{C}(M)\right|_{M=1}$, where 
$\mathbf{C}(M)$ is the chromatic polynomial of the diagram, which is the 
number of distinct ways of colouring the windows of the diagram with $M$ 
colours available, subject to the map-colouring rule that no two windows 
that meet along a propagator, are coloured the same colour.  This 
coefficient \emph{vanishes} for the vast majority of diagrams, because their 
chromatic polynomial, $\mathbf{C}(M)$, contains two or more factors of 
$(M-1)$.

The long-distance behaviour of the vacuum expectation value of one 
Wilson loop, and the correlation functions of two or more Wilson loops, 
is completely controlled by the diagrams that have an island, or in 
other words, a connected group of propagators, none of which joins onto 
any of the Wilson loops in the left-hand side of the equation.  The 
requirement that 
$\left.\frac{\mathrm{d}}{\mathrm{d}M}\mathbf{C}(M)\right|_{M=1}$ be 
non-vanishing, implies that a diagram can have at most one island, and, 
if it has an island, it can have \emph{no} propagator that does not form 
part of the island.

In particular, an island diagram, in the Group-Variation Equation for 
the vacuum expectation value of one Wilson loop, has one window of 
cylinder topology, that represents the correlation function of the 
left-hand side Wilson loop, and the Wilson loop defined on the closed 
loop of paths that forms the outer border of the island, and one or more 
simply connected windows inside the island, that represent the vacuum 
expectation values of the Wilson loops defined on the closed loops of 
paths that form their borders.  This structure results in the emergence 
of the Wilson area law in a simple and robust way: the window weights 
suppress long propagators, essentially by generating an effective mass 
for the propagators in their borders.  Thus, once some semblance of an 
area law has appeared, the window weights, for the internal windows of 
the island, suppress the contributions of all islands of size larger 
than $\frac{1}{\sqrt{\sigma}}$, where $\sigma$ is the coefficient of the 
area, in the emerging area law.  An island diagram will then give a
contribution proportional to the area of the minimal-area spanning
surface of the left-hand side Wilson loop, because we may expect a
comparable contribution to the island diagram, from an island of size
about $\frac{1}{\sqrt{\sigma}}$, that is situated close enough to any
point of the minimal-area spanning surface of the left-hand side Wilson
loop.

Applying $L\frac{\partial}{\partial L}$ to the left-hand side Wilson loop
also results in a factor of the area of the minimal-area spanning surface
of the loop, thus verifying the consistency of the area law.  This is
called the ``island diagram mechanism''.  A similar argument shows that
the island diagrams are also responsible for the massive glueball saturation
of the correlation functions, provided the mass of the lightest glueball
is less than twice the effective mass generated for the propagators by
the window weights.

The group-variation equations are derived by expressing the vacuum
expectation values and correlation functions, for $\textrm{SU}(NM)$, 
first in terms of those for $(\textrm{SU}(N))^M$, then in terms of those
for $\textrm{SU}(N)$, 
differentiating with respect to $M$ and setting $M = 1$, and in the 
left-hand sides, expressing the derivative with respect to $M$, in terms 
of a derivative with respect to $g^2$, then in terms of a derivative with respect to the sizes and separations of the Wilson loops involved,
via the renormalization group.

To motivate the Group-Variation Equations, crudely, as a way of summing
the planar diagrams of large-$N$ Yang-Mills theory, consider the sum of
all the planar Feynman diagrams that contribute, at leading non-vanishing
order in $\frac{1}{N}$, to the vacuum expectation value of one Wilson
loop, or to the correlation function of two or more Wilson loops.  They
are the planar diagrams that can be drawn on a sphere with $n$ holes, or
a disk with $n-1$ holes, where $n$ is the number of Wilson loops
involved.  Imagine colouring each window, of each diagram, independently,
with any of $M$ possible colours.  This results in taking each diagram a
total of $M^w$ times, where $w$ is the number of windows of the diagram,
which means that, up to an overall power of $M$, that is
the same for all the diagrams, we are now calculating the vacuum expectation value, or correlation function, for $\textrm{SU}(NM)$, with
coupling constant $g$, or alternatively, for $\textrm{SU}(N)$, with
coupling constant $g\sqrt{M}$, rather than for $\textrm{SU}(N)$, with
coupling constant $g$.  Now draw thick lines, or ``borders'', along all
propagators that run between two windows coloured in two \emph{different}
colours, and group together all coloured diagrams that have the same
pattern of coloured ``countries'', if we ignore the internal lines inside
the coloured ``countries''.  We then see that, if we look at the internal
lines inside the coloured ``countries'', in such a group of coloured
diagrams, and ignore distinctions between gluon lines and Fadeev-Popov
lines, (which are dealt with in detail in the paper), we have, for each
simply-connected
coloured ``country'', all the planar Feynman diagrams that contribute,
in leading non-vanishing order in the $\frac{1}{N}$ expansion, to the
vacuum expectation value of the Wilson loop defined on the ``path''
around the border of the ``country'', and for each non-simply-connected
coloured ``country'', all the planar Feynman diagrams that contribute,
in leading non-vanishing order in the $\frac{1}{N}$ expansion, to the
correlation function of the Wilson loops defined on the ``paths'' around
the two, or more, connected components of the border of that ``country''.
Here we have focussed on specific paths, in the sums over paths that
build up the propagators around the borders of the ``countries'', and we
then have to sum over these paths, weighted by the appropriate kinetic
factors, which are given in the paper.  The window weights for these
``countries'' are in $\textrm{SU}(N)$ Yang-Mills theory, with coupling
constant $g$.

Now add together all the sets of coloured Feynman diagrams that correspond to the same
pattern of ``countries'', but coloured in different ways, with $M$
different colours available, subject to the map-colouring rule that no
two ``countries'', that share a common border, are coloured in the same
colour.  This simply results in multiplying the contribution of any one
of the possible colourings, by $\mathbf{C}(M)$, where $\mathbf{C}(M)$,
the chromatic polynomial of the pattern of ``countries'', is the number
of different possible colourings of the pattern of countries, with $M$
different colours available, subject to the map-colouring rule.  We can
now differentiate with respect to $M$ and set $M$ equal to 1, which gives
the factor 
$\left.\frac{\mathrm{d}}{\mathrm{d}M}\mathbf{C}(M)\right|_{M=1}$.  Each
pattern of ``countries'', for which this factor is non-vanishing, gives a
diagram that contributes to the right-hand side, of the group-variation
equation, for the vacuum expectation value, or correlation function, that
we started with, provided we again ignore distinctions between gluon and
Fadeev-Popov lines, which are treated in detail in the paper.

It is not hard to see, in a crude way, that if, in $\textrm{SU}(NM)$ 
Yang-Mills theory, we integrate over the gauge fields \emph{outside} the 
$(\textrm{SU}(N))^M$
subgroup, for fixed values of the fields \emph{in} the $(\textrm{SU}(N))^M$ subgroup, 
which are treated as background fields, then for large $N$, at fixed $M$,
we get planar diagrams, each window of which is associated with one of the
$M$ different $\textrm{SU}(N)$ subgroups of $(\textrm{SU}(N))^M$, such that any two windows
that share a common border, (i.e. meet along a propagator), must be in
\emph{different} $\textrm{SU}(N)$ subgroups of $(\textrm{SU}(N))^M$.  This is because, if we
view the $\textrm{SU}(NM)$ gauge fields as entries in an $NM\times NM$ matrix, each
gauge field, outside the block-diagonal $(\textrm{SU}(N))^M$ subgroup, interacts with precisely two \emph{different} $\textrm{SU}(N)$ subgroups, of the 
block-diagonal $(\textrm{SU}(N))^M$ subgroup.  In the planar diagram picture, the
two distinct $\textrm{SU}(N)$ subgroups of $(\textrm{SU}(N))^M$, that a non-block-diagonal gauge field interacts with, can be viewed as associated with the two
``sides'' of the propagator, of that non-block-diagonal gauge field.

The windows of these planar diagrams do not have to be simply-connected.
In particular, a window can look like a ``lake'', with one or more
``islands'' in it.  The only requirement is that the windows, on the two
sides of a propagator, must be in \emph{different} $\textrm{SU}(N)$ subgroups of
$(\textrm{SU}(N))^M$, and this implies, in particular, that no ``island'' can be
connected to the ``shore'' of a ``lake'' by a single ``causeway'', (i.e. a
single propagator), because both sides of that propagator, being two 
different parts of the edge of a single window, would then be in the
\emph{same} $\textrm{SU}(N)$ subgroup of $(\textrm{SU}(N))^M$.

If we now use the general realization of propagators in background gauge
fields, as sums over paths, weighted by path-ordered phase factors,
calculated in those background gauge fields, we then see that, still
considering the block-diagonal $(\textrm{SU}(N))^M$ gauge fields as fixed background
fields, we get planar diagrams, in which the propagators represent sums
over paths, each weighted by the product of two path-ordered phase 
factors, one in each of the two different $\textrm{SU}(N)$ subgroups of 
$(\textrm{SU}(N))^M$, associated with the windows, on the two sides of that 
propagator, in the diagram.  The path-ordered phase factors, around the
border of any window, are all in the same $\textrm{SU}(N)$ subgroup of $(\textrm{SU}(N))^M$,
and the traces of their products, around the connected components of the
border of that window, are gauge-invariant.

We now, in the limit of large $N$, at fixed $M$, integrate over the 
block-diagonal $(\textrm{SU}(N))^M$ gauge fields.  Then in view of the 
factorization of vacuum expectation values of products of Wilson loops at
leading order in large $N$, we will find that for each simply-connected
window, and fixed paths in the sums over paths in the propagators at the
edges of that window, we simply get the vacuum expectation value, in
$\textrm{SU}(N)$, of the Wilson loop formed by the paths round the edge of that window.  This is of course completely independent of \emph{which} of the
$M$ $\textrm{SU}(N)$ subgroups of $(\textrm{SU}(N))^M$ is associated with that window.

It is not so obvious that for a non-simply-connected window, firstly, all
the two or more Wilson loops around its perimeter must be in the
\emph{same} $\textrm{SU}(N)$ subgroup of $(\textrm{SU}(N))^M$, and secondly, we get the 
leading non-vanishing, (at large $N$), contribution to the 
\emph{correlation function}, in $\textrm{SU}(N)$, of the Wilson loops that form its
perimeter.  That this is so, is shown in detail in the first half of this
paper.  A crude way of seeing that the Wilson loops around the perimeter
of a non-simply connected window must be in the same $\textrm{SU}(N)$ subgroup of
$(\textrm{SU}(N))^M$, is to consider the planar \emph{Feynman} diagrams that can contribute to that window, as discussed above,
and note that the distinct $\textrm{SU}(N)$ subgroups of
$(\textrm{SU}(N))^M$ do not interact with one another.

Finally, we sum over all distinct assignments of the different $\textrm{SU}(N)$
subgroups of $(\textrm{SU}(N))^M$ to the windows of our diagram, subject to the 
``map-colouring'' rule that no two adjacent windows share the same one of
the $M$ different ``colours''.  The total number of distinct assignments
of $M$ different ``colours'' to the windows of the diagram, subject to the
map-colouring rule, is given by the ``chromatic polynomial'' of the 
diagram, $\mathbf{C}(M)$, which is easily calculated.

We thus find that, at leading non-vanishing order at large $N$, the vacuum
expectation values and correlation functions of Wilson loops, in $\textrm{SU}(NM)$
Yang-Mills theory, are expressed in terms of the same vacuum expectation
values and correlation functions, in $\textrm{SU}(N)$ Yang Mills theory.  But at
leading non-vanishing order at large $N$, these same vacuum expectation
values and correlation functions, in $\textrm{SU}(NM)$ Yang Mills theory, are also
related to those of $\textrm{SU}(N)$ Yang Mills theory, by a rescaling of the
$\textrm{SU}(N)$ coupling constant: $g\rightarrow g\sqrt{M}$.  Thus we have
expressed the leading non-vanishing contributions, at large $N$, to the
vacuum expectation values and correlation functions of Wilson loops, in $\textrm{SU}(N)$
Yang Mills theory, with the coupling constant $g\sqrt{M}$, in terms of 
those same vacuum expectation values and correlation functions, in $\textrm{SU}(N)$
Yang Mills theory, with the coupling constant $g$, in such a way that the
only dependence of the right-hand sides of the equations on $M$, is 
through the chromatic polynomials $\mathbf{C}(M)$.

We can thus differentiate with respect to $M$, then set $M$ equal to 1.
The effect on the left-hand sides of the equations is that we now have the
derivative, with respect to $g$, of the vacuum expectation values and 
correlation functions, which may be further expressed, via the 
renormalization group, in terms of the derivative of those vacuum
expectation values and correlation functions, with respect to an overall
length-scale factor, if the sizes and separations of all the Wilson loops
involved, are re-scaled by a common scaling factor.

The effect on the right-hand sides of the equations is to produce a vast
reduction in the number of diagrams involved, because for huge classes of
diagrams, the chromatic polynomial $\mathbf{C}(M)$ has two or more factors
of $(M-1)$.  The reduction in diagrams is comparable to reducing a 
sum over ladders, to a Bethe-Salpeter kernel, although the way it works
is, of course, totally different.

In particular, in the equation for the vacuum expectation value of a 
single Wilson loop, only two types of diagrams survive in the right-hand 
side.  In the first type, every window is simply connected, and if the
left-hand side Wilson loop is rubbed out, the resulting diagram is still
connected.  In the second type, there is precisely one
non-simply-connected window, which looks like a ``lake'', whose outer
perimeter is precisely the left-hand side Wilson loop, and which contains
exactly one ``island'' of propagators.

These equations, called the Group-Variation Equations, are complete, 
because the Feynman-diagram expansions can be recovered from them order by
order in perturbation theory, by developing their solutions in powers of
$g$.  They have the advantage, over perturbation theory, that their 
solutions manifestly have the correct behaviour at large distances, namely
the Wilson area law, and massive glueball saturation of the correlation
functions.

To observe that this behaviour is a self-consistent solution of the 
group-variation equations, as discussed briefly above,
note that the principal effect of area-law
window weights, in the two windows beside a propagator, will be to give
that propagator an effective mass, which can be crudely estimated to be at
least $1.3\sqrt{\sigma}$, where $\sigma$ is the area-law parameter.  Thus
in the sums over paths, paths whose length is greater than $1/\sqrt{
\sigma}$ will be suppressed.  Thus when the area of the minimal-area
spanning surface of the left-hand side Wilson loop is greater than
$1/\sigma$, the non-island diagrams in the right-hand side will give a
contribution proportional to the perimeter of the loop, whereas the island
diagrams will give a contribution proportional to the area of the 
minimal-area spanning surface of the loop, and thus give the dominant
contribution, because we may expect a comparable contribution to the 
island diagram, from an island of size about $1/\sqrt{\sigma}$, that is
situated close enough to any point of the minimal-area spanning surface of
the left-hand side loop.

The correlation function of the 
left-hand side Wilson loop, (of size large compared to
$1/\sqrt{\sigma}$),
and the Wilson loop defined on the paths that
form the outer boundary of the island, will be largest when the island is
close to the minimal-area spanning surface $S$ of the left-hand side Wilson
loop, and it will approximately factorize into a factor $e^{-\sigma A}$,
where $A$ is the area of $S$, and a factor dependent on the orientation of
the island with respect to $S$, and on the perpendicular
distance of the island from $S$.  The result of performing the sums over the
island paths, subject to a fixed mean position of
the island paths, will be approximately
independent of the fixed mean position of the island
paths, other than through the overall factor that depends on
the perpendicular distance of the island from $S$.

Thus, for a crude first
estimate, we expect that the contribution of any island diagram, to the
right-hand side of the Group-Variation Equation, for the vacuum
expectation value of a Wilson loop, whose size is large compared to
$1/\sqrt{\sigma}$, will be equal to a constant, times $Ae^{-\sigma A}$,
where $A$ is the area of the minimal-area spanning surface, of the 
left-hand side Wilson loop.

And, as noted briefly above, applying $L\frac{\partial}{\partial L}$, to
the left-hand side Wilson loop,
also produces a factor of $A$, thus verifying the consistency of the
area law, within a crude first estimate.

A more careful study, in Chapter 7 of the paper,
shows that the term $e^{-\sigma A_c}$, in the correlation function of two
Wilson loops, where $A_c$ is the area of the cylinder-topology 
minimal-area orientable spanning surface of the two loops, must develop,
when it exists, a pre-exponential factor that decreases, at large $A_c$,
at least as fast as $1/\ln(\sigma A_c)$, because otherwise island
diagrams give contributions that are too \emph{large}: in addition to the
factor of $A$, they also get a factor of $\ln(\sigma A)$, which must be
cancelled by a pre-exponential factor,
that decreases at least as fast as $1/\ln(\sigma A_c)$, at large $A_c$.

If this pre-exponential factor decreases $\emph{faster}$ than
$1/\ln(\sigma A_c)$, at large $A_c$, then the cylinder-topology
minimal-area spanning surface term, in the correlation
function of two Wilson loops, gives \emph{no} contribution to the asymptotic form, in the limit of large $A$, of the 
right-hand side of the Group-Variation Equation, for the vacuum
expectation value of a single Wilson loop,
and the entire asymptotic contribution comes
from the term, in the correlation function of two loops, that has the form of
the lightest glueball, propagating by the shortest possible path, between
the separate minimal-area spanning surfaces of the two loops.  This term
always gives exactly the correct contribution, to the right-hand sides, of
the Group-Variation Equations.

As mentioned briefly above, a similar study shows that the island
diagrams are also responsible for the self-consistency of the massive
glueball saturation
of the correlation functions, provided the mass of the lightest glueball 
is less than twice the effective mass generated for the propagators by the
window weights.  This is a stringent consistency
condition: while it is satisfied for the zeroth-order approximation to the
lightest glueball mass, namely $2.38\sqrt{\sigma}$, that results if the
pre-exponential factor, in the cylinder-topology minimal-area orientable
spanning surface term, in the correlation function of two Wilson loops, decreases
\emph{faster} than $1/\ln(\sigma A_c)$, at large $A_c$, this zeroth-order
approximation
to the lightest glueball mass is some
33 percent smaller than the best lattice value of $3.56\sqrt{\sigma}$
\cite{Teper}.  Thus more refined estimates, of the effective mass of the
propagators, generated by the window weights, will have to give a value of
at least $1.78\sqrt{\sigma}$, in comparison with the simplest estimate,
obtained in Chapter 4 of the paper, 
that the value is at least $1.3\sqrt{\sigma}$.  (The nature of the 
calculation in Chapter 4 shows that the value of $1.3\sqrt{\sigma}$ is
conservative, in the sense that it is more likely to be an under-estimate
than an over-estimate.)

The island diagram mechanism is robust, in the sense that it works in
exactly the same way for the entire sum of island diagrams, as it does for
the leading order island diagrams, or for any finite sum of island diagrams.  The higher-order island diagrams only give adjustments to the
mass of the lightest glueball, as well as adjusting the pre-asymptotic
large-distance behaviour of the correlation functions, so that they 
correspond to a spectrum of massive glueballs with sharp masses.

Furthermore, the island diagram mechanism does not require a particularly
large value of $g$.  Indeed, when the renormalization group is used to
replace $\frac{\partial}{\partial g}$, in the left-hand side of the 
group-variation equations, by $L\frac{\partial}{\partial L}$, where $L$ is
an overall scaling parameter of the sizes and separations of the Wilson
loops involved, the right-hand side becomes multiplied by 
$\frac{\beta(g)}{g}$.  If
the right-hand sides are then restricted to the leading island diagrams, 
in which the islands are simple loops with no vertices, then this factor 
of $\frac{\beta(g)}{g}$, 
in the right-hand side, is the \emph{only} explicit 
dependence on $g$ of the equations: all the remaining dependence on $g$ is
through the dependence on $g$ of the vacuum expectation values and
correlation functions.

What this means is that, once $g$ reaches a critical value, determined by
the group-variation equations, at which the island diagram mechanism
operates to produce the correct long-distance behaviour of the vacuum
expectation values and correlation functions, there is no need to
consider any larger value of $g$.  This is due to the fact that the
long-distance behaviour, of all the vacuum expectation values and
correlation functions, is completely determined by islands whose size is
approximately \emph{fixed}, at about $1/\sqrt{\sigma}$.  The critical
value of $g$ will have some dependence on the renormalization scheme,
but not a lot, since the first two coefficients in $\frac{\beta(g)}{g}$
are independent of the renormalization scheme.  Comparison with
experimental results suggests that the critical value of
$\frac{g^2}{4\pi}$, as normalized in the second half of this paper, will
be larger than 0.43, (which corresponds to $\alpha_s(1784 \textrm{ MeV})=
0.35$, observed in $\tau$ decay \cite{beta in MS bar 1}),
but the most significant point is not the
actual value, of the critical value of $g$, but the fact it is determined
by the point, where the absolute value, of $\frac{\beta(g)}{g}$, reaches a
certain critical value.

't Hooft has demonstrated, in reference \cite{'t Hooft a}, that the sums
of the planar Feynman diagrams, in large-$N_c$ QCD,
converge geometrically, if one throws away all the divergent subdiagrams,
and furthermore, in reference \cite{'t Hooft b}, that
a similar result holds in the presence of the divergent
subdiagrams, if one uses a suitably generalized running coupling, and gives the gluons a mass, to cut off the long-distance growth of the running coupling.  The lower bound on the radius of
convergence, in the complex $g^2$ plane, proved by 't Hooft in these
references, is several orders of magnitude smaller than the expected
critical value of $g^2$.

It would seem reasonable to make the hypothesis, that in a ``natural''
renormalization scheme, such as $\overline{\mathrm{MS}}$ \cite{MS bar}, 
the convergence
behaviour of $\frac{\beta(g)}{g}$, as a power series in $g^2$, is neither
better, nor worse, than the convergence behaviour of other physical
quantities, such as the sums in the right-hand sides of the 
Group-Variation Equations.  If this is so, then we may expect the
large-$N$ limit of $\frac{\beta(g)}{g}$, as a power series in $g^2$, to
converge geometrically, for sufficiently small $g^2$.  Then since, in 
$\overline{\mathrm{MS}}$, all the coefficients, in the power series for
$\frac{\beta(g)}{g}$, seem to have the same sign 
\cite{beta in MS bar 1}, 
\cite{beta in MS bar 2}, the fastest direction of growth of 
$\left|\frac{\beta(g)}{g}\right|$, in the complex $g^2$ plane, will be along
the positive $g^2$ axis.  This implies that the critical value of $g^2$
will be \emph{strictly
smaller} than the radius of convergence of the power series for
$\frac{\beta(g)}{g}$, so the power series for $\frac{\beta(g)}{g}$ will
\emph{converge geometrically}, at the critical value of $g^2$.

The same hypothesis then implies that the sums, in the right-hand sides
of the Group-Variation Equations, will also \emph{converge
geometrically}, at the critical value of $g^2$.

Study of the large-$N$ limit of the four-loop $\beta$-function in
$\overline{\mathrm{MS}}$, obtained from the general result given in
reference \cite{beta in MS bar 2}, shows that
the ratios of successive pairs of coefficients in the expansion are
increasing, but at a decreasing rate, and indicates that the series is
likely to diverge for $\frac{g^2}{4\pi}$, as normalized in the second
half of this paper, somewhere in the range 0.53 to 1.05, and most likely,
near the lower end of this range.  (The four-loop term is essential to
reach these conclusions.)  The critical value of $\frac{g^2}{4\pi}$ is
expected to be strictly smaller than the value where the series diverges.

But as noted earlier, the experimental result, from observations of 
$\tau$ decay, that $\alpha_s(1748 \textrm{ MeV})=0.35$, indicates that 
the critical value, of this $\frac{g^2}{4\pi}$, is greater than 0.43.
This is because this $\frac{g^2}{4\pi}$ is equal to $\frac{3}{2}$ times
the value $\alpha_s$ would have in the absence of quarks, which at
1748 MeV is approximately $\frac{3}{2}\times\frac{1}{1.22}\times\alpha_s=1.23
\alpha_s$.  Thus $\alpha_s(m_\tau)$, which is the largest value of
$\alpha_s$ for which there is experimental evidence, must be very close
to the critical value.

It is interesting to note, in Figure 9.2, on page 19, of Chapter 9,
\textit{Quantum Chromodynamics}, of reference \cite{beta in MS bar 1},
that the experimental value, of $\alpha_s(m_\tau)$, lies about 1.6 standard
deviations above the best fit curve, to all measurements, of $\alpha_s$.
This suggests that the curve of $\alpha_s(\mu)$ is indeed starting to
curve upwards towards a vertical slope, as $\mu$ approaches $m_\tau$ from
above, which is what is expected as $\frac{\beta(g)}{g}$ starts to 
diverge.

Since the experimental value of $\sqrt{\sigma}$ is about 0.44 GeV
\cite{Michael}, this experimental evidence, that $\alpha_s$ is
approaching the critical value as $\mu$ approaches $1748\textrm{ MeV}=
4.0\sqrt{\sigma}$, is a further indication that the estimate in Chapter
4 of this paper, that the effective mass generated for the propagators by
the window weights, is at least $1.3\sqrt{\sigma}$, is an underestimate.
Indeed, the typical size, of the islands, of approximately fixed size,
that determine the long distance behaviour of all the vacuum expectation
values and correlation functions, is approximately equal to the 
reciprocal, of twice the effective mass generated for the propagators by the
window weights, since when an island elongates in
any direction, at least two of the propagators in the island get
elongated.  Thus since the mass at which $\alpha_s$ stops evolving, is
equal to the reciprocal of the size of these typical islands, the mass at
which $\alpha_s$ stops evolving, is equal to about twice the effective
mass generated for the propagators, by the window weights.  Thus the
experimental evidence that $\alpha_s$ is stopping evolving, by reaching
the critical value, at about
$\mu=4.0\sqrt{\sigma}$, indicates that the effective mass, generated for
the propagators by the window weights, is about
$2.0\sqrt{\sigma}$, rather than $1.3\sqrt{\sigma}$.

Since the
long-distance behaviour of vacuum expectation values and correlation
functions is determined, via the island diagram mechanism, by islands 
whose size is approximately fixed at $\frac{1}{4.0\sqrt{\sigma}}$, and this
mechanism works in exactly the same way for the full sum of island
diagrams, as it does for the leading-order island diagrams, or the sum of
any finite number of island diagrams, there is no physical reason why the
sums, in the right-hand sides of the Group-Variation Equations, should
not converge.

There is also a possibility that the Group-Variation Equations might
be systematically solvable, by the iterative substitution of their left-hand
sides into their right-hand sides, starting from a reasonable ansatz.

Whether the $1/N$ expansions themselves can converge, is more dubious.
For example, each time $N$ decreases from a positive integer, to the next
smaller positive integer, some of the states must drop out of the 
spectrum.  Whether this occurs by their masses becoming infinite, or by
their coefficients, in the expansion of any correlation function in terms
of ``eigenstates'', somehow vanishing for all integer values of $N$ below
a certain integer, or by some mechanism connected with the fact that most
of the states are also developing widths, is not known.

The fact that, within the context of the Group-Variation Equations, 
$\alpha_s$ does not evolve, at distances larger than 
the typical island size, might be compared with the fact that, within
quantum electrodynamics, the fine structure constant does not evolve, at
distances larger than the reciprocal of the mass of the lightest charged
particle, namely the electron.

The foregoing introduction has omitted many important points, and some
difficulties, that are considered in detail in the paper.  The general
idea of a ``two-stage'' integration over the gauge fields of some large
gauge group, (which in practice will be $\textrm{SU}(NM)$), in which the 
gauge 
fields that are \emph{not} members of some subgroup, (which in practice
will be $(\textrm{SU}(N))^M$), are integrated over first, treating the 
gauge fields
in the subgroup as background fields, after which the gauge fields in the
subgroup are integrated over, enabling the vacuum expectation values and
correlation functions, of gauge-invariant quantities in the larger group,
to be expressed in terms of sums over diagrams, involving sums
over paths, weighted by the vacuum expectation values and correlation
functions, of gauge-invariant quantities in the smaller group, is
most simply studied in the context of a general larger group $H$, and 
subgroup $G$, resulting in equations called the Group-Changing Equations
for $H$ and $G$.  A special choice of gauge-fixing and Fadeev-Popov terms
must be made, which is, however, renormalizable, and such that the sum of
the gauge-fixing and Fadeev-Popov terms is a BRST variation in the normal
way, with completely standard BRST variations.  There is a normal gauge
parameter, but the requirement that Fadeev-Popov loops stay either in the
subgroup or out of it, rather than wandering in and out of the subgroup,
requires that the analogue of Landau gauge must be chosen.

Starting from this systematic basis, the group-variation equations can be
derived, not only for the leading non-vanishing terms in the $1/N$
expansions, but for all the terms in the $1/N$ expansions.  The 
coefficient of every term in, the $1/N$ expansion, of the vacuum 
expectation
value or correlation function, of a product of Wilson loops, has its own
Group-Variation Equation, and the Group-Variation Equations for the
$n^\mathrm{th}$ non-vanishing terms, in the vacuum expectation value of 
one
Wilson loop, and the correlation functions of two or more Wilson loops,
close among themselves, and are complete, in the sense that the 
Feynman-diagram expansions can be recovered, by developing them, as power
series, in the coupling constant.

Explicitly, the equations for the
leading non-vanishing term, in the vacuum expectation value, of one 
Wilson
loop, and the leading non-vanishing terms, in the correlation functions, 
of
two or more Wilson loops, are complete, and close among themselves, the
equations for the next-to-leading non-vanishing term, in the vacuum
expectation value, of one Wilson loop, and the next-to-leading 
non-vanishing terms, in the correlation functions, of two or more Wilson
loops, are complete, and close among themselves, but with the leading
non-vanishing terms also in the right-hand sides, and so on.  Thus the 
equations for the leading non-vanishing contributions can be solved, then
these solutions can be used, in the right-hand sides, of the equations 
for 
the next-to-leading non-vanishing contributions, and so on.
Thus the Group-Variation Equations can be used to calculate all the
Yang Mills vacuum expectation values, and correlation functions, required
for the study of mesons and glueballs.
\footnote{The determination of the form of the Group-Variation Equations,
to all orders in $1/N$, was carried out in response to a question from
G. Ross.}
Witten has given evidence, in reference \cite{baryons}, that baryons are
monopoles, or solitons, of large-$N_c$ QCD, to be studied in a
Hartree-Fock manner.  It remains to be determined whether the
Group-Variation Equations can be used to calculate the Yang Mills vacuum
expectation values, and correlation functions, required for the study of
baryons.  Some recent results, in the application of the $1/N_c$ 
expansion to baryons, are given in references \cite{recent baryons a},
\cite{recent baryons b}, \cite{recent baryons c}, and
\cite{recent baryons d}.

The determination of the form of the group-variation equations to all
orders in $1/N$, and in particular, determining what happens to the 
linear
combinations of the diagonal elements of an $\textrm{SU}(NM)$ matrix, 
that are
\emph{not} elements of the $(\textrm{SU}(N))^M$ subgroup, requires 
choosing a
specific basis for the generators of $\textrm{SU}(NM)$, that is suited 
to the
$(\textrm{SU}(N))^M$ subgroup.  In the basis used in this paper, the 
off-diagonal
$\textrm{SU}(NM)$ generators, that are not members of the 
$(\textrm{SU}(N))^M$ subgroup, are
called $t_4$'s and $t_5$'s, and the diagonal $\textrm{SU}(NM)$ 
generators, that are
not members of the block-diagonal $(\textrm{SU}(N))^M$ subgroup, are 
called 
$t_6$'s.  The ``6-fields'' do not couple at all to the fields in the
$(\textrm{SU}(N))^M$ subgroup, and thus have free propagators, even in 
the presence
of background fields in the $(\textrm{SU}(N))^M$ subgroup.  They have no 
occurrence
at all, in the Group-Variation Equations for the leading non-vanishing
terms, in the vacuum expectation values and correlation functions.

Determining the detailed form of the Group-Variation Equations, and in
particular, verifying that the equations for the leading non-vanishing
terms, in the $1/N$ expansions, have the expected simple form, involves
identifying some eight ``selection rules,'' that restrict the diagrams 
that can
occur, in the Group-Changing Equations, for $\textrm{SU}(NM)$ and 
$(\textrm{SU}(N))^M$.

As mentioned briefly above,
the detailed study, of the island diagram mechanism, shows that the 
introductory description, given above, must be modified for the 
following reason.  If the
correlation function, of a Wilson loop of size approximately
$1/\sqrt{\sigma}$, close to the minimal-area spanning surface of a much
larger Wilson loop, is proportional to $e^{-\sigma A_c}$, with no
pre-exponential factor,
where $A_c$ is
the area of the \emph{cylinder-topology} minimal-area spanning surface,
whose boundary is the two loops, properly oriented, then the contribution
of the island diagrams is \emph{too large}: it grows in proportion to
$L^2\ln L$, as the dimensions of the left-hand side Wilson loop are 
scaled
by a factor $L$, rather than in proportion to $L^2$, as required.  This
would require the vacuum expectation value, of the left-hand side Wilson
loop, to depend on $L$ as $ae^{-bL^2\ln(cL)}$, which is impossible, 
because
it violates the Seiler bound \cite{Seiler}\cite{Simon Yaffe}.

The simplest resolution, of this problem, is that the cylinder-topology
minimal-area spanning surface term, in the correlation function of two
Wilson loops, must be multiplied, when it exists, by a pre-exponential
factor, that decreases at least as fast as $\frac{1}{\ln(\sigma A_c)}$,
at large $A_c$.

This behaviour occurs because, if the area of the minimal-area spanning
surface of the small loop is $1/\sigma$, and the area of the 
minimal-area spanning surface, $S$, of the large loop, is $A$, then in 
the limit
of large $\sigma A$, if the perpendicular distance, $z$, from the small
loop, to $S$, is not too 
large, and the perpendicular projection, of the small loop, onto $S$, is 
not too near the edge
of $S$, or in other words, not too near the large loop itself,
and the small loop is oriented, so as to minimize the area, of the 
cylinder-topology, minimal-area, spanning surface, $S_c$, of the two 
loops, then the area, $A_c$, of $S_c$,
depends on the perpendicular distance, $z$, from the small loop to $S$, 
only through the additive
term $\frac{2\pi z^2}{\ln(\sigma A)}$.  The $\ln(\sigma A)$, in the 
denominator, means that the larger the area $A$, of $S$, the more weakly 
the small loop is
attracted to $S$.  It is 
this denominator factor, of $\ln(\sigma A)$, that produces the 
unacceptable
factor of $\ln(L)$, in the island diagram contributions, if the
cylinder-topology term, in the correlation function of two Wilson loops,
behaves, when it exists, simply as $e^{-\sigma A_c}$, with no 
pre-exponential factor.

As mentioned briefly above, if the pre-exponential factor, in the 
cylinder-topology term, in the correlation function of two Wilson loops,
decreases \emph{faster} than $\frac{1}{\ln(\sigma A_c)}$, at large $A_c$,
then this term makes \emph{no} contribution to the asymptotic form, at
large $A$, of the right-hand side, of the Group-Variation Equation, for
the vacuum expectation value, of a Wilson loop, the area of whose
minimal-area spanning surface, is $A$.  If this is the case, then the 
asymptotic form, at large $A$, of the right-hand side of this
Group-Variation Equation, comes entirely from the term
$f^2\sqrt{\frac{m}{32\pi^3z^3}}e^{-mz}e^{-\sigma A_1}e^{-\sigma A_2}$, in
the correlation function of two Wilson loops, whose separate minimal-area
spanning surfaces, $S_1$, and $S_2$, have areas $A_1$, and $A_2$, where
$z$ is the shortest distance between any point of $S_1$, and any point of
$S_2$, $m$ is the mass of the lightest glueball, and $f$ is the
glueball-to-surface coupling constant.  This immediately gives the 
correct form of 
the Wilson area law, and also, in the approximation that the sums, over 
the
island diagrams, are dominated by their leading terms, gives the 
zeroth-order approximation, $m=2.38\sqrt{\sigma}$, for the mass of the 
lightest glueball, which is about 33 percent less
than the best lattice value, of $3.56\sqrt{\sigma}$ \cite{Teper}.

In Section \ref{Dimensional Regularization}, I give some evidence that
the Group-Variation Equations can be regularized by dimensional
regularization \cite{'t Hooft Veltman}, \cite{Bollini Giambiagi},
\cite{Ashmore}, in a gauge-invariant manner, which also preserves the 
property
that Fadeev-Popov loops stay either in the subgroup or out of it, and in
Section \ref{BPHZ PVRHDT Lattice}, I give some evidence that this can 
also
by achieved, by the
method of adding gauge-invariant higher derivative terms to the action,
plus Pauli-Villars scalar and spinor regulator fields, to cancel the 
one-loop divergences \cite{Lee Zinn-Justin}.

The Group
Variation Equations represent a minimum
resummation of the Feynman diagrams contributing to the coefficients, 
in the
1/N expansions, of physical quantities in \textrm{SU}(N) gauge 
theories, in that each
side, of a Group Variation Equation, is simply equal to
${g^2} {{\partial} \over {\partial{g^2}}}$, plus a finite integer 
constant,
(1 or 0 in the simplest cases), acting on the corresponding sum of 
Feynman
diagrams.

Shortly after the discovery of asymptotic freedom, Gross and Wilczek 
\cite{Gross Wilczek} suggested that the endless increase of the coupling 
constant, at long distances, might be responsible for quark confinement, 
and Weinberg \cite{Weinberg infra-red} suggested an explanation based 
on infra-red divergences, and the masslessness of the gluons, while 
't Hooft, in references \cite{'t Hooft a} and \cite{'t Hooft b}, later 
showed that neither of these effects, in the absence of the other, can 
produce confinement, within the direct sum of planar Feynman diagrams.  
It has turned out that all the coefficients in $\beta(g)$ appear to have 
the same sign, so that $\left\vert\frac{\beta(g)}{g}\right\vert$ 
runs to infinity at a finite 
distance, less than $\frac{1}{\Lambda_s}\simeq\frac{1}{200
\mathrm{ MeV}}$, so that perturbation theory cannot be used at long distances.  In this paper we see that the two effects together produce confinement by the island diagram mechanism, and that within the
Group-Variation Equations, the window weights produce an effective mass
for the gluon paths, that cuts off the infra-red divergences, and ensures
that the long-distance behaviour, of vacuum expectation values and
correlation functions, is determined by islands of fixed size $\simeq
\frac{1}{4.0\sqrt{\sigma}}\simeq\frac{1}{1800\mathrm{ MeV}}$, so that the
running coupling never increases beyond a critical value $\alpha_s\simeq
0.35$, at which the sums in the right-hand sides of the Group-Variation
Equations probably converge, so that a minimal resummation of 
perturbation theory, as given by the Group-Variation Equations, can be
used for the calculation of all strong interaction quantities, that do 
not involve baryons.

\chapter{Group-Changing Equations}
\section{Background and Conventions}

We consider Yang Mills theory \cite{IZ} for a compact Lie algebra $G$ in 4
Euclidean dimensions, with the conventions that the matrices $t_a$ of a
representation of $G$ satisfy $(t_a)^\dag =-t_a$, and $[t_a,t_b]=f_{abc}t_c$,
where $f_{abc}$ are the real
and totally antisymmetric structure constants, (so that the adjoint
representation of $G$ is given by $(t_c)_{ab}=f_{acb}$), and that the gauge
variation of the gauge field $A_{\mu a}$ is
$A_{\mu a} \to A_{\mu a}+\partial_\mu\epsilon_a+A_{\mu b}f_{abc}\epsilon_c$,
so that the general covariant derivative is $D_{\mu
ij}\psi_j=\partial_\mu\psi_i+A_{\mu b}(t_b)_{ij}\psi_j$, and the gauge
variation of the matter field $\psi_i$ is $\psi_i\to\psi_i-\epsilon_c(t_c)_{ij}\psi_j$.  
The Yang Mills action density is
$\frac{1}{4g^2}F_{\mu\nu a}F_{\mu\nu a}$, where $F_{\mu\nu a}=\partial_\mu A_{\nu
a}-\partial_\nu A_{\mu a}+f_{abc}A_{\mu b}A_{\nu c}$, and $g$ is the coupling
constant.

We wish to determine the vacuum expectation values and correlation functions
of gauge-invariant quantities constructed from the gauge-invariant
path-ordered\\phase-factor:
\begin{displaymath}
W(A,x(s))_{ij}=\sum_{n=0}^\infty\int_0^1ds_1\dots\int_0^1ds_n\theta(s_2-s_1)\dots
\theta(s_n-s_{n-1})\frac{dx_{\mu_1}(s_1)}{ds_1}\dots
\frac{dx_{\mu_n}(s_n)}{ds_n}\quad\times
\end{displaymath}
\begin{equation}\label{E1}
\times\quad A_{\mu_1a_1}(x(s_1))\dots A_{\mu_na_n}(x(s_n))(t_{a_1}\dots
t_{a_n})_{ij}
\end{equation}
where the continuous path $x(s)$, $0\le s\le1$, either consists of a finite
number of straight segments, or else is a smooth curve, and $\theta(s)$ is the step 
function, $\theta(s) = 1$ for $s\ge0$, $\theta(s)=0$ for $s<0$.

The gauge-variation of $W$ is given by:
\begin{eqnarray}\label{E2}
W(A+D\epsilon,x(x))_{ij} & = &
W(A,x(s))_{ij}-\epsilon_b(x(0))(t_b)_{ik}W(A,x(s))_{kj}+{}
\nonumber\\
& & {}+W(A,x(s))_{ik}\epsilon_b(x(1))(t_b)_{kj}
\end{eqnarray}
The simplest form of gauge-invariant quantity formed from $W$ is obtained by
taking the trace of $W$ when $x(s)$ is a closed path, so that $x(0)$ is equal
to $x(1)$.  This is called a Wilson loop.  More general gauge-invariant
quantities may be formed from the $W$'s, possibly in different representations, by taking 
a network of
paths meeting at junctions, and contracting the $W$'s at the junctions with
invariant tensors whose indices are in the appropriate representations.

\section{Gauge Fixing and BRS Invariance}

We consider a compact Lie algebra $H$ and its compact Lie sub-algebra $G$,
and obtain integral equations expressing the vacuum expectation values, in
the Yang Mills theory for $H$, of the gauge-invariant quantities of $H$, in
terms of the vacuum expectation values, in the Yang Mills theory for $G$, of the 
gauge-invariant
quantities of $G$.  We let lower case letters $a$, $b$, $c$, \dots run over
the elements of the Lie algebra $G$, and upper-case letters $A$, $B$, $C$,
\dots run over the elements of the Lie algebra $H$ that are \emph{not} elements of $G$, 
and we use the
summation convention that repeated indices run over the domains just defined.
Then the fact that $G$ is a sub-\emph{algebra} of $H$ is expressed by the
vanishing of the structure constants $f_{abC}$ with two lower-case indices and one 
upper-case index, so that
there is no $f_{abC}t_C$ term in the right-hand side of the commutation
relation $[t_a,t_b]=f_{abc}t_c$.  Now with Greek indices running over the
\emph{whole} of $H$, the Jacobi
 identity for $H$ may be written:
\begin{equation}\label{E3}
f_{\gamma\alpha\epsilon}f_{\epsilon\beta\delta}-
f_{\gamma\beta\epsilon}f_{\epsilon\alpha\delta}=
f_{\alpha\beta\epsilon}f_{\gamma\epsilon\delta}
\end{equation}
and taking $\gamma$ as $C$, $\beta$ as $b$, and $\delta$ as $D$ in this
equation, and noting that $f_{Cae}$ and $f_{abE}$ are equal to zero, we
obtain:
\begin{equation}\label{E4}
f_{CaE}f_{EbD}-f_{CbE}f_{EaD}=f_{abe}f_{CeD}
\end{equation}
Thus the matrices $(t_a)_{CD}=f_{CaD}$ form a representation, (possibly
reducible), of the Lie algebra $G$.

We denote the $H$-covariant derivative by $D_\mu$, and the $G$-covariant
derivative by $\bar{D}_\mu$.  The fields $A_{\mu A}$ transform as matter
fields under gauge variations in $G$, and their $G$-covariant derivative is
given by:
\begin{equation}\label{E5}
(\bar{D}_\mu A_\nu)_A=\partial_\mu A_{\nu A}+A_{\mu a}f_{AaB}A_{\nu B}
\end{equation}
We introduce gauge-fixing auxiliary fields $B_a$ and $B_A$, and choose the
gauge-fixing action to be:
\begin{equation}\label{E6}
\frac{1}{g^2}\left(iB_a(\partial_\mu A_{\mu a})+\frac{\alpha}{2}B_a B_a +
iB_A(\bar{D}_\mu A_\mu)_A+\frac{\beta}{2}B_A B_A\right)
\end{equation}
We then find in the usual way that we must add the Fadeev-Popov action
density:
\begin{equation}\label{E7}
\frac{1}{g^2}\left(\psi_a(\partial_\mu(D_\mu\phi)_a)+\psi_A(\bar{D}_\mu(D_\mu\phi)_A)+
\psi_A(D_\mu\phi)_af_{AaB}A_{\mu B}\right)
\end{equation}
where $\psi_a$, $\phi_a$, $\psi_A$, and $\phi_A$ are the Fadeev-Popov fields.

Now if we again allow Greek indices $\alpha$, $\beta$, $\gamma$, \dots to run
over the \emph{whole} of $H$, and indicate position arguments by subscripts
$x$, $y$, \dots, then the standard BRS operator \cite{IZ} $\delta$ for $H$
may be written:
\begin{equation}\label{E8}
\delta=\int d^4x\left((D_\mu\phi)_{\alpha x}\frac{\delta}{\delta A_{\mu\alpha x}} -
\frac{1}{2}f_{\alpha\beta\gamma}\phi_{\beta x}\phi_{\gamma x}\frac{\delta}
{\delta\phi_{\alpha x}}+iB_{\alpha x}\frac{\delta}{\delta\psi_{\alpha x}}\right)
\end{equation}
We note that, in consequence of our use of the gauge-fixing auxiliary fields
$B_\alpha$, $\delta^2$ vanishes \emph{exactly}, $\delta^2=0$.

Now the sum of our gauge-fixing and Fadeev-Popov action densities (\ref{E6})
and (\ref{E7}) is equal to the action of $\delta$ on:
\begin{equation}\label{E9}
\frac{1}{g^2}\left(\psi_a(\partial_\mu A_{\mu a})
-i\frac{\alpha}{2}\psi_aB_a
+\psi_A(\bar{D}_\mu A_\mu)_A
-i\frac{\beta}{2}\psi_AB_A\right)
\end{equation}
Hence it directly follows from the nilpotence of $\delta$ that our full
action has the standard BRS invariance for $H$, hence that the effective action
$\Gamma$, in the presence of sources for $(\bar{D}_\mu A_\mu)_A$ and for the
BRS variations of $A_{\mu\alpha}$ and $\phi_\alpha$, satisfies Ward
identities of the usual form
\cite{IZ page 597}.

Indeed, if this structure can be preserved by a
gauge-invariant regularization, we could conclude
directly, from the general result, that two actions
which differ by a BRS-variation, give the same results
for the vacuum expectation values of BRS-invariant
quantities, that this gauge-fixing procedure will give
the same results for the vacuum expectation values of
BRS-invariant quantities, as the usual one.

Now $g^2$ times the Fadeev-Popov action density (\ref{E7}) is equal to:
\begin{displaymath}
\biggl(\psi_a(\partial^2\phi_a+\partial_\mu(A_{\mu b}f_{abc}\phi_c)+\partial_\mu(A_{\mu B}
f_{aBC}\phi_C))+
\end{displaymath}
\begin{displaymath}
+\psi_A(\partial^2\phi_A+\partial_\mu(A_{\mu b}f_{AbC}\phi_C)+\partial_\mu(A_{\mu B}
f_{ABc}\phi_c)+\partial_\mu(A_{\mu B}f_{ABC}\phi_C))+
\end{displaymath}
\begin{displaymath}
+\psi_A A_{\mu a}f_{AaE}(\partial_\mu\phi_E+A_{\mu b}f_{EbC}\phi_C+A_{\mu B}
f_{EBc}\phi_c+A_{\mu B}f_{EBC}\phi_C)+
\end{displaymath}
\begin{equation}\label{E10}
+\psi_A(\partial_\mu\phi_a+A_{\mu c}f_{acd}\phi_d+A_{\mu C}f_{aCD}\phi_D)f_{AaB}
A_{\mu B}\biggr)
\end{equation}
We examine the terms that contain both $\psi$ with an \emph{upper}-case index
and $\phi$ with a \emph{lower}-case index.  The first such term is
$\psi_A(\partial_\mu(A_{\mu B}f_{ABc}\phi_c))$.  The second such term is\\$\psi_A
A_{\mu a}f_{AaE}(A_{\mu B}f_{EB
c}\phi_c)$, which is equal to $-\psi_A A_{\mu c}\phi_d A_{\mu
B}f_{AcE}f_{EdB}$.  The third such te-rm is
$\psi_A\!\!\;(\partial_\mu\phi_a)f_{AaB}A_{\mu B}$, which is equal to $-\psi_A A_{\mu
B}f_{ABc}(\partial_\mu\phi_c)$.  And the fourth and final such term is $\psi_
A A_{\mu c}f_{acd}\phi_d f_{AaB}A_{\mu B}$, which by (\ref{E4}) is equal to
$\psi_A A_{\mu c}\phi_d A_{\mu B}(f_{AcE}f_{EdB}-f_{AdE}f_{EcB})$.  Thus the
sum of the first and third such terms is equal to $\psi_A(\partial_\mu A_{\mu
B})f_{ABc}\phi_c$, and the sum of the second and fourth such terms is equal to:
\begin{displaymath}
-\psi_A A_{\mu c}\phi_d A_{\mu B}f_{AdE}f_{EcB}=\psi_A(A_{\mu c}f_{EcB}A_{\mu
B})f_{AEd}\phi_d=\psi_A(A_{\mu a}f_{BaC}A_{\mu C})f_{ABc}\phi_c
\end{displaymath}
Hence the sum of all four such terms is equal to:
\begin{equation}\label{E11}
\psi_A(\bar{D}_\mu A_\mu)_Bf_{ABc}\phi_c
\end{equation}

\section{Propagators for Fields not in the Subgroup, in the presence of Background Fields in the Subgroup}

We now define $\bar{F}_{\mu\nu a}=\partial_\mu A_{\nu a}-\partial_\nu
A_{\mu a} + f_{a b c} A_{\mu b} A_{\nu c} $, so that
$F_{\mu\nu a}$ is equal to $\bar{F}_{\mu\nu a}+f_{aBC}A_{\mu B}A_{\nu C}$,
and we note that $F_{\mu\nu A}$ is equal to $(\bar{D}_\mu
A_\nu)_A-(\bar{D}_\nu A_\mu)_A+f_{ABC}A_{\mu B}
A_{\nu C}$.  Thus there are \emph{no} terms in $F_{\mu\nu a}F_{\mu\nu
a}+F_{\mu\nu A}F_{\mu\nu A}$ that contain exactly one $A_\mu$ with an
upper-case group index, and the sum of all the terms in $F_{\mu\nu
a}F_{\mu\nu a}+F_{\mu\nu A}F_{\mu\nu A}$ that contain exactly two $A_\mu$'s with 
upper-case group indices, is:
\begin{equation}\label{E12}
2((\bar{D}_\mu A_\nu)_A(\bar{D}_\mu A_\nu)_A-(\bar{D}_\nu
A_\mu)_A(\bar{D}_\mu A_\nu)_A+\bar{F}_{\mu\nu a}f_{aBC}A_{\mu B}A_{\nu C})
\end{equation}
Thus if we consider $A_{\mu A}$ and $B_A$ to be propagating in a
\emph{background} field given by $A_{\mu a}$, and if we again denote position
arguments by subscripts $x$, $y$, \dots, and if we define the propagator
matrix for $A_{\mu A}$ and $B_A$ propagating in the background field $A_{\mu a}$ by:
\begin{equation}\label{E13}
\left(A_{\mu Ax},B_{Ax}\right)\left(\begin{array}{cc}G_{\mu Ax,\nu By} &
\tilde{G}_{\mu Ax,By} \\ G_{Ax,\nu By} & G_{Ax,By} \end{array}\right)
\left(\begin{array}{c}A_{\nu By} \\ B_{By} \end{array}\right)
\end{equation}
then we find that this propagator matrix satisfies the equation:
\begin{displaymath}
\left(\begin{array}{c|c}\!\!-(\bar{D}^2)_{ACx}\delta_{\mu\sigma}+
(\bar{D}_\mu
\bar{D}_\sigma)_{ACx}-2\bar{F}_{\mu\sigma ax}f_{AaC} \!& -i(\bar{D}_\mu)_{ACx} \\
i(\bar{D}_\sigma)_{ACx} & \beta \delta_{AC}\end{array}\!\right)\!\left(
\begin{array}{cc}G_{\sigma Cx,\nu By} & \tilde{G}_{\sigma Cx,By} \\ G_{Cx,\nu By} &
G_{Cx,By} \end{array}\right)\!=
\end{displaymath}
\begin{equation}\label{E14}
=g^2\left(\begin{array}{c|c}\delta_{\mu\nu}\delta_{AB}\delta^4(x-y) & 0 \\
0 & \delta_{AB}\delta^4(x-y)\end{array}\right)
\end{equation}
which we abbreviate as:
\begin{equation}\label{E15}
\left(\begin{array}{c|c}-\bar{D}^2\delta_{\mu\sigma}+\bar{D}_\mu\bar{D}_\sigma
-2\bar{F}_{\mu\sigma} & -i\bar{D}_\mu \\ i\bar{D}_\sigma & \beta\end{array}\right)
\left(\begin{array}{cc}G_{\sigma\nu} & \tilde{G}_\sigma \\ G_\nu & G\end{array}
\right) = g^2\left(\begin{array}{cc}\delta_{\mu\nu} & 0 \\ 0 & 1\end{array}
\right)\end{equation}

We define the operator $E$ by:
\begin{equation}\label{E16}E=\bar{D}_\mu\frac{1}{\bar{D}^2}\bar{D}_\mu
\end{equation}
and we note that $E_{Ax,By}$ is equal to $\delta_{AB}\delta^4(x-y)$ plus
terms of degree one and higher in $A_{\mu a}$.

Now the exact solution of the equation obtained from (\ref{E15}) by deleting
the term $-2\bar{F}_{\mu\sigma}$, is given by:
\begin{equation}\label{E17}
g^2\left(\begin{array}{c|c}-\frac{1}{\bar{D}^2}\delta_{\sigma\nu}+\frac{1}
{\bar{D}^2}\bar{D}_\sigma\left(\frac{1-\beta}{(1-\beta)E+\beta}\right)\bar{D}_\nu
\frac{1}{\bar{D}^2} & \frac{1}{\bar{D}^2}\bar{D}_\sigma\left(\frac{-i}
{(1-\beta)E+\beta}\right) \\
\left(\frac{i}{(1-\beta)E+\beta}\right)\bar{D}_\nu
\frac{1}{\bar{D}^2} & \left(\frac{1-E}{(1-\beta)E+\beta}\right)\end{array}
\right)\end{equation}

And if we denote (\ref{E17}) by $g^2G_0$, and the exact solution of
(\ref{E15}) by $g^2G_1$, and the matrix
$\left(\begin{array}{cc}2\bar{F}_{\mu\sigma} & 0 \\ 0 & 0\end{array}\right)$ by
$F$, then the exact solution of (\ref{E15}) is given by:
\begin{equation}\label{E18}
g^2G_1=g^2\left(G_0+G_0FG_0+G_0FG_0FG_0+G_0FG_0FG_0FG_0+\dots\right)
\end{equation}

And furthermore, if we denote the $AA$ component of (\ref{E17}) by $g^2G_0$,
and the $AA$ component of the exact solution of (\ref{E15}) by $g^2G_1$, then
in consequence of the vanishing of the $AB$, $BA$, and $BB$ components of
$F$, the $AA$ component of the exact solution of (\ref{E15}) is also given by (\ref{E18}), 
with $F$
now interpreted simply as $\bar{F}_{\mu\sigma}$.

\subsection{Landau gauge}

We observe that, for $\beta\ne0$, it follows from (\ref{E17}) and (\ref{E18})
that the $AA$ component of the exact solution of (\ref{E15}) is also the
exact solution of the equation for $G_{\mu Ax,\nu By}$ alone that is obtained
if the gauge-fixing auxiliary field has been integrated out:
\begin{equation}\label{E19}
\left(-(\bar{D}^2)_{ACx}\delta_{\mu\sigma}+\left(1-\frac{1}{\beta}\right)
(\bar{D}_\mu\bar{D}_\sigma)_{ACx}-2\bar{F}_{\mu\sigma ax}f_{AaC}\right)G_{\sigma Cx,\nu
By}=g^2\delta_{\mu\nu}\delta_{AB}\delta^4(x-y)\end{equation}

And we observe that, as $\beta$ tends to 0, the propagator matrix (\ref{E17})
tends smoothly to the Landau gauge form:
\begin{equation}\label{E20}
g^2\left(\begin{array}{c|c}-\frac{1}{\bar{D}^2}\delta_\sigma\nu+\frac{1}
{\bar{D}^2}\bar{D}_\sigma\frac{1}{E}\bar{D}_\nu\frac{1}{\bar{D}^2} & -i\frac{1}
{\bar{D}^2}\bar{D}_\sigma\frac{1}{E} \\ i \frac{1}{E}\bar{D}_\nu\frac{1}
{\bar{D}^2} & \frac{1}{E}-1\end{array}\right)\end{equation}

Thus we see that the Landau gauge case $\beta=0$ may be treated without any
problems, without any need for any limiting process from $\beta\ne0$, and
without any reference at all to the gauges with $\beta\ne0$, by means of the
gauge-fixing auxiliary field $B_A$, and that the results obtained by this method are in 
exact agreement
with the results obtained by letting $\beta$ tend to 0 in the results
obtained, for general $\beta\ne0$, without the use of the gauge-fixing
auxiliary field.

We note furthermore \cite{DJGLH} that the Landau gauge condition $\beta=0$
is preserved under changes of renormalization point, so that in Landau gauge
there is no $\frac{\partial}{\partial\beta}$ term in the renormalization group
equation, even for non-gauge-invariant quantities.

\subsection{Fadeev-Popov loops stay either in or out of the subgroup in Landau gauge}

Now the $AA$ component of (\ref{E20}) satisfies the identity:
\begin{equation}\label{E21}
\bar{D}_\sigma\left(-\frac{1}{\bar{D}^2}\delta_{\sigma\nu}+\frac{1}{\bar{D}^2}
\bar{D}_\sigma\frac{1}{E}\bar{D}_\nu\frac{1}{\bar{D}^2}\right)=0\end{equation}
Hence it immediately follows from (\ref{E18}) that in Landau gauge $G_{\mu
Ax,\nu By}$, which is the exact $A_{\mu Ax}A_{\nu By}$ propagator in the
background field $A_{\mu a}$, satisfies the indentity:
\begin{equation}\label{E22}
\bar{D}_\mu G_{\mu\nu}=0\end{equation}
Hence it immediately follows from the form of (\ref{E11}) that in Landau
gauge for $A_{\mu A}$, that is, for $\beta=0$ in (\ref{E6}), and for the
vacuum expectation value of any quantity that includes no $B_A$'s, the terms
in the Fadeev-Popov action density (\ref{E10}) that contain both $\psi$ with an 
\emph{upper}-case index and
$\phi$ with a \emph{lower}-case index, make \emph{no contribution at all}.
Furthermore, in the vacuum expectation value of any quantity that includes no
Fadeev-Popov fields, the Fadeev-Popov propagators occur only in closed loops, and, since 
the
Fadeev-Popov propagators do not mix upper-case indices and lower-case
indices, any Fadeev-Popov loop that includes any vertex that contains both
$\psi$ with a \emph{lower}-case index and $\phi$ with an \emph{upper}-case index, must 
also include at least one
vertex that contains both $\psi$ with an \emph{upper}-case index and $\phi$
with a \emph{lower}-case index.  Hence if we use Landau gauge for $A_{\mu
A}$, or in other words, if we
set $\beta=0$ in (\ref{E6}), and if we are calculating the vacuum expectation
value of any quantity that includes no $B_A$'s and no Fadeev-Popov fields,
then we may completely neglect the five terms in the Fadeev-Popov action
density (\ref{E10}) that contain both a Fadeev-Popov field with an upper-case index and a 
Fadeev-Popov
field with a lower-case index.  The remaining terms in (\ref{E10}) consist of
the standard Fadeev-Popov action density $\psi_a(\partial^2\phi_a+\partial_\mu(A_{\mu
b}f_{abc}\phi_c))$ for
$G$, plus seven terms that may be put in the manifestly $G$-gauge-invariant
form:
\begin{equation}\label{E23}
\psi_A(\bar{D}^2\phi)_A+\psi_A(\bar{D}_{\mu AE}(A_{\mu B}f_{EBC}\phi_C))+
\psi_Af_{AaB}A_{\mu B}\phi_Cf_{CaD}A_{\mu D}\end{equation}

\subsection{The propagators expressed as sums over paths, weighted by Wilson lines of fields in the subgroup}

We next note that $\left(\frac{1}{\bar{D}^2}\right)_{Ax,By}$ may be expressed
as a sum over paths from $x$ to $y$, each weighted by the path-ordered phase
factor (\ref{E1}), in the representation of $G$ given by
$(t_a)_{CD}=f_{CaD}$.  Indeed, we may write:
\begin{equation}\label{E24}
\left(\frac{-1}{\bar{D}^2}\right)_{Ax,By}=\int_0^\infty ds\left(e^{s\bar{D}^2
}\right)_{Ax,By}\end{equation}
We choose a value $\sigma$, which is to represent the ``maximum tolerable'' width
of a Gaussian, and write:
\begin{equation}\label{E25}
\int_0^\infty ds\left(e^{s\bar{D}^2}\right)_{Ax,By}=\sum_{n=0}^\infty
\int_{n\sigma}^{(n+1)\sigma}ds\left(e^{s\bar{D}^2}\right)_{Ax,By}\end{equation}
Then for $n\ge1$, with $n\sigma\le s\le(n+1)\sigma$, we write:
\begin{equation}\label{E26}
\left(e^{s\bar{D}^2}\right)_{Ax,By}=\int d^4z_1\dots\int d^4z_n\left(e^
{\frac{s\bar{D}^2}{n+1}}\right)_{Ax,C_1z_1}\left(e^{\frac{s\bar{D}^2}{n+1}}
\right)_{C_1z_1,C_2z_2}\dots\left(e^{\frac{s\bar{D}^2}{n+1}}\right)_
{C_nz_n,By}\end{equation}
Then in this exact expression, noting that $\frac{n\sigma}{n+1}\le\frac{s}{n+1}
\le\sigma$ holds, we approximate each $\left(e^{\frac{s\bar{D}^2}{n+1}}\right)_
{Ep,Fq}$ by $\left(e^{\sigma\partial^2}\right)_{p,q}=\frac{e^{-\frac{(p-q)^2}{4\sigma}}}
{(4\pi\sigma)^2}$ multiplied by the path-ordered phase factor $W_{Ep,Fq}$ for the
\emph{straight line} from $p$ to $q$.

Thus if we denote by $\left(W_{xz_1z_2\dots z_ny}\right)_{AB}$ the
path-ordered phase factor for the path consisting of the straight line from
$x$ to $z_1$, then the straight line from $z_1$ to $z_2$, and so on, and then
finally the straight line from $z_
n$ to $y$, then the $\sigma$-approximation to
$\left(\frac{-1}{\bar{D}^2}\right)_{Ax,By}$ is given by:
\begin{equation}\label{E27}
\sigma\sum_{n=0}^\infty\int d^4z_1\dots\int d^4z_n\left(W_{xz_1z_2\dots z_ny}
\right)_{AB}\frac{e^{-\frac{(x-z_1)^2}{4\sigma}}}{(4\pi\sigma)^2}\frac{e^{-\frac
{(z_1-z_2)^2}{4\sigma}}}{(4\pi\sigma)^2}\dots\frac{e^{-\frac{(z_n-y)^2}{4\sigma}}}
{(4\pi\sigma)^2}\end{equation}
In Section \ref{Final Section} we sketch the derivation, directly from 
the $\sigma\to0$ limit of this expression, of the standard expansion:
\begin{equation}\label{E28}
\frac{-1}{\bar{D}^2}=\frac{-1}{\partial^2}+\frac{-1}{\partial^2}(\partial
A+A\partial+AA)\frac{-1}{\partial^2}
+\frac{-1}{\partial^2}(\partial A+A\partial+AA)\frac{-1}{\partial^2}(\partial A+
A\partial+AA)\frac{-1}{\partial^2}+\dots\end{equation}
where $\left(\frac{-1}{\partial^2}\right)_{xy}=\frac{1}{4\pi^2(x-y)^2}$.

The $G$-covariant derivative, acting on $A_{\nu Ax}$, for example, may be
approximated by:
\begin{equation}\label{E29}
(\bar{D}_\mu A_\nu)_{Ax}\simeq\int d^4y\frac{(y-x)_\mu}{2\sigma}\frac{e^{-\frac
{(y-x)^2}{4\sigma}}}{(4\pi\sigma)^2}W_{Ax,By}A_{\nu By}\end{equation}

Now as noted after equation (\ref{E16}), $E_{Ax,By}$ is equal to $\delta_{AB}
\delta^4(x-y)$ plus terms of degree one and higher in $A_{\mu a}$.  Thus
$\frac{1}{E}$ may be expressed as:
\begin{equation}\label{E30}
\frac{1}{E}=\frac{1}{1-(1-E)}=1+(1-E)+(1-E)^2+\dots\end{equation}
Hence, by (\ref{E16}), (\ref{E27}), and (\ref{E29}), $\frac{1}{E}$ may also
be expressed in terms of sums over paths, weighted by the path-ordered phase
factor.  Hence both the Landau-gauge $A_{\mu A}A_{\mu A}$ propagator in the
background field $A_{\mu
a}$, and the $\psi_A\phi_A$ propagator in the background field $A_{\mu a}$,
are fully expressed as sums over paths, weighted by the path-ordered
phase-factor in the representation $(t_a)_{CD}=
f_{CaD}$ of $G$.

We note furthermore that $\frac{1}{4g^2}(F_{\mu\nu a}F_{\mu\nu a}+
F_{\mu\nu A}F_{\mu\nu A})$ is equal to the sum of the $A_{\mu a}$'s Yang
Mills action density $\frac{1}{4g^2}\bar{F}_{\mu\nu a}\bar{F}_{\mu\nu a}$,
plus $\frac{1}{4g^2}$ times the $A_{\mu A}$'s kinetic terms (\ref{E12}) in
the background field $A_{\mu a}
$, plus the manifestly $G$-gauge-invariant interaction terms:
\begin{equation}\label{E31}
\frac{1}{4g^2}(4(\bar{D}_\mu A_\nu)_Af_{ABC}A_{\mu B}A_{\nu C}+(f_{aBC}
f_{aEF}+f_{ABC}f_{AEF})A_{\mu B}A_{\nu C}A_{\mu E}A_{\nu F})\end{equation}

\section{Group-Changing Equations for th$\textrm{e }\!\!\textrm{V}$acuum Expectation Values and Correlation Functions for the Group, in terms of those for the Subgroup}
\label{GCE section}

We now consider the calculation of the vacuum expectation value, in the Yang
Mills theory for the compact Lie algebra $H$, of a general
$H$-gauge-invariant quantity, formed from the path-ordered phase factors, in
various representations of $H$, that correspond to the paths in some network of paths, 
where the paths in the network
meet at junctions, and the path-ordered phase factors are contracted at the
junctions with $H$-invariant tensors with indices in the appropriate
representations of $H$.  We use
the following procedure: we \emph{first} functionally integrate over the
$A_{\mu A}$, $B_A$, $\psi_A$, and $\phi_A$ fields, in the presence of a
general background $A_{\mu a}$ field configuration, which we do \emph{not}
functionally integrate over at this
 stage.  We use Landau gauge for the $A_{\mu A}$ and $B_A$ fields, or in
other words, we set $\beta$ equal to 0 in (\ref{E6}).  Then, as shown above,
we may use the manifestly $G$-gauge-invariant form (\ref{E23}) for the
$\phi_A\psi_B$ action.  There are now \emph{no} terms in the action that contain both one 
or more of the
fields $A_{\mu A}$, $B_A$, $\psi_A$, and $\phi_A$, and one or more of the
fields $B_a$, $\psi_a$, and $\phi_a$, (thus it does not matter whether or not
the $B_a$, $\psi_a$, and $\phi
_a$ fields are integrated over at this stage).  The exact $A_{\mu A}A_{\nu
B}$ and $\psi_A\phi_B$ propagators in the background field $A_{\mu a}$ are
expressed as sums over paths, weighted by the $G$-covariant path-ordered
phase factors in the representation $(t_a)_{CD}=f_{CaD}$ of $G$, by (\ref{E18}), 
(\ref{E20}), (\ref{E27}),
(\ref{E29}), and (\ref{E30}), while the $A_{\mu A}B_B$ and $B_AB_B$
propagator components are irrelevant, since the $B_A$ field is not involved
at all in this vacuum expectation value.

In each $H$-covariant path-ordered phase factor in the $H$-invariant quantity
whose vacuum expectation value we are calculating, we put $A_{\mu\alpha}
(t_\alpha)_{ij}=A_{\mu a}(t_a)_{ij}+A_{\mu A}(t_A)_{ij}$, and expand our
$H$-invariant quantity in powers of $A_{\mu A}$.  We then see immediately
from (\ref{E1}) that between each occurrence of $A_{\mu A}(t_A)_{ij}$ along
the path-ordered phase factor in the representation $t_\alpha$ of $H$, we have a 
\emph{G-covariant}
path-ordered phase-factor in the representation $t_a$ of $G$.

Furthermore, $(t_A)_{ij}$, considered as a tensor in its three indices $A$,
$i$, and $j$, is an invariant tensor of $G$, where the index $A$ is in the
representation $f_{AaB}$ of $G$, the index $i$ is in the representation $t_a$
of $G$, and the index $j$ is in the complex conjugate representation $(t_a)^\ast$ of $G$.  
Indeed, the
commutation relation $[t_a,t_A]=f_{aAB}t_B$ implies immediately that for any
infinitesimal parameters $\epsilon_a$, $a\in G$, we have:
\begin{equation}
	\label{E31a}
(\delta_{ik}+(\epsilon_at_a)_{ik})(\delta_{jm}+\epsilon_b((t_b)_{jm})^\ast)
(\delta_{AB}+\epsilon_cf_{AcB})(t_B)_{km}=(t_A)_{ij}
\end{equation}
And similarly, we find from the commutation relation (\ref{E4}) that
	\[
\epsilon_af_{CaE}f_{EbD}+\epsilon_af_{DaE}f_{CbE}+\epsilon_af_{bae}f_{CeD}
=0
\]
hence that $f_{CbD}$ is an invariant tensor of $G$, and by taking, in
the general Jacobi identity $f_{\beta\alpha\epsilon}f_{\epsilon\gamma\delta}+
f_{\gamma\alpha\epsilon}f_{\beta\epsilon\delta}+f_{\delta\alpha\epsilon}
f_{\beta\gamma\epsilon}=0$ of $H$, $\alpha=a$, $\beta=B$, $\gamma=C$, and
$\delta=D$, and recalling that the structure constants with one upper-case
index and two lower-case indices are all equal to zero, we find that $f_{BaE}
f_{ECD}+f_{CaE}f_{BED}+f_{DaE}f_{BCE}=0$, hence that $f_{BCD}$ is an
invariant tensor of $G$.

We now develop the standard perturbation expansion for the functional
integral over the $A_{\mu A}$, $B_A$, $\psi_A$, and $\phi_A$ fields, with the
pre-exponential factor given by our $H$-invariant quantity, and in the
presence of the general $A_{\mu a}$ ``background field'' configuration, which we do 
\emph{not} yet functionally
integrate over, and we find immediately from the foregoing that we have a sum
over ``decorations'' of our $H$-gauge-invariant quantity by new paths and
junctions, where all the
paths are now in the appropriate representations of $G$, and contracted at
the junctions by the appropriate $G$-invariant tensors, and all new paths are
to be summed over with the appropriate position-space weight according to
which of our propagators in the background field $A_{\mu a}$ they represent, and all new 
junctions are
to be integrated, as appropriate, either over all four space dimensions or,
if they belong to a path of our $H$-invariant quantity, along that path,
respecting path-ordering along that path with any other new junctions that belong on that 
path.

Let us denote the vacuum functional integral over \emph{all} our fields,
(both upper-case index and lower-case index), by $Z_H$.  We do \emph{not} yet
divide by $Z_H$.  Therefore we must also include all vacuum bubbles, in the
presence of the general background field configuration $A_{\mu a}$.  Each vacuum bubble 
consists of a
$G$-invariant network of paths and junctions formed from $G$-covariant
path-ordered phase factors and $G$-invariant tensors just as before, with the
paths being summed over and the junction positions integrated over, with the only 
difference being that
there are now \emph{no} fixed paths or junctions, (and \emph{all} paths are
now in the representation $f_{AaB}$ of $G$).  And we treat $Z_H$ in exactly
the same fashion, first functionally integrating just over the $A_{\mu A}$, $B_A$, 
$\psi_A$, and $\phi_A$
fields, in the presence of a general background field $A_{\mu a}$, and
develop $Z_H$ as a sum over vacuum bubbles in the presence of $A_{\mu a}$.

\emph{Then, for each separate term in each of the above expansions}, we
functionally integrate over the $A_{\mu a}$ field, (and also over the $B_a$,
$\psi_a$, and $\phi_a$ fields, if that has not already been done).
Furthermore, we divide every term by
the vacuum functional integral over the fields $A_{\mu a}$, $B_a$, $\psi_a$,
and $\phi_a$, which we denote by $Z_G$.  Thus all vacuum bubbles due to the
\emph{lower}-case index fields propagating as quantum fields in loops are
cancelled out, while all the vacuum bubbles due to the \emph{upper}-case index fields 
propagating as
quantum fields in loops, remain.

\subsection{Emergence of correlation functions involving vacuum bubbles in the subgroup}

Let $W$ represent our initial $H$-invariant quantity.  Let $[W]_H$ denote
$\frac{1}{Z_G}$ times the functional integral over \emph{all} the fields,
with the pre-exponential factor given by $W$, where the subscript $H$
indicates that we have done the functional integral over \emph{all} the fields, and the 
square brackets are to
remind us that we have divided by $Z_G$, not by $Z_H$.  And let $\tilde{Z}_H$
denote $\frac{Z_H}{Z_G}$.

We then see that for every term in each of our expansions, and for every
position-space configuration of the new paths and junctions that occur in
that term, we have the \emph{vacuum expectation value in the Yang Mills
theory for $G$} of the $G$-invariant quantity that corresponds to that term, and to that 
configuration of the
new paths and junctions of that term.

Now the vacuum expectation value of $W$ \emph{in the Yang Mills theory for}
$H$ is given by $\frac{[W]_H}{\tilde{Z}_H}$.

Suppose $[W]_H$ contains a term $\langle W_1W_2\rangle_G$, where $W_1$ is a
$G$-invariant, decorated version of $W$, and $W_2$ is a $G$-invariant vacuum
bubble, and the subscript $G$ indicates the vacuum expectation value in the
Yang Mills theory for $G$.  Then $[W]_H$ also contains the term $\langle W_1\rangle_G$, 
and
$\tilde{Z}_H$ includes the term $\langle W_2\rangle_G$, so that
$\frac{1}{\tilde{Z}_H}$, expanded in powers of $(\tilde{Z}_H-1)$, includes
the term $-\langle W_2\rangle_G$.  Thus the total of all the terms in 
$\frac{[W]_H}{\tilde{Z}_H}$ that contain precisely the
two $G$-invariants $W_1$ and $W_2$, is $\langle W_1W_2\rangle_G-\langle
W_1\rangle_G\langle W_2\rangle_G$, which is the \emph{correlation function}
of $W_1$ and $W_2$ in the Yang Mills theory for $G$.  And similarly, suppose $[W]_H$ 
contains a term
$\langle W_1W_2W_3\rangle_G$, where of  $W$, and $W_2$ and $W_3$ are
$G$-invariant vacuum bubbles.  Then $[W]_H$ also includes terms $\langle
W_1\rangle_G, \langle W_1W_2\rangle_G$,
and $\langle W_1W_3\rangle_G$, and $\tilde{Z}_H$ contains terms $\langle
W_2\rangle_G$, $\langle W_3\rangle_G$, and $\langle W_2W_3\rangle_G$, hence
$\frac{1}{\tilde{Z}_H}$ includes terms $+2\langle W_2\rangle_G\langle
W_3\rangle_G$, (from $+(\tilde{Z}_H-1)^2$), and $-\langle W_2W_3\rangle_G$, (from 
$-(\tilde{Z}_H-1)$).  Hence
the total of all the terms in $\frac{[W]_H}{\tilde{Z}_H}$ that contain
precisely the three $G$-invariants $W_1$, $W_2$, and $W_3$, is:
\begin{displaymath}
\langle W_1W_2W_3\rangle_G-\langle W_1W_2\rangle_G\langle
W_3\rangle_G-\langle W_1W_3\rangle_G\langle W_2\rangle_G-\langle
W_1\rangle_G\langle W_2W_3\rangle_G+\end{displaymath}
\begin{displaymath}
+2\langle W_1\rangle_G\langle
W_2\rangle_G\langle W_3\rangle_G\end{displaymath}
which is the \emph{correlation function} of $W_1$, $W_2$, and $W_3$, in the
Yang Mills theory for $G$.

And in general, if $W_1$ is a $G$-invariant decoration of $W$, and
$W_2$,\dots,$W_n$ are $G$-invariant vacuum bubbles, then the total of all the
terms in $\frac{[W]_H}{\tilde{Z}_H}$ that contain precisely the $n$
$G$-invariants $W_1$, $W_2$,\dots,$W_n$,
is equal to the sum, over all \emph{partitions} of the set
$\{W_1,W_2,\dots,W_n\}$, of $(-1)^{m-1}(m-1)!$, where $m$ is the number of
parts of the partition, times the product, overthe parts of the partition, of
the vacuum expectation value, in the Yang
Mills theory for $G$, of the $W_i$'s in that part of the partition.  And this
is precisely the correlation function, in the Yang Mills theory for $G$, of
the $n$ $G$-invariants $W_1$, $W_2$,\dots,$W_n$.  (We note that each
partition into $m$ parts comes
from a term in $[W]_H$ times a term in $(-1)^{m-1}(\tilde{Z}_H-1)^{m-1}$.)

Now if $W$ is itself a product of two or more $H$-invariant quantities, then
we may wish to calculate the correlation function of those quantities, in the
Yang Mills theory for $H$.  For example, if $W$ is equal to the product of
$W_1$ and $W_2$, where $W_1$ and $W_2$ are $H$-invariant quantities, then the correlation 
function of
$W_1$ and $W_2$, in the Yang Mills theory for $H$, is equal to
$\frac{[W_1W_2]_H}{\tilde{Z}_H}-\frac{[W_1]_H}{\tilde{Z}_H}\frac
{[W_2]_H}{\tilde{Z}_H}$.  Each of the three vacuum expectation values, in the
Yang Mills theory for $H$, that occurs here, may be expressed in terms of
vacuum expectation values, in the Yang Mills theory for $G$, of $G$-invariant
decorations of $W_1W_2$, by the results already obtained.  Now a decoration of $W_1W_2$ 
may connect
$W_1$ and $W_2$ into a single $G$-invariant quantity, say $W_3$, or $W_1$ and
$W_2$ may be decorated into two separate $G$-invariant quantities, say $W_3$
and $W_4$, and in the latter case, $[W_1]_H$ contains the term $\langle W_3\rangle_G$ and 
$[W_2]_H$
contains the term $\langle W_4\rangle_G$, so that we obtain the correlation
function of $W_3$ and $W_4$ in the Yang Mills theory for $G$.  Now suppose,
for example, that $[W_1W_2]_H$ contains the term $\langle W_3W_4W_5\rangle_G$, where $W_3$ 
is a
$G$-invariant decoration of $W_1$, $W_4$ is a $G$-invariant decoration of
$W_2$, and $W_5$ is a $G$-invariant vacuum bubble.  Then $[W_1]_H$ contains
the terms $\langle W_3\rangle_G$
and $\langle W_3W_5\rangle_G$, $[W_2]_H$ contains the terms $\langle
W_4\rangle_G$ and $\langle W_4W_5\rangle_G$, and $\tilde{Z}_H$ contains the
term $\langle W_5\rangle_G$, from which we find immediately that the total of
all the terms in $\frac{[W_1W_2
]_H}{\tilde{Z}_H}-\frac{[W_1]_H}{\tilde{Z}_H}\frac{[W_2]_H}
{\tilde{Z}_H}$ that contain precisely the three $G$-invariants $W_3$, $W_4$,
and $W_5$, is the correlation function of $W_3$, $W_4$, and $W_5$, in the
Yang Mills theory for $G$.

And in general, if we calculate the correlation function, in the Yang Mills
theory for $H$, of $n$ separate $H$-invariant quantities $W_1$,\dots,$W_n$,
and use our previous results to express all the vacuum expectation values, in
the Yang Mills theory for $H$, of subsets of $\{W_1,\dots,W_n\}$, in terms of correlation 
functions,
in the Yang Mills theory for $G$, of $G$-invariant decorations of those
subsets of $\{W_1,\dots,W_n\}$, and $G$-invariant vacuum bubbles, then we
find that if $P$ is a partition into $m$ parts of the set $\{W_1,\dots,W_n\}$, (so that 
$1\le m\le n$
holds), and $\tilde{W}_1$, \dots, $\tilde{W}_m$ are $G$-invariant
\emph{connected} decorations of the parts of $P$, and $\tilde{W}_{m+1}$,
\dots, $\tilde{W}_{m+r}$ are $G$-invariant vacuum bubbles, then the total of all the terms 
in the correlation function of
$\{W_1,\dots,W_n\}$ in the Yang Mills theory for $H$, that involve precisely
the $(m+r)$ $G$-invariant quantities $\tilde{W}_1$, \dots, $\tilde{W}_{m+r}$,
is equal to the correlation function, in the Yang Mills theory for $G$, of
$\{\tilde{W}_1,\dots,\tilde{W}_{m+r}\}$.

\subsection{One-loop vacuum bubbles in the subgroup}
The one-loop vacuum bubbles require special treatment, because they have no
junctions.  For the Fadeev-Popov one-loop vacuum bubble we have $\mathrm{tr}\ 
\mathrm{ln}(\bar{D}^2)=
\mathrm{ln}\ \mathrm{det}(\bar{D}^2)$, while for the
$A_{\mu A}$ and $B_A$ one-loop vacuum bubble we have $-\frac{1}{2}$ times the trace of the 
logarithm
of the matrix of $G$-covariant derivatives that occurs at the left of
(\ref{E14}) and (\ref{E15}), with $\beta$ set equal to $0$.

For $\mathrm{tr}\ \mathrm{ln}(\bar{D}^2)$ we use:
\begin{equation}\label{E32}
\int_0^\infty \frac{ds}{s}\left(e^{s\bar{D}^2}-e^{s\partial^2}\right)=
-\mathrm{ln}\left(\bar{D}^2\left(\frac{1}{\partial^2}\right)\right)\end{equation}
We then treat $e^{s\bar{D}^2}$ in the same manner as before, with the result
being obtained from (\ref{E27}) by removing the overall factor of $\sigma$,
dividing term $n$ by $(n+1)$, setting $x$ equal to $y$ and integrating, and
taking the trace on the group indices.

For the $A_{\mu A}$ and $B_A$ one-loop vacuum bubble the calculation is
facilitated by use of the identity:
\begin{displaymath}
\mathrm{tr}\ \mathrm{ln}(MN)=\mathrm{ln}\ \mathrm{det}(MN)=\mathrm{ln}(
\mathrm{det}M\mathrm{det}N)=\end{displaymath}
\begin{equation}\label{E33}
=(\mathrm{ln}\ \mathrm{det}M)+(\mathrm{ln}\ \mathrm{det}N)=
(\mathrm{tr}\ \mathrm{ln}M)+(\mathrm{tr}\ \mathrm{ln}N)\end{equation}
which holds for general $M$ and $N$.

Now let $M$ denote the matrix of $G$-covariant derivatives that occurs at the
left of (\ref{E14}) and (\ref{E15}), with $\beta$ set equal to 0, so that the
one-loop vacuum bubble for $A{\mu A}$ and $B_A$ is given by $-\frac{1}{2}
\mathrm{tr}\ \mathrm{ln}M$.  Let $M_0$ denote $M$ with the term
$-2\bar{F}_{\mu\sigma}$ removed, let $g^2G_0$ denote the propagator matrix
(\ref{E20}), and let $F$ denote the matrix $\left(\begin{array}{c|c}
2\bar{F}_{\mu\sigma} & 0 \\ 0 & 0 \end{array}\right)$.

Then
\begin{displaymath}
\mathrm{tr}\ \mathrm{ln}M=\mathrm{tr}\ \mathrm{ln}(M_0G_0M) = \mathrm{tr}
\ \mathrm{ln}(M_0G_0(M_0-F))=\mathrm{tr}\ \mathrm{ln}(M_0(1-G_0F))=
\end{displaymath}
\begin{equation}\label{E34}
=(\mathrm{tr}\ \mathrm{ln}M_0)-\mathrm{tr}(G_0F)-\frac{1}{2}\mathrm{tr}
(G_0FG_0F)-\frac{1}{3}\mathrm{tr}(G_0FG_0FG_0F)-\dots\end{equation}
To calculate $\mathrm{tr}\ \mathrm{ln}M_0$ we express $M_0$ as:
\begin{equation}\label{E35}
\left(\begin{array}{c|c}-\bar{D}^2\delta_{\mu\alpha} & 0 \\ 0 &
\frac{1}{\theta}\end{array}\right)\left(\begin{array}{c|c}-\frac{1}
{\bar{D}^2}\delta_{\alpha\gamma} & 0 \\ 0 & \theta\end{array}\right)\left(
\begin{array}{c|c}-\bar{D}^2\delta_{\gamma\sigma}+\bar{D}_\gamma\bar{D}_\sigma &
-i\bar{D}_\gamma \\ i\bar{D}_\sigma & 0\end{array}\right)\end{equation}
where $\theta$ will be chosen for convenience.  Now the product of the second
and third matrices in (\ref{E35}) is:
\begin{equation}\label{E36}
\left(\begin{array}{c|c}\delta_{\alpha\sigma}-\frac{1}{\bar{D}^2}\bar{D}_\alpha
\bar{D}_\sigma & i\frac{1}{\bar{D}^2}\bar{D}_\alpha \\ i\theta\bar{D}_\sigma & 0
\end{array}\right)=\left(\begin{array}{c|c}\delta_{\alpha\sigma} & 0 \\ 0 & 1
\end{array}\right)-\left(\begin{array}{c|c}\frac{1}{\bar{D}^2}\bar{D}_\alpha
\bar{D}_\sigma & -i\frac{1}{\bar{D}^2}\bar{D}_\alpha \\ -i\theta\bar{D}_\sigma & 1
\end{array}\right)\end{equation}
We denote the right-hand side of (\ref{E36}) by $1-N$, and we now choose
$\theta=-1$ so that $N$ factorizes as:
\begin{equation}\label{E37}
\left(\begin{array}{c|c}\frac{1}{\bar{D}^2}\bar{D}_\alpha\bar{D}_\sigma & -i
\frac{1}{bar{D}^2}\bar{D}_\alpha \\ i\bar{D}_\sigma & 1\end{array}\right)=
\left(\begin{array}{c}\frac{1}{\bar{D}^2}\bar{D}_\alpha \\ i\end{array}
\right)\left(\bar{D}_\sigma\ ,\ -i\right)\end{equation}
We then find immediately that for $n\ge1$, $N^n$ is equal to:
\begin{equation}\label{E38}
\left(\begin{array}{c}\frac{1}{\bar{D}^2}\bar{D}_\alpha \\ i \end{array}
\right)(E+1)^{-1}(E+1)^n\left(\bar{D}_\sigma\ ,\ -i\right)\end{equation}
hence
\begin{displaymath}
\mathrm{ln}(1-N)=-N-\frac{1}{2}N^2-\frac{1}{3}N^2-\dots=\end{displaymath}
\begin{displaymath}
=\left(\begin{array}{c}\frac{1}{\bar{D}^2}\bar{D}_\alpha\\i\end{array}\right)
(E+1)^{-1}\biggl(-(E+1)-\frac{1}{2}(E+1)^2-\frac{1}{3}(E+1)^3-\dots\biggr)
\left(\bar{D}_\sigma\ ,\ -i\right)\end{displaymath}
\begin{equation}\label{E39}
=\left(\begin{array}{c}\frac{1}{\bar{D}^2}\bar{D}_\alpha\\i\end{array}\right)
(E+1)^{-1}(\mathrm{ln}(-E))\biggl(\bar{D}_\sigma\ ,\ -i\biggr)\end{equation}
hence
\begin{equation}\label{E40}
\mathrm{tr}\ \mathrm{ln}(1-N)=\mathrm{tr}\ \mathrm{ln}(-E)\end{equation}
hence
\begin{equation}\label{E41}
\mathrm{tr}\ \mathrm{ln}M_0=4\mathrm{tr}\ \mathrm{ln}(\bar{D}^2)+\mathrm{tr}
\ \mathrm{ln}E+\mathrm{constant}\end{equation}
where the $A_{\mu A}$-independent constant cancels between $[W]_H$ and
$\tilde{Z}_H$ for any $W$, due to exponentiation of the vacuum bubbles.

We call the equations we have derived in this section, which express the
vacuum expectation values and correlation functions of $H$-invariant
quantities, in the Yang Mills theory for $H$, in terms of the vacuum
expectation values and correlation functions of $G$-invariant quantities, in
the Yang Mills theory for the Lie subalgebra $G$ of $H$, the
\emph{group-changing equations for} $G$ \emph{and} $H$.

In the applications we need to consider path-ordered phase factors having an
insertion of $F_{\mu\nu a}(t_a)_{ij}$ at a finite number of points along the
path, where $t_a$ is the representation appropriate to that path, but these
can all be treated by the principles already given.

\section{The Group-Changing Equations for $\textrm{SU}(NM)$ and $(\textrm{SU}(N))^M$}

In this section we apply the group-changing equations to the groups
$G=(\textrm{SU}(N))^M$ and $H=\textrm{SU}(NM)$, then take the derivative with respect to $M$ at
$M=1$, to obtain a complete and closed set of equations among the
coefficients of the $\frac{1}{N}$ expansions of the one-Wilson-loop vacuum expectation 
value and the multi-Wilson-loop
correlation functions in $\textrm{SU}(N)$ Yang Mills theory.  In the next section we
shall see that the equations for the leading terms in the $\frac{1}{N}$
expansions have a very simple structure, and that their solution will manifestly satisfy 
the Wilson area
law for the one-Wilson-loop vacuum expectation value, and the requirement of
massive glueball saturation for the multi-Wilson-loop correlation functions.

\subsection{Review of the $\frac{1}{N}$ expansion}

We begin be recalling some well-known facts about $\textrm{SU}(N)$ Yang Mills theory
\cite{Planar}.  For any integers $\alpha$, $\beta$, we define $\theta_{\alpha\beta}$
to be equal to 1 if $\alpha<\beta$ and equal to 0 if $\alpha\ge\beta$.  Then as a basis
for the fundamental representation of $\mathop{\mathrm{SU}}(N)$ we may take, with $1\le 
j\le N$,
$1\le k\le N$:
\begin{eqnarray}\label{E42}
(t_{1ab})_{jk}&=&\frac{i}{\sqrt{2}}(\delta_{aj}\delta_{bk}+\delta_{ak}\delta_{bj})
\hspace{3.5cm}(1\le a<b\le N)\nonumber\\
(t_{2ab})_{jk}&=&\frac{1}{\sqrt{2}}(\delta_{aj}\delta_{bk}-\delta_{ak}\delta_{bj})
\hspace{3.5cm}(1\le a<b\le N)\nonumber\\
(t_{3a})_{jk}&=&\frac{i}{\sqrt{a(a-1)}}\delta_{jk}(\theta_{ja}-(a-1)\delta_{ja})
\hspace{2.0cm}(2\le a\le N)\end{eqnarray}
where the $t_1$'s, $t_2$'s, and $t_3$'s are repectively generalizations of
$\frac{i}{\sqrt{2}}$ times the Pauli matrices $\sigma_1$, $\sigma_2$, and $\sigma_3$.

We let greek indices run over all the $N^2-1$ generators of
$\mathop{\mathrm{SU}}(N)$, that is, over the $\frac{1}{2}N(N-1)$ $t_1$'s, the
$\frac{1}{2}N(N-1)$ $t_2$'s, and the $N-1$ $t_3$'s.  We then find that the
generators satisfy:
\begin{equation}\label{E43}
\mathop{\mathrm{tr}}(t_\alpha 
t_\beta)=-\delta_{\alpha\beta}\end{equation}
which immediately implies, from the commutation relation 
$[t_\alpha,t_\beta]=
f_{\alpha\beta\gamma}t_\gamma$, that:
\begin{equation}\label{E44}
f_{\alpha\beta\gamma}=-\mathop{\mathrm{tr}}([t_\alpha,t_\beta]t_\gamma)=-
\mathop{\mathrm{tr}}(t_\alpha t_\beta t_\gamma-t_\gamma t_\beta t_\alpha)
\end{equation}
hence that the structure constants are totally antisymmetric.

We now find, with the summation convention temporarily suspended, that:
\begin{equation}\label{E45}
\sum_{1\le a<b\le N}(t_{1ab})_{jk}(t_{1ab})_{pq}=-\frac{1}{2}(\delta_{jp}
\delta_{kq}+\delta_{jq}\delta_{kp})(\theta_{jk}+\theta_{kj})
\end{equation}
\begin{equation}\label{E46}
\sum_{1\le a<b\le N}(t_{2ab})_{jk}(t_{2ab})_{pq}=-\frac{1}{2}(\delta_{jp}
\delta_{kq}-\delta_{jq}\delta_{kp})(\theta_{jk}+\theta_{kj})\end{equation}
\begin{eqnarray}\label{E47}
\sum_{1\le a<b\le N}((t_{1ab})_{jk}(t_{1ab})_{pq}+(t_{2ab})_{jk}(t_{2ab})_
{pq})&=&-\delta_{jq}\delta_{kp}(\theta_{jk}+\theta_{kj})\nonumber\\ &=&-\delta_{jq}\delta_
{pk}(1-\delta_{jk})\end{eqnarray}
\begin{displaymath}\label{E48}
\sum_{2\le a\le N}(t_{3a})_{jk}(t_{3a})_{pq}=-\delta_{jk}\delta_{pq}\sum_
{2\le a\le N}\frac{1}{a}\left(\frac{\theta_{ja}\theta_{pa}}{(a-1)}-\delta_{ja}\theta_{pa}
-\delta_{pa}\theta_{ja}+(a-1)\delta_{ja}\delta_{pa}\right)\end{displaymath}
\begin{displaymath}
=-\delta_{jk}\delta_{pq}\left(\left(\sum_{\begin{array}{c}\max(j,p)+1\le a\\a\le N
\end{array}}\left(\frac{1}{a-1}-\frac{1}{a}\right)\right)-\frac{\theta_{pj}}{j}
-\frac{\theta_{jp}}{p}+\left(\frac{j-1}{j}\right)\delta_{jp}\right)
\end{displaymath}
\begin{equation}
=-\delta_{jk}\delta_{pq}\delta_{jp}-\delta_{jk}\delta_{pq}\left(\left(\frac{1}{\max(j,p)}-
\frac{1}{N}\right)-\frac{1}{\max(j,p)}\right)=-\delta_{jq}\delta_{pk}\delta_{jk}+
\frac{1}{N}\delta_{jk}\delta_{pq}\end{equation}
hence, with the usual summation convention, and recalling our convention that
the greek indices run over \emph{all} the generators of $\mathop{\mathrm{SU}}(N)$:
\begin{equation}\label{E49}
(t_\alpha)_{jk}(t_\alpha)_{pq}=-\delta_{jq}\delta_{pk}+\frac{1}{N}\delta_{jk}\delta_{pq}
\end{equation}
(We note that $(t_\alpha)_{jk}(t_\alpha)_{pq}$ is an invariant tensor of 
$\mathop{\mathrm{SU}}(N)$, and that once it is known that $\delta_{jq}\delta_{pk}$ and
$\delta_{jk}\delta_{pq}$ are the only invariant tensors that occur in the
right-hand side of (\ref{E49}), then the coefficients may be determined
directly from (\ref{E43}) and the tracelessness of the $t_\alpha$'s.  But for other groups 
and representations, further
linearly independent invariant tensors might occur in the right-hand side of
(\ref{E49}).)

We replace the coupling constant $g$ of the preceding sections by
$\frac{g}{\sqrt{N}}$, so that the Yang Mills action density is now given by:
\begin{equation}\label{E50}
\frac{N}{4g^2}F_{\mu\nu\alpha}F_{\mu\nu\alpha}\end{equation}
Then, in each Feynman diagram that contributes to a vacuum expectation value
or correlation function of Wilson loops, we use the relation (\ref{E44}) to
express all the structure constants in terms of the $t_\alpha$'s, then use the
relation (\ref{E49}) and
the $\delta_{\alpha\beta}$ colour structure of the propagators, (which has the
consequence that the $t_\alpha$'s all occur in pairs with the adjoint
representation indices contracted as in the left-hand side of (\ref{E49})),
to eliminate all the $t_\alpha$'s in terms
of fundamental representation Kronecker deltas.  Then for each independent
choice of one of the two terms in (\ref{E44}) for each structure constant,
and one of the two terms in (\ref{E49}) for each propagator and for the
$\delta_{\alpha\beta}$ in the middle of
each four-gluon vertex, we obtain a collection of closed loops of fundamental
representation Kronecker deltas, each of which simply gives a factor $N$.
The total $N$-dependence of each such term is given by a factor $N$ for each
vertex \emph{that comes
from the action}, a factor $\frac{1}{N}$ for each propagator, a factor
$\frac{1}{N}$ for each choice of the second term in the right-hand side of
(\ref{E49}), and a factor $N$ for each closed loop of fundamental
representation Kronecker deltas.  (There is \emph{no} factor associated with each vertex 
where a gluon line ends at a
Wilson loop.)

We note that replacing the first term in (\ref{E49}) by the second term in
(\ref{E49}) in any one location, (i.e. in any one propagator or four-gluon
vertex), either leaves the total power of $N$ unaltered, or else decreases it
by 2.  For if the two Kronecker deltas in the first term are in the \emph{same} closed 
loop of
Kronecker deltas, then replacing the first term of (\ref{E49}) by the second
term in that location, \emph{increases} the number of closed loops of
Kronecker deltas by 1, so that the total power of $N$ is unaltered, while if the two 
Kronecker deltas in the first
term are in two \emph{different} closed loops of Kronecker deltas, then
replacing the first term by the second in that location \emph{decreases} the
number of closed loops of Kronecker deltas by 1, so that the total power of $N$ decrease 
by 2.  Thus we
may determine the maximum possible power of $N$ corresponding to any Feynman
diagram by choosing the first term in (\ref{E49}) at every location.

Now every Feynman diagram that contributes to the vacuum expectation value of
$n$ Wilson loops may be built up from the diagram with no propagators by the
successive addition of ``new'' propagators to lower-order diagrams.  Each end
of the ``new'' propagator may be either on a Wilson loop, or at an ``old'' cubic vertex, 
(which
becomes in quartic vertex in the ``new'' diagram), or in the ``middle'' of an
``old'' propagator, (which results in a new cubic vertex, and an increase by
1 in the number of ``old'' propagators).  Thus each such addition of one ``new'' 
propagator results
in the difference between the \emph{total} number of propagators, and the
total number of vertices \emph{coming from the action}, increasing by 1, thus
each addition of one ``new'' propagator brings a factor $\frac{g^2}{N}$ from the explicit 
factor
$\frac{N}{g^2}$ in (\ref{E50}).  We assume first that we choose the first
term in (\ref{E49}) in every propagator and every quartic vertex, so that we
just have a sum of terms corresponding to which of the two terms in (\ref{E43}) is chosen 
for each structure
constant.  Then each end of the ``new'' propagator breaks into an ``old''
closed loop of Kronecker deltas, and if both ends of the ``new'' propagator
break into the \emph{same} old loop, then the total number of closed loops 
\emph{increases} by 1, hence
the total power of $N$ is \emph{unaltered}, while if the two ends of the
``new'' propagator break into \emph{different} old loops, then the total
number of closed loops of Kronecker deltas \emph{decreases} by 1, hence the total power of 
$N$ decreases by
2.  Hence we see immediately by induction on the power of $g^2$ that every
term has the power of $N$ of the zeroth order term, multiplied by an integer
power $\ge0$ of $\frac{1}{N^2}$, and by the preceding paragraph we see immediately that 
this remains
true when we allow the second term in (\ref{E49}) as well.  We furthermore
see by induction, allowing again just the first term in (\ref{E49}), that in
any term that contributes at the leading power of $N$, i.e. at the same power of $N$ as in 
the zeroth
order term, every occurrence of the first term in (\ref{E49}), has its two
Kronecker deltas in \emph{different} closed loops of Kronecker deltas.  For
the only way to obtain the leading power of $N$ is to ensure that, as the diagram and the 
term of
interest is built up by successive additions of ``new'' propagators to
lower-order diagrams, the two ends of each ``new'' propagator break into 
the \emph{same} closed loop of Kronecker deltas,
so that each addition of a ``new'' propagator increases the total number of closed loops 
of Kronecker deltas by 1,
and this means firstly that the two Kronecker deltas in the ``new''
propagator are in two different closed loops of Kronecker deltas, and
secondly that no propagator of the ``new'' diagram that formed the whole or part of a 
propagator of the ``old''
diagram, has both its Kronecker deltas in a single closed loop of Kronecker
deltas, unless that was the case in the ``old'' diagram, which by the
induction assumption was not so.
Hence by the preceding paragraph again, we see that when we again allow both
terms in (\ref{E49}), \emph{no} term that contributes at the leading power of
$N$ contains any occurrence of the second term in (\ref{E49}).

Now it is impossible to add a ``new'' propagator to a diagram with two or
more connected components, in such a way that each end of the ``new''
propagator is in a different connected component of the ``old'' diagram, (and
thus the ``new'' propagator decreases the number of connected components of the diagram by 
1), and retain the
same power of $N$ as the ``old'' diagram, for the fact that the two ends of
the ``new'' propagator are in different connected components of the ``old''
diagram, means that the
two ends of the ``new'' propagator must break into \emph{different} closed
loops of Kronecker deltas of the old diagram.  Thus we find that the leading
power of $N$ of the diagrams that contribute to the vacuum expectation value
of $n$ Wilson loops, and
have $m$ or fewer connected components, is:
\begin{equation}\label{E51}n-2(n-m)=2m-n\end{equation}
And in particular, the leading power of $N$ of the diagrams that contribute
to the vacuum expectation value of $n$ Wilson loops and have \emph{one}
connected component, or in other words, of the diagrams that contribute to
the correlation function of $n$ Wilson loops, is:
\begin{equation}\label{E52}2-n\end{equation}

Now by repeating the same arguments as before, we find that in any term that contributes
to the correlation function of n Wilson loops with the largest possible power of $N$, or
in other words, with $(2-n)$ powers of $N$, there are \emph{no} occurrences of the second
term in (\ref{E49}), with one exception: if any propagator, or any $\delta_{\alpha\beta}$ in
the middle of a four-gluon vertex, is such that by removing it, the number of connected
components of the diagram is increased by 1, then when the first term in (\ref{E49}) is
taken in that propagator or $\delta_{\alpha\beta}$, the two Kronecker deltas from part of
a \emph{single} closed loop of Kronecker deltas, (for any closed loop tha intersects both
the ``remainder'' connected components must have at least two of its Kronecker deltas
intersecting both those connected components, and the two Kronecker deltas concerned are
the \emph{only} ones that intersect both those connected components), hence when we 
replace the
first term in (\ref{E19}) by the second in just that one propagator or 
$\delta_{\alpha\beta}$,
we get a term that exactly cancels the previous term.  In fact this argument shows that no
Feynman diagram whose number of connected components may be increased by 1 by removing a
single propagator or the $\delta_{\alpha\beta}$ from the middle of a single four-gluon
vertex, makes any contribution to any correlation function or vacuum expectation value of
gauge-invariant quantities, since replacing the first term in (\ref{E49}) by the second in
that one ``key'' propagator or $\delta_{\alpha\beta}$, and leaving everything else 
unchanged,
always gives a term that exactly cancels the previous term.

And with this one exception, where, as just noted, replacing the first term in (\ref{E49})
by the second term in just one ``key'' propagator or $\delta_{\alpha\beta}$, results in a
term that exactly cancels the previous term, we again find by induction on the power of 
$g^2$
that the second term in (\ref{E49}) makes \emph{no} contribution to any Feynman diagram term
that contributes at the leading power of $N$, i.e. with $(2-n)$ powers of $N$, to the 
correlation
function of $n$ Wilson loops.

Now if we consider any Feynman diagram term that contributes at the leading power of $N$,
 i.e. with $(2-n)$ powers of $N$, to the correlation function of $n$ Wilson loops, we may
 ``fill'' each closed loop of Kronecker deltas with an oriented topological 2-disk, and we
 see by induction on the power of $g^2$ that these oriented topological 2-disks, one for 
each
 closed loop of Kronecker deltas, join up to form an oriented 2-sphere with $n$oriented 
holes
 in it, where the boundaries of the holes are the $n$ Wilson loops, and the boundaries of 
the
 holes are all oriented the same way.  In fact, as we build up each diagram from 
lower-order
 diagrams by successive addition of ``new'' propagators, we see that the condition that 
the
 two ends of each ``new'' propagator break into a \emph{single} ``old'' closed loop of 
 Kroneker deltas, means that each ``new'' propagator is drawn inside one ``window'' of the
 lower-order diagram, or in other words, on the oriented 2-disk that fills the single 
``old''
 closed loop of Kronecker deltas into which the ``new'' propagator breaks.  And conversely
 we see again by induction on the power of $g^2$, starting from the leading terms, that
 every connected Feynman diagram that contributes to the correlation function of $n$ 
Wilson
 loops and can be drawn on the oriented 2-sphere with $n$ oriented holes of the same 
orientation,
 the boundaries of the holes being the Wilson loops, gives a contribution to that 
correlation
 function at the leading power of $N$, i.e. with $(2-n)$ powers of $N$, provided that it
 has no ``key'' propagator or $\delta_{\alpha\beta}$ in a four-gluon vertex, whose removal
 increases the number of connected components of the diagram.
 
 For $n\geq 1$ the leading contributions to the correlation function of $n$ Wilson loops 
have
 two closed loops of Kronecker deltas and $n$ powers of $g^2$, and have the general form shown here:
\vspace{1mm}
\unitlength 5pt
\linethickness{0.61pt}
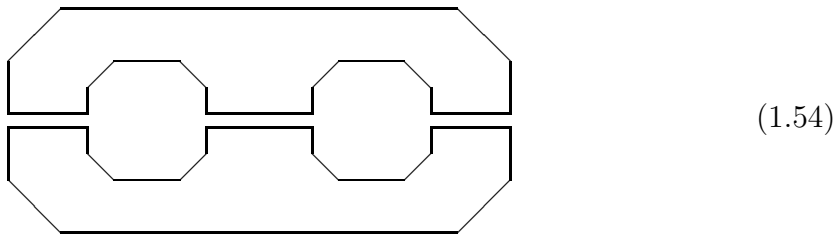
\begin{equation}\label{E53}
\raisebox{-45pt}{
\begin{picture}(39.00,18.00)
\put(35.00,1.00){\line(1,1){4.00}}
\put(39.00,5.00){\line(0,1){4.00}}
\put(39.00,9.00){\line(-1,0){6.00}}
\put(33.00,9.00){\line(0,-1){2.00}}
\put(33.00,7.00){\line(-1,-1){2.00}}
\put(31.00,5.00){\line(-1,0){5.00}}
\put(26.00,5.00){\line(0,0){0.00}}
\put(26.00,5.00){\line(-1,1){2.00}}
\put(24.00,7.00){\line(0,1){2.00}}
\put(9.00,5.00){\line(1,0){5.00}}
\put(14.00,5.00){\line(1,1){2.00}}
\put(16.00,7.00){\line(0,1){2.00}}
\put(16.00,9.00){\line(1,0){8.00}}
\put(35.00,18.00){\line(1,-1){4.00}}
\put(39.00,14.00){\line(0,-1){4.00}}
\put(39.00,10.00){\line(-1,0){6.00}}
\put(33.00,10.00){\line(0,1){2.00}}
\put(33.00,12.00){\line(-1,1){2.00}}
\put(24.00,12.00){\line(0,-1){2.00}}
\put(9.00,14.00){\line(1,0){5.00}}
\put(14.00,14.00){\line(1,-1){2.00}}
\put(16.00,12.00){\line(0,-1){2.00}}
\put(16.00,10.00){\line(1,0){8.00}}
\put(35.00,1.00){\line(-1,0){30.00}}
\put(5.00,1.00){\line(-1,1){4.00}}
\put(7.00,7.00){\line(1,-1){2.00}}
\put(1.00,5.00){\line(0,1){4.00}}
\put(1.00,9.00){\line(1,0){6.00}}
\put(7.00,9.00){\line(0,-1){2.00}}
\put(9.00,14.00){\line(-1,-1){2.00}}
\put(7.00,12.00){\line(0,-1){2.00}}
\put(7.00,10.00){\line(-1,0){6.00}}
\put(1.00,10.00){\line(0,1){4.00}}
\put(1.00,14.00){\line(1,1){4.00}}
\put(5.00,18.00){\line(1,0){30.00}}
\put(24.00,12.00){\line(1,1){2.00}}
\put(26.00,14.00){\line(1,0){5.00}}
\end{picture}
}
\end{equation}

\vspace{1mm}

\noindent where we have shown the closed loops of Kronecker deltas for the
case $n=3$, and the pairs
of parallel lines represent gluon propagators, while the single lines represent Wilson
loop segments.

Now by the foregoing, the $N$-dependence of the correlation function of $n$ Wilson loops 
in
$\textrm{SU}(N)$ Yang Mills theory is given, for all $n\geq 1$, by:
\begin{equation}\label{E54}
N^{2-n}(f_0(W_1,\dots,W_n,g^2)+\frac{1}{N^2}f_1(W_1,\dots,W_n,g^2)+\frac{1}{N^4}f_2
(W_1,\dots,W_n,g^2)+\dots)\end{equation}
where $W_1,\dots,W_n$ are the $n$ Wilson loops, and we recall that we define the 
"correlation
function" of \emph{one} Wilson loop to be equal to the one-Wilson-loop vacuum expectation
value.

We note that if we denote the correlation function of $n$ Wilson loops by $[W_1\dots 
W_n]$,
(where this use of square brackets should not be confused with the square brackets used 
for
a different purpose in Section 2), then the $n$-Wilson-loop vacuum expectation values may 
be
expressed in terms of the $n$-Wilson loop correlation functions by:
\begin{displaymath}
<W_1>=[W_1]
\end{displaymath}
\begin{displaymath}
<W_1W_2>=[W_1W_2]+[W_1][W_2]
\end{displaymath} 
\begin{equation}\label{E55}
<W_1W_2W_3>=[W_1W_2W_3]+[W_1W_2][W_3]+[W_1W_3][W_2]+[W_1][W_2W_3]+[W_1][W_2][W_3]
\end{equation}
hence (\ref{E54}) also defines the $\frac{1}{N^2}$ expansion of the $n$-Wilson-loop vacuum
expectation values for all $n\geq 1$.

\subsection{The $\frac{1}{N}$ expansion in the presence of quarks}

The expansion coefficients $f_r(W_1,\dots,W_n,g^2)$ in (\ref{E54}) contain all the 
dynamics
of gluons and glueballs.  We note that if quarks had been present, we would have had to
write an expansion in powers of $\frac{1}{N}$ in (\ref{E54}), rather than an expansion in 
powers of $\frac{1}{N^2}$.  However all QCD calculation can be done by first calculating 
the 
expansion coefficients in (\ref{E54}) in pure Yang Mills theory, then using these 
expansion
coefficients to weight the sums over quark and antiquark paths when quarks are included.
This is because each Wilson loop can play either the role of a glueball state, or the role
of a quark/antiquark path.  Thus to calculate the effect of an extra quark/antiquark 
vacuum
bubble on any process, for example, we include an extra Wilson loop in the relevant pure 
Yang Mills theory correlation function, where that extra Wilson loop follows the 
quark/antiquark
path of that extra vacuum bubble, and sum over the paths followed by that extra Wilson 
loop
with the appropriate kinematic weight, which is obtained from the quark propagator in a
general background Yang Mills field by a procedure analogous to that used in Section 2:
\begin{equation}\label{E56}
\frac{1}{\gamma.D+m}=(\gamma.D-m)\frac{1}{D^2+\frac{1}{2}\sigma_{\mu\nu}F_{\mu\nu}-m^2}=
-(\gamma.D-m)\int_0^\infty ds\ e^{s(D^2+\frac{1}{2}\sigma_{\mu\nu}F_{\mu\nu}-m^2)}
\end{equation}
where $\sigma_{\mu\nu}=\frac{1}{2}[\gamma_\mu,\gamma_\nu]$.  The 
$\frac{1}{2}\sigma_{\mu\nu}
F_{\mu\nu}$ term in the exponent in the right-hand side of (\ref{E56}) requires insertions 
of
$F_{\mu\nu}$ at finite numbers of points along the path-ordered phase factors, just as we
found in Section 2 for the $A_{\mu A}A_{\mu A}$ propagator in a general $A_{\mu a}$ 
``background'' field, in the context of the group-changing equations.  We note that it is 
the
combined effect of a $\frac{1}{2}\sigma_{\mu\nu}F_{\mu\nu}$ term on one quark or antiquark
line and a $\frac{1}{2}\sigma_{\mu\nu}F_{\mu\nu}$ term on another quark or antiquark line
that produces the ``hyperfine'' splitting that separates the $\pi$ from the $\rho$ and the 
$N$
from the $\Delta$ \cite{Perkins}.  The large $u$ and $d$ quark masses (about 350 MeV) used
in that calculation come from the Wilson loop factors weighting the paths, in the manner
indicated later in this paper.

\subsection{A basis for $\textrm{SU}(NM)$ suited to the $(\textrm{SU}(N))^M$ subgroup}

We now apply the group-changing equations to the groups $G=(\textrm{SU}(N))^M$ and $H=\textrm{SU}(NM)$, to
obtain a complete and closed set of equations among the expansion coefficients
$f_r(W_1,\dots,W_n,g^2)$ in (\ref{E54}).  When we have obtained these equations we will 
find
that we can differentiate with respect to $M$ and then set $M=1$, (as may be expected from
the analytic dependence on $N$ of the correlation functions (\ref{E54})), but for now we
assume that $M$ is an integer.

We choose the following basis for the generators of $\textrm{SU}(NM)$.  The rows and columns of the
generators in the fundamental representation of $\textrm{SU}(NM)$ are labelled by a lower-case 
index
that runs from 1 to $N$, and an upper-case index that runs from 1 to $M$, and we recall 
that for any integers $\alpha$, $\beta$, we define $\theta_{\alpha\beta}$ to be equal to 1
if $\alpha<\beta$ and equal to 0 if $\alpha\geq\beta$.  The generators are:
\begin{displaymath}
(t_{1Aab})_{JjKk}=\frac{i}{\sqrt{2}}\delta_{JA}\delta_{KA}(\delta_{aj}\delta_{bk}+
\delta_{ak}\delta_{bj})\qquad\qquad(1\leq A\leq M,\quad 1\leq a<b\leq N)
\end{displaymath}
\begin{displaymath}
(t_{2Aab})_{JjKk}=\frac{1}{\sqrt{2}}\delta_{JA}\delta_{KA}(\delta_{aj}\delta_{bk}-
\delta_{ak}\delta_{bj})\qquad\qquad(1\leq A\leq M,\quad 1\leq a<b\leq N)
\end{displaymath}
\begin{displaymath}
(t_{3Aa})_{JjKk}=\frac{i}{\sqrt{a(a-1)}}\delta_{JA}\delta_{KA}\delta_{jk}(\theta_{ja}-
(a-1)\delta_{ja})\qquad\qquad(1\leq A\leq M,\quad 2\leq a\leq N)
\end{displaymath}
\begin{displaymath}\begin{array}{cc}
(t_{4AaBb})_{JjKk}
=\frac{i}{\sqrt{2}}(\delta_{AJ}\delta_{aj}\delta_{BK}\delta_{bk}+
\delta_{AK}\delta_{ak}\delta_{BJ}\delta_{bj}) & \quad(1\leq A<B\leq M,\quad\! 1\leq a\leq N, \\ &
\quad 1\leq b\leq N)\end{array}
\end{displaymath}
\begin{displaymath}\begin{array}{cc}
(t_{5AaBb})_{JjKk}
=\frac{1}{\sqrt{2}}(\delta_{AJ}\delta_{aj}\delta_{BK}\delta_{bk}-
\delta_{AK}\delta_{ak}\delta_{BJ}\delta_{bj}) & \quad(1\leq A<B\leq M,\quad\! 1\leq a\leq N, \\ &
\quad 1\leq b\leq N)\end{array}
\end{displaymath}
\begin{equation}\label{E57}
(t_{6A})_{JjKk}
=\frac{i}{\sqrt{A(A-1)N}}\delta_{JK}\delta_{jk}
(\theta_{JA}-(A-1)\delta_{JA})
\qquad\qquad(2\leq A\leq M)\end{equation}
We observe that the $t_1$'s, $t_2$'s, $t_4$'s, and $t_5$'s here are simply a re-labelling 
of the
``off-diagonal'' generators, i.e. the $t_1$'s and $t_2$'s, of (\ref{E42}), when the $N$ of 
(\ref{E42}) is replaced by $NM$, while the $t_3$'s and $t_6$'s here are related by an 
orthogonal linear transformation to the ``diagonal'' generators, i.e. the $t_3$'s, of
(\ref{E42}), when the $N$ of (\ref{E42}) is replaced by $NM$.

We now use the convention that Greek indices run over \emph{all} the $(N^2M^2-1)$ 
generators
of $\mathrm{SU}(NM)$, that is, over the $\frac{1}{2}MN(N-1)$ $t_1$'s, the 
$\frac{1}{2}MN(N-1)$
$t_2$'s, the $M(N-1)$ $t_3$'s, the $\frac{1}{2}M(M-1)N^2$ $t_4$'s, the 
$\frac{1}{2}M(M-1)N^2$
$t_5$'s, and the $(M-1)$ $t_6$'s.  Then we find, just as before, that the $t_\alpha$'s
satisfy the equation (\ref{E43}), which again immediately implies the result (\ref{E44}),
hence that the structure constants are totally antisymmetric.

Now for each fixed value of $A$, $1\leq A\leq M$, the $t_{1Aab}$'s, $t_{2Aab}$'s, and 
$t_{3Aa}$'s
generate a distinct $\mathrm{SU}(N)$ subalgebra of $\mathrm{SU}(NM)$, hence the set of all
the $t_1$'s, $t_2$'s, and $t_3$'s generates an $(\mathrm{SU}(N))^M$ subalgebra of 
$\mathrm{SU}(NM)$.  Hence when we apply the results of Section 2, the lower-case indices 
of
Section 2 run over all the sets $\{(1Aab)\mid 1\leq A\leq M, 1\leq a<b\leq N\}$,
$\{(2Aab)\mid 1\leq A\leq M, 1\leq a<b\leq N\}$, and $\{(3Aa)\mid 1\leq A\leq M, 2\leq 
a\leq N\}$, and the upper-case indices of Section 2 run over all the sets $\{(4AaBb)\mid 
1\leq A<B\leq M, 1\leq a\leq N, 1\leq b\leq N\}$, $\{(5AaBb)\mid 1\leq A<B\leq M, 1\leq 
a\leq N, 1\leq b\leq N\}$, and $\{(6A)\mid 2\leq A\leq M\}$.

We now find, with the summation convention temporarily suspended, that for each value of 
$A$, $1\leq A\leq M$, we have:
\begin{displaymath}
\Bigg\{\sum_{1\leq a<b\leq N}\left((t_{1Aab})_{JjKk}(t_{1Aab})_{PpQq}+(t_{2Aab})_{JjKk}
(t_{2Aab})_{PpQq}\right)+\end{displaymath}
\begin{equation}\label{E58}
+\sum_{2\leq A\leq N}(t_{3Aa})_{JjKk}(t_{3Aa})_{PpQq}\Bigg\}=\delta_{JA}\delta_{KA}\delta
_{PA}\delta_{QA}\left(-\delta_{jq}\delta_{kp}+\frac{1}{N}\delta_{jk}\delta_{pq}\right)
\end{equation}
(which is the relation (\ref{E49}) for the number $A$ $\textrm{SU}(N)$ subgroup), 
and that we also have:
	\[\sum_{\begin{array}{c}1\leq A<B\leq M\\1\leq a\leq N\\1\leq b\leq N
	\end{array}}\left(\left(t_{4AaBb}\right)_{JjKk}\left(t_{4AaBb}
	\right)_{PpQq}+\left(t_{5AaBb}\right)_{JjKk}\left(t_{5AaBb}
	\right)_{PpQq}\right)=
\]
\begin{equation}
	\label{E59}
	=-\delta_{JQ}\delta_{jq}\delta_{PK}\delta_{pk}\left(1-\delta_{JK}\right)
\end{equation}
and also:
	\[\sum_{2\leq A\leq M}\left(t_{6A}\right)_{JjKk}\left(t_{6A}\right)
	_{PpQq}=\frac{1}{N}\delta_{JK}\delta_{jk}\delta_{PQ}\delta_{pq}
	\left(-\delta_{JQ}+\frac{1}{M}\right)=
\]
\begin{equation}
	\label{E60}
	=-\frac{1}{N}\delta_{JQ}\delta_{PK}\delta_{JK}\delta_{jk}\delta_{pq}+
	\frac{1}{NM}\delta_{JK}\delta_{jk}\delta_{PQ}\delta_{pq}
\end{equation}
hence, with the summation convention restored, and recalling that greek
indices now run over \emph{all} the generators of $\textrm{SU}(NM)$, we find that:
\begin{equation}
	\label{E61}
	\left(t_\alpha\right)_{JjKk}\left(t_\alpha\right)_{PpQq}=-\delta_{JQ}
	\delta_{jq}\delta_{KP}\delta_{kp}+\frac{1}{NM}\delta_{JK}\delta_{jk}
	\delta{PQ}\delta_{pq}
\end{equation}
which is the analogue of (\ref{E49}) for $\textrm{SU}(NM)$ in the present basis.

\chapter{Reduction of the Group-Changing Equations for $\textrm{SU}(NM)$ and \\$(\textrm{SU}(N))^M$, to Equations Expressing the Vacuum Expectation Values and Correlation Functions for $\textrm{SU}(NM)$, in terms of those for $\textrm{SU}(N)$}

Now let $W_1,\ldots,W_n$ be $n$ Wilson loops in the fundamental
representation (\ref{E57}) of $\textrm{SU}(NM)$.  Then by Section \ref{GCE section}, the 
group-changing equations for $(\textrm{SU}(N))^M$ and $\textrm{SU}(NM)$ express the 
correlation function $\left[W_1\ldots W_n\right]_{\mathrm{SU}(NM)}$, in the Yang
Mills theory for $\textrm{SU}(NM)$, in terms of sums of correlation functions
$\left[\tilde{W}_1\ldots\tilde{W}_s\right]_{(\mathrm{SU}(N))^M}$ of connected,
$(\textrm{SU}(N))^M$-gauge-invariant quantities $\tilde{W}_1,\ldots,\tilde{W}_s$, 
in the Yang Mills theory for $(\textrm{SU}(N))^M$.  We note that
$\tilde{W}_1,\ldots,\tilde{W}_s$ are \emph{not} in general Wilson loops,
but rather are more general $(\textrm{SU}(N))^M$-gauge-invariant quantities
obtained from the $\textrm{SU}(NM)$ Wilson loops $W_1\ldots W_n$ as follows.

We ``decorate'' the closed paths defining the Wilson loops $W_1\ldots W_n$
 by the addition of new paths and junctions, such that the $n$ separate closed paths defining $W_1\ldots W_n$ are replaced by $m$ connected
``networks'' $\tilde{W}_1,\ldots,\tilde{W}_m$ of paths and junctions,
where $1\leq m\leq n$, and for each $1\leq i\leq m$, the network defining
$\tilde{W}_i$ contains at least one of the $n$ original closed paths, and
we also include $r\geq 0$ ``vacuum bubbles''
$\tilde{W}_{m+1},\ldots,\tilde{W}_{m+r}$, each of which is a connected
network of new paths and junctions that does \emph{not} contain any of the
$n$ original closed paths.

Each $\tilde{W}_i$ is an $(\textrm{SU}(N))^M$-gauge-invariant quantity formed from
$(\textrm{SU}(N))^M$- covariant path-ordered phase factors and 
$(\textrm{SU}(N))^M$-invariant tensors as follows.  Each path that comes from the 
whole or part of one of the $n$ original closed paths, is in the 
representation of $(\textrm{SU}(N))^M$ obtained by restricting the representation
(\ref{E57}) of $\textrm{SU}(NM)$ to the subgroup $(\textrm{SU}(N))^M$, i.e. to the $t_1$'s,
$t_2$'s, and $t_3$'s.  And each \emph{new} path is in the representation
 of $(\textrm{SU}(N))^M$ which in the notation of Section 2 is given by:
\begin{equation}
	\label{E62}
	\left(t_a\right)_{AB}=f_{AaB}=-\mathrm{tr}\left(t_At_at_B-t_Bt_at_A
	\right)
\end{equation}
where the $t_\alpha$'s in the right-hand side of (\ref{E62}) are given by
(\ref{E57}), and $a$ in (\ref{E62}) runs over the sets $\{(1Eef)\vert1\leq
E\leq M, 1\leq e<f\leq N\}$, $\{(2Eef)\vert1\leq E\leq M,1\leq e<f\leq N
\}$, and $\{(3Ee)\vert1\leq E\leq M,2\leq e\leq N\}$, (i.e. over the 
generators of $(\textrm{SU}(N))^M$), and $A$ and $B$ in (\ref{E62}) each runs over
all the sets $\{(4EeFf)\vert1\leq E<F\leq M,1\leq e\leq N,1\leq f\leq
N\}$, $\{(5EeFf)\vert1\leq E<F\leq M,1\leq e\leq N,1\leq f\leq N\}$, and
$\{(6E)\vert2\leq E\leq M\}$.

An at each junction where a \emph{new} path ends at one of the $n$ original
closed paths, we have the $(\textrm{SU}(N))^M$-invariant tensor obtained by 
restricting the fundamental representation (\ref{E57}) of $\textrm{SU}(NM))$ to the
$t_4$'s, $t_5$'s, and $t_6$'s, (which gives an $(\textrm{SU}(N))^M$-invariant tensor
by equation (\ref{E31a})), and at each junction at which \emph{three} new
paths end, we have the $(\textrm{SU}(N))^M$-invariant tensor:
\begin{equation}
	\label{E63}
	f_{ABC}=-\mathrm{tr}\left(t_At_Bt_C-t_Ct_Bt_A\right)
\end{equation}
where the $t_\alpha$'s in (\ref{E63}) are given by (\ref{E57}), and $A$, 
$B$, and $C$ in (\ref{E63}) each runs over the same domain as $A$ and $B$
in (\ref{E62}), and at each junction at which \emph{four} new paths end we
have either the $(\textrm{SU}(N))^M$-invariant tensor $f_{AaB}f_{CaD}$, where
$f_{AaB}$ is given by (\ref{E62}), the sum on $a$ runs over the generators
of $(\textrm{SU}(N))^M$, and $A$, $B$, $C$, and $D$ each runs over the same domain 
as $A$ and $B$ in (\ref{E62}), or else the $(\textrm{SU}(N))^M$-invariant tensor
$f_{ABC}f_{DBE}$, where $f_{ABC}$ is given by (\ref{E62}), the sum on $B$
runs over the same domain as $A$ and $B$ in (\ref{E63}), and $A$, $C$,
$D$, and $E$ each runs over the same domain as $A$ and $B$ in (\ref{E62}),
and no junction-types are permitted in the networks $\tilde{W}_i$,
$1\leq i\leq (m+r)$, other than those just described.

Now, as just described, the path-ordered phase factor for each \emph{new}
path has at each end an index that runs over all the sets
$\{(4EeFf)\vert1\leq E<F\leq M, 1\leq e\leq N,1\leq f\leq N\}$,
$\{(5EeFf)\vert1\leq E<F\leq M, 1\leq e\leq N,1\leq f\leq N\}$,
and $\{(6E)\vert2\leq E\leq M\}$.  We observe that $f_{AaB}$ of
(\ref{E62}) \emph{vanishes} whenever $A$ or $B$ or both is a member of
$\{(6E)\vert2\leq E\leq M\}$, which is a consequence of the fact that, for
each separate $\textrm{SU}(N)$ subgroup, each $t_6$ is a multiple of the unit
matrix for that $\textrm{SU}(N)$ subgroup, so that each $t_6$ commutes with every
generator of the $(\textrm{SU}(N))^M$ subgroup of $\textrm{SU}(NM)$.  (Hence each 
$A_{\mu(6E)}$ field interacts with \emph{none} of the gauge fields in the 
$(\textrm{SU}(N))^M$ subgroup.)  It follows immmediately from this that, firstly,
the path-ordered phase factor for a new path has \emph{no} matrix elements
between any $(6E)$ at one end and any $(4GgHh)$ or $(5GgHh)$ at its other
end, and, secondly, the matrix elelments of the path-ordered phase factor
for a new path between any $(6E)$ at one end and any $(6F)$ at the other 
end, are simply given by $\delta_{EF}$.  Thus we may treat the $t_6$'s
completely separately from the $t_4$'s and $t_5$'s.

We now use (\ref{E62}) and (\ref{E63}) to express all the 
$(\textrm{SU}(N))^M$-invariant tensors that occur at junctions at which three or 
four new paths end, in terms of traces of the matrices of the $\textrm{SU}(NM)$
fundamental representation (\ref{E57}).  This has the consequence that,
since the $(\textrm{SU}(N))^M$-invariant tensor at a junction where a new path ends
at one of the $n$ original closed paths, is also a $t_4$, $t_5$, or $t_6$,
the path-ordered phase factor $W_{AB}$ for each new path is contracted with
a $t_A$ and a $t_B$ in the form
$W_{AB}\left(t_A\right)_{JjKk}\left(t_B\right)_{PpQq}$, where $A$ and $B$
are each summed over the sets 
$\{(4EeFf)\vert1\leq E<F\leq M, 1\leq e\leq N,1\leq f\leq N\}$,
$\{(5EeFf)\vert1\leq E<F\leq M, 1\leq e\leq N,1\leq f\leq N\}$,
and $\{(6E)\vert2\leq E\leq M\}$.  Now by the preceding paragraph,
$W_{AB}$ vanishes if one of $A$ and $B$ is a $(6E)$ and the other is a
$(4GgHh)$ or a $(5GgHh)$, and $W_{(6E)(6F)}$ is equal to $\delta_{EF}$, so
the total contribution to 
$W_{AB}\left(t_A\right)_{JjKk}\left(t_B\right)_{PpQq}$, of terms involving
one or more $t_6$'s, is $\displaystyle\sum_{2\leq E\leq M}\left(t_{6E}
\right)_{JjKk}\left(t_{6E}\right)_{PpQq}$, which by (\ref{E60}) is equal
to $\frac{1}{N}\delta_{JK}\delta_{jk}\delta_{PQ}\delta_{pq}\left(
-\delta_{JQ}+\frac{1}{M}\right)$.

\section{Reduction of the new Wilson lines that involve $(\textrm{SU}(N))^M$ gauge fields}

We now change our convention for greek indices again, and define greek
indices to run just over the generators of $\textrm{SU}(N)$, as given by 
(\ref{E42}).  Then we note that the generators $\left(t_{1Aab}\right)
_{JjKk}$, $\left(t_{2Aab}\right)_{JjKk}$, and
$\left(t_{3Aa}\right)_{JjKk}$ of (\ref{E57}) can all be expressed in the
form:
\begin{equation}
	\label{E64}
	\left(t_{A\alpha}\right)_{JjKk}=\delta_{JA}\delta_{KA}\left(t_\alpha
	\right)_{jk}=\delta_{JA}\delta_{JK}\left(t_\alpha\right)_{jk}
\end{equation}
where $\alpha$ runs over the sets $\{(1ab)\vert1\leq a<b\leq N\}$,
$\{(2ab)\vert1\leq a<b\leq N\}$, and $\{(3a)\vert2\leq a\leq N\}$, and
 the $\textrm{SU}(N)$ generators $\left(t_\alpha\right)_{jk}$ are defined in
 (\ref{E42}).  We note that in (\ref{E64}), $A$, $J$, and $K$ satisfy
 $1\leq A\leq M$, $1\leq J\leq M$, and $1\leq K\leq M$, and that the 
 summation convention does \emph{not} apply in the right-hand side.
 
 Then the group index and $(\textrm{SU}(N))^M$-gauge-field structure of the term in
 \\
 $W_{AB}\left(t_A\right)_{JjKk}\left(t_B\right)_{PpQq}$ that is of degree
 $u$ in the $(\textrm{SU}(N))^M$-gauge-fields may be expressed in the form, for 
 $u\geq1$:
 	\[(-1)^u\bigg\{A\raisebox{-4pt}{$\stackrel{\displaystyle{(x(s_1))}}
 	{\scriptstyle{\mu_1(E_1\alpha_1)}}$}
 	A\raisebox{-4pt}{$\stackrel{\displaystyle{(x(s_2))}}
 	{\scriptstyle{\mu_2(E_2\alpha_2)}}$}
 	\dots A\raisebox{-4pt}{$\stackrel{\displaystyle{(x(s_u))}}
 	{\scriptstyle{\mu_u(E_u\alpha_u)}}$}
 	\left(t_A\right)_{JjKk}\mathrm{tr}\left(t_At_{E_1\alpha_1}t_{C_1}-
 	t_{C_1}t_{E_1\alpha_1}t_A\right)\times
\]
\begin{equation}
	\label{E65}
	\times\mathrm{tr}\left(t_{C_1}t_{E_2\alpha_2}t_{C_2}-t_{C_2}t_{E_2
	\alpha_2}t_{C_1}\right)\dots\mathrm{tr}\left(t_{C_{u-1}}t_{E_u\alpha_u}
	t_B-t_Bt_{E_u\alpha_u}t_{C_{u-1}}\right)\left(t_B\right)_{PpQq}\bigg\}
\end{equation}
where (\ref{E62}) has been used, and the $E_i$'s, $1\leq i\leq u$, are to
be summed from 1 to $M$, the $\alpha_i$'s, $1\leq i\leq u$, are to be 
summed over the generators of $\textrm{SU}(N)$, (as specified after (\ref{E64})),
and the $A$, $B$, and the $C_i$'s, $1\leq i\leq (u-1)$, are to be summed
over the sets $\{(4GgHh)\vert1\leq G<h\leq M,1\leq g\leq N,1\leq h\leq
N\}$, and $\{(5GgHh)\vert1\leq G<h\leq M,1\leq g\leq N,1\leq h\leq N\}$.
(We note that, due to the assumption $u\geq1$, it immediately follows from
the preceding paragraphs that no terms involving any $t_6$'s make any
contribution.)

We now apply (\ref{E64}) to the $t_{E_i\alpha_i}$'s in (\ref{E65}), then
use the result (\ref{E59}) to perform the sums over $A$, $B$, and the 
$C_i$'s, $1\leq i\leq (u-1)$, to obtain:
 	\[(-1)^u\Bigg\{A\raisebox{-4pt}{$\stackrel{\displaystyle{(x(s_1))}}
 	{\scriptstyle{\mu_1(E_1\alpha_1)}}$}
 	A\raisebox{-4pt}{$\stackrel{\displaystyle{(x(s_2))}}
 	{\scriptstyle{\mu_2(E_2\alpha_2)}}$}
 	\dots A\raisebox{-4pt}{$\stackrel{\displaystyle{(x(s_u))}}
 	{\scriptstyle{\mu_u(E_u\alpha_u)}}$}
 	\left(t_A\right)_{JjKk}\quad\times
\]
	\[\times\bigg(\left(t_A\right)_{R_1r_1S_1s_1}\left(t_{C_1}\right)
	_{V_1v_1W_1w_1}\left(\left(t_{E_1\alpha_1}\right)_{S_1s_1V_1v_1}\delta
	_{W_1R_1}\delta_{w_1r_1}-\delta_{S_1V_1}\delta_{s_1v_1}\left(t_{E_1
	\alpha_1}\right)_{W_1w_1R_1r_1}\right)\bigg)\times
\]
	\[\times\bigg(\left(t_{C_1}\right)_{R_2r_2S_2s_2}\left(t_{C_2}\right)
	_{V_2v_2W_2w_2}\left(\left(t_{E_2\alpha_2}\right)_{S_2s_2V_2v_2}\delta
	_{W_2R_2}\delta_{w_2r_2}-\delta_{S_2V_2}\delta_{s_2v_2}\left(t_{E_2
	\alpha_2}\right)_{W_2w_2R_2r_2}\right)\bigg)\times
\]
	\[\times\quad\ldots\quad\times
\]
	\[\times\!\bigg(\!\!\left(t_{C_{u-1}}\right)_{R_ur_uS_us_u}\!
	\!\left(t_B\right)
	_{V_uv_uW_uw_u}\!\left(\!\left(t_{E_u\alpha_u}\right)_{S_us_uV_uv_u}
	\delta
	_{W_uR_u}\!\!\delta_{w_ur_u}\!\!-\!\delta_{S_uV_u}\delta_{s_uv_u}\!\left(
	t_{E_u
	\alpha_u}\!\right)_{W_uw_uR_ur_u}\!\right)\!\!\bigg)\!\times
\]
	\[\times\quad\left(t_B\right)_{PpQq}\Bigg\}\quad=
\]
 	\[=\quad(-1)^u\Bigg\{A\raisebox{-4pt}{$\stackrel{\displaystyle{(x(s_1))}}
 	{\scriptstyle{\mu_1(E_1\alpha_1)}}$}
 	A\raisebox{-4pt}{$\stackrel{\displaystyle{(x(s_2))}}
 	{\scriptstyle{\mu_2(E_2\alpha_2)}}$}
 	\dots A\raisebox{-4pt}{$\stackrel{\displaystyle{(x(s_u))}}
 	{\scriptstyle{\mu_u(E_u\alpha_u)}}$}
 	\left(t_A\right)_{JjKk}\quad\times
\]
	\[\times\bigg(\left(t_A\right)_{R_1r_1S_1s_1}\left(t_{C_1}\right)
	_{V_1v_1W_1w_1}\delta_{S_1V_1}\delta_{W_1R_1}
	\left(\delta_{E_1S_1}\left(t_{\alpha_1}\right)_{s_1v_1}
	\delta_{w_1r_1}-\delta_{s_1v_1}\delta_{E_1W_1}\left(t_{
	\alpha_1}\right)_{w_1r_1}\right)\bigg)\times
\]
	\[\times\bigg(\left(t_{C_1}\right)_{R_2r_2S_2s_2}\left(t_{C_2}\right)
	_{V_2v_2W_2w_2}\delta_{S_2V_2}\delta_{W_2R_2}
	\left(\delta_{E_2S_2}\left(t_{\alpha_2}\right)_{s_2v_2}
	\delta_{w_2r_2}-\delta_{s_2v_2}\delta_{E_2W_2}\left(t_{
	\alpha_2}\right)_{w_2r_2}\right)\bigg)\times
\]
	\[\times\quad\ldots\quad\times
\]
	\[\times\!\bigg(\!\left(t_{C_{u-1}}\right)_{R_ur_uS_us_u}\!\left(t_B\right)
	_{V_uv_uW_uw_u}\delta_{S_uV_u}\delta_{W_uR_u}\!
	\left(\delta_{E_uS_u}\!\left(t_{\alpha_u}\right)_{s_uv_u}\!
	\delta_{w_ur_u}\!\!-\!\delta_{s_uv_u}\delta_{E_uW_u}\!\left(t_{
	\alpha_u}\right)_{w_ur_u}\right)\!\!\bigg)\!\times
\]
	\[\times\quad\left(t_B\right)_{PpQq}\Bigg\}\quad=
\]
 	\[=\quad-\ \ \Bigg\{A\raisebox{-4pt}{$\stackrel{\displaystyle{(x(s_1))}}
 	{\scriptstyle{\mu_1(E_1\alpha_1)}}$}
 	A\raisebox{-4pt}{$\stackrel{\displaystyle{(x(s_2))}}
 	{\scriptstyle{\mu_2(E_2\alpha_2)}}$}
 	\dots A\raisebox{-4pt}{$\stackrel{\displaystyle{(x(s_u))}}
 	{\scriptstyle{\mu_u(E_u\alpha_u)}}$}
 	\delta_{JS_1}\delta_{js_1}\delta_{R_1K}\delta_{r_1k}\left(1-
 	\delta_{JK}\right)\ \ \times
\]
	\[\times\delta_{V_1S_2}\delta_{v_1s_2}\delta_{R_2W_1}\delta_{r_2w_1}
	\left(1-\delta_{V_1W_1}\right)\ldots\delta_{V_{u-1}S_u}\delta_{v_{u-1}
	s_u}\delta_{R_uW_{u-1}}\delta_{r_uw_{u-1}}\left(1-\delta_{V_{u-1}W_{u-1}}
	\right)\times
\]
	\[\times\delta_{V_uQ}\delta_{v_uq}\delta_{PW_u}\delta_{pw_u}\left(1-
	\delta_{V_uW_u}\right)\delta_{S_1V_1}\delta_{W_1R_1}
	\left(\delta_{E_1S_1}\left(t_{\alpha_1}\right)_{s_1v_1}
	\delta_{w_1r_1}-\delta_{s_1v_1}\delta_{E_1W_1}\left(t_{
	\alpha_1}\right)_{w_1r_1}\right)\times
\]
	\[\times\delta_{S_2V_2}\delta_{W_2R_2}
	\left(\delta_{E_2S_2}\left(t_{\alpha_2}\right)_{s_2v_2}
	\delta_{w_2r_2}-\delta_{s_2v_2}\delta_{E_2W_2}\left(t_{
	\alpha_2}\right)_{w_2r_2}\right)\ \ \times\ \ \ldots\ \ \times
\]
	\[\times\delta_{S_uV_u}\delta_{W_uR_u}
	\left(\delta_{E_uS_u}\left(t_{\alpha_u}\right)_{s_uv_u}
	\delta_{w_ur_u}-\delta_{s_uv_u}\delta_{E_uW_u}\left(t_{
	\alpha_u}\right)_{w_ur_u}\right)\Bigg\}\quad=
\]
 	\[=\quad-\ \ \Bigg\{A\raisebox{-4pt}{$\stackrel{\displaystyle{(x(s_1))}}
 	{\scriptstyle{\mu_1(E_1\alpha_1)}}$}
 	A\raisebox{-4pt}{$\stackrel{\displaystyle{(x(s_2))}}
 	{\scriptstyle{\mu_2(E_2\alpha_2)}}$}
 	\dots A\raisebox{-4pt}{$\stackrel{\displaystyle{(x(s_u))}}
 	{\scriptstyle{\mu_u(E_u\alpha_u)}}$}
 	\delta_{JQ}\delta_{PK}\left(1-\delta_{JK}\right)\ \ \times
\]
	\[\times\left(\delta_{E_1J}\left(t_{\alpha_1}\right)_{jv_1}
	\delta_{w_1k}-\delta_{jv_1}\delta_{E_1K}\left(t_{
	\alpha_1}\right)_{w_1k}\right)
	\left(\delta_{E_2J}\left(t_{\alpha_2}\right)_{v_1v_2}
	\delta_{w_2w_1}-\delta_{v_1v_2}\delta_{E_2K}\left(t_{
	\alpha_2}\right)_{w_2w_1}\right)\times
\]
	\[\times\ \ \ldots\ \ \times
	\ \ \left(\delta_{E_uJ}\left(t_{\alpha_u}\right)_{v_{u-1}q}
	\delta_{pw_{u-1}}-\delta_{v_{u-1}q}\delta_{E_uK}\left(t_{
	\alpha_u}\right)_{pw_{u-1}}\right)\Bigg\}\quad=
\]
 	\[=\quad-\delta_{JQ}\delta_{PK}\left(1-\delta_{JK}\right)
 	\left(A\raisebox{-4pt}{$\stackrel{\displaystyle{(x(s_1))}}
 	{\scriptstyle{\mu_1(J\alpha_1)}}$}
 	\left(t_{\alpha_1}\right)_{jv_1}
	\delta_{w_1k}-\delta_{jv_1}
	A\raisebox{-4pt}{$\stackrel{\displaystyle{(x(s_1))}}
 	{\scriptstyle{\mu_1(K\alpha_1)}}$}
	\left(t_{\alpha_1}\right)_{w_1k}\right)\times
\]
	\[\times
	\left(A\raisebox{-4pt}{$\stackrel{\displaystyle{(x(s_2))}}
 	{\scriptstyle{\mu_2(J\alpha_2)}}$}
	\left(t_{\alpha_2}\right)_{v_1v_2}
	\delta_{w_2w_1}-\delta_{v_1v_2}
	A\raisebox{-4pt}{$\stackrel{\displaystyle{(x(s_2))}}
 	{\scriptstyle{\mu_2(K\alpha_2)}}$}
	\left(t_{
	\alpha_2}\right)_{w_2w_1}\right)\times
\]
\begin{equation}
	\label{E66}
\times\ \ldots\ \times
	\ \left(A\raisebox{-4pt}{$\stackrel{\displaystyle{(x(s_u))}}
 	{\scriptstyle{\mu_u(J\alpha_u)}}$}
	\left(t_{\alpha_u}\right)_{v_{u-1}q}
	\delta_{pw_{u-1}}-\delta_{v_{u-1}q}
	A\raisebox{-4pt}{$\stackrel{\displaystyle{(x(s_u))}}
 	{\scriptstyle{\mu_u(K\alpha_u)}}$}
	\left(t_{
	\alpha_u}\right)_{pw_{u-1}}\right)
\end{equation}
where the $E_i$'s, $1\leq i\leq u$, the $\alpha_i$'s, $1\leq i\leq u$, 
$A$, $B$, and the $C_i$'s, $1\leq i\leq (u-1)$, are to be summed over the 
same domains as in (\ref{E65}), the $R_i$'s, $S_i$'s, $V_i$'s, and
$W_i$'s, $1\leq i\leq u$, are to be summed from 1 to $M$, and the $r_i$'s,
$s_i$'s, $v_i$'s, and $w_i$'s, $1\leq i\leq u$, are to be summed from 1 to
$N$.  (We note that after the application of (\ref{E59}), the sums on the
$S_i$'s and $V_i$'s collapsed to 
	\[J=S_1=V_1=S_2=V_2=\ldots=S_u=V_u(=Q)
\]
and the sums on the $R_i$'s and $W_i$'s collapsed to 
	\[K=R_1=W_1=R_2=W_2=\ldots=R_u=W_u(=P)
\]
and that the summation convention is \emph{not} to be applied to $J$ and 
$K$ in the right-hand side of (\ref{E66}).)

We now substitute the result (\ref{E66}) for each $u\geq1$ into the 
definition (\ref{E1}) of the path-ordered phase factor in the 
representation of $(\textrm{SU}(N))^M$ given by (\ref{E62}), where the indices in
(\ref{E62}) run over the sets specified immediately after (\ref{E62}),
(and $n$ in (\ref{E1}) is re-written as $u$), and sum over $u$ from 1 to
$\infty$, and also add the $u=0$ term as given by (\ref{E59}), to conclude
that if $x(x)$, $0\leq s\leq1$, is any new path, and $W((\textrm{SU}(N))^M,x(s))
_{AB}$ denotes its path-ordered phase factor as specified above, (so that
$A$ and $B$ each run over the sets
$\{(4EeFf)\vert1\leq E\leq F\leq M,1\leq e\leq N,1\leq f\leq N\}$,
$\{(5EeFf)\vert1\leq E\leq F\leq M,1\leq e\leq N,1\leq f\leq N\}$, and
$\{(6E)\vert2\leq E\leq M\}$), then the total contribution to 
$W((\textrm{SU}(N))^M,x(s))_{AB}\left(t_A\right)_{JjKk}\left(t_B\right)_{PpQq}$ 
from terms that involve no $t_6$'s, is simply equal to $-\delta_{JQ}
\delta_{PK}\left(1-\delta_{JK}\right)$ times the \emph{product} of the 
path-ordered phase factor $W\left(A_{\mu(J\alpha)},x(s)\right)_{jq}$, in
the \emph{fundamental} representation of $\textrm{SU}(N)$, \emph{for the gauge
fields of the number} $J$ $\textrm{SU}(N)$ \emph{subgroup}, and the path-ordered
phase factor $W\left(A_{\mu(K\alpha)},x(1-s)\right)_{pk}$, in the
\emph{fundamental} representation of $\textrm{SU}(N)$, \emph{for the gauge fields 
of the number} $K$ $\textrm{SU}(N)$ \emph{subgroup}, where $x(1-s)$ denotes the 
given path $x(s)$, traversed in the opposite direction.  Indeed, when we
write out the product of these two path-ordered phase factors, with each
being given by equation (\ref{E1}), we may collect together, for each
$u\geq1$, all the terms of \emph{total} degree $u$ in the gauge fields,
(where the general such term has $r$ $A_{\mu(J\alpha)}$'s, for some 
$0\leq r\leq u$, and $(u-r)$ $A_{\mu(K\alpha)}$'s).  Then in each such 
term we replace the integration variables $s_i$ coming from the second
path-ordered phase factor by new integration variables $\tilde{s}_i=(1-s_i
)$, then break up the product of the path-ordered integrals coming from 
the two separate phase factors into a sum of \emph{totally} path-ordered
terms, (i.e. such that the set of all the integration variables $s_i$
coming from the first phase factor, and all the integration variables
$\tilde{s}_i$ coming from the second phase factor, is totally ordered),
and relabel the integration variables in each such term in accordance with
the total path ordering in that term.  We then see that we have a sum over
the $2^u$ independent choices of specifying, independently for each gauge
field along the path, whether that gauge field is an $A_{\mu(J\alpha)}$ or
an $A_{\mu(K\alpha)}$, which is precisely what wee have in the right-hand
side of (\ref{E66}), when we note that each $A_{\mu(K\alpha)}$ also 
brings a minus sign, from
	\[\frac{dx_{\mu_i}(1-s_i)}{ds_i}=-\frac{dx_{\mu_i}(\tilde{s}_i)}
	{d\tilde{s}_i}
\]

\subsection{The $\textrm{SU}(N)$ Wilson lines, in a new $(\textrm{SU}(N))^M$ Wilson line, belong to \emph{different} $\textrm{SU}(N)$'s}

Hence since, as noted above, the total contribution to 
  \[W((\textrm{SU}(N))^M,x(s))_{AB}\left(t_A\right)_{JjKk}\left(t_B\right)_{PpQq}
\]
from terms that involve one or more $t_6$'s, is simply equal to 
$\frac{1}{N}\delta_{JK}\delta_{jk}\delta_{PQ}\delta_{pq}\left(-\delta_{JQ}
+\frac{1}{M}\right)$, we finally find that, for any new path $x(s)$:
	\[W((\textrm{SU}(N))^M,x(s))_{AB}\left(t_A\right)_{JjKk}\left(t_B\right)_{PpQq}=
\]
	\[=\ \bigg\{-\delta_{JQ}\delta_{PK}\left(1-\delta_{JK}\right)
	W(A_{\mu(J\alpha)},x(s))_{jq}W(A_{\mu(K\alpha)},x(1-s))_{pk}+
\]
\begin{equation}
	\label{E67}
	+\frac{1}{N}\delta_{JK}\delta_{jk}\delta_{PQ}\delta_{pq}\left(-\delta
	_{JQ}+\frac{1}{M}\right)\bigg\}
\end{equation}
where in the left-hand side here, $A$ and $B$ are each summed over the 
sets
$\{(4EeFf)\vert1\leq E<F\leq M,1\leq e\leq N,1\leq f\leq N\}$,
$\{(5EeFf)\vert1\leq E<F\leq M,1\leq e\leq N,1\leq f\leq N\}$, and
$\{(6E)\vert2\leq E\leq M\}$, and the meanings of all the terms in the 
right-hand side have been explained above.

(We note that in equation (\ref{E67}), as in equation (\ref{E1}), $s$ is 
not an argument of either side of the equation - we wrote $x(s)$ simply to
display the fact that the argument $x$ of $W$ is a path, not a point.)

We shall refer to the first term in the right-hand side of (\ref{E67}),
which comes from all the terms in the left-hand side which involve
\emph{no} $t_6$'s, as the \emph{45-term}, and the second term in the 
right-hand side of (\ref{E67}), which comes from all the terms in the 
left-hand side which involve one or more $t_6$'s, as the \emph{6-term}.

We shall also refer to the $A_{\mu(4EeFf)}$ fields, the $A_{\mu(5EeFf)}$
fields, and the $A_{\mu(6E)}$ fields, respectively, and also the 
corresponding Fadeev-Popov fields, as the 4-fields, the 5-fields, and the 
6-fields, respectively.

We note that the crucial $\left(1-\delta_{JK}\right)$ factor in the
45-term, which results in the 45-term vanishing whenever $J=K$, comes from
the fact that the 4-fields and 5-fields interact with two \emph{distinct}
$\textrm{SU}(N)$ subgroups.

And we note that the summation convention does \emph{not} apply to the
right-hand side of equation (\ref{E67}).

\section{Reduction of the $(\textrm{SU}(N))^M$ quantities, to sums of products of Wilson loops in the $\textrm{SU}(N)$'s}

We now return to the discussion, which we began after equation
(\ref{E61}), of the particular group-changing equaton for $\textrm{SU}(N))^M$ and
$\textrm{SU}(NM)$, that applies to the correlation function $[W_1\ldots W_n]_
{\textrm{SU}(NM)}$, in the Yang Mills theory for $\textrm{SU}(NM)$, of $n$ Wilson loops
$W_1,\dots,W_n$, each in the fundamental representation (\ref{E57}) of
$\textrm{SU}(NM)$, and we consider a term $[\tilde{W}_1\dots\tilde{W}_s]_{\mathrm{SU}(N))^M}
$ in the right-hand side of this equation, where each $\tilde{W}_i$ is a
connected $\textrm{SU}(N))^M$-gauge-invariant quantity formed, as described in the
discussion following equation (\ref{E61}), from $(\textrm{SU}(N))^M$-covariant
path-ordered phase factors corresponding to paths that form parts or
wholes of the $n$ original closed paths, (each taken in the representation
(\ref{E64}) of $(\textrm{SU}(N))^M$), form $(\textrm{SU}(N))^M$-covariant path-ordered phase
factors corresponding to ``new'' paths, (each takenin the representation
(\ref{E62}) of $(\textrm{SU}(N))^M$), and from the particular
$\textrm{SU}(N))^M$-invariant tensors, at the junctions of the paths, specified in
the discussion accompanying equation (\ref{E63}).

We note that each vertex involving new paths \emph{only}, comes either
from a term in (\ref{E31}) or from a term in (\ref{E23}).  We have to sum
over assignments of which of the new paths are gauge-field paths and 
which are Fadeev-Popov paths, subject to the requirements that there are
either 0 or 2 Fadeev-Popov path-ends at any vertex that involves new paths
ony, and no Fadeev-Popov path ends on any original path.  We consider some
specific such assignment.

We now, as described in the paragraphs preceding equation (\ref{E64}), use
equations (\ref{E62}) and (\ref{E63}) to express all the 
$\textrm{SU}(N))^M$-invariant tensors that occur at junctions at which three or 
four new paths end, in terms of traces of the matrices of the $\textrm{SU}(NM)$
fundamental representation (\ref{E57}), which, as noted before, has the
consequence that every new path has its path-ordered phase factor
$W((\textrm{SU}(N))^M,x(s))_{AB}$ contracted with a $t_A$ and a $t_B$ in the form
$W((\textrm{SU}(N))^M,x(s))_{AB}\left(t_A\right)_{JjKk}\left(t_B\right)_{PpQq}$,
where $A$ and $B$ are each summed over the sets
$\{(4EeFf)\vert1\leq E<F\leq M,1\leq e\leq N,1\leq f\leq N\}$,
$\{(5EeFf)\vert1\leq E<F\leq M,1\leq e\leq N,1\leq f\leq N\}$, and
$\{(6E)\vert2\leq E\leq M\}$, which is exactly what we have in the 
left-hand side of equation (\ref{E67}).  We therefore apply (\ref{E67}) to
every new path.

To see what we get, let us first take the \emph{first} term in the 
right-hand side of (\ref{E67}), i.e. the 45-term, in every new path.  We
then see immediately that we get a finite number fo terms, each 
corresponding to an independent choice, at each vertex at which three new
paths meet, of one of the two terms in the right-hand side of (\ref{E63}),
and also to an independent choice, at each vertex with four gauge-field
path-ends, of either the $f_{aBC}f_{aEF}$ term or the 
$f_{ABC}f_{AEF}$ term in the ``quartic vertex'' term in (\ref{E31}), 
(while at a vertex with two gauge-field path-ends and two Fadeev-Popov
path-ends, we must take the $f_{AaB}f_{CaD}$ quartic vertex term in
(\ref{E23}), and alsto to independent choices for each of the structure 
constants occurring in the chosen terms in the quartic vertices, of either
of the two terms in (\ref{E62}) or in (\ref{E63}).  Each such choice
corresponds to a different routeing, through the relevant vertex, of the
index connections of the two separate path-ordered phase factors, each in
the fundamental representation of $SUN(N)$, but involving the fields of 
two \emph{different} $\textrm{SU}(N)$ subgroups of $(\textrm{SU}(N))^M$, (and traversing
their path inn opposite directions), that are multiplied together in the
45-term.  We consider any one particular such routeing, which corresponds
to the $\textrm{SU}(N)$ fundamental representation path-ordered phase factros 
joining across the vertices to form a particular collection of closed 
loops, and we immediately see, that when we contract the upper-case
indices through a vertex, the fields in each $\textrm{SU}(N)$ fundamental 
representation path-ordered phase factor are in the \emph{same} $\textrm{SU}(N)$
subgroup of $(\textrm{SU}(N))^M$ in both members of each pair of $\textrm{SU}(N)$
fundamental representation path-ordered phase factors that join onto one
another across a vertex.  It immediately follows from this, that in each 
closed loop of $\textrm{SU}(N)$ fundamental representation path-ordered phase
factors, all the fields occurring in all the phase factors in that closed
loop, are in the \emph{same} $\textrm{SU}(N)$ subgroup of $\textrm{SU}(N))^M$.  \emph{Thus
each closed loop of} $\textrm{SU}(N)$ \emph{fundamental representation 
path-ordered phase factors, (formed by joining} $\textrm{SU}(N)$ \emph{fundamental
representation path-ordered phase factors in the individual paths, across
the vertices, in accordance with the chosen routeing), is a Wilson loop in
one of the} $M$ $\textrm{SU}(N)$ \emph{subgroups of} $(\textrm{SU}(N))^M$.  We note,
furthermore, that if any $\textrm{SU}(N)$ fundamental representation Wilson loop
passes along any new path \emph{in both directions}, then both the $\textrm{SU}(N)$
fundamental representation path-ordered phase factors passing along that
path will be in the \emph{same} one of the $M$ $\textrm{SU}(N)$'s hence the $\left(
1-\delta_{JK}\right)$ factor in the 45-term for that path will vanish.
Hence any routeing of the $(\textrm{SU}(N))^M$ indices through the vertices which 
results in any of the $\textrm{SU}(N)$ fundamental representation Wilson loops
produced by that routeing, passing along any new path in both directions,
gives no contribution at all, hence we may restrict the sum over routeings
of the $(\textrm{SU}(N))^M$ indices through the vertices to routeings which satisfy
the ``selection rule'' that \emph{none} of the $\textrm{SU}(N)$ fundamental
representation Wilson loops which they produce, passes along any new path
in both directions.

\subsection{Summing over partitions of the set of the $\textrm{SU}(N)$ Wilson loops, into parts whose members belong to the same $\textrm{SU}(N)$, subject to the selection rule}

We now have to sum over the upper-case indices, (which distinguish the $M$
separate $\textrm{SU}(N)$ subgroups of $(\textrm{SU}(N))^M$), subject to the constraint,
which follows immediately from the foregoing, that all the upper-case
indices around any $\textrm{SU}(N)$ fundamental representation Wilson loop be
equal, with their common value identifying the particular one of the $M$
$\textrm{SU}(N)$ subgroups of $(\textrm{SU}(N))^M$ to which all the fields occurring in that
$\textrm{SU}(N)$ fundamental representation Wilson loop belong, and also to the 
constraint, which follows from the  $\left(1-\delta_{JK}\right)$ factor in
the 45-term, that \emph{no new path has the two} $\textrm{SU}(N)$ \emph{fundamental
representation path-ordered phase factors that pass along it, being in
the} same $\textrm{SU}(N)$ \emph{subgroup of} $(\textrm{SU}(N))^M$.  Thus for each choice of
the routeings through the vertices, we have to sum over all
\emph{partitions}, into not more than $M$ parts, of the set of all the 
$\textrm{SU}(N)$ fundamental representation closed loops that result from that 
routeing, subject to the constraint that no two loops that are members of
the same part of the partition, pass along any common new path.  And for 
each such partition, we sum over all distinct ways of assigning a distinct
one of the $M$ $\textrm{SU}(N)$'s to each part of the partition, (so that all the 
members of any given part of the partition, are in the \emph{same} 
$\textrm{SU}(N)$, which is different from the $\textrm{SU}(N)$ assigned to every other part
of the partition).  We note that we are here considering together 
\emph{all} the $\tilde{W}_i$'s that occur in the $(\textrm{SU}(N))^M$ correlation
function $[\tilde{W}_1\ldots\tilde{W}_s]_{(\mathrm{SU}(N))^M}$, \emph{not} just one
of the $\tilde{W}_i$'s on its own, and that the set of $\textrm{SU}(N)$ 
fundamental representation Wilson loops that is being partitioned is the 
set of \emph{all} the $\textrm{SU}(N)$ Wilson loops that arise in the manner 
described, in \emph{all} the $\tilde{W}_i$'s.

Now the $(\textrm{SU}(N))^M$ correlation function
$[\tilde{W}_1\ldots\tilde{W}_s]_{(\mathrm{SU}(N))^M}$ is itself equal to the sum,
over all partitions of the set $\{\tilde{W}_1,\ldots,\tilde{W}_s\}$, of 
$(-1)^{m-1}(m-1)!$, where $m$ is the number of parts of the partition,
times the product, over the parts of the partition, of the vacuum 
expectation value, in the Yang Mills theory for $(\textrm{SU}(N))^M$, of the 
$\tilde{W}_i$'s in that part of the partition.  (This is simply the
inverse of the general expression (\ref{E55}) for vacuum expectation
values in terms of correlation functions, and has already been used in
Section 2.)

But the vacuum expectation value, in the Yang Mills theory for 
$(\textrm{SU}(N))^M$, of any product of gauge-invariant quantities, each individual
one of which involves just \emph{one} of the $\textrm{SU}(N)$'s, factorizes exactly
into the product, over the individual $\textrm{SU}(N)$'s, of the vacuum
expectation value, in that $\textrm{SU}(N)$, of the product of all those factors
in the original product of gauge-invariant quantities, that belong to that
$\textrm{SU}(N)$.  Thus when we consider any term in our expansion of
$[\tilde{W}_1\ldots\tilde{W}_s]_{(\mathrm{SU}(N))^M}$ as a sum of numerical
coefficients times products of vacuum expectation values, in the Yang 
Mills theory for $(\textrm{SU}(N))^M$, of subsets of the set $\{\tilde{W}_1,\ldots,
\tilde{W}_s\}$, (where each such product runs over the parts of some 
partition of $\{\tilde{W}_1,\ldots,\tilde{W}_s\}$), and we also consider
a particular term in our sum over all partitions of the set of all the
$\textrm{SU}(N)$ fundamental representation Wilson loops that have been generated
from the paths in the $\tilde{W}_i$'s in the manner described, and a 
particular assignment of each part of \emph{this} partition to a distinct
one of the $M$ $\textrm{SU}(N)$'s, (so that all the $\textrm{SU}(N)$ Wilson loops belonging
to any one part of this partition,are in the \emph{same} $\textrm{SU}(N)$, and
$\textrm{SU}(N)$ Wilson loops belonging to \emph{distinct} parts of this partition,
are in \emph{distinct} $\textrm{SU}(N)$'s), we see that every term factorizes into
a product of vacuum expectation values, in the Yang Mills theory for 
$\textrm{SU}(N)$, of products of Wilson loops in the fundamental representation of
$\textrm{SU}(N)$.

\subsection{No two Wilson loops, in any of the $\textrm{SU}(N)$ vacuum expectation values, touch one another along any path}

Furthermore, the $\left(1-\delta_{JK}\right)$ factor in the 
45-term in (\ref{E67}) vanishes whenever two of our $\textrm{SU}(N)$ fundamental
representation Wilson loops that share a common ``new'' path, are in the 
\emph{same} one of the $M$ $\textrm{SU}(N)$'s.  Hence, in every term in our double
sum over partitions, (i.e. over all allowed partitions of the set of all
our $\textrm{SU}(N)$ fundamental representation Wilson loops, and over all 
partitions of the set $\{\tilde{W}_1,\ldots,\tilde{W}_s\}$), any two of 
our $\textrm{SU}(N)$ fundamental representation Wilson loops that pass alon any
common new path, lie in \emph{different} $\textrm{SU}(N)$ vacuum expectation 
values.  \emph{Hence, due to the} $\left(1-\delta_{JK}\right)$ 
\emph{factor in the 45-term in (\ref{E67}), we} never \emph{have two
Wilson loops in any one of thes vacuum expectation values, touching one
another along any path.}  (We recall that, as already noted, the 
$\left(1-\delta_{JK}\right)$ factor also results in any routeing of the
$(\textrm{SU}(N))^M$ indices through the vertices that results in any of the 
$\textrm{SU}(N)$ fundamental representation Wilson loops passingalong any new path
in both directions, giving vanishing contribution, so that none of our
individual $\textrm{SU}(N)$ Wilson loops passes along any new path in both 
directions, hence none of our individual $\textrm{SU}(N)$ Wilson loops touches
itself along any path.)  Furthermore, we see immediately that at any 
vertex at which three new paths meet, the constraint that the two $\textrm{SU}(N)$
fundamental representation path-ordered phase factors that run along each
new path, must be in \emph{different} $\textrm{SU}(N)$'s, means that \emph{none} of
the $\textrm{SU}(N)$ vacuum expectation values in any of our products, contains
more than one of the three $\textrm{SU}(N)$ fundamental representation Wilson loops
that meet at any such vertex.  Thus in each vacuum expectation value in
each of our products, the only ``essential'' intersection or touching, (or
self-intersection or self-touching), of one or more of the $\textrm{SU}(N)$ 
fundamental representation Wilson loops in that vacuum expectation value,
that can occur, (where by ``essential'', we mean that is not a consequence
of some accidental intersection or touching in the configuration space of
the paths and vertices), is when two loops intersect, (or one loop
intersects itself), at a quartic vertexin (\ref{E31}) or (\ref{E23}).
And furthermore, this kind of ``essential'' intersection resulting from a
quartic verex, is itself only an ``occasional'' occurrence, occurring
either when the partition of the set of our $\textrm{SU}(N)$ fundamental
representation Wilson loops is such that two different loops that meet at
some quartic vertex, are in the \emph{same} part of that partition, (which
results in two different loops in some vacuum expectation value 
intersecting one another), or else when the routeings of the $(\textrm{SU}(N))^M$
indices through the vertices are such that some $\textrm{SU}(N)$ fundamental
representation Wilson loop intersects itself at some quaric vertex.

We thus see that, at least for the terms where we take the 45-term in
(\ref{E67}) for every new path, an with the exception of the possible
``essential'' intersections which can arise from quartic vertices, (which
will require special treatment in renormalization, but are not
substantially worse than the ``accidental'' intersections which can occur
due to intersections or coincidences of paths or vertices in configuration
space), the right-hand sides of the group-changing equations for 
$(\textrm{SU}(N))^M$ and $\textrm{SU}(NM)$, applied to correlation functions, in the Yang
Mills theory for $\textrm{SU}(NM)$, of products of Wilson loops in the fundamental
representantion of $\textrm{SU}(NM)$, reduce to sums of products of vacuum
expectation values, and hence, by (\ref{E55}), to sums of products of 
correlation functions, in the Yang Mills theory for $\textrm{SU}(N)$, of products 
of Wilson loops in the fundamental representation of $\textrm{SU}(N)$, and 
furthermore, the $\textrm{SU}(N)$ fundamental representation Wilson loops that 
occur in any individual $\textrm{SU}(N)$ vacuum expectation value or correlation
function that occurs in the right-hand side have, in the absence of any
accidental intersections or touchings associated with particular
configurations in position space of the paths and vertices in the
$\tilde{W}_i$'s, \emph{no} intersections or touchings other than the 
possible ``occasional'' simple intersections described above arising from
quartic vertices.  In other words, at least for all the terms where we
choose the 45-term in (\ref{E67}) for every new path, all the 
complicated gauge-invariant ``networks'' with junctions that occur in the
right-hand sides of the group-changing equations for general $G$ and $H$,
(even when we only have simple products of Wilson loops in the left-hand
sides), have, in the present case of $G=(\textrm{SU}(N))^M$ and $H=\textrm{SU}(NM)$, with
correlation functions of products of $\textrm{SU}(NM)$ fundamental representation
Wilson loop in the left-hand sides, \emph{reduced away}, leaving only
sums of products of correlation functions of products of $\textrm{SU}(N)$
fundamental representation Wilson loops in the right-hand sides.  We note
that these results apply for \emph{all} $N$.

\subsection{Summing over assignments of the $\textrm{SU}(N)$'s to the parts of the partition gives a polynomial factor in $M$}

We still have to sum over all the distinct ways of assigning a distinct
one of the $M$ $\textrm{SU}(N)$ subgroups of $(\textrm{SU}(N))^M$ to each of the $r$ parts
of our partition of the st of all our $\textrm{SU}(N)$ fundamental representation
Wilson loops, but we now see that this sum simply gives a factor
	\[M(M-1)(M-2)\ldots(M-r+1)=\frac{M!}{(M-r)!}
\]
Now this is a polynomial in $M$ that vanishes for $r\geq(M+1)$, hence we
may extend the sum over the partitions of the set of all our $\textrm{SU}(N)$
fundamental representation Wilson loops, (which was previously restricted
to partitions with not more than $M$ parts), into a sum over \emph{all}
partitions of the set of all our $\textrm{SU}(N)$ fundamental representation
Wilson loops, with the sum over all the distinct assignments of a distinct
one of the $M$ $\textrm{SU}(N)$ subgroups of $(\textrm{SU}(N))^M$ to each part of a
partition into $r$ parts, giving a factor 
	\[M(M-1)(M-2)\ldots(M-r+1)=\frac{M!}{(M-r)!}
\]
which vanishes for $r\geq(M+1)$.  Hence we now see that, at least for the 
terms where we take the 45-term in (\ref{E67}) for every new path, the
\emph{only} dependence on $M$ of the right-hand sides of our equations is
through these simple factors $M(M-1)(M-2)\ldots(M-r+1)$, there being
precisely one such factor in each term, and that factor having its 
``$r$'' equal to the number of parts of the particular partition of the
set of all our $\textrm{SU}(N)$ fundamental representation Wilson loops, that
occurs in the definition of that term.

\section{Inclusion of the 6-terms does not alter the main results}

Let us now consider what happens when we choose the 6-term in (\ref{E67}),
rather than the 45-term, in some of the new paths,  We first note that 
there are \emph{no} path-ordered phase factors in the 6-term, which is an
immediate consequence of the fact that each 6-field interacts with 
\emph{none} of the $\textrm{SU}(N)$'s.  Furthermore, the $t_6$'s are all diagonal
matrices hence commute with one another, hence any $\textrm{SU}(NM)$ structure
constant with two or more of its indices in the set $\{(6E)\vert
2\leq E\leq M\}$, vanishes identically, hence at any vertex at which
three new paths meet, \emph{at most one of them} can get the 6-term rather
than the 45-term in (\ref{E67}), while at any vertex at which four new
paths meet, \emph{at most two of them} can get the 6-term rather than the
45-term in (\ref{E67}).  Furthermore, the $\delta_{JK}\delta_{jk}
\delta_{PQ}\delta_{pq}$ index structure of the 6-term means that, as far
as the index routeings and path-ordered phase factors are concerned, it is
just as if any new path on which we choose the 6-term rather than the 
45-term in (\ref{E67}), is simply not there at all.  Thus we may
completely analyse the effects of the 6-term in (\ref{E67}), by starting
from ``core'' diagrams in which the 45-term in (\ref{E67}) is taken for 
every new path, and in which we have already specified, (for each 
particular ``core'' diagram term), what the index routeings through the 
vertices are, and what partition we take of the set of all the $\textrm{SU}(N)$
fundamental representation Wilson loops produced by that routeing, and
considering additions to each ``core'' diagram term, of additional new
paths, each having both its ends on paths or vertices of the core diagram,
and subject to the constraint that a total of at most four new paths can
end at any vertex of the resulting diagram, with the 6-term in (\ref{E67})
being taken on each new path.  Now $f_{AaB}$ in (\ref{E62}) vanishes if
either $S$ or $B$ is a ``6'' (i.e. a member of $\{(6E)\vert2\leq E\leq M
\}$), since every $t_6$ commutes with every $t_1$, $t_2$, and $t_3$, hence
all the couplings of ``6-paths'', (i.e. new paths on which we take the 
6-term in (\ref{E67})), to vertices \emph{not} on any of the original $n$
$\textrm{SU}(NM)$ fundamental representation Wilson loop closed paths, are via
$f_{ABC}$ in (\ref{E63}), with \emph{one} of $A$, $B$, and $C$ being a
``6'', and the other two being ``45's'', (i.e. members of 
$\{(4EeFf)\vert1\leq E<F\leq M, 1\leq e\leq N,1\leq f\leq N\}$ or
$\{(5EeFf)\vert1\leq E<F\leq M, 1\leq e\leq N,1\leq f\leq N\}$).  And we
immediately see that when one of $A$, $B$, and $C$ is a ``6'' and the 
other two are ``45's'', each of the two terms in the right-hand side of 
(\ref{E63}) corresponds, when a new ``6-path'' ends on a new path of the
``core'' diagram to form a new vertex at which two ``45'' paths, (i.e. 
new paths on which we take the 45-term in (\ref{E67})), and one ``6'' path
end, to the upper-case indices ``$J$'' and ``$K$'' at that end of that 
``6'' path, both being equal to one, or both being equal to the other, of
the two different numbers wihch identify the two different $\textrm{SU}(N)$'s to 
whcih the two $\textrm{SU}(N)$ fundamental representation path-ordered phase 
factors that run along that ``45'' path in the ``core'' diagram, belong,
(in that particular term in our sum over all partitions of the set of all
the $\textrm{SU}(N)$ fundamental representation Wilson loops in that particular
core diagram term).  Now when a ``6'' path ``attaches'' to the 
\emph{same} one of the $M$ $\textrm{SU}(N)$'s at both its ends, the $\left(
-\delta_{JQ}+\frac{1}{M}\right)$ factor in the 6-term in (\ref{E67}) takes
the value $-\frac{(M-1)}{M}$, while when a ``6'' path ``attaches'' to a
\emph{different} $\textrm{SU}(N)$ at each end, the $\left(
-\delta_{JQ}+\frac{1}{M}\right)$ factor in the 6-term takes the value
$\frac{1}{M}$.  And whether a ``6''-path ``attaches'' to the \emph{same}
$\textrm{SU}(N)$ at each end or to a \emph{different} $\textrm{SU}(N)$ at each end is
determined, firstly, by whether it ``attaches'' to the \emph{same} 
$\textrm{SU}(N)$ fundamental representation Wilson loop at each end or to a 
\emph{different} $\textrm{SU}(N)$ fundamental representation Wilson loop at each
end, and secondly, if it attaches to a \emph{different} $\textrm{SU}(N)$ 
fundamental representation Wilson loop at each end, whether or not those
two different $\textrm{SU}(N)$ fundamental representation Wilson loops are members
of the same, or different, parts of the partition of the set of \emph{all}
the $\textrm{SU}(N)$ fundamental representation Wilson loops of the present 
particular ``core'' diagram term, to which the present ``partition-term''
corresponds.  Hence, for each particular ``partition-term'' in each
particular ``core'' diagram term, the \emph{only} modification to the
$M$-dependence of that term produced by the addition of a new ``6''-path,
is by multiplication by a factor $-\frac{(M-1)}{M}$ or $\frac{1}{M}$, as
appropriate.

Furthermore, due to the absence of any path-ordered phase factor 
associated with any ``6''-path, we can do the configuration-space sum over
paths for any ``6''-path \emph{immediately}, to obtain simply a 
\emph{free} Landau-gauge vector boson propagator or a \emph{free}
Fadeev-Popov propagator, as appropriate.

We must, however, be careful to note that, when we identify the 
individual $\tilde{W}_i$'s, (which are the ``inseparable units'' with 
respect to which the correlation function
$[\tilde{W}_1\ldots\tilde{W}_s]_{(\mathrm{SU}(N))^M}$ is expressed in terms of
$(\textrm{SU}(N))^M$ vacuum expectation values by the inverse of (\ref{E55})), as
the connected parts of any decoration of our initial $n$ $\textrm{SU}(NM)$
fundamental representation Wilson loops, (as is required by the general
derivation of the group-changing equations given in Section 2), we must
consider \emph{all} the new paths, i.e. all the ``45'' paths \emph{and}
all the ``6'' paths, in case two or more parts of the decoration, which
would appear to be disconnected from one another if we only look at the 
``45'' paths, (i.e. if we only look at the ``core'' decoration, which
contains all the ``45'' paths but none of the ``6'' paths), are in fact
connected to one another by one or more ``6'' paths in the full
decoration.

We should also note that when the full decoration contains some ``6''
paths, there will be cases when the Fadeev-Popov paths in the ``core''
decoration do not form closed lops, due to some of the paths in the 
Fadeev-Popov loops in the full decoration being ``6'' paths.

We also note here that, due to the $\left(1-\delta_{JK}\right)$ factor in 
the 45-term in (\ref{E67}), any \emph{``core''} decoration whose number of
connected components can be increased by the deletion of just \emph{one}
``45''-path, gives no contribution at all, since no matter what routeing
of the indices through the vertices is chosen, the two $\textrm{SU}(N)$
fundamental representation path-ordered phase factors that run (in
opposite directions) along that ``key'' ``45''-path, must be in the 
\emph{same} $\textrm{SU}(N)$ fundamental representation Wilson loop, (for
otherwise, when the key ``45''-path was deleted, any two points of the 
``core'' diagram that were previously connected to one another via a
sequence of paths that included the ``key'' ``45''-path, would remain
connected to one another by an alternative sequence of paths that 
replaced the ``key'' ``45''-path in the original sequence of paths by a
sequence of paths that traced the unbroken remainder part of either of the
two different $\textrm{SU}(N)$ fundamental representation Wilson loops that passed
along the ``key'' ``45''-path, so that deleting the ``key'' ``45''-path
would \emph{not} increase the number of connected components of the 
``core'' decoration, contrary to assumption), hence for every routeing of 
the indices through the vertices, the $\left(1-\delta_{JK}\right)$ factor
for the ``key'' ``45''-path, vanishes.

Thus, in summary, we see that our previous conclusion, namely that the
right-hand sides of the group-changing equations for $(\textrm{SU}(N))^M$ and
$\textrm{SU}(NM)$, applied to correlation functions, in the Yang Mills theory for
$\textrm{SU}(NM)$, of products of Wilson loops in the fundamental representation of
$\textrm{SU}(NM)$, reduce to sums of products of correlation functions, in the 
Yang Mills theory for $\textrm{SU}(N)$, of products of Wilson loops in the
fundamental representation of $\textrm{SU}(N)$, and furthermore, that the $\textrm{SU}(N)$
fundamental representation Wilson loops that occur in any individual
$\textrm{SU}(N)$ correlation function that occurs in the right-hand side, have, in
the absence of any accidental intersections or touchings associated with
particular configurations in position space of the paths and vertices in
the $\tilde{W}_i$'s, \emph{no} intersections or touchings, other than the
possible ``occasional'' simple intersections arising from quartic vertices
as described before, (these conclusions being proved previously only for
the case where we choose the 45-term in (\ref{E67}) for every new path),
remain completely valid in the general case, where both the 45-term in
(\ref{E67}) and the 6-term in (\ref{E67}) are allowed.

Furthermore, the entire $M$-dependence of the right-hand sides is given by
simple coefficients, where the coefficient that applies for a given 
routeing of the indices through the vertices of the ``core'' decoration,
(i.e. the decoration with all the ``6''-paths removed), and a given
partition $P$ of the set of all the $\textrm{SU}(N)$ fundamental representation
Wilson loops that result from that routeing, is given by
	\[M(M-1)(M-2)\ldots(M-r+1)=\frac{M!}{(M-r)!}
\]
where $r$ is the number of parts of the partition $P$, and there is in 
addition, for each ``6''-path, a factor of either $-\frac{(M-1)}{M}$ or
$\frac{1}{M}$, where the factor for a given ``6''-path is
$-\frac{(M-1)}{M}$ if both ends of that ``6''-path ``attach'' to the 
\emph{same} $\textrm{SU}(N)$ fundamental representation Wilson loop, or to two 
different $\textrm{SU}(N)$ fundamental representation Wilson loops that are members
of the same part of the partition $P$, and is $\frac{1}{M}$ if each end of
that ``6''-path ``attaches'' to a \emph{different} $\textrm{SU}(N)$ fundamental
representation Wilson loop, and those two $\textrm{SU}(N)$ fundamental 
representation Wilson loops are members of \emph{different} parts of the
partition $P$, where we note that for every ``6''-path end that is 
\emph{not} simply on one of the $n$ original $\textrm{SU}(NM)$ fundamental
representation Wilson loops, we also have to sum over the two terms in the
right-hand side of (\ref{E63}), (in the case where exactly one of $A$, 
$B$, and $C$ is a ``6'', and the other two are ``45's''), which 
corresponds, due to the $\delta_{JK}\delta_{jk}$ index structure at the 
$JjKk$ ``end'' of the 6-term in (\ref{E67}), to summing over two 
alternative choices of which of two alternative possible $\textrm{SU}(N)$
fundamental representation path-ordered phase factor ``lines'' that 
``6''-path ``attaches'' to.

And we note, furthermore, that due to the absence of any path-ordered
phase factor associated with any ``6''-path, we can do the 
configuration-space sum over paths for any ``6''-path
\emph{immediately}, to obtain simply a \emph{free} Landau-gauge vector
boson propagator or a \emph{free} Fadeev-Popov propagator, as
appropriate.

\section{Closed equations for the expansion coefficients in the $\frac{1}{N}$ expansions of vacuum expectation values and correlation functions}

We are now ready to apply the group-changing equations for $SU\!(N))^M$ and
$SU\!(N\!M)$ to obtain a complete and closed set of equations among the 
expansion coefficients $f_r(W_1,\ldots,W_n,g^2)$ in (\ref{E54}).

Now if we were to take the Yang Mills action density for $\textrm{SU}(NM)$ to be
	\[\frac{NM}{4g^2}F_{\mu\nu\alpha}F_{\mu\nu\alpha}
\]
(where $\alpha$ runs over all the generators of $\textrm{SU}(NM)$), then the 
expansions for the Wilson loop correlation functions of $\textrm{SU}(NM)$ would be
obtained from (\ref{E54}) simply by replacing $N$ by $NM$.  But the 
group-changing equations apply when the external factor of the action
density is the \emph{same} for both the ``large group'' $H$ and the 
``small group'' $G$, hence we take the action density as
	\[\frac{N}{4g^2}F_{\mu\nu\alpha}F_{\mu\nu\alpha}
\]
for \emph{both} $H=\textrm{SU}(NM)$ and $G=(\textrm{SU}(N))^M$, with the only difference
being that the sum on $\alpha$ runs over the generators of $\textrm{SU}(NM)$ for
$\textrm{SU}(NM)$, and over the generators of $(\textrm{SU}(N))^M$ for $\textrm{SU}(N))^M$.

Hence the $\textrm{SU}(NM)$ fundamental representation correlation functions, in 
the Yang Mills theory for $\textrm{SU}(NM)$, are given in terms of the expansion
coefficients \\$f_r(W_1,\ldots,W_n,g^2)$ in (\ref{E54}), by:
	\[(NM)^{2-n}\left(f_0(W_1,\ldots,W_n,g^2M)+\frac{1}{(NM)^2}f_1(W_1,
	\ldots,W_n,g^2M)\ +\right.
\]
\begin{equation}
	\label{E68}
	\left.+\ \frac{1}{(NM)^4}f_2(W_1,\ldots,W_n,g^2M)+\ldots\right)
\end{equation}
where the replacement of the argument $g^2$ of the $f_r$'s by $g^2M$ is
due to having the external factor $\frac{N}{4g^2}$ rather than the 
external factor $\frac{NM}{4g^2}$ in the Yang Mills action density for
$\textrm{SU}(NM)$.

We now proceed to derive closed equations for the expansion coefficients
in (\ref{E54}) and (\ref{E68}), by substituting the expansions (\ref{E68})
into the left-hand sides, and the expansions (\ref{E54}) into the 
right-hand sides, of the group-changing equations for $(\textrm{SU}(N))^M$ and
$\textrm{SU}(NM)$, applied to the correlation functions, in the Yang Mills theory
for $\textrm{SU}(NM)$, of $n$ Wilson loops $W_1,\ldots,W_n$, each in the 
fundamental representation of $\textrm{SU}(NM)$

We need a Lemma which expresses the correlation functions of $s$ 
``bunches'' of quantities, (where in practice each ``bunch'' of quantities
will be the product of all the $\textrm{SU}(N)$ fundamental representation Wilson
loops, in various $\textrm{SU}(N)$ subgroups of $(\textrm{SU}(N))^M$, that occur in a 
particular connected component $\tilde{W}_i$ of some decoration of our
initial $n$ Wilson loops, for a particular routeing of the $(\textrm{SU}(N))^M$
indices through the vertices of that connected component of that 
decoration), in terms of the correlation functions of the individual
quantities.

We recall \cite{BPHZ} that a partition is a set $P$ such that every 
member of $P$ is a set, no member of $P$ is empty, and if $i$ and $j$ are
any two distinct members of $P$, then $i\cap j=\emptyset$, where
$\emptyset$ is the empty set.

For any finite set $X$, we define $\mathbf{P}(X)$ to be the set whose
members are all the partitions of $X$, (so that, in other words,
$\mathbf{P}(X)$ is the set whose members are all the partitions $P$ such
that the union of all the members of $P$ is equal to $X$), and we define
$\mathbf{N}(X)$ to be equal to the number of members of $X$.

And we also recall \cite{BPHZ} that if $X$ is a set, and $Y$ is a set
whose members are sets, then we say that $X$ is $Y$-connected if and only
if, for every partition of $X$ into two nonempty parts, there exists a 
member of $Y$ that has nonempty intersection with both those parts, and we
also recall that a $Y$-connected component of $X$ is a nonempty
$Y$-connected subset of $X$ that is \emph{not} a strict subset of any
$Y$-connected subset of $X$.  (We also recall our convention that the
statement, ``$Z$ is a subset of $X$,'' includes the possibility that 
$Z=X$, and that we say, ``Z is a strict subset of X,'' if we wish to 
exclude the possibility that $Z=X$.)

And we also recall \cite{BPHZ} that if $Y$ is a set whose members are 
sets, then we define $\mathcal{U}(Y)$ to be the union of all the members
of $Y$.

\subsection{A Lemma about correlation functions}

\emph{Lemma}  Let $K$ be a finite set of indices, (for example, the 
integers 1 to $k$), and let $\{W_i\vert i\in K\}$ be a set of quantities
indexed by $K$, (for example, $W_1,\ldots,W_k$ could be a set of $k$
Wilson loops.)  Let $S$ be a partition of $K$.

For each member $i$ of $S$ we define:
\begin{equation}
	\label{E69}
	B_i=\prod_{j\in i}W_j
\end{equation}

Then the correlation function of the $B_i$'s, in which each $B_i$ is
treated as an ``indivisible unit'', namely:
  \[\left[\prod_{i\in S}B_i\right]=\sum_{P\in\mathbf{P}(S)}(-1)^{(\mathbf
  {N}(P)-1)}(\mathbf{N}(P)-1)!\prod_{i\in P}\left\langle \prod_{j\in i}B_j
\right\rangle
\]
	\[=\sum_{P\in\mathbf{P}(S)}(-1)^{(\mathbf
  {N}(P)-1)}(\mathbf{N}(P)-1)!\prod_{i\in P}\left\langle \prod_{j\in i}
  \prod_{l\in j}W_l\right\rangle
\]
\begin{equation}
	\label{E70}
	=\sum_{P\in\mathbf{P}(S)}(-1)^{(\mathbf
  {N}(P)-1)}(\mathbf{N}(P)-1)!\prod_{i\in P}\left\langle
  \prod_{l\in \mathcal{U}(i)}W_l\right\rangle
\end{equation}
is equal to a sum:
\begin{equation}
	\label{E71}
	\sum_R\prod_{i\in R}\left[\prod_{j\in i}W_j\right]
\end{equation}
where the sum runs over all partitions $R$ of $K$ such that $K$ is 
$(R\cup S)$-connected.

\emph{Proof.}  We substitute into the right-hand side of (\ref{E70}) the
general identities (\ref{E55}), which in our present notation, are 
expressed as:
\begin{equation}
	\label{E72}
	\left\langle \prod_{i\in N}W_i\right\rangle=\sum_{Q\in\mathbf{P}(N)}
	\prod_{i\in Q}\left[\prod_{j\in i}W_j\right]
\end{equation}
for a general finite set $N$, to obtain the correlation function of the 
$B_i$'s in terms of the correlation functions of the $W_j$'s as:
\begin{equation}
	\label{E73}
	\sum_{P\in\mathbf{P}(S)}(-1)^{\mathbf{N}(P)-1)}(\mathbf{N}(P)-1)!
	\prod_{i\in P}\sum_{Q\in \mathbf{P}(\mathcal{U}(i))}\prod_{j\in Q}
	\left[\prod_{l\in j}W_l\right]
\end{equation}

This has the general form of (\ref{E71}), namely a sum of products of 
correlation functions of the $W_j$'s, with each individual product
corresponding to a partition of $K$, (in the sense that the individual
correlation-function factors in that product are in one-to-one
correspondence with the parts of that partition of $K$), and the partition
$R$ of $K$ that corresponds to a particular partition $P$ of $S$, and
particular partitions $Q_i$, (one for each part $i$ of $P$), of the unions
of the parts of $P$, being given by:
\begin{equation}
	\label{E74}
	R=\bigcup_{i\in P}Q_i
\end{equation}

It remains to determine the net coefficient of the product corresponding
to each partition $R$ of $K$, and check that it is equal to 1 if $K$ is
$R\cup S$-connected, and equal to 0 otherwise.

We begin by asking, for a general partition $R$ of $K$, what partitions
$P$ of $S$ can produce $R$ by (\ref{E74}), for suitable partitions 
$Q_i$ of the unions of the parts $i$ of $P$, and we see immediately that a
partition $P$ of $S$ is able to produce $R$ in this way if and only if, 
for every member $i$ of $P$ and every member $j$ of $R$, j is either a
\emph{subset} of $\mathcal{U}(i)$, or else $j\cap\mathcal{U}(i)$ is empty,
or in other words, a partition $P$ of $S$ is able to produce $R$ in this
way if and only if, for every member $j$ of $R$, $j$ intersects the union
of \emph{exactly one} member $i$ of $P$, (and hence is a subset of the 
union of that member $i$ of $P$).  And furthermore, if a partition $P$ of
$S$ \emph{is} able to produce $R$ in this way, then there is 
\emph{exactly one} partition $Q_i$ of the union of each member $i$ of $P$
that produces $R$ in this way from $P$: namely, for each member $i$ of 
$P$, $Q_i$ is the set of all the members of $R$ that are subsets of 
$\mathcal{U}(i)$.

Thus for a general partition $R$ of $K$, the net coefficient in 
(\ref{E73}) of the product corresponding to $R$, is equal to a sum:
\begin{equation}
	\label{E75}
	\sum_P(-1)^{(\mathbf{N}(P)-1)}(\mathbf{N}(P)-1)!
\end{equation}
where the sum runs over all partitions $P$ of $S$ such that each member of
$R$ intersects the union of exactly one member of $P$, (and hence is a 
subset of the union of that member of $P$).

We next note that the set of all partitions $P$ of $S$ such that each
member of $R$ intersects the union of exactly one member of $P$, (and 
hence is a subset of the union of that member of $P$), is in one-to-one
correspondence with the set of all partitions $T$ of the set of all the
$R\cup S$-connected components of $K$, and that furthermore, each such
partition $P$ of $S$ has the same number of members as the corresponding
partition $T$ of the set of all the $R\cup S$-connected components of $K$.
For if $P$ is any such partition of $S$, and $j$ is any 
$R\cup S$-connected component of $K$, then $j$ is certainly a subset of 
the union of some member of $P$, since every member of $S$ is a subset of
the union of some member of $P$, and every member of $R$ is a subset of the
union of some member of $P$.  (We recall \cite{BPHZ} that the set of all
the $R\cup S$-connected components of $K$, is itself a partition of $K$.)
And in fact, if $P$ is any such partition of $S$, and we define the 
corresponding partition $U$ \emph{of} $K$ to be the set whose members are
the unions $\mathcal{U}(i)$ of the members $i$ of $P$, then the 
corresponding partition $T$ of the set of all the $R\cup S$-connected
components of $K$, is the set whose members are in one-to-one 
correspondence with the members of $U$, and such that the member of $T$
that corresponds to a given member $m$ of $U$, is the set whose members 
are all the $R\cup S$-connected components of $K$ that are subsets of $m$,
(where we note that there is at least one such $R\cup S$-connected
component of $K$, since $m$ is nonempty, and the set of all the
$R\cup S$-connected components of $K$ is a partition of $K$ such that each
individual $R\cup S$-connected component of $K$ is either a subset of $m$
or else does not intersect $m$).  And given any partition $T$ of the set of
all the $R\cup S$-connected components of $K$, we immediately obtain the
corresponding ``allowed'' partition $P$ of $S$ by reversing this 
construction, and we note that each of the three corresponding partitions
$P$, $U$, and $T$ has the same number of members.  Hence for any partition
$R$ of $K$, the net coefficient in (\ref{E73}) of the product of 
correlation functions of the $W_j$'s that corresponds to $R$, is given by:
\begin{equation}
	\label{E76}
	\sum_{T\in\mathbf{P}(V)}(-1)^{(\mathbf{N}(T)-1)}(\mathbf{N}(T)-1)!
\end{equation}
where we have defined $V$ to be the set of all the $R\cup S$-connected
components of $K$.

Now the sum (\ref{E76}) is well known to have the value 1 if $\mathbf{N}
(V)=1$, and the value 0 if $\mathbf{N}(V)\geq2$, and indeed, this very
same sum, with $V$ replaced by a general finite set $N$, is exactly what
occurs in the verification fo the standard inversion formula:
\begin{equation}
	\label{E77}
	\left[\prod_{i\in N}W_i\right]=\sum_{Q\in\mathbf{P}(N)}(-1)^{(\mathbf{N}
	(Q)-1)}(\mathbf{N}(Q)-1)!\prod_{i\in Q}\left\langle \prod_{j\in i}W_j
	\right\rangle
\end{equation}
of the standard expression (\ref{E72}) for the vacuum expectation values in
terms of the correlation functions,(i.e. when (\ref{E72}) is substituted
into the right-hand side of (\ref{E77})).

This concludes the proof of the Lemma, for we have now shown that the
coefficient in (\ref{E72}) of the product of correlation functions of the
$W_j$'s that corresponds to a partition $R$ of $K$, is equal to 1 if $R$ is
such that $K$ has exactly \emph{one} $R\cup S$-connected component, (or in
other words, if $R$ is such that $K$ is $R\cup S$-connected), and equal to 
0 if $R$ is such that $K$ has \emph{more} than one $R\cup S$-connected
component.

For completeness, we note here that the fact that (\ref{E76}) is equal to
1 if $\mathbf{N}(V)=1$, and equal to 0 if $\mathbf{N}(V)\geq2$, may be
proved by classifying the partitions $T$ of $V$ by their numbers $m_q$,
$q\geq1$, of $q$-membered parts, subject to the constraint that:
\begin{equation}
	\label{E78}
	m_1+2m_2+3m_3+\ldots=\mathbf{N}(V)
\end{equation}
there being:
\begin{equation}
	\label{E79}
	(\mathbf{N}(V))!\prod_{q\geq1}\frac{1}{(q!)^{m_q}m_q!}
\end{equation}
such partitions $T$ of $V$.  It is then plain that the value of 
(\ref{E76}) depends only on $\mathbf{N}(V)\equiv v$.  We multiply 
(\ref{E76}) by $\frac{\lambda^v}{v!}$ and sum from $v=0$ to $\infty$ to
get:
	\[\sum_{\begin{array}{c}m_1\geq0\\m_2\geq0\\ \ldots\end{array}}(-1)
	\left(\left(m_1+m_2+m_3+\ldots\right)-1\right)!\prod_{q\geq1}
	\frac{1}{m_q!}\left(\frac{-\lambda^q}{q!}\right)^{m_q}\qquad=
\]
	\[=\ \sum_{z=0}^\infty\frac{(-1)}{z}\!\!\!\!\!\!\!\!\!\!\!\!\!\!\!\!
	\!\!\!\!\!\!\!\!\!\!\!\!\!\!\!\!\!\!\!\!\!\!\!\!
	\sum_{\qquad\qquad\qquad\left.\begin{array}{c}m_1\geq0\\
	m_2\geq0\\ \ldots\end{array}\right|m_1+m_2+m_3+\ldots=z}\!\!\!\!\!\!\!\!
	\!\!\!\!\!\!\!\!\!\!\!\!\!\!\!\!\!\!\!\!\!\!\!\!\!
	z!\prod_{q\geq1}
	\frac{1}{m_q!}\left(\frac{-\lambda^q}{q!}\right)^{m_q}=
	\ \sum_{z=0}^\infty\frac{(-1)}{z}\left(\sum_{q=1}^\infty\left(
	\frac{-\lambda^q}{q!}\right)\right)^z=
\]
\begin{equation}
	\label{E80}
	=\sum_{z=0}^\infty\frac{(-1)}{z}\left(1-e^\lambda\right)^z=\ln\left(1-
	\left(1-e^\lambda\right)\right)=\lambda
\end{equation}

\subsection[Substitution of the $\frac{1}{N}$ expansions into the Group-Changing Equations for $\textrm{SU}(NM)$ and
$(\textrm{SU}(N))^M$]{Substitution of the $\frac{1}{N}$ expansions into the \\Group-Changing Equations for $\textrm{SU}(NM)$ and $(\textrm{SU}(N))^M$}

We now substitute the expansions (\ref{E68}) into the left-hand sides, and
the expansions (\ref{E54}) into the right-hand sides, of the 
group-changing equations for $(\textrm{SU}(N))^M$ and $\textrm{SU}(NM)$, applied to the 
correlation functions,in the Yang Mills theory for $\textrm{SU}(NM)$, of $n$ Wilson
loops $W_1,\ldots,W_n$, each in the fundamental representation of 
$\textrm{SU}(NM)$, and equate coefficients of powers of $N$, in order to derive
closed equations for the $f_r$'s in (\ref{E54}) and (\ref{E68}), bearing
in mind the reduction of the right-hadn sides of the group-changing
equations for this case, to sums of products of correlation functions of
$\textrm{SU}(N)$ fundamental representation Wilson loops, in various $\textrm{SU}(N)$
subgroups of $(\textrm{SU}(N))^M$, as derived in the foregoing.

We adopt the following procedure.  We write out the right-hand side of the
group-changing equation for $[W_1\ldots W_n]_{\mathrm{SU}(NM)}$ as a sum of 
decorations of the $n$ closed paths defining the $n$ Wilson loops 
$W_1,\ldots,W_n$, as given by the general group-changing equations as
derived in Section 2.  We consider a particular term $[\tilde{W}_1\ldots
\tilde{W}_s]_{(\mathrm{SU}(N))^M}$ in this sum over decorations, where $\tilde{W}
_1,\ldots,\tilde{W}_s$ are the connected components of this particular
decoration of $W_1,\ldots,W_n$, (and some of the $\tilde{W}_i$'s may be
vacuum bubbles).  We also consider a specific assignment of which of the
``new paths'' of the decorations $\tilde{W}_1,\ldots\tilde{W}_s$, (i.e.
paths that are not wholes or parts of any of the $n$ original closed
paths), are gauge-field paths, (to be summed over with kinematic weights
given by (\ref{E16}), (\ref{E18}), (\ref{E20}), and (\ref{E27}), and which
are Fadeev-Popov paths, (to be summed over with kinematic weights given by
(\ref{E24}) and (\ref{E27})), consistent with the ``action vertices'' as
given by (\ref{E31}), and the second and third terms in (\ref{E23}), in
consequence of which the Fadeev-Popov paths form closed loops, which 
involve new paths and action vertices only.  We note that, as usual, there
is a factor (-1) for each closed loop of Fadeev-Popov paths, corresponding
to the Fermi statistics of the Fadeev-Popov fields.  We also note that the
gauge-covariant derivatives in the second term in (\ref{E23}) and the first
term in (\ref{E31}) result in the modification of the kinematic weights of
the appropriate paths by the addition of an extra segment to each term in
(\ref{E27}), with a weight fro this segment as given by(\ref{E29}).  We
also note that, as mentioned between equations (\ref{E62}) and (\ref{E63}),
at each vertex where a new path ends at one of the $n$ original closed
paths, we have either a $t_4$, a $t_5$, or a $t_6$, and we note that any
new path that ends at one of the $n$ original closed paths, is a 
gauge-field path, not a Fadeev-Popov path, (since the definition 
(\ref{E1}) of a path-ordered phase factor involve no Fadeev-Popov fields).

We call a path in a decoration an ``original'' path if it is the whole, or 
a part, of any of the $n$ original closed paths, and a ``new'' path if it
is not an original path, or in other words, if it is a gauge-field path or
a Fadeev-Popov path.  And we call a vertex in a decoration an ``action''
vertex if it arises from a term in (\ref{E31}) or from the second or third
term in (\ref{E23}), and an ``original-path vertex'' if it is a vertex
where a gauge-field path ends at one of the $n$ original closed loops.

We also consider a specific assignment of which of the new paths are
``45''-paths and which are ``6''-paths, subject to the selection rules,
derived in the foregoing, that at most one 6-path can end at any cubic
action vertex, and at most two 6-paths can end at any quartic action 
vertex, and that it must not be possible to increase the number of
connected components of the ``core'' part of any $\tilde{W}_i$, (i.e. of 
that $\tilde{W}_i$ with all its 6-paths removed), by removing just
\emph{one} 45-path.  (We note that it \emph{is} possible, however, for the
core part of a $\tilde{W}_i$ to have more than one connected component.)

We express all the structure constants in the action vertices, (i.e. in
(\ref{E31}), and the second and third term in (\ref{E23})), in terms of
traces of the $t_i$'s ($1\leq i\leq 6$), by (\ref{E62}) and (\ref{E63}), 
and consider a specific choice, independently for each structure constant
in each vertex, of one of the two terms in (\ref{E62}) or (\ref{E63}).

We also consider separately, at each quartic action vertex at which four
gauge-field 45-paths meet, the term
	\[\frac{1}{4g^2}f_{aBC}f_{aEF}A_{\mu B}A_{\nu C}A_{\mu E}A_{\nu F}
\]
in (\ref{E31}), (which corresponds to $a$ being summed over the 1's, the
2's, and the 3's), and the part of the
	\[\frac{1}{4g^2}f_{ABC}f_{AEF}A_{\mu B}A_{\nu C}A_{\mu E}A_{\nu F}
\]
term in (\ref{E31}) where $A$ is summed over the 4's and the 5's, and the
part of the 
	\[\frac{1}{4g^2}f_{ABC}f_{AEF}A_{\mu B}A_{\nu C}A_{\mu E}A_{\nu F}
\]
term in (\ref{E31}) where $A$ is summed over the 6's.

We note that at any quartic action vertex where one or two 6-paths end,
all four new paths ending at that vertex must be gauge-field paths, since
in the 
	\[\psi_Af_{AaB}A_{\mu B}\phi_Cf_{CaD}A_{\mu D}
\]
in (\ref{E23}), the sum on $a$ runs only over the 1's, the 2's, and the 
3's, and $f_{AaB}$ and $f_{CaD}$ vanish when any of their upper-case 
indices is a ``6'', and furthermore the only term in (\ref{E31}) that
contributes in these cases is the 
	\[f_{ABC}f_{AEF}A_{\mu B}A_{\nu C}A_{\mu E}A_{\nu F}
\]
term, with the sum on $A$ running just over the 4's and the 5's, Since
$f_{ABC}$ vanishes when two or more of its indices are 6's, and likewise
for $f_{AEF}$.

We use (\ref{E58}), (\ref{E59}), (\ref{E60}), and (\ref{E67}), to express
the particular term we are considering, as a sum of a finite number of 
terms, each of which is a product of $\textrm{SU}(N)$ fundamental representation
Wilson loops, with the individual Wilson loops being in various $\textrm{SU}(N)$
subgroups of $(\textrm{SU}(N))^M$, and where the sum is over the distinct allowed
assignments of the individual Wilson loops to the different $\textrm{SU}(N)$
subgroups of $(\textrm{SU}(N))^M$, subject to the constraint, (due to the $\left(
1-\delta_{JK}\right)$ factor in the 45-term in (\ref{E67})), that any two
Wilson loops that share a common 45-path, must be in \emph{different}
$\textrm{SU}(N)$ subgroups of $(\textrm{SU}(N))^M$.  (There is also an analogous constraint,
coming from the $\left(1-\delta_{JK}\right)$ factor in (\ref{E59}), at
each quartic vertex at which four gauge-field 45-pats meet, and at which
we have chosen the $f_{ABC}f_{AEF}A_{\mu B}A_{\nu C}A_{\mu E}A_{\nu F}$
term in (\ref{E31}), with $A$ being summed over the 4's and the 5's.)

\noindent $\ \ \ $We note that, at any quartic vertex that corresponds to
 the 
$f_{aBC}f_{aEF}A_{\mu B}A_{\nu C}A_{\mu E}A_{\nu F}$ term in (\ref{E31}) or
to the $\psi_Af_{AaB}A_{\mu B}\phi_Cf_{CaD}A_{\mu D}$ term in (\ref{E23}),
two of the four $\textrm{SU}(N)$ fundamental representation path-ordered phase
factors that pass through the vertex are forced, by (\ref{E58}), to be in
the \emph{same} $\textrm{SU}(N)$ subgroup of $(\textrm{SU}(N))^M$.

(We note that the $\delta_{JA}\delta_{KA}\delta_{PA}\delta_{QA}$ factor in
(\ref{E58}), when summed over $A$, gives $\delta_{JQ}\delta_{PK}\delta
_{JK}$, which is also equal to $\delta_{JK}\delta_{PQ}\delta_{JQ}$, and 
that the summation convention is of course \emph{not} to be applied to $J$,
$K$, or $Q$ here.)  We also consider separately, at every such quartic
vertex, the $-\delta_{JA}\delta_{KA}\delta_{PA}\delta_{QA}\delta_{jq}
\delta_{kp}$ and the $\frac{1}{N}\delta_{JA}
\delta_{KA}\delta_{PA}\delta_{QA}\delta_{jk}\delta_{pq}$ term in the 
right-hand side of (\ref{E58}), (making an independent choice of one or the
other of these two terms at every such quartic vertex).

Now our choice, independently at each cubic action vertex, of one of the
two terms in (\ref{E63}), and independently at each structure constant in
each quartic action vertex, of one of the two terms in (\ref{E62}) or in
(\ref{E63}), as appropriate, and independently, at each quartic vertex at
which four gauge-field 45-paths meet, of either the ``123-terms'', the
``45-terms'', or the ``6-terms'' in $(f_{aBC}f_{aEF}+f_{ABC}f_{AEF})$ in
(\ref{E31}), and independently, at each ``123-type'' quartic vertex, (i.e.
the $f_{aBC}f_{aEF}$ term in (\ref{E31}) or the quartic interaction term in
(\ref{E23})), of either the $-\delta_{jq}\delta_{kp}$ term or the 
$\frac{1}{N}\delta_{jk}\delta_{pq}$ term in (\ref{E58}), corresponds to 
making a specific choice, independently at each vertex, of the routeings of
all the $\textrm{SU}(N)$ fundamental representation path-ordered phase factors that
pass through that vertex, and hence to a specific choice of the way that 
the pairs of (oppositely-directed) $\textrm{SU}(N)$ fundamental representation
path-ordered phase factors that run alon the 45-path, and the single
$\textrm{SU}(N)$ fundamental representation path-ordered phase factors that run 
along each original path, are connected up into $\textrm{SU}(N)$ fundamental
representation Wilson loops, and it also corresponds, wherever a 6-path
ends at an action vertex,(and hence at a structure constant), rather than
at one of the $n$ original closed paths, to making a specific choice of 
which of the two $\textrm{SU}(N)$ fundamental representation path-ordered phase
factors that ``pass through'' that structure constant, (when (\ref{E67})
or (\ref{E59}) is applied to each of the other two indices of that
structure constant, each of which must be either a ``4'' or a ``5''), that
6-path end ``attaches to'', in the sense that the two upper-case indices at
that end of that 6-path, are both equal to the integer $A$, $1\leq A\leq
M$, which specifies to which of the $M$ $\textrm{SU}(N)$ subgroups of $(\textrm{SU}(N))^M$,
the $\textrm{SU}(N)$ fundamental representation path-ordered phase factor to which
that end of that 6-path ``attaches'', belongs.  (For when that structure
constant is expressed, by (\ref{E63}), as:
\begin{equation}
	\label{E81}
	f_{ABC}=-\left(\left(t_A\right)_{LlJj}\left(t_B\right)_{JjKk}\left(t_C
	\right)_{KkLl}-\left(t_C\right)_{LlJj}\left(t_B\right)_{JjKk}\left(t_A
	\right)_{KkLl}\right)
\end{equation}
we find, if $B$ is the ``6'', and $A$ and $C$ are the 4's and 5's, that 
when we contract with $\left(t_B\right)_{PpQq}$, say, with $B$ being
summed over all the ``6's'', and use (\ref{E67}), remembering that the 
6-term in the right-hand side of (\ref{E67}) is the contribution of all the
terms in the left-hand side that involve one or more $t_6$'s, and that the
only nonvanishing such term in the left hand side is precisely $\delta
_{AB}\left(t_A\right)_{JjKk}\left(t_B\right)_{PpQq}$, with $A$ and $B$
being summed over all the ``6's'', we get:
\begin{equation}
	\label{E82}
	-\left(\left(t_A\right)_{LlJj}\left(t_C\right)_{KkLl}-\left(t_C\right)
	_{LlJj}\left(t_A\right)_{KkLl}\right)\frac{1}{N}\delta_{JK}\delta_{jk}
	\delta_{PQ}\delta_{pq}\left(-\delta_{JQ}+\frac{1}{M}\right)
\end{equation}
where $J$, $K$, and $L$ are to be summed from 1 to $M$, and $j$, $k$, and
$l$ are to be summed from 1 to $N$, and this is equal to:
\begin{equation}
	\label{E83}
	-\left(t_A\right)_{LlJj}\left(t_C\right)_{JjLl}\frac{1}{N}\delta_{PQ}
	\delta_{pq}\left(\left(-\delta_{JQ}+\frac{1}{M}\right)-\left(-\delta_
	{LQ}+\frac{1}{M}\right)\right)
\end{equation}
where $J$ and $L$ are to be summed from 1 to $M$, and $j$ and $l$ are to be
summed from 1 to $N$.  The stated result follows from this if we note that
if we multiply by $W((\textrm{SU}(N))^M,x(s))_{EA}\left(t_E\right)_{GgHh}$, say,
 sum $E$ and $A$ over the 4's and 5's, and use the ``45 part'' of 
(\ref{E67}), then $J$ and $L$, (which become equal to $G$ and $H$,
respectively, by the $\delta_{GJ}$ and $\delta_{LH}$ factors in the 45-term
in the form of (\ref{E67}) with the appropriate indices), identify the two
(distinct) $\textrm{SU}(N)$ subgroups of $(\textrm{SU}(N))^M$ to which the gauge fields
occurring in the two $\textrm{SU}(N)$ fundamental representation path-ordered phase
factors which run, in opposite directions, along the path $x(s)$, belong.)

\chapter{The Group-Variation Equations for the $\textrm{SU}(N)$ Groups}

\section{The Group-Variation Equations to all orders in $\frac{1}{N}$}

\subsection{Diagram notation}

We introduce the following diagram notation for identifying individual
terms of the type we have described.  45-paths will be denoted by a pair
of parallel lines, representing the two $\textrm{SU}(N)$ fundamental representation
path-ordered phase factors which run, in opposite directions, along them.
We adopt the rule, ``drive on the left'', for specifying the directions of
the two phase factors, and arrows will only be put on the lines when their
directions cannot be unambiguously identified by this rule.  6-paths will
be indicated by single dashed lines.  Original paths, (i.e. paths that
form parts or wholes of any of the $n$ original closed paths), will be
indicated by single solid lines.  The main purpose of the diagrammatic
notation is to display the routeings of the $\textrm{SU}(N)$ fundamental
representation path-ordered phase factors through the vertices.  We also
indicate, wherever a 6-path ends at an action vertex, (and hence at a
structure constant), which of the two $\textrm{SU}(N)$ fundamental representation
path-ordered phase factors passing through that structure constant, that
6-path end ``attaches to'', in the sense explained above, by ensuring that 
the broken line that represents the 6-path, ends with a dash that
terminates on the appropriate $\textrm{SU}(N)$ fundamental representation 
path-ordered phase factor.

The possible triple action vertices are:
\unitlength 1mm
\linethickness{0.61pt}
\begin{equation}
	\label{E84}
	\raisebox{-7mm}{
\begin{picture}(57.00,17.00)
\put(9.00,8.00){\line(1,0){10.00}}
\put(9.00,10.00){\line(1,0){10.00}}
\put(9.00,10.00){\line(-1,1){7.00}}
\put(1.00,15.00){\line(1,-1){6.00}}
\put(7.00,9.00){\line(-1,-1){6.00}}
\put(9.00,8.00){\line(-1,-1){7.00}}
\put(47.00,10.00){\line(-1,1){7.00}}
\put(47.00,8.00){\line(-1,-1){7.00}}
\put(39.00,3.00){\line(1,1){7.00}}
\put(39.00,15.00){\line(1,-1){7.00}}
\put(47.00,8.00){\line(0,1){2.00}}
\put(57.00,8.00){\line(-1,0){11.00}}
\put(57.00,10.00){\line(-1,0){11.00}}
\end{picture}
}
\end{equation}
and:
	\unitlength 1mm
\linethickness{0.61pt}
\begin{equation}
	\label{E85}
	\raisebox{-7mm}{
\begin{picture}(54.00,17.00)
\put(1.00,17.00){\line(0,-1){16.00}}
\put(3.00,17.00){\line(0,-1){16.00}}
\put(3.00,9.00){\line(1,0){2.00}}
\put(7.00,9.00){\line(1,0){2.00}}
\put(11.00,9.00){\line(1,0){2.00}}
\put(15.00,9.00){\line(1,0){2.00}}
\put(39.00,17.00){\line(0,-1){16.00}}
\put(41.00,17.00){\line(0,-1){16.00}}
\put(44.00,9.00){\line(1,0){2.00}}
\put(48.00,9.00){\line(1,0){2.00}}
\put(52.00,9.00){\line(1,0){2.00}}
\put(39.00,9.00){\line(1,0){3.00}}
\end{picture}
}
\end{equation}

There is always a relative minus sign between the two terms in 
(\ref{E84}), corresponding to the relative minus sign between the two terms
in (\ref{E63}), and there is similarly always a relative minus sign between
the two terms in (\ref{E85}), again corresponding to the relative minus 
sign between the two terms in (\ref{E63}).

The original-path vertices are:
\unitlength 1pt
\linethickness{0.61pt}
\begin{equation}
	\label{E86}
	\raisebox{-20pt}{
\begin{picture}(67.00,41.00)
\put(1.00,11.00){\line(1,0){30.00}}
\put(31.00,11.00){\line(0,1){30.00}}
\put(37.00,41.00){\line(0,-1){30.00}}
\put(37.00,11.00){\line(1,0){30.00}}
\put(20.00,11.00){\line(-1,1){10.00}}
\put(10.00,1.00){\line(1,1){10.00}}
\put(56.00,11.00){\line(-1,1){10.00}}
\put(46.00,1.00){\line(1,1){10.00}}
\end{picture}
}\qquad\textrm{or}\qquad\raisebox{-20pt}{
\begin{picture}(71.00,41.00)
\put(62.00,1.00){\line(-1,1){10.00}}
\put(12.00,11.00){\line(1,1){10.00}}
\put(22.00,1.00){\line(-1,1){10.00}}
\put(52.00,11.00){\line(1,1){10.00}}
\put(33.00,41.00){\line(0,-1){20.00}}
\put(33.00,21.00){\line(1,-1){10.00}}
\put(39.00,41.00){\line(0,-1){20.00}}
\put(39.00,21.00){\line(-1,-1){10.00}}
\put(71.00,11.00){\line(-1,0){28.00}}
\put(1.00,11.00){\line(1,0){28.00}}
\end{picture}
}
\end{equation}
where the cross-over in the second form of (\ref{E86}) occurs simply in
order to follow the rule: ``drive on the left'', (and we note that the two
forms in (\ref{E86}) are two different ways of drawing the same thing,
\emph{not} two different terms to be added or subtracted from one
another), and :
	\unitlength 1.00pt
\linethickness{0.61pt}
\begin{equation}
	\label{E87}
	\raisebox{-20pt}{
\begin{picture}(61.00,46.00)
\put(1.00,11.00){\line(1,0){30.00}}
\put(31.00,11.00){\line(1,0){30.00}}
\put(20.00,11.00){\line(-1,1){10.00}}
\put(10.00,1.00){\line(1,1){10.00}}
\put(50.00,11.00){\line(-1,1){10.00}}
\put(40.00,1.00){\line(1,1){10.00}}
\put(31.00,11.00){\line(0,1){5.00}}
\put(31.00,21.00){\line(0,1){5.00}}
\put(31.00,31.00){\line(0,1){5.00}}
\put(31.00,41.00){\line(0,1){5.00}}
\end{picture}
}
\end{equation}

There are two distinct types of phase-factor routeing through quartic
action vertices at which four 45-paths end:\\
\emph{Type 1}
\unitlength 1.00pt
\linethickness{0.61pt}
\begin{equation}
	\label{E88}
	\raisebox{-15pt}{
\begin{picture}(43.00,43.00)
\put(7.00,43.00){\line(1,-1){12.00}}
\put(13.00,25.00){\line(-1,1){12.00}}
\put(25.00,31.00){\line(1,1){12.00}}
\put(43.00,37.00){\line(-1,-1){12.00}}
\put(19.00,31.00){\line(1,0){6.00}}
\put(31.00,25.00){\line(0,-1){6.00}}
\put(31.00,19.00){\line(1,-1){12.00}}
\put(37.00,1.00){\line(-1,1){12.00}}
\put(25.00,13.00){\line(-1,0){6.00}}
\put(19.00,13.00){\line(-1,-1){12.00}}
\put(1.00,7.00){\line(1,1){12.00}}
\put(13.00,19.00){\line(0,1){6.00}}
\end{picture}
}\ \quad\raisebox{-15pt}{
\begin{picture}(43.00,43.00)
\put(7.00,43.00){\line(1,-1){12.00}}
\put(13.00,25.00){\line(-1,1){12.00}}
\put(25.00,31.00){\line(1,1){12.00}}
\put(43.00,37.00){\line(-1,-1){12.00}}
\put(31.00,19.00){\line(1,-1){12.00}}
\put(37.00,1.00){\line(-1,1){12.00}}
\put(19.00,13.00){\line(-1,-1){12.00}}
\put(1.00,7.00){\line(1,1){12.00}}
\put(19.00,31.00){\line(0,-1){18.00}}
\put(25.00,13.00){\line(0,1){18.00}}
\put(13.00,25.00){\line(1,0){18.00}}
\put(31.00,19.00){\line(-1,0){18.00}}
\end{picture}
}\ \quad\raisebox{-15pt}{
\begin{picture}(43.00,43.00)
\put(43.00,37.00){\line(-1,-1){12.00}}
\put(31.00,19.00){\line(1,-1){12.00}}
\put(31.00,25.00){\line(0,-1){6.00}}
\put(7.00,1.00){\line(1,1){18.00}}
\put(25.00,19.00){\line(0,1){6.00}}
\put(25.00,25.00){\line(-1,1){18.00}}
\put(1.00,37.00){\line(1,-1){36.00}}
\put(1.00,7.00){\line(1,1){36.00}}
\end{picture}
}\ \quad\raisebox{-15pt}{
\begin{picture}(43.00,43.00)
\put(1.00,37.00){\line(1,-1){12.00}}
\put(13.00,19.00){\line(-1,-1){12.00}}
\put(13.00,25.00){\line(0,-1){6.00}}
\put(37.00,1.00){\line(-1,1){18.00}}
\put(19.00,19.00){\line(0,1){6.00}}
\put(19.00,25.00){\line(1,1){18.00}}
\put(43.00,37.00){\line(-1,-1){36.00}}
\put(43.00,7.00){\line(-1,1){36.00}}
\end{picture}
}\ \quad\raisebox{-15pt}{
\begin{picture}(43.00,43.00)
\put(7.00,1.00){\line(1,1){12.00}}
\put(25.00,13.00){\line(1,-1){12.00}}
\put(19.00,13.00){\line(1,0){6.00}}
\put(43.00,37.00){\line(-1,-1){18.00}}
\put(25.00,19.00){\line(-1,0){6.00}}
\put(19.00,19.00){\line(-1,1){18.00}}
\put(7.00,43.00){\line(1,-1){36.00}}
\put(37.00,43.00){\line(-1,-1){36.00}}
\end{picture}
}\ \quad\raisebox{-15pt}{
\begin{picture}(43.00,43.00)
\put(7.00,43.00){\line(1,-1){12.00}}
\put(25.00,31.00){\line(1,1){12.00}}
\put(19.00,31.00){\line(1,0){6.00}}
\put(43.00,7.00){\line(-1,1){18.00}}
\put(25.00,25.00){\line(-1,0){6.00}}
\put(19.00,25.00){\line(-1,-1){18.00}}
\put(7.00,1.00){\line(1,1){36.00}}
\put(37.00,1.00){\line(-1,1){36.00}}
\end{picture}
}
\end{equation}
\emph{Type 2}
	\unitlength 1.00pt
\linethickness{0.61pt}
\begin{equation}
	\label{E89}
	\raisebox{-15pt}{
\begin{picture}(43.00,43.00)
\put(7.00,1.00){\line(1,1){36.00}}
\put(37.00,1.00){\line(-1,1){36.00}}
\put(1.00,7.00){\line(1,1){36.00}}
\put(7.00,43.00){\line(1,-1){36.00}}
\end{picture}
}\qquad\raisebox{-15pt}{
\begin{picture}(43.00,43.00)
\put(7.00,43.00){\line(1,-1){12.00}}
\put(13.00,25.00){\line(-1,1){12.00}}
\put(25.00,31.00){\line(1,1){12.00}}
\put(43.00,37.00){\line(-1,-1){12.00}}
\put(31.00,25.00){\line(0,-1){6.00}}
\put(31.00,19.00){\line(1,-1){12.00}}
\put(37.00,1.00){\line(-1,1){12.00}}
\put(19.00,13.00){\line(-1,-1){12.00}}
\put(1.00,7.00){\line(1,1){12.00}}
\put(13.00,19.00){\line(0,1){6.00}}
\put(19.00,31.00){\line(0,-1){18.00}}
\put(25.00,13.00){\line(0,1){18.00}}
\end{picture}
}\qquad\raisebox{-15pt}{
\begin{picture}(43.00,43.00)
\put(1.00,7.00){\line(1,1){12.00}}
\put(19.00,13.00){\line(-1,-1){12.00}}
\put(13.00,25.00){\line(-1,1){12.00}}
\put(7.00,43.00){\line(1,-1){12.00}}
\put(19.00,31.00){\line(1,0){6.00}}
\put(25.00,31.00){\line(1,1){12.00}}
\put(43.00,37.00){\line(-1,-1){12.00}}
\put(31.00,19.00){\line(1,-1){12.00}}
\put(37.00,1.00){\line(-1,1){12.00}}
\put(25.00,13.00){\line(-1,0){6.00}}
\put(13.00,19.00){\line(1,0){18.00}}
\put(13.00,25.00){\line(1,0){18.00}}
\end{picture}
}
\end{equation}

\vspace{4mm}

Type 1 quartic vertices arise from the ``45-terms'' in $f_{ABC}f_{ADE}$ in
(\ref{E31}), (when (\ref{E59}) is used), and from the $-\delta_{jq}
\delta_{kp}$ term in (\ref{E58}), when (\ref{E58}) is applied to the 
``123-terms'', namely $f_{aBC}f_{aEF}$ in (\ref{E31}), and the quartic
interaction term \\$\psi_Af_{AaB}A_{\mu B}\phi_Cf_{CaD}A_{\mu D}$ in
(\ref{E23}).

Type 2 quartic vertices arise from the ``6-terms'' in $f_{ABC}f_{ADE}$ in
(\ref{E31}), (when (\ref{E60}) is used), and from the $\frac{1}{N}\delta_
{jk}\delta{pq}$ term in (\ref{E58}), when (\ref{E58}) is applied to the 
``123-terms''.

Type 1 quartic vertices are characterized by the property that in two of 
the three ``2/2 channels'', (i.e. the three channels, traditionally called
$s$, $t$, and $u$, in which the four legs of the quartic vertex are
divided into two sets of two legs each), \emph{two} of the four $\textrm{SU}(N)$
fundamental representation path-ordered phase factors that pass through
that vertex, pass from one to the other of the two sets of two legs that
define that channel, (or in other words, ``pass through that channel''),
while in the third 2/2 channel, all four path-ordered phase factors pass
through that channel, and the Type 2 quartic vertices are characterized by
the property that, in one of the three 2/2 channels, \emph{none} of the 
path-ordered phase factors pass throught that channel, while in the other
two 2/2 channels, all four of the path-ordered phase factors pass through
those channels.

The other two types of quartic action vertex, which have either one or two
6-paths ending at them, and which we call Type 3 and Type 4 respectively,
(and which arise only from the $f_{ABC}f_{AEF}$ term in (\ref{E31}), and 
not at all from (\ref{E23}), due to the vanishing of $f_{aBC}$ whenever 
$B$ or $C$ is a ``6''), look respectively like an (\ref{E84}) and an
(\ref{E85}) in close proximity, or like two (\ref{E85})'s in close 
proximity.

To have each diagram formed from these components correspond to precisely
one term of the type described above, it would be necessary, for each Type
1 quartic vertex, to indicate whether that Type 1 quartic vertex arises
from a ``123-term'' via equation (\ref{E58}), or from a ``45-term'' via
equation (\ref{E59}), and also to indicate which of the two 2/2 channel
through which \emph{two} phase factors pass, corresponds to the ``source''
of that vertex, in the sense that there is one of the two structure 
constants from which that vertex arises, at each ``end'' of that channel,
and it would also be necessary, for each Type 2 quartic vertex, to
indicate whether that Type 2 quartic vertex arises from a ``123-term'' via
equation (\ref{E58}), or from a ``6-term'' via equation (\ref{E60}), and 
also to indicate which of the four possible pairs of phase factors, (there
being one member of each pair at each end of the ``source'' channel, which
in this case is the one through which \emph{none} of the phase factors
pass), is the pair ``involved'' in that term, in the sense that whether
those two phase factors are in the same, or different, $\textrm{SU}(N)$ subgroups 
of $(\textrm{SU}(N))^M$, determines whether or not the contribution from 
(\ref{E58}) vanishes, and whether the contribution from (\ref{E60}) 
includes the factor $\frac{1}{M}$ or the factor $-\frac{(M-1)}{M}$.

Certain simplifications occur.  Firstly, if it is known that two of the 
paths that meet at a quartic vertex are Fadeev-Popov paths, rather than
gauge-field paths, (which is not shown on our diagrams), then that vertex
can only arise from a ``123-term'', namely the term $\psi_Af_{AaB}A_{\mu B
}\phi_Cf_{CaD}A_{\mu D}$ in (\ref{E23}).  And secondly, if it is known 
that all four paths that meet at a quartic vertex are gauge-field paths,
(so that that vertex arises from the $\left(f_{aBC}f_{aEF}+f_{ABC}f_{AEF}
\right)$ term in (\ref{E31})), then if that vertex is a Type-1 quartic
vertex, we may add the contributions from (\ref{E58}) and from 
(\ref{E59}), to obtain the $-\delta_{JQ}\delta_{jq}\delta_{KP}\delta_{kp}$
term in (\ref{E61}), which is completely independent of which $\textrm{SU}(N)$
subgroups of $(\textrm{SU}(N))^M$ the four phase factors are in, (i.e. independent
of whether or not any of them are in the same $\textrm{SU}(N)$ subgroups of 
$(\textrm{SU}(N))^M$), and if that vertex is a Type-2 quartic vertex, we may add 
the contributions from (\ref{E58}) and from (\ref{E60}) to obtain the term
$\frac{1}{NM}\delta_{JK}\delta_{jk}\delta_{PQ}\delta_{pq}$ in (\ref{E61}),
which is again completely independent of which $\textrm{SU}(N)$ subgroups of
$(\textrm{SU}(N))^M$ the four phase factors are in, and which consequently
\emph{cancels out} when we sum over the four possible pairs of phase 
factors which can be ``involved'', due to the relative minus sign between
the two terms in (\ref{E62}) and in (\ref{E63}).  (This corresponds to a
similar cancellation, in the Feynman diagrams for $\textrm{SU}(N)$ Yang Mills 
theory, of the contributions from the $\frac{1}{N}\delta_{jk}\delta_{pq}$
term in (\ref{E49}) whenever that term occurs between two structure
constants, i.e. inside a quartic vertex or between any two action 
vertices.  Indeed, if the $\frac{1}{N}\delta_{jk}\delta_{pq}$ term in 
(\ref{E49}) is substituted for $\left(t_\alpha\right)_{jk}\left(t_\alpha
\right)_{pq}$ in:
	\[\mathrm{tr}\left(t_\alpha t_\beta t_\gamma-t_\gamma t_\beta t_\alpha
	\right)\mathrm{tr}\left(t_\alpha t_\epsilon t_\phi-t_\phi t_\epsilon
	t_\alpha\right)=
\]
\begin{equation}
	\label{E90}
	=\left(t_\beta t_\gamma-t_\gamma t_\beta\right)_{kj}\left(t_\alpha
	\right)_{jk}\left(t_\alpha\right)_{pq}\left(t_\epsilon t_\phi-t_\phi
	t_\epsilon\right)_{qp}
\end{equation}
we immediately obtain $\frac{1}{N}\mathrm{tr}\left(t_\beta t_\gamma-
t_\gamma t_\beta\right)\mathrm{tr}\left(t_\epsilon t_\phi-t_\phi
t_\epsilon\right)=0$.)

But it is still necessary to indicate, for each Type-1 quartic vertex, 
which of the two 2/2 channels through which two phase factors pass, is the
``source'' of that quartic vertex.

However for our immediate purpose, namely to determine, for each $f_r$ in 
the expansions (\ref{E54}) and (\ref{E68}), which of the $f_s$'s can occur
in the right-hand side of the group-changing equation for that $f_r$, 
there is no need to include any more detail in our diagrams than is given
by the diagram rules as defined above.  The important point to note is
that, as follows immediately from (\ref{E58}) and (\ref{E60}), \emph{there
is an explicit factor of} $\frac{1}{N}$ \emph{associated with every Type-2
quartic vertex.}  We note here that, as follows directly from the 6-term
in (\ref{E67}), \emph{there is also an explicit factor of} $\frac{1}{N}$
\emph{associated with each 6-path.}  We also note here that, as pointed 
out previously, the $\left(1-\delta_{JK}\right)$ factor in the 45-term in
(\ref{E67}) has the consequence, that any routeing of the phase factors
through the vertices, that disobeys the ``selection rule'' that no $\textrm{SU}(N)$
fundamental representation Wilson loop may pass in both directions along
any 45-path, gives no contribution at all.

\subsection{Application of the Lemma}

Now the break up of the $\tilde{W}_i$'s into sums of terms that correspond
to specific routeings of the $\textrm{SU}(N)$ fundamental representation 
path-ordered phase factors through the vertices, operates completely
independently within each separate $\tilde{W}_i$ in the $(\textrm{SU}(N))^M$
correlation function $\left[\tilde{W}_1\ldots\tilde{W}_s\right]_{(\mathrm{SU}(N))^
M}$, and consequently this $(\textrm{SU}(N))^M$ correlation function is equal to a
sum of $(\textrm{SU}(N))^M$ correlation functions of the form \newline
$\left[B_1\ldots B_s
\right]_{(\mathrm{SU}(N))^M}$, where each $B_i$ is a term in the corresponding
$\tilde{W}_i$ that corresponds to specific routeings through the vertices
of the $\textrm{SU}(N)$ fundamental representation path-ordered phase factors that
occur in that $\tilde{W}_i$, and consequently to a specific set $K_i$ of
$\textrm{SU}(N)$ fundamental representation Wilson loops, in various $\textrm{SU}(N)$
subgroups of $(\textrm{SU}(N))^M$, in that $\tilde{W}_i$, and also to specific
attachment points of 6-path ends, where appropriate.

Furthermore, the assignment of the $\textrm{SU}(N)$ fundamental representation
Wilson loops in the $B_i$'s to various $\textrm{SU}(N)$ subgroups of $(\textrm{SU}(N))^M$,
subject to the selection rules explained in the foregoing, and in 
particular to the crucial selection rule that \emph{no two} $\textrm{SU}(N)$ 
\emph{fundamental representation Wilson loops that pass, (in opposite
directions), along a common 45-path, can be in the same} $\textrm{SU}(N)$ 
\emph{subgroup of} $(\textrm{SU}(N))^M$, (which again follows directly from the 
$\left(1-\delta_{JK}\right)$ factor in the 45-term in (\ref{E67})), again
operates completely independently within the separate $B_i$'s in the
$(\textrm{SU}(N))^M$ correlation function $\left[B_1\ldots B_s\right]_{(\mathrm{SU}(N))^M}$,
hence each such $(\textrm{SU}(N))^M$ correlation function
$\left[B_1\ldots B_s\right]_{(\mathrm{SU}(N))^M}$ is itself equal to a sum of 
$(\textrm{SU}(N))^M$ correlation functions $\left[\tilde{B}_1\ldots\tilde{B}_s
\right]_{(\mathrm{SU}(N))^M}$, where each $\tilde{B}_i$ corresponds to a specific
assignment of the $\textrm{SU}(N)$ fundamental representation Wilson loops in the
corresponding $B_i$, to various specific $\textrm{SU}(N)$ subgroups of $(\textrm{SU}(N))^M$,
consistent with the selection rules as described above.

For any such term we define, for each $\tilde{B}_i$, $K_i$ to be the set
of all the $\textrm{SU}(N)$ fundamental representation Wilson loops in $\tilde{B}_
i$, and $S$ to be the set whose members are all the $K_i$'s, (so that $S$
is a partition of the set $K$ of \emph{all} the $\textrm{SU}(N)$ fundamental
representation Wilson loops in all the $\tilde{B}_i$'s).  We note that
$s=\mathbf{N}(S)$.

We now apply our Lemma to each such $(\textrm{SU}(N))^M$ correlation function \\
$\left[\tilde{B}_1\ldots\tilde{B}_s\right]_{(\mathrm{SU}(N))^M}$, to express it as
a sum of produts of $(\textrm{SU}(N))^M$ correlation functions of the individual
$\textrm{SU}(N)$ fundamental representation Wilson loops that occur in the $\tilde{
B}_i$'s, as in (\ref{E71}), where the sum on $R$ in (\ref{E71}) runs over
all partitions $R$ of the set $K$ of all the $\textrm{SU}(N)$ fundamental 
representation Wilson loops in all the $\tilde{B}_i$'s, such that $K$ is
$(R\cup S)$-connected.  We note here that the requirement that $R$ be such
that $K$ is $(R\cup S)$-connected, has the consequence that \emph{the 
maximum possible number of parts of the partition} $R$ \emph{is equal to}
$k+1-s$\emph{, where} $k=\mathbf{N}(K)$ \emph{is the total number of}
$\textrm{SU}(N)$ \emph{fundamental representation Wilson loops in all the} $\tilde
{B}_i$\emph{'s.}

Now an $(\textrm{SU}(N))^M$ correlation function $\left[\displaystyle\prod_{j\in i}
W_j\right]_{(\mathrm{SU}(N))^M}$, where $i$ is a part of the partition $R$, 
\emph{vanishes} unless all the $W_j$'s, $j\in i$, are in the \emph{same}
$\textrm{SU}(N)$ subgroup of $(\textrm{SU}(N))^M$, and we thus obtain the following 
additional selection rule on the assignments of the $\textrm{SU}(N)$ fundamental
representation Wilson loops to various $\textrm{SU}(N)$ subgroups of $(\textrm{SU}(N))^M$,
for each partition $R$, as generated by the Lemma, of the set $K$ of all 
the $\textrm{SU}(N)$ fundamental representation Wilson loops in all the $\tilde{B}
_i$'s, the only contributing assignments of the members of $K$ to various
$\textrm{SU}(N)$ subgroups of $(\textrm{SU}(N))^M$, that contribute to the ``Lemma term''
defined by $R$, are those which satisfy the requirement, that for each 
member $i$ of $R$, all the $W_j$'s, $j\in i$, are assigned to the 
\emph{same} $\textrm{SU}(N)$ subgroup of $(\textrm{SU}(N))^M$.  This means that, if for each
assignment of all the $W_j$'s, $j\in K$, to various $\textrm{SU}(N)$ subgroups of
$(\textrm{SU}(N))^M$, we define the corresponding partition $P$ of $K$ to be the 
unique partition of $K$ such that members $i$ and $j$ of $K$ are members 
of the \emph{same} member of $P$ if and only if they are assigned to the
\emph{same} $\textrm{SU}(N)$ subgroup of $(\textrm{SU}(N))^M$, then the only such 
assignments that contribute to the ``Lemma term'' defined by the partition
$R$ of $K$, are those whose corresponding partition $P$ satisfies the
requirement that, for each member $i$ of $R$, $i$ is a \emph{subset} of
some member $j$ of $P$.

\subsection{The derivative with respect to $M$, at $M=1$, becomes a derivative with respect to $g^2$}

Furthermore, if all the $W_j$ in the $(\textrm{SU}(N))^M$ correlation function
$\left[\displaystyle\prod_{j\in i}W_j\right]_{(\mathrm{SU}(N))^M}$ \emph{are}
assigned to the same $\textrm{SU}(N)$ subgroup of $(\textrm{SU}(N))^M$, then this
$(\textrm{SU}(N))^M$ correlation function is equal to the corresponding $\textrm{SU}(N)$
correlation function $\left[\displaystyle\prod_{j\in i}W_j\right]_{\mathrm{SU}(N)
}$, and it immediately follows from this that, as noted earlier, (in the 
discussion between equations (\ref{E67}) and (\ref{E68})), the entire
dependence of any term on its assignments of the members of $K$, to the
various $\textrm{SU}(N)$ subgroups of $(\textrm{SU}(N))^M$, is through the corresponding
partition $P$, and that, furthermore, when we sum over all those 
assignments that given the same $P$, the entire dependence on $M$ is
given, for this term, by the simple factor
	\[M(M-1)(M-2)\ldots(M-p+1)=\frac{M!}{(M-p)!}
\]
(where $p=\mathbf{N}(P)$ is the number of parts of the partition $P$),
times a factor $-\frac{(M-1)}{M}$ or $\frac{1}{M}$ for each 6-path, (where
the choice depends on $P$ and on the attachment points of the ends of that
6-path), times a similar factor for each Type-2 quartic vertex, as 
discussed above.

\emph{It follows that we may differentiate with respect to} $M$ \emph{and
then set} $M$ \emph{equal to 1.}  We call the resulting equations the
\emph{group-variation equations for the} $\textrm{SU}(N)$ \emph{groups.}

Now
	\[\left.\frac{d}{dM}\left((NM)^{2-n-2r}f_r(W_1,\ldots,W_n,g^2M)\right)
	\right|_{M=1}\qquad=
\]
\begin{equation}
	\label{E91}
	=(2-n-2r)N^{2-n-2r}f_r(W_1,\ldots,W_n,g^2)+N^{2-n-2r}g^2\frac{d}{dg^2}
	f_r(W_1,\ldots,W_n,g^2)
\end{equation}
hence we immediately see, on substituting the expansions (\ref{E68}) into
the left hand sides, that the group-variation equations for the $\textrm{SU}(N)$
groups express the derivatives with respect to $g^2$ of the expansion
coefficients in (\ref{E54}), in terms of those expansion coefficients
themselves.

\subsection{The Group-Variation Equation for a coefficient in the $\frac{1}{N}$ expansion of a vacuum expectation value or correlation function,
involves only expansion coefficients of the same, or lower, non-vanishing order in the $\frac{1}{N}$ expansions}

We shall now determine which of the $f_s$'s in (\ref{E54}) can occur in 
the right-hand side of the group-variation equation for an expansion
coefficient $f_r(W_1,\ldots,W_n,g^2)$ in (\ref{E54}).

For any right-hand side term of the type described above, we define $l$ to
be the total number of new paths minus the total number of action 
vertices, $v$ to be the total number of 6-paths plus the total number of
Type-2 quartic action vertices, $k$ to be the total number of $\textrm{SU}(N)$
fundamental representation Wilson loops, and $s$ to be the total number of
connected components of the decoration $\tilde{W}_1,\ldots,\tilde{W}_s$ of
the original $n$ Wilson loops $W_1,\ldots,W_n$.  We also define, for each
individual connected component $\tilde{W}_i$ of that decoration, $l_i$ to
be the total number of new paths minus the total number of action vertices
of that connected component, $v_i$ to be the total number of 6-paths plus
the total number of Type-2 quartic action vertices of that connected
component, $k_i$ to be the total number of $\textrm{SU}(N)$ fundamental 
representation Wilson loops of that connected component, and $n_i$ to be
the total number of the \emph{original} $n$ Wilson loops that form part of
that connected component.  Thus $l=\sum l_i$, $v=\sum v_i$, $k=\sum k_i$,
and $n=\sum n_i$, where the sums all run from $i=1$ to $s$.

One-loop vacuum bubbles are assigned $l=0$  We note that there is no need
to consider the ``6-path'' one-loop vacuum bubble, since it is completely
independent of the $(\textrm{SU}(N))^M$ gauge fields $A_{\mu a}$, and therefore
cancels out.

Now every allowed decoration can be built up from the original $n$ closed
paths which define the $n$ $\textrm{SU}(NM)$ Wilson loops $W_1,\ldots,W_n$, plus
one one-loop 45-path vacuum bubble for each $\tilde{W}_i$ that contains
\emph{none} of the original $n$ Wilson loops, (and hence has $n_i=0$, or
in other words, is a vacuum bubble), by repeated applicatons of an ``add a
path'' operation, similar to what we used for $\textrm{SU}(N)$ Feynman diagrams.
Each end of the path that is ``added'' at each stage may be either on a
previously existing new path, or at a previously existing triple action
vertex, or on an original path.  We may build up the ``core'' decoration
first, (that is, the decoration with all its 6-paths removed), and then
add the 6-paths, (all of which have both their ends on the core 
decoration).  Each added path increases the \emph{total} number of new 
paths be either 3, 2, or 1, depending on where the ends of the added path
go, and corresponding to these cases, the added path increases the
\emph{total} number of action vertices by 2, 1, or 0, respectively, thus
every ``add a path'' operation increases $l$ by 1, and indeed $l$ is equal
to the total number of times the ``add a path'' operation has been
performed.  And furthermore, each $l_i$ is equal to the number of ``add a
path'' operations that have gone into the building of that $\tilde{W}_i$,
where we note that a $\tilde{W}_i$, which by definition is a connected
component of the full decoration, may, at early stages of the building
process, have had more than one connected component.

We define one additional building operation, that consists of pinching
together two previously existing 45-paths to form a Type-2 quartic action
vertex, or pinching together two different points on a single previously
existing 45-path, again to form a quartic action vertex.  This operation
increases the total number of new paths by 2, and increases the total 
number of action vertices by 1, hence again corresponds to increasing $l$
by 1.

At each stage of the building of a valid decoration, a valid decoration is
obtained, except that in some cases some of our selection rules, in
particular the selection rule that it must not be possible to increase the
number of connected components of a core decoration, by the removal of 
just one 45-path, may be disobeyed.  But that doesn't matter, since our
purpose is to derive, for each $\tilde{W}_i$, an inequality between $k_i$,
$l_i$, $v_i$, and $n_i$, and this inequality is true even for those cases
where these selection rules are disobeyed.  We now find, by induction on
$l$ and on the individual $l_i$'s, that for every $\tilde{W}_i$, the 
following inequality is satisfied:
\begin{equation}
	\label{E92}
	k_i\leq2+l_i+v_i-n_i
\end{equation}
and this immediately implies, on summing over $i$, that:
\begin{equation}
	\label{E93}
	k\leq2s+l+v-n
\end{equation}

Indeed, (\ref{E92}) and (\ref{E93}) are satisfied by a single original
closed path, which has $k=1$, $l=v=0$, $n=1$, and a single one-loop
45-path vacuum bubble, which has $k=2$, $l=v=0$, $n=0$, and also by any
number of original closed paths and one-loop 45-path vacuum bubbles.  And
when a 45-path is added to a core decoration, $k$ \emph{increases} by 1 if
both ends of the added 45-path ``break into'' the \emph{same} previously
existing $\textrm{SU}(N)$ Wilson loop, and $k$ \emph{decreases} by 1 if each end of
the added 45-path ``breaks into'' a \emph{different} previously existing
$\textrm{SU}(N)$ Wilson loop.  Furthermore, if the added 45-path has the effect of
decreasing the number of connected components of the core decoration by 1,
then each of its ends is on a different previously existing connected 
component, hence each of its ends must ``break into'' a \emph{different}
previously existing Wilson loop, hence it results in $k$ 
\emph{decreasing} by 1.  Hence (\ref{E92}) and (\ref{E93}) are preserved
when the induction step consists of adding a 45-path to a core 
decoration.  And when two previously existing 45-paths are pinched 
together to form a Type-2 quartic vertex, or two different points on a 
single previously existing 45-path are pinched together to form a Type-2
quartic vertex, $k$ is unaltered, $l$ and $v$ each increase by 1, and $s$
is either unaltered or else decreases by 1,(since this operation may pinch
together two 45-paths, that are on different connected components of the
previously existing core decoration), hence (\ref{E92}) and (\ref{E93})
are also preserved when the induction stip consists of forming a Type-2
quartic vertex by pinching togther two different points on the previously
existing core diagram.  And finally, adding a 6-path leaves $k$ unaltered,
increases $l$ by 1, increases $v$ by 1, and either leaves $s$ unaltered or
else decreases $s$ by 1, (since each end of the added 6-path may be on a
different connected component of the previously existing decoration),
hence (\ref{E92}) and (\ref{E93}) are also preserved when the induction
stip consists of adding a 6-path.

We note, as an immediate corollary to this proof, that if there is any
6-path or Type-2 quartic vertex whose removal, (in the case of a 6-path),
or ``unpinching'', (in the case of a Type-2 quartic vertex), does 
\emph{not} result in an increase of the number of connected components of
the decoration, then the right-hand side of (\ref{E93}) may be replaced 
by:
\begin{equation}
	\label{E94}
	2s+l+v-n-2
\end{equation}
(Any decoration may be built up in the sequence: first build the core
decoration with all its Type-2 quartic vertices ``unpinched'', then do the
``pinchings'' to form the Type-2 quartic vertices, then add the 6-paths.)

We note here that there is in fact an additional selection rule, namely
that no decoration that contributes to the right-hand side of any of our
$\textrm{SU}(NM)/(\textrm{SU}(N))^M$ group-changing equations or our $\textrm{SU}(N)$ 
group-variation equations, can be such that its number of connected 
components can be increased by ``unpinching'' a single Type-2 quartic
action vertex.  This is because, as observed just before equation 
(\ref{E90}), the various terms associated with any Type-2 quartic action
vertex at which four gauge-field paths meet, always exactly cancel out, so
that we may assume that two of the paths that meet at any Type-2 quartic
action vertex are Fadeev-Popov paths, and that that vertex arises from the
$\psi_Af_{AaB}A_{\mu B}\phi_Cf_{CaD}A_{\mu D}$ term in (\ref{E23}).  And 
it immediately follows from the form of this term, that there is 
\emph{one} Fadeev-Popov path ending at that vertex at each ``end'' of the
``source channel'' of that vertex, i.e. at each end of the channel of 
that vertex through which \emph{no} path-ordered phase factors pass.  This
means that when a Type-2 quartic action vertex is formed by pinching
together two previously existing 45-paths, half of each of those two
previously existing paths must become a Fadeev-Popov path, (the building
process does not in general preserve the separate identities of 
gauge-field paths and Fadeev-Popov paths), and, in particular, if a Type-2
quartic action vertex is such that ``unpinching'' it results in increasing
the number of connected ocmponents of the decoration, there would have to
be one Fadeev-Popov path ending at that quartic action vertex in each of 
the two separate connected components that are produced by unpinching that
Type-2 quartic action vertex.  But Fadeev-Popov paths only occur in closed
loops of Fadeev-Popov paths, hence the situation just described is 
impossible, since there is no way the two Fadeev-Popov paths ending at
that ``key'' Type-2 quartic action vertex could form a closed loop of
Fadeev-Popov paths.  We note, however, that we \emph{do} have to allow 
such ``key'' Type-2 quartic action vertices at intermediate stages in the
process of building-up a decoration.  (We do not distinguish gauge-field
paths from Fadeev-Popov paths until the ``building'' of the decoration has
been completed.)

We now continue with the determination of which of the $f_s$'s in 
(\ref{E54}) can occur in the right-hand side of the group-variation
equation for an expansion coefficient $f_r(W_1,\ldots,W_n,g^2)$ in
(\ref{E54}), with the definitions of $l$, $l_i$, $v$, $v_i$, $k$, $k_i$,
$n$, $n_i$, and $s$, and the sets $K$ and $S$, as already introduced.  Let
us consider a term generated by our Lemma where we have $r$ 
correlation-function factors, corresponding to the partition $R$ of $K$,
so that $r=\mathbf{N}(R)$.  Now as we have already noted, (between 
equations (\ref{E90}) and (\ref{E91})), the requirement that $R$ be such
that $K$ is $(R\cup S)$-connected, implies that:
\begin{equation}
	\label{E95}
	r\leq k+1-s
\end{equation}

For each $j$, $1\leq j\leq r$, let $m_j$ be the number of members of the 
$j^\mathrm{th}$ member of $R$.  We note that:
\begin{equation}
	\label{E96}
	m_1+m_2+\ldots+m_r=k=\mathbf{N}(K)
\end{equation}

We substitute in the expansion (\ref{E54}) for each of these
correlation-function factors.  Let us consider the term where, for the
$j^\mathrm{th}$ correlation-function factor, $1\leq j\leq r$, we take the
$b_j^\mathrm{th}$ term in the expansion (\ref{E54}), or in other words, 
the term:
\begin{equation}
	\label{E97}
	N^{2-m_j-2b_j}f_{b_j}(W_1,\ldots,W_{m_j},g^2)
\end{equation}
where $W_1,\ldots,W_{m_j}$ here are some of the $\textrm{SU}(N)$ fundamental
representation Wilson loops occurring in this decoration term, 
(specifically, those that are members of the $j^\mathrm{th}$ member of
$R$), \emph{not} the original $n$ Wilson loops.

Then the total power of $N$, for this term, coming from the 
correlation-function factors, is:
	\[(2-m_1-2b_1)+(2-m_2-2b_2)+\ldots+(2-m_r-2b_r)\qquad=
\]
	\[=\qquad2r-k-2(b_1+b_2+\ldots+b_r)
\]
	\[\leq\qquad2(k+1-s)-k-2(b_1+b_2+\ldots+b_r)
\]
	\[=\qquad k+2-2s-2(b_1+b_2+\ldots+b_4)
\]
\begin{equation}
	\label{E98}
	\leq2+l+v-n-2(b_1+b_2+\ldots+b_r)
\end{equation}
where at the last step here we used (\ref{E93}).

Now the factor $\frac{N}{4g^2}$ outside the action produces a factor
$\frac{1}{N}$ for each new path and a factor $N$ for each action vertex,
hence an overall factor $N^{-1}$, and there is also, as already noted, an
additional, explicit factor of $\frac{1}{N}$ associated with each 6-path,
and with each Type-2 quartic action vertex, (coming from the explicit
factors of $\frac{1}{N}$ in the appropriate terms in (\ref{E58}) and 
(\ref{E67})), which gives an additional factor of $N^{-v}$.

Hence the \emph{total} power of $N$ for this term is:
\begin{equation}
	\label{E99}
	\leq\qquad2-n-2(b_1+b_2+\ldots+b_r)
\end{equation}
where we note that the inequality arises from the use of the upper bounds
(\ref{E93}) and (\ref{E95}) on $k$ and $r$.

Now $f_a(W_1,\ldots,W_n,g^2)$ in the left-hand side, (where $W_1,\ldots,
W_n$ here are the $n$ original Wilson loops), corresponds to a power of 
$N$ of:
\begin{equation}
	\label{E100}
	2-n-2a
\end{equation}
(see (\ref{E91}), with $r$ in (\ref{E91}) replaced by $a$).

Hence, equating coefficients of powers of $N$ in the original 
group-variation equations, we see that the term defined above can only
contribute to the group-variation equation for $f_a(W_1,\ldots,W_n,g^2)$,
if:
\begin{equation}
	\label{E101}
	b_1+b_2+\ldots+b_r\leq a
\end{equation}
But all the $b_j$'s are $\geq0$, hence \emph{an} $f_b$ \emph{can only
contribute to the right-hand side of the group-variation equation for}
$f_a$ \emph{if} $b\leq a$.

\subsection{The Group-Variation Equations are complete}

Now the group-variation equations are \emph{complete}.  Indeed, we may
directly recover the Feynman diagram expansions of the $f_a$'s in powers
of $g^2$ by developing the sums over paths in powers of $g$, (as we show
how to do later in this paper, and as corresponds to developing the 
original background-field propagators in Section 2 in powers of $g$ rather
than as sums over paths), using the boundary condition that the 
one-Wilson-loop correlation function, (i.e. the one-Wilson-loop vacuum
expectation value), is equal to 1 at $g^2=0$, and that all the other 
correlation functions are equal to 0 at $g^2=0$.  (This procedure doesn't
directly give the Feynman diagrams in one of the usual linear covariant
gauges, but the usual Feynman diagrams can be recovered by the use of 
``propagator gauge invariance'' identities, similar to those used by 
Mills \cite{Mills}, as we will show in detail in our next paper.)

The group-variation equations have, of course, to be renormalized, and
also to be re-written to take account of the fact that we must divide a 
short-distance factor out of each Wilson loop.  We indicate briefly how to
do this later in this paper, and will give the full details in our next
paper.

The important point for now is that, due to the result just obtained, we
may first solve the equations for the $f_0$'s, (which give the solution of
the glueball sector of large-$N_c$ QCD), \emph{then} solve the equations
for the $f_1$'s, (using the $f_0$'s already calculated), and so on.

\subsection{Use of the renormalization group to express the derivative with respect to $g^2$}

We note here that, once the group-variation equations have been 
renormalized, we will be able to express the derivative  $\frac{d}{dg^2}
f_a(W_1,\ldots,W_n,g^2)$ that occurs in (\ref{E91}), in terms of a 
derivative with respect to an overall linear scale factor $L$ of the sizes
and separations of the Wilson loops $W_1,\ldots,W_n$, by means of the 
renormalization group equation \cite{RGE}:
\begin{equation}
	\label{E102}
	\left(L\frac{\partial}{\partial L}+\beta(g)\frac{\partial}{\partial g}
	\right)f_a(W_1,\ldots,W_n,g^2)=0
\end{equation}
(In practice there will be additional terms in this equation due to the
short-distance factor which we have to divide out.)

When (\ref{E102}) is used, the boundary conditions at $g^2=0$ quoted above
will be replaced by the behaviours of the $f_a$'s as $L\rightarrow0$, as
determined by renomalization-group-improved perturbation theory.

\section{Simplifications at leading non-vanishing order in $\frac{1}{N}$}

We shall devote the remainder of this paper to the analysis of the
group-variation equations for the $f_0$'s, which, as we shall soon see, 
have the advantage, over the direct sum of Feynman diagrams, that their
solution manifestly has the correct qualitative behaviour, namely, the
Wilson area law for the one-Wilson-loop vacuum expectation value, and
massive glueball saturation for the multi-Wilson-loop correlation 
functions, and that, moreover, this correct qualitative behaviour is
obtained for the very simplest approximation to the solution, and is 
maintained in all higher approximations.

Let $l$, $l_i$, $v$, $v_i$, $k$, $k_i$, $n$, $n_i$, and $s$, and the sets
$K$, $S$, and $R$, be defined, for each right-hand side term, as in the
preceding discussions.

We begin by noting some further simplifications that occur in the 
right-hand sides of the group-variation equations for the $f_0$'s.

We first note that for $a=0$, the inequality in (\ref{E101}) becomes an
equality, (and of course, all the $b_j$ are equal to 0), which immediately
implies that $k$k and $r$ must be equal to their upper limits as given by
(\ref{E93}) and (\ref{E95}) respectively.

Now the requirement that $k$ be equal to its upper limit as given by
(\ref{E93}) implies, firstly, that at each step of building up the core of
an allowed decoration $\tilde{W}_1\ldots\tilde{W}_s$ from the initial $n$
Wilson loops, plus as many one-loop 45-path vacuum bubbles as there are
vacuum bubbles among the $\tilde{W}_i$'s, the added 45-path must satisfy
the requirement that both its ends ``break into'' the \emph{same}
previously existing $\textrm{SU}(N)$ fundamental representation Wilson loop, except
at those steps where this is \emph{impossible}, due to the added 45-path
having each end in a different previously existing connected component,
(and thus reducing the total number of connected components by 1).  And
secondly, as immediately follows from the discussion accompanying formula
(\ref{E94}), there must be \emph{no} 6-pat or Type-2 quartic vertex whose
removal, (in the case of a 6-path), or ``unpinching'', (in the case of a
Type-2 quartic vertex), does \emph{not} result in an increase of the 
number of connected components of the decoration.  But by the discussion
following formula (\ref{E94}), there can be \emph{no} such Type-2 quartic
vertex in the full decoration, hence there must be \emph{no Type-2 quartic
vertices at all}.

Now $r$ is to be equal to its maximum possible value, as given by 
(\ref{E95}), and this implies, in particular, that if $i$ is any member of
$S$, and $j$ is any member of $R$, then $i\cap j$ has at most one member.
For if $i\cap j$ had two or more members then we could remove all but one 
of the members of $i\cap j$ from $j$, and for each member $u$ of $i\cap j$
that we remove from $j$, add the one-member set $\{u\}$ to $R$, (thus
increasing $r=\mathbf{N}(R)$), and $K$ would remain $(R\cup S)$-connected,
contradicting the assumption that $r$ has its maximum possible value 
consistent with $K$ being $(R\cup S)$-connected.

(We note that the upper bound (\ref{E95}) on $r$ follows from the 
following consideration: for given $K$ and $S$, we form $R$ by a sequence
of steps starting from an ``initial'' $R$ which is defined to be the set 
of all the one-members subsets of $K$.  Thus the ``initial'' $R$ has $k=
\mathbf{N}(K)$ members, and $K$ has $s=\mathbf{N}(S)$ 
$(R\cup S)$-connected components.  We transform this ``initial'' $R$ to
the given $R$ by a sequence of steps, each of which consists of replacing
two members of ``$R$ at that stage'', by their union.  Thus each step
results in reducing $r=\mathbf{N}(R)$ by 1.  Now each step may or may not
reduce the number of $(R\cup S)$-connected components of $K$ by 1, but 
certainly no step reduces the number of $(R\cup S)$-connected components 
of $K$ by \emph{more} than 1, since each step consists of replacing just
two members of $R$ by their union, and no member of $R$ intersects more
than one $(R\cup S)$-connected component of $K$, hence no two members of
$R$ intersect more than two $(R\cup S)$-connected components of $K$.  
Hence to reach an $R$ such that the number of $(R\cup S)$-connected
components of $K$ has been reduced from $s$ to 1, there must be at least
$s-1$ of these steps.)

We now return for a moment to the group-changing equations for $(\textrm{SU}(N))^M$
and $\textrm{SU}(NM)$, and consider a 6-pat, with its ends ``attached'', (in the
usual sense - see e.g. the discussion before and after (\ref{E81})), to
specific $\textrm{SU}(N)$ fundamental representation Wilson loops.  We consider the
sum over the assignments of which of the $M$ $\textrm{SU}(N)$ subgroups of 
$(\textrm{SU}(N))^M$ the $\textrm{SU}(N)$ fundamental representation Wilson loops to which
that 6-path attaches, are ``in'', (or in other words, that their 
gauge-fields belong to).  Let us \emph{suppose} that we can sum over the
$M$ possibilities completely independently at each end of that 6-path, (so
that we assume, in particular, that each end of that 6-path attaches to a
\emph{different} $\textrm{SU}(N)$ fundamental representation Wilson loop).  We then
see immediately , from the 6-term in (\ref{E67}), that from the $M$
possibilities where we have the \emph{same} $\textrm{SU}(N)$ subgroup of 
$(\textrm{SU}(N))^M$ at each end of the 6-path, we get a contribution that 
includes, from the overall factor $M$ and from the $\left(-\delta_{JQ}+
\frac{1}{M}\right)$ factor in the 6-term, the factor $M\left(-\frac{(M-1)}
{M}\right)=-(M-1)$, while from the $M(M-1)$ possibilities where we have a
\emph{different} $\textrm{SU}(N)$ subgroup of $(\textrm{SU}(N))^M$ at each end of the
6-path, we get the same contribution, except that from the overall factor
$M(M-1)$ and from the $\left(-\delta_{JQ}+\frac{1}{M}\right)$ factor in
the 6-term, we now get the factor $M(M-1)\left(\frac{1}{M}\right)=(M-1)$.
Thus \emph{the total is zero}: we have yet another ``selection rule''.  We
note that this holds for general $M$, and is thus also true for the 
group-\emph{variation} equations for the $\textrm{SU}(N)$ groups.

We did not mention this selection rule before, because the circumstances
under which it can be applied are extremely restricted: we must be able to
sum completely independently over the ``which $\textrm{SU}(N)$'' assignments of the
two $\textrm{SU}(N)$ fundamental representation Wilson loops to which that 6-path
attaches, and every other factor in the integrand of the term under 
consideration, must be completely independent of whether or not those two
$\textrm{SU}(N)$ fundamental representation Wilson loops are in the same part, or 
in different parts of the partition $P$ of $K$ that is defined, as 
described between (\ref{E90}) and (\ref{E91}), for each assignment of the
$\textrm{SU}(N)$ fundamental representation Wilson loops to various $\textrm{SU}(N)$
subgroups of $(\textrm{SU}(N))^M$.

We now return to the analysis of the terms that occur in the right-hand
sides of the group-variation equations for the $f_0$'s.

We have already seen that there can be no Type-2 quartic vertices at all,
and that any 6-path that occurs must be such that its removal results in
an increase in the number of connected components of the decoration.  And
we have also seen that the fact that for each right-hand side term, 
$r=\mathbf{N}(R)$ is equal to its maximum possible value, as given by
(\ref{E95}), implies that for each right-hand side term, if $i$ is any
member of $S$ and $j$ is any member of $R$, then $i\cap j$ has at most one
member.  Let us consider a right-hand side term that has a 6-path whose
removal results in an increase in the number of connected components of
that decoration.  Then each end of that 6-path certainly attaches to a
\emph{different} $\textrm{SU}(N)$ fundamental representation Wilson loop, indeed 
the $\textrm{SU}(N)$ fundamental representation Wilson loops to which the ends of 
that 6-path attach are in different connected components of the decoration
that remains when that 6-path is removed.  And furthermore, by the result
just mentioned, these two $\textrm{SU}(N)$ fundamental representation Wilson loops
are members of \emph{different} members of $R$, since they are members of
the same member of $S$.  Furthermore, it also follows from the fact that
$r$ is equal to its maximum possible value, that if we define $\tilde{S}$
to be the partition of $K$ that is obtained from $S$ by removing the 
member of $S$ that contains the $\textrm{SU}(N)$ fundamental representation Wilson
loops of the connected component of our decoration which contains the 
6-path we are considering, and replacing it by the two (nonempty) sets of
$\textrm{SU}(N)$ fundamental representation Wilson loops that correspond to the two
connected components into which that connected component of our decoration
separates when we remove that 6-path, (so that the members of $\tilde{S}$
are in one-to-one correspondence with the connected components of the
decoration we obtain by removing that 6-path, and the members of each 
member of $\tilde{S}$ are the $\textrm{SU}(N)$ fundamental representation Wilson
loops in the corresponding connected component of that decoration), then
$K$ is \emph{not} $(R\cup\tilde{S})$-connected.  For $\mathbf{N}(\tilde{S}
)=\mathbf{N}(S)+1=s+1$, hence by the same proof as before, if $K$ was
$(R\cup\tilde{S})$-connected, then $R$ could have at most $k+1-s-1=k-s$
members, which contradicts the assumption that $R$ has $k+1-s$ members.
Hence there exists a partition of $K$ into two nonempty parts, say $U$ and
$V$, such that \emph{no} member of $(R\cup\tilde{S})$ intersects both $U$
and $V$.  Now if $i$ is the member of $S$ that we removed in forming
$\tilde{S}$ from $S$, and $u$ and $v$ are the two members of $\tilde{S}$
into which $i$ has ``split'', then $u$ is a subset of one member of $\{U,V
\}$, and $v$ is a subset of the other member of $\{U,V\}$, (for $u$ and 
$v$ are members of $\tilde{S}$, hence neither $u$ nor $v$ intersects both
$U$ and $V$, while if both $u$ and $v$ were subsets of the \emph{same}
member of $\{U,V\}$, then the fact that $K$ is $(R\cup S)$-connected, 
hence that some member of $(R\cup S)$ intersects both $U$ and $V$, and the
fact that $i=u\cup v$ is a subset of \emph{one} member of $\{U,V\}$, hence
does \emph{not} intersect both members of $\{U,V\}$, would imply that 
some member of $(R\cup\tilde{S})$ intersects both $U$ and $V$, contrary to
assumption).  Suppose for definiteness that $u$ is a subset of $U$ and $v$
is a subset of $V$.  Then one end of our 6-path attaches to a member of
$u$, hence a member of $U$, while the other end of our 6-path attaches to
a member of $v$, hence a member of $V$.  Now the fact that no member of
$R$ intersects both $U$ and $V$ implies that the product of the 
correlation functions generated by our Lemma for this term, completely
factorizes into a factor that depends only on the members of $U$, and a 
factor that depends only on the members of $V$.  And furthermore, the fact
that no member of $\tilde{S}$ intersects both $U$ and $V$ implies that 
apart from our 6-path itself, the integrand for this term completely
factors into a factor associated with $U$, and a factor associated with 
$V$.  Hence the conditions for the applicability of the selection rule
which we derived just above, are satisfied: apart from the $\left(\delta_
{JQ}+\frac{1}{M}\right)$ factor in the 6-term in (\ref{E67}), the 
integrand for this term is completely independent of whether the two
$\textrm{SU}(N)$ fundamental representation Wilson loops to which the ends of our
6-path attach, are in the same, or different, $\textrm{SU}(N)$ subgroups of
$(\textrm{SU}(N))^M$, so that we sum over all possibilities, the result is zero.

It immediately follows from this thata no decoration that includes a 
6-path whose removal results in increasing the number of connected
components of that decoration, gives any contribution, hence, combining
this with the previous results, we see that \emph{there are no 6-paths and
no Type-2 quartic vertices at all in the right-hand sides of the 
group-variation equations for the} $f_0$\emph{'s.}

Hence every contributing decoration term can be obtained from one-loop
45-path vacuum bubbles and the $n$ original Wilson loops by repeated
applications of the simple ``add a path'' procedure, with every added path
being a 45-path, so that all the new paths are 45-paths.

We can now give a simple topological interpretation of each term, as 
defined above, in the right-hand side of the group-changing equation for
any $f_0$.  Let the integer $n\geq1$ denote, as usual, the number of 
Wilson loop arguments of the $f_0$ whose group-changing equation we are
considering, and let the integers $l$, $l_i$, $k$, $k_i$, $n_i$, and $s$,
and the sets $K$, $S$, and $R$, have their usual meanings of the term 
under consideration, so that the suffix $i$ runs over the connected 
components of the decoration under considerataion of the initial $n$
Wilson loops, $l$ is the total number of new paths minus the total number
of action vertices, $l_i$ is the corresponding quantity for the $i^
\textrm{th}$ connected component of the decoration, $K$ is the set of all
the $\textrm{SU}(N)$ fundamental representation Wilson loops into which the $\textrm{SU}(N)$
fundamental representation path-ordered phase factors in the 45-paths and
the original paths are connected by the specific routeings through the 
action vertices defined by the term under consideration, 
$k=\mathbf{N}(K)$, $S$ is the partition of $K$ whose parts are in 
one-to-one correspondence with the $s$ connected components of the
decoration, such that the $k_i$ members of the $i^\mathrm{th}$ part of 
$S$ are precisely those members of $K$ that belong to the $i^\mathrm{th}$
connected component fo the decoration, $n_i$ is the number of the $n$
original Wilson loops that form part of the $i^\mathrm{th}$ connected
component of the decoration, and $R$ is the partition of $K$ which 
defines, in accordance with the Lemma, the product of correlation
functions of members of $K$, to which the particular ``Lemma term'' that
we are considering, corresponds, (so that $R$ satisfies the requirement
that $K$ is $(R\cup S)$-connected).  The integers $v$ and $v_i$ defined
previously are of course zero in the present situation, by the results 
just derived.

We first note that the requirement that $k$ be equal to its maximum
possible value, as given by (\ref{E93}), implies that if we ``fill'' each
member of $K$, (i.e. each $\textrm{SU}(N)$ fundamental representation Wilson loop),
with an oriented topological 2-disk, then for each individual connected
component of the decoration, the $k_i$ such oriented topological 2-disks
associated with that connected component join up to form an oriented
topological 2-sphere with $n_i$ holes, where the boundaries of the holes
are the $n_i$ original Wilson loops that form part of that connected
component, and the boundaries of the holes are all oriented the same way.
The ``joining up'' process consists of ``sewing together'', along each
45-path, the two oriented topological 2-disks whose boundaries meet along
that 45-path.  (We recall that one of our selection rules is that no 
member of $K$ passes in both directions along any 45-path.)  The reasoning
is exactly as given for Feynman diagrams before (\ref{E53}), with ``closed
loop of Kronecker deltas'' now replaced by ``$\textrm{SU}(N)$ fundamental
representation Wilson loop'', or in other words, ``member of $K$''.

We note that the oriented topological 2-sphere with $n_i$ holes is 
equivalent to the oriented topological 2-disk with $n_i-1$ holes, so that
(\ref{E53}) shows a possible example with $n_i=3$.

\subsection{Topological representation of the right-hand side terms}

We shall now construct a topological representation of our right-hand side
term, in which each member $T_j$ of $R$, $1\leq j\leq r$, where $r=
\mathbf{N}(R)$, is represented by an oriented topological 2-sphere with
$t_j=\mathbf{N}(T_j)$ holes, with the boundaries of the holes being the
$t_j$ members of $K$ that are members of $T_j$, and all the boundaries of
the holes being oriented the same way, i.e. such that their orientations 
as defined by their path-orderings are consistent with the orientation of
that topological 2-sphere.  (We recall that each member $T_j$ of $R$
corresponds to a factor, in the term under consideration, equal to the
correlation function of the members of $T_j$.)

We now see that the topological construction that we have already given,
where we filled \emph{every} member of $K$ with an oriented topological
2-disk,(or in other words, an oriented topological 2-sphere with one 
hole), irrespective of whether that member of $K$ was as member of a
one-member member of $R$, or a member of a member $T_j$ of $R$ with two or
more members, corresponds to carrying out the above construction for the
``initial'' $R$ that is defined to be the set of all the one-member 
subsets of $K$, as we described in the proof given above, (some paragraphs
after (\ref{E102})), of the upper limit (\ref{E95}) on $r$.  We now
proceed to ``correct'' this ``initial'' topological construction into a
topological construction that exactly corresponds to the ``true'' $R$ by
the above rule, in a series of steps that exactly correspond to the steps
that we used, in the above-mentioned proof of the upper limit on $r$, to
transform the ``initial'' $R$ into the ``true'' $R$.

We note first that in this proof given above, (some paragraphs after
(\ref{E102})), of the upper limit (\ref{E95}) on $r$, each step in the 
process of transforming the ``initial'' $R$, (i.e. the set of all the 
one-member subsets of $K$), to the ``true'' $R$, consists of replacing two
members of the ``$R$ at that stage'' by their union, and thus reducing the
number of members of $R$ by 1.  Now such a step reduces by 1 the number of
$(R\cup S)$-connected components of $K$ if the two members of the ``$R$ at
that stage'', which we replace by their union, were subsets of two
\emph{different} $(R\cup S)$-connected components of $K$, while it has no
effect on the number of $(R\cup S)$-connected components of $K$ if the two
members of the ``$R$ at that stage'' which we replace by their union, were
subsets of the \emph{same} $(R\cup S)$-connected component of $K$.  Now 
the upper bound (\ref{E95}) on $r$ comes from noting that to reduce the
number of $(R\cup S)$-connected components of $K$ from its initial value
of $s$, to 1, there must be \emph{at least} $s-1$ such steps, and we now
see that to have $r=\mathbf{N}(R)$ equal to its maximum possible value of 
$k+1-s$, as given by (\ref{E95}), (which, as we have already seen, is a
necessary requirement for our term to contribute to the right-hand side of
the group-variation equation for an $f_0$), we must be able to construct
the ``true'' $R$ from the ``initial'' $R$ in \emph{exactly} $s-1$ such
steps, which immediately implies that at each step, the two members of the
``$R$ at that stage'' which we replace by their union, must be subsets of
\emph{different} $(R\cup S)$-connected components of $K$, (so that the 
number of $(R\cup S)$-connected components of $K$ decreases by 1 at each
step.)

Now the topological equivalent of replacing two members of the ``$R$ at
that stage'' by their union is given by the following consideration: at
the first step, since all the members of the ``initial'' $R$ are 
one-member subsets of $K$, corresponding to 2-spheres with one hole each,
with the boundary of each being a member of $K$, we are to form the union
of two one-member subsets of $K$.  We cut a small hole in each of the 
``2-spheres with one hole'' whose boundaries are the two relevant members
of $K$, and join those two ``2-spheres with one hole'' to one another by a
long tube, (which is itself a topological 2-sphere with two holes), sewing
each end of the tube to one of the ``small'' holes which we have just made
in each of those two ``2-spheres with one hole'', in such a way that the
orientations of the two ``2-spheres with one hole'' are consistent with
one another, and also with the orientation of the tube, or cylinder.  We
see immediately that the result of this operation is an oriented 
topological 2-sphere with two holes, exactly as required by our rule for a
two-member member of $R$.  Now it is a general property of oriented
topological 2-spheres with holes \cite{Henle} that if we have two oriented
topological 2-spheres, one with $u$ holes and one with $v$ holes, and we
make a small additional hole in each sphere, and join the two spheres
together by a long cylinder, sewing each end of the cylinder to one of the
small additional holes we just made in each of the spheres, in such a way
that the orientations of the two spheres are consistent with one another
and also consistent with the orientation of the cylinder, then the result
is an oriented topological 2-sphere with $u+v$ holes.  (It is not, of
course, necessary for the additional holes to be ``small'', or for the 
cylinder to be ``long'': it simply helps with the visualization of the
process.  In fact the cylinder or tube could be dispensed with, and the 
two spheres sewn directly to ne another at the boundaries of the 
additional holes, but we retain it to help with the visualization.)  Thus
every step of the process by which we transform the ``initial $R$'' into 
the ``true $R$'', (each step consisting of replacing two members of the 
``$R$ at that stage'' by their union), can be reproduced exactly in the
topological picture, by joining the appropriate two ``oriented
topological 2-spheres with holes'' (which correspond to the two members of
the ``$R$ at that stage'' whose union we are forming), into a new oriented
topological 2-sphere with holes, whose set of holes is the union of the 
sets of holes of the original two 2-spheres.

Now what happens to the $s$ oriented topological 2-spheres with holes, the
$i^\mathrm{th}$ having $n_i$ holes, that correspond to the $s$ connected
components of the decoration, (and which we formed as described above, by
filling \emph{every} member of $K$ by an oriented topological 2-sphere 
with one hole), when we carry out these operations step by step, 
corresponding exactly to the transformation of the ``initial $R$'', in
exactly $s-1$ steps, to the ``true $R$''?  We see immediately that
because, (in order to reach the ``true $R$'' in \emph{exactly} $s-1$ 
steps, as is necessary for $r=\mathbf{N}(R)$ to be equal to the upper 
limit as given by (\ref{E95})), we must, as just shown, at every step, 
form the union of two members of the ``$R$ at that stage'' that are 
subsets of two \emph{different} $(R\cup S)$-connected components of $K$,
every step reduces the number of connected components of our ``topological
model'' of our term by 1, and every step consists of joining two oriented
topological 2-spheres with holes, into a single oriented topological
2-sphere with holes, by the tube procedure, where this last conclusion
follows by induction on the number of steps so far performed.  In fact at
each step, the connected components of our topological model at that 
stage, exactly correspond to the $(R\cup S)$-connected components of $K$
at that stage.  \emph{Thus at the end of the procedure, (i.e. when we have
reached the ``true''} $R$\emph{, after having performed} $s-1$ 
\emph{steps), our topological model for this term consists of exactly one
oriented topological 2-sphere, with} $n$ \emph{holes, with the boundaries
of the holes being the initial $n$ Wilson loops, each oriented, (in the
sense of its path-ordering), consistently with the orientation of this one
oriented topological 2-sphere.}

We note that to reach this conclusion that the topological model of our
term is an oriented topological 2-\emph{sphere}, as opposed to a surface
of higher genus, (i.e. with handles), it is crucial that, at each step of
the above procedure, the two members of the ``$R$ at that stage'' whose
union we form, are subsets of two \emph{different} $(R\cup S)$-connected
components of $K$, (and thus members of two \emph{different} connected
components of the topological model at that stage), for if we were to make
the ``two small holes'', at any step, in a \emph{single} connected 
component of the topological model at that stage, the result of sewing 
the ends of the cylinder to the two small holes would be a surface of 
higher genus, i.e. with a handle.

Now as we have already remarked, an oriented topological 2-sphere with $n$
holes is equivalent to an oriented topological 2-disk with $n-1$ holes,
(since of course we assume $n\geq1$), so we see that we may represent the
topological model of our term by a diagram where we choose one of our
original $n$ Wilson loops as the outer boundary of a 2-disk, and draw the
other $n-1$ Wilson loops inside this outer boundary, (but \emph{not}
within one another), and draw the 45-paths of the decoration on the 
oriented 2-disk, with $n-1$ holes, thus formed.  We immediately see that
we can draw all the 45-paths without any crossings of their $\textrm{SU}(N)$
fundamental representation path-ordered phase factors, and respecting our
rule ``drive on the left'', provided that the \emph{outermost} boundary of
the disk is oriented \emph{anti-clockwise}, (i.e. has its path-ordering
arrows pointing \emph{anti-clockwise}), while each of the $n-1$ ``inner
boundaries'' are oriented \emph{clockwise}.  (This is precisely what is 
required for all $n$ boundaries of this disk, when it is stretched to look
like a 2-sphere with $n$ holes, to be oriented the same way.)  We can now
immediately read off this diagram what correlation function factors we
have in this term: a simply-connected ``window'', (or ``region''), of this
diagram corresponds to the vacuum expectation value of the single $\textrm{SU}(N)$
fundamental representation Wilson loop which forms its perimeter, a 
doubly-connected ``window'' corresponds to the correlation function of the
two $\textrm{SU}(N)$ fundamental representation Wilson loops which form its 
perimeter, and so on.

(For historical reasons, we note that ``window-weighted'' path integrals
have previously been considered in the context of large-$N_c$ QCD by 
Migdal and Makeenko \cite{MigMak}.  However our equations have absolutely
no connection at all with theirs.)

And furthermore, we can go the other way: given the oriented 2-disk with
$n-1$ holes, and its boundaries oriented as described, and any planar
diagram of 45-paths drawn on this disk, with \emph{no} crossings of 
$\textrm{SU}(N)$ fundamental representation path-ordered phase factors, and the 
only vertices being:
	\unitlength 1pt
\linethickness{0.61pt}
\begin{equation}
	\label{E103}
	\raisebox{-26pt}{
	\begin{picture}(58.00,53.00)
\put(28.00,30.00){\line(1,0){30.00}}
\put(28.00,24.00){\line(1,0){30.00}}
\put(28.00,30.00){\line(-1,1){23.00}}
\put(28.00,24.00){\line(-1,-1){23.00}}
\put(22.00,27.00){\line(-1,-1){21.00}}
\put(22.00,27.00){\line(-1,1){21.00}}
\end{picture}
}\qquad\qquad
	\raisebox{-33pt}{
	\begin{picture}(67.00,67.00)
\put(1.00,37.00){\line(1,0){30.00}}
\put(31.00,37.00){\line(0,1){30.00}}
\put(37.00,67.00){\line(0,-1){30.00}}
\put(37.00,37.00){\line(1,0){30.00}}
\put(1.00,31.00){\line(1,0){30.00}}
\put(31.00,31.00){\line(0,-1){30.00}}
\put(37.00,1.00){\line(0,1){30.00}}
\put(37.00,31.00){\line(1,0){30.00}}
\end{picture}
}\qquad\qquad
	\raisebox{-20pt}{
\begin{picture}(67.00,41.00)
\put(1.00,11.00){\line(1,0){30.00}}
\put(31.00,11.00){\line(0,1){30.00}}
\put(37.00,41.00){\line(0,-1){30.00}}
\put(37.00,11.00){\line(1,0){30.00}}
\put(20.00,11.00){\line(-1,1){10.00}}
\put(10.00,1.00){\line(1,1){10.00}}
\put(56.00,11.00){\line(-1,1){10.00}}
\put(46.00,1.00){\line(1,1){10.00}}
\end{picture}
}
\end{equation}
(where the single line with arrows represents part of one of the 
boundaries of the disk), we obtain a diagram which precisely corresponds
to the topological model of one of our terms, provided the number of 
connected components of the \emph{drawing}, (i.e. of the 
\emph{decoration}, \emph{not} the topological model), cannot be increased
by deleting just one 45-path, (this corresponds t one of our selection
rules).  The diagram \emph{can}, of course, have ``islands'', such as:
	\unitlength 1.00pt
\linethickness{0.61pt}
\begin{equation}
	\label{E104}
	\raisebox{-28pt}{
\begin{picture}(63.00,56.00)
\put(57.00,40.00){\line(-1,1){10.00}}
\put(47.00,50.00){\line(-1,0){12.00}}
\put(35.00,50.00){\line(0,-1){43.00}}
\put(35.00,7.00){\line(1,0){12.00}}
\put(47.00,7.00){\line(1,1){10.00}}
\put(57.00,17.00){\line(0,1){23.00}}
\put(7.00,40.00){\line(1,1){10.00}}
\put(17.00,50.00){\line(1,0){12.00}}
\put(29.00,50.00){\line(0,-1){43.00}}
\put(29.00,7.00){\line(-1,0){12.00}}
\put(17.00,7.00){\line(-1,1){10.00}}
\put(7.00,17.00){\line(0,1){23.00}}
\put(50.00,1.00){\line(1,1){13.00}}
\put(1.00,43.00){\line(1,1){13.00}}
\put(1.00,43.00){\line(0,-1){29.00}}
\put(1.00,14.00){\line(1,-1){13.00}}
\put(14.00,1.00){\line(1,0){36.00}}
\put(14.00,56.00){\line(1,0){36.00}}
\put(50.00,56.00){\line(1,-1){13.00}}
\put(63.00,43.00){\line(0,-1){29.00}}
\end{picture}
}
\end{equation}

In such a case the window surrounding the islands corresponds, by our
general rule, to the correlation function of the $\textrm{SU}(N)$ fundamental
representation Wilson loops that form the outer boundaries of the islands
and the outer perimeter of that window.

It is appropriate to note here that planar diagrams don't usually have
symmetry factors associated with specific subdiagrams.  This is because if
a subdiagram of a Feynman diagram or decoration is left invariant by a
certain group of permutations of its lines and vertices, there is usually
precisely the right number of ``leading large-$N$'' routeings of its
Kronecker deltas or path-ordered phase factors through its vertices, to
cancel its symmetry factor.  However some planar diagrams contributing to
Wilson loop vacuum expectation values, and similarly some of our present
diagrams, have symmetry factors corresponding to ``rotational'' symmetries
of the entire diagram.  These symmetries, and the corresponding symmetry
factors, vanish if we arbitrarily assign one point on each Wilson loop as
the ``base point'' of that loop, past which no vertex on that loop may go.
But if we do that, we immediately get a larger number of the less 
symmetric diagrams, which would be equivalent to one another but for 
inequivalent assignments of which adjacent pair of their vertices on the 
loop the base point goes between.  We will display our diagrams assuming
that such base points are \emph{not} introduced, so that in some cases we
will need to introduce symmetry factors.

Furthermore, in our diagrams we sometimes need symmetry factors associated
with individual ``islands'', due to rotational symmetries of those 
islands.  For example, the island (\ref{E104}) needs the symmetry factor
$\frac{1}{2}$, (provided none of its paths are Fadeev-Popov paths).  The
corresponding vacuum bubble, (with three gauge-field paths), has the 
symmetry factor $\frac{1}{6}$, but there are three possible choices of 
which one of the three $\textrm{SU}(N)$ fundamental representation Wilson loops of
that vacuum bubble, to make into the outside of the island, (i.e. to 
include in the same member of $R$ as some member of $K$ from another
connected component of the decoration).

It is now appropriate to introduce a notation to distinguish Fadeev-Popov
45-paths from gauge-field 45-paths, and we do this by putting a big arrow
on each Fadeev-Popov 45-path:
	\unitlength 1.00pt
\linethickness{0.61pt}
\begin{equation}
	\label{E105}
	\raisebox{-10pt}{
\begin{picture}(67.00,21.00)
\put(1.00,14.00){\line(1,0){66.00}}
\put(67.00,8.00){\line(-1,0){66.00}}
\put(39.00,11.00){\line(-1,1){10.00}}
\put(29.00,1.00){\line(1,1){10.00}}
\end{picture}
}
\end{equation}
We may consider the arrow, on a Fadeev-Popov propagator, as pointing from
the $\psi_A$ field to the $\phi_A$ field, hence all these arrows should
point in the same cyclic direction around any closed loop of Fadeev-Popov
paths.

For the remainder of this paper, any 45-path without such a big arrow, is
to be interpreted as a gauge-field path.

We now return for a moment to the group-changing equations for $(\textrm{SU}(N))^M$
and $\textrm{SU}(NM)$, in order to determine the $M$-dependence of the terms
corresponding to our diagrams, so that we can differentiate with respect 
to $M$, then set $M=1$.  We have two fundamental rules:
\begin{enumerate}
	\item If two $\textrm{SU}(N)$ fundamental representation Wilson loops share any
	45-path, they must be in \emph{different} $\textrm{SU}(N)$ subgroups of
	$(\textrm{SU}(N))^M$.  (This follows from the $\left(1-\delta{JK}\right)$ factor
	in the 45-term in (\ref{E67}).)
	\item All the $\textrm{SU}(N)$ fundamental representation Wilson loops involved 
	in any correlation function must be in the \emph{same} $\textrm{SU}(N)$ subgroup
	of $(\textrm{SU}(N))^M$.  (This follows from the fact that the $(\textrm{SU}(N))^M$
	correlation function of a set of $\textrm{SU}(N)$ fundamental representation
	Wilson loops vanishes unless all the $\textrm{SU}(N)$ fundamental representation
	Wilson loops in that set are in the \emph{same} $\textrm{SU}(N)$ subgroup of
	$(\textrm{SU}(N))^M$.  And if they \emph{are} all in the same $\textrm{SU}(N)$ subgroup of
	$(\textrm{SU}(N))^M$, then that $(\textrm{SU}(N))^M$ correlation function is equal to the
	corresponding $\textrm{SU}(N)$ correlation function.)
\end{enumerate}
With reference to (1) here, we note that our ``planar diagram'' rules,
namely that we use only the vertices (\ref{E103}), and that there must be
no crossings of $\textrm{SU}(N)$ fundamental representation path-ordered phase
factors, and our rule that it must not be possible to increase the number
of connected components of the decoration by deleting a single 45-path,
together imply, as required, that no $\textrm{SU}(N)$ fundamental representation
Wilson loop can pass in both directions along any 45-path.  This follows
by induction on $l$, exactly as for the corresponding result for planar
Feynman diagrams, (see the discussion between (\ref{E50}) and 
(\ref{E51})).

The implication of (2) for our diagrams is simple:  For each window of our
diagram, irrespective of whether that window is simply-connected or 
multiply-connected, \emph{all} the $\textrm{SU}(N)$ fundamental representation
Wilson loops that form the boundary of that window, must be in the 
\emph{same} $\textrm{SU}(N)$ subgroup of $(\textrm{SU}(N))^M$.

\subsection{Chromatic Polynomials}

Thus we see, if we disregard for a moment any possible complications 
associated with the vertices, that we have an analogue of the famous
``map-colouring'' problem.  Any allowed assignment of the members of $K$,
i.e. the $\textrm{SU}(N)$ fundamental representation Wilson loops, ot the various
$\textrm{SU}(N)$ subgroups of $(\textrm{SU}(N))^M$, consists of an assignment, consistent
with (1) and (2) above, of an integer in $\{1,2,\ldots,M\}$ to each 
$\textrm{SU}(N)$ fundamental representation Wilson loop.  We have to calculate the
$M$-dependence of the set of all such allowed assignments, defined for all
$M$ as the lowest-degree polynomial in $M$ that gives the correct result
for all integers $M\geq0$, (since this is what is given by the general
form of the $M$-dependence as discussed, e.g., before equation 
(\ref{E91}), and is also consistent with the ``simplest sum of powers'' by
which the $N$-dependence of the Feynman diagrams is generalized to all
$N$), then we have to calculate the derivative with respect to $M$ at 
$M=1$.

Instead of $M$ \emph{numbers} we may think of $M$ ``\emph{colours}''.  We
then see immediately that (2) implies that instead of colouring the 
\emph{lines}, i.e. the $\textrm{SU}(N)$ fundamental representation Wilson loops, we
may colour the windows, (since (2) says that all the lines that form the
boundary of any window, are to be coloured the same colour).  Then (1) is
the classic map-colouring requirement that ``countries'' that share a 
border must be coloured in different colours.  The polynomial 
$\mathbf{C}(M)$, defined as above to be the lowest-degree polynomial that
gives the correct number of distinct colourings, with $M$ available 
colours, for all integer $M\geq0$, is sometimes called the ``chromatic
polynomial''.  We note that some of our ``countries'' are of course 
allowed to be non-simply connected, i.e. to enclose or surround other
countries, or one or more of the $n-1$ ``interior'' holes, or both.
However we do \emph{not} have any ``disconnected'' countries, i.e.
countries with two or more connected components, as is sometimes allowed
in the map-colouring problem.

Now, are there any complications coming from the vertices?

We note first that the triple vertices certainly cause no problems.  For
only two windows meet at an original-path vertex, and those two windows
are forced by their common 45-path to be different colours, and similarly
only three windows meet at an action vertex, and they are again forced by
their common 45-paths to have three different colours.

However four windows meet at a quartic vertex, and the constraint (1)
allows four different ``partitions'' (into like-coloured subsets) of the
set of these four windows: all four may be different colours, one opposite
pair may be like-coloured, the other opposite pair being unlike-coloured,
(two different choices here, corresponding to which is the like-coloured
opposite pair), or both opposite pairs may be like-coloured.  However it
immediately follows from the discussion in the paragraph beginning
``Certain simplifications occur'', (after(\ref{E89})), that if all four 
paths ending at a quartic vertex are gauge-field paths, then when we add
the contributions of the $f_{aBC}f_{aEF}$ and the $f_{ABC}f_{AEF}$ terms
in (\ref{E31}), which give respectively the $-\delta_{jq}\delta_{kp}$ term
in (\ref{E58}), and the whole of (\ref{E59}), the result is simply the
$-\delta_{JQ}\delta_{jq}\delta_{KP}\delta_{kp}$ term in (\ref{E61}), which
is completely independent of which of the $M$ colours each of the four
$\textrm{SU}(N)$ path-ordered phase factors passing through the vertex is in.  In
fact, the $f_{aBC}f_{aEF}$ term in (\ref{E31}) gives the contribution of
all the ``colour assignments'' where both the phase-factor lines passing
through the ``source channel'' of that vertex, (i.e. the 2/2 channel which
has one structure constant at each ``end''), have the \emph{same} colour,
while the $f_{ABC}f_{AEF}$ term in (\ref{E31}) gives the contribution of
all the ``colour assignments'' where the two phase-factor lines passing
through the ``source channel'' of the vertex are assigned two
\emph{different} colours.  Thus we see that a quartic vertex at which
four gauge-field paths end gives \emph{no complications at all}: the
simple map-colouring rules given above, which take no account at all of
whether opposite pairs of windows at quartic vertices are like or unlike
coloured, are exactly correct.

However if two of the paths that end at a quartic vertex are Fadeev-Popov
paths, then the only possible source of that vertex is the term
$\psi_Af_{AaB}A_{\mu B}\phi_Cf_{CaD}A_{\mu D}$ in (\ref{E23}), which is a
``123-term'', in the sense that the suffix $a$ is summed only over the
1's, 2's, and 3's in (\ref{E57}).  Thus this Type-1 quartic vertex only
gets the Type-1 contribution from (\ref{E58}), namely the
$-\delta_{JA}\delta_{KA}\delta_{PA}\delta_{QA}\delta_{jq}\delta_{kp}$ term
in (\ref{E58}), summed over $A$ from 1 to $M$, to give $-\delta_{JQ}
\delta_{PK}\delta_{JK}\delta_{jq}\delta_{pk}$, (where the summation 
convention is of course \emph{not} applied to $J$ or $K$).  Thus this
vertex only contributes ``colour assignments'' where the two phase-factor
lines passing through it ``source channel'' are the \emph{same} colour.
This means that at such a quartic vertex in our diagrams, the opposite
pair of windows at that vertex which lie on either \emph{side} of the 
``source channel'' of that vertex, are to be coloured the \emph{same}
colour.  In other words, for the purposes of window colourings, it is just
as though that vertex has been sliced into two across its ``source 
channel'', so that the opposite pair of windows, which lie on either side
of the ``source channel'' of that vertex, are ``merged'' into a single
window.

Now the structure of the $\psi_Af_{AaB}A_{\mu B}\phi_Cf_{CaD}A_{\mu D}$
term in (\ref{E23}) shows that such a vertex has one Fadeev-Popov ``leg''
at each end of its ``source channel''.  Thus if the two Fadeev-Popov 
``legs'' of the vertex share a phase-factor line, or in other words, if
the two Fadeev-Popov ``legs'' of that vertex are ``neighbours'' in the 
planar diagram, for example:
	\unitlength 1.00pt
\linethickness{0.61pt}
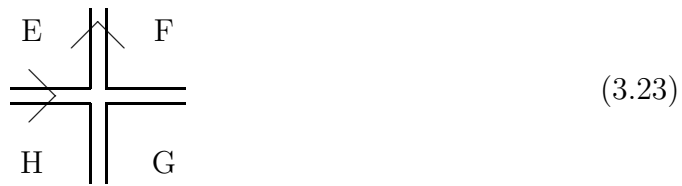
\begin{equation}
	\label{E106}
	\raisebox{-33pt}{
\begin{picture}(67.00,67.00)
\put(1.00,37.00){\line(1,0){30.00}}
\put(31.00,37.00){\line(0,1){30.00}}
\put(37.00,67.00){\line(0,-1){30.00}}
\put(37.00,37.00){\line(1,0){30.00}}
\put(1.00,31.00){\line(1,0){30.00}}
\put(31.00,31.00){\line(0,-1){30.00}}
\put(37.00,1.00){\line(0,1){30.00}}
\put(37.00,31.00){\line(1,0){30.00}}
\put(8.00,44.00){\line(1,-1){10.00}}
\put(18.00,34.00){\line(-1,-1){10.00}}
\put(24.00,52.00){\line(1,1){10.00}}
\put(34.00,62.00){\line(1,-1){10.00}}
\put(9.00,59.00){\makebox(0,0)[cc]{E}}
\put(59.00,59.00){\makebox(0,0)[cc]{F}}
\put(59.00,9.00){\makebox(0,0)[cc]{G}}
\put(9.00,9.00){\makebox(0,0)[cc]{H}}
\end{picture}
}
\end{equation}
then the ``source channel'' of the vertex is fixed uniquely.  In example
(\ref{E106}) the windows at the \emph{ends} of the source channel are $F$
and $H$, and the windows at the \emph{sides} of the source channel are $E$
and $G$.  Therefore, for purposes of calculating the chromatic polynomial
$\mathbf{C}(M)$ for one of our diagrams containing the vertex
(\ref{E106}), this vertex is to be ``sliced into two'' by a cut from
window $E$ to window $G$, so that windows $E$ and $G$ are merged into a
single window:
	\unitlength 1.00pt
\linethickness{0.61pt}
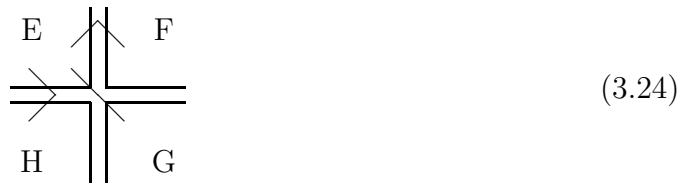
\begin{equation}
	\label{E107}
	\raisebox{-33pt}{
\begin{picture}(67.00,67.00)
\put(1.00,37.00){\line(1,0){30.00}}
\put(31.00,37.00){\line(0,1){30.00}}
\put(37.00,67.00){\line(0,-1){30.00}}
\put(37.00,37.00){\line(1,0){30.00}}
\put(1.00,31.00){\line(1,0){30.00}}
\put(31.00,31.00){\line(0,-1){30.00}}
\put(37.00,1.00){\line(0,1){30.00}}
\put(37.00,31.00){\line(1,0){30.00}}
\put(8.00,44.00){\line(1,-1){10.00}}
\put(18.00,34.00){\line(-1,-1){10.00}}
\put(24.00,52.00){\line(1,1){10.00}}
\put(34.00,62.00){\line(1,-1){10.00}}
\put(9.00,59.00){\makebox(0,0)[cc]{E}}
\put(59.00,59.00){\makebox(0,0)[cc]{F}}
\put(59.00,9.00){\makebox(0,0)[cc]{G}}
\put(9.00,9.00){\makebox(0,0)[cc]{H}}
\put(24.00,44.00){\line(1,-1){20.00}}
\end{picture}
}
\end{equation}

However if the two Fadeev-Popov ``legs'' of a quartic vertex at which two
Fadeev-Popov paths end, do \emph{not} share any phase-factor line, or in
other words, if the two Fadeev-Popov ``legs'' of the vertex are ``opposite
legs'' in the planar diagram, for example:
	\unitlength 1.00pt
\linethickness{0.61pt}
\begin{equation}
	\label{E108}
	\raisebox{-33pt}{
\begin{picture}(67.00,67.00)
\put(1.00,37.00){\line(1,0){30.00}}
\put(31.00,37.00){\line(0,1){30.00}}
\put(37.00,67.00){\line(0,-1){30.00}}
\put(37.00,37.00){\line(1,0){30.00}}
\put(1.00,31.00){\line(1,0){30.00}}
\put(31.00,31.00){\line(0,-1){30.00}}
\put(37.00,1.00){\line(0,1){30.00}}
\put(37.00,31.00){\line(1,0){30.00}}
\put(8.00,44.00){\line(1,-1){10.00}}
\put(18.00,34.00){\line(-1,-1){10.00}}
\put(9.00,59.00){\makebox(0,0)[cc]{E}}
\put(59.00,59.00){\makebox(0,0)[cc]{F}}
\put(59.00,9.00){\makebox(0,0)[cc]{G}}
\put(9.00,9.00){\makebox(0,0)[cc]{H}}
\put(62.00,34.00){\line(-1,1){10.00}}
\put(62.00,34.00){\line(-1,-1){10.00}}
\end{picture}
}
\end{equation}
then the ``source channel'' of the vertex is \emph{not} fixed uniquely:
its ends could be windows $E$ and $G$, or windows $F$ and $H$.  Now use of
(\ref{E62}) shows that $f_{AaB}f_{CaD}$, with the conventions of our 
diagrams, has the structure:
	\unitlength 1.00pt
\linethickness{0.61pt}
\begin{equation}
	\label{E109}
	-\raisebox{-26pt}{
\begin{picture}(53.00,53.00)
\put(47.00,47.00){\line(0,-1){11.00}}
\put(47.00,47.00){\line(-1,0){11.00}}
\put(14.00,14.00){\line(0,-1){11.00}}
\put(14.00,14.00){\line(-1,0){11.00}}
\put(21.00,29.00){\line(0,-1){4.00}}
\put(25.00,21.00){\line(1,0){4.00}}
\put(33.00,25.00){\line(0,1){4.00}}
\put(29.00,33.00){\line(-1,0){4.00}}
\put(41.00,27.00){\line(-1,0){28.00}}
\put(53.00,5.00){\line(-1,1){20.00}}
\put(33.00,29.00){\line(1,1){20.00}}
\put(49.00,53.00){\line(-1,-1){20.00}}
\put(5.00,53.00){\line(1,-1){20.00}}
\put(1.00,49.00){\line(1,-1){20.00}}
\put(1.00,5.00){\line(1,1){20.00}}
\put(5.00,1.00){\line(1,1){20.00}}
\put(49.00,1.00){\line(-1,1){20.00}}
\put(1.00,53.00){\makebox(0,0)[cb]{B}}
\put(53.00,53.00){\makebox(0,0)[cb]{A}}
\put(53.00,1.00){\makebox(0,0)[ct]{D}}
\put(1.00,1.00){\makebox(0,0)[ct]{C}}
\end{picture}
}+\raisebox{-26pt}{
\begin{picture}(53.00,53.00)
\put(14.00,14.00){\line(0,-1){11.00}}
\put(14.00,14.00){\line(-1,0){11.00}}
\put(21.00,29.00){\line(0,-1){4.00}}
\put(25.00,21.00){\line(1,0){4.00}}
\put(33.00,25.00){\line(0,1){4.00}}
\put(29.00,33.00){\line(-1,0){4.00}}
\put(41.00,27.00){\line(-1,0){28.00}}
\put(53.00,5.00){\line(-1,1){20.00}}
\put(33.00,29.00){\line(1,1){20.00}}
\put(49.00,53.00){\line(-1,-1){20.00}}
\put(5.00,53.00){\line(1,-1){20.00}}
\put(1.00,49.00){\line(1,-1){20.00}}
\put(1.00,5.00){\line(1,1){20.00}}
\put(5.00,1.00){\line(1,1){20.00}}
\put(49.00,1.00){\line(-1,1){20.00}}
\put(1.00,53.00){\makebox(0,0)[cb]{A}}
\put(53.00,53.00){\makebox(0,0)[cb]{B}}
\put(53.00,1.00){\makebox(0,0)[ct]{D}}
\put(1.00,1.00){\makebox(0,0)[ct]{C}}
\put(7.00,47.00){\line(1,0){11.00}}
\put(7.00,47.00){\line(0,-1){11.00}}
\end{picture}
}+\raisebox{-26pt}{
\begin{picture}(53.00,53.00)
\put(40.00,14.00){\line(0,-1){11.00}}
\put(40.00,14.00){\line(1,0){11.00}}
\put(33.00,29.00){\line(0,-1){4.00}}
\put(29.00,21.00){\line(-1,0){4.00}}
\put(21.00,25.00){\line(0,1){4.00}}
\put(25.00,33.00){\line(1,0){4.00}}
\put(13.00,27.00){\line(1,0){28.00}}
\put(1.00,5.00){\line(1,1){20.00}}
\put(21.00,29.00){\line(-1,1){20.00}}
\put(5.00,53.00){\line(1,-1){20.00}}
\put(49.00,53.00){\line(-1,-1){20.00}}
\put(53.00,49.00){\line(-1,-1){20.00}}
\put(53.00,5.00){\line(-1,1){20.00}}
\put(49.00,1.00){\line(-1,1){20.00}}
\put(5.00,1.00){\line(1,1){20.00}}
\put(53.00,53.00){\makebox(0,0)[cb]{A}}
\put(1.00,53.00){\makebox(0,0)[cb]{B}}
\put(1.00,1.00){\makebox(0,0)[ct]{D}}
\put(53.00,1.00){\makebox(0,0)[ct]{C}}
\put(47.00,47.00){\line(-1,0){11.00}}
\put(47.00,47.00){\line(0,-1){11.00}}
\end{picture}
}-\raisebox{-26pt}{
\begin{picture}(53.00,53.00)
\put(7.00,47.00){\line(0,-1){11.00}}
\put(7.00,47.00){\line(1,0){11.00}}
\put(40.00,14.00){\line(0,-1){11.00}}
\put(40.00,14.00){\line(1,0){11.00}}
\put(33.00,29.00){\line(0,-1){4.00}}
\put(29.00,21.00){\line(-1,0){4.00}}
\put(21.00,25.00){\line(0,1){4.00}}
\put(25.00,33.00){\line(1,0){4.00}}
\put(13.00,27.00){\line(1,0){28.00}}
\put(1.00,5.00){\line(1,1){20.00}}
\put(21.00,29.00){\line(-1,1){20.00}}
\put(5.00,53.00){\line(1,-1){20.00}}
\put(49.00,53.00){\line(-1,-1){20.00}}
\put(53.00,49.00){\line(-1,-1){20.00}}
\put(53.00,5.00){\line(-1,1){20.00}}
\put(49.00,1.00){\line(-1,1){20.00}}
\put(5.00,1.00){\line(1,1){20.00}}
\put(53.00,53.00){\makebox(0,0)[cb]{B}}
\put(1.00,53.00){\makebox(0,0)[cb]{A}}
\put(1.00,1.00){\makebox(0,0)[ct]{D}}
\put(53.00,1.00){\makebox(0,0)[ct]{C}}
\end{picture}
}
\end{equation}
The first and last terms here show that whenever the vertex (\ref{E108})
occurs in our diagrams, we are, for the purposes of calculating the 
chromatic polynomial $\mathbf{C}(M)$ of the diagram, to sum over the two
possible identifications of the ``source channel'' of the vertex.  Thus we
adopt the diagrammatic identity:
	\unitlength 1.00pt
\linethickness{0.61pt}
\begin{equation}
	\label{E110}
	\raisebox{-33pt}{
\begin{picture}(67.00,67.00)
\put(1.00,37.00){\line(1,0){30.00}}
\put(31.00,37.00){\line(0,1){30.00}}
\put(37.00,67.00){\line(0,-1){30.00}}
\put(37.00,37.00){\line(1,0){30.00}}
\put(1.00,31.00){\line(1,0){30.00}}
\put(31.00,31.00){\line(0,-1){30.00}}
\put(37.00,1.00){\line(0,1){30.00}}
\put(37.00,31.00){\line(1,0){30.00}}
\put(8.00,44.00){\line(1,-1){10.00}}
\put(18.00,34.00){\line(-1,-1){10.00}}
\put(9.00,59.00){\makebox(0,0)[cc]{E}}
\put(59.00,59.00){\makebox(0,0)[cc]{F}}
\put(59.00,9.00){\makebox(0,0)[cc]{G}}
\put(9.00,9.00){\makebox(0,0)[cc]{H}}
\put(62.00,34.00){\line(-1,1){10.00}}
\put(62.00,34.00){\line(-1,-1){10.00}}
\end{picture}
}=\raisebox{-33pt}{
\begin{picture}(67.00,67.00)
\put(1.00,37.00){\line(1,0){30.00}}
\put(31.00,37.00){\line(0,1){30.00}}
\put(37.00,67.00){\line(0,-1){30.00}}
\put(37.00,37.00){\line(1,0){30.00}}
\put(1.00,31.00){\line(1,0){30.00}}
\put(31.00,31.00){\line(0,-1){30.00}}
\put(37.00,1.00){\line(0,1){30.00}}
\put(37.00,31.00){\line(1,0){30.00}}
\put(8.00,44.00){\line(1,-1){10.00}}
\put(18.00,34.00){\line(-1,-1){10.00}}
\put(9.00,59.00){\makebox(0,0)[cc]{E}}
\put(59.00,59.00){\makebox(0,0)[cc]{F}}
\put(59.00,9.00){\makebox(0,0)[cc]{G}}
\put(9.00,9.00){\makebox(0,0)[cc]{H}}
\put(62.00,34.00){\line(-1,1){10.00}}
\put(62.00,34.00){\line(-1,-1){10.00}}
\put(24.00,44.00){\line(1,-1){20.00}}
\end{picture}
}+\raisebox{-33pt}{
\begin{picture}(67.00,67.00)
\put(1.00,37.00){\line(1,0){30.00}}
\put(31.00,37.00){\line(0,1){30.00}}
\put(37.00,67.00){\line(0,-1){30.00}}
\put(37.00,37.00){\line(1,0){30.00}}
\put(1.00,31.00){\line(1,0){30.00}}
\put(31.00,31.00){\line(0,-1){30.00}}
\put(37.00,1.00){\line(0,1){30.00}}
\put(37.00,31.00){\line(1,0){30.00}}
\put(8.00,44.00){\line(1,-1){10.00}}
\put(18.00,34.00){\line(-1,-1){10.00}}
\put(9.00,59.00){\makebox(0,0)[cc]{E}}
\put(59.00,59.00){\makebox(0,0)[cc]{F}}
\put(59.00,9.00){\makebox(0,0)[cc]{G}}
\put(9.00,9.00){\makebox(0,0)[cc]{H}}
\put(62.00,34.00){\line(-1,1){10.00}}
\put(62.00,34.00){\line(-1,-1){10.00}}
\put(24.00,24.00){\line(1,1){20.00}}
\end{picture}
}
\end{equation}
(We may confirm that this is correct by symmetrizing $f_{AaB}f_{CaD}$ 
under $B\rightleftharpoons D$.)

For practical purposes, we may consider the ``slice'' through such a 
vertex as simply connecting together the two windows which are to be 
coloured in the same colour.  (The other opposite pair of windows, i.e.
the pair \emph{separated} by the ``slice'', may be either like-coloured or
unlike-coloured.)

Thus a vertex like (\ref{E108}), shown without a ``slice'', is to be 
interpreted by (\ref{E110}).  We don't usually display the ``slice'' for a
vertex like (\ref{E106}), since the only place it can go is as shown in
(\ref{E107}).

Thus, in summary, the chromatic polynomial $\mathbf{C}(M)$ is to be
calculated, for each of our diagrams, as the lowest-degree polynomial in
$M$ which gives correctly, for every integer $M\geq0$, the number of 
distinct ways of colouring the windows of that diagram with $M$ available
colours, subject to the requirement that if two windows share a common
``boundary'', (i.e. a common 45-path), they are to be coloured in two
\emph{different} colours, and subject to the special rules, as just
explained, for windows that meet at a quartic vertex that has two
Fadeev-Popov ``legs''.

We then differentiate the chromatic polynomial $\mathbf{C}(M)$ with 
respect to $M$, and evaluate the derivative $\frac{d}{dM}\mathbf{C}(M)$ at
$M=1$.  This gives the coefficient with which that diagram occurs in the
right-hand side of the group-variation equation for the appropriate $f_0$.
(For the purposes of the present discussion, we consider the possible
``symmetry factors'', associated with ``rotational symmetries'' of the
diagram, as discussed above, to be an ``intrinsic'' part of the
mathematical expression corresponding to the diagram.)

\chapter{Vanishing Of The Chromatic Polynomial Factor When There Is More Than One Island, Effective Mass For The 45-Paths From The Window Weights, and Minimal-Length Spanning Trees}

\section{Examples of the diagrams, and their chromatic polynomials}

We now consider some examples of our diagrams, and their associated 
chromatic polynomials $\mathbf{C}(M)$, first for $n=1$:
	\unitlength 1.00pt
\linethickness{0.61pt}
\begin{equation}
	\label{E111}
	\raisebox{-14pt}{

}\qquad\quad\mathbf{C}(M)=M(M-1)^3\qquad\quad
\end{equation}

We note that the chromatic polynomial $\mathbf{C}(M)$ may be calculated
for any of our diagrams as the sum, over all permitted partitions of the
set of all the windows into like-coloured sets, of 
$M(M-1)\ldots(M-p+1)=\frac{M!}{(M-p)!}$, where $p$ is the number of parts
of the partition.  However it may frequently be calculated more quickly by
other methods.

We next consider some examples for $n=2$:
	\unitlength 1.00pt
\linethickness{0.61pt}
\begin{equation}
	\label{E158}
	\raisebox{-51pt}{
%
}\qquad\quad\mathbf{C}(M)=M(M-1)^2\qquad\quad
\end{equation}
and some examples for $n=3$:
	\unitlength 1.00pt
\linethickness{0.61pt}
\begin{equation}
	\label{E174}
	\raisebox{-51pt}{
%
}\qquad\mathbf{C}(M)=M(M-1)(M-2)\qquad
\end{equation}

\section{Circumstances that guarantee that $\mathbf{C}(M)$ has at least two factors of $(M-1)$}

We now observe, firstly, that for \emph{every} diagram, apart from those,
namely (\ref{E111}), (\ref{E158}), and (\ref{E174}), that have \emph{no}
45-paths, $\mathbf{C}(M)$ includes at least one factor of $(M-1)$.  This is
indeed exactly what we expect, since it is impossible to colour a diagram
that includes 45-paths with just one colour.

But we also observe that for a large number of the diagrams, $\mathbf{C}(
M)$ includes two or more factors of $(M-1)$, so that $\left.\frac{d}{dM}
\mathbf{C}(M)\right|_{M=1}$ is equal to zero, \emph{so that that diagram
makes no contribution at all to the right-hand side of the relevant
group-variation equation.}

Indeed for \emph{any} diagram, such that the set $U$ of all the windows of
that diagram, can be partitioned into three nonempty sets $X$, $Y$, and
$Z$, such that $X$ contains exactly one window, and no member of $Y$ 
shares a 45-path with any member of $Z$, $\mathbf{C}(M)$ will have at 
least two factors of $(M-1)$.  For $\mathbf{C}(M)$ in such a case factors
into a factor $M$ for the one window in $X$, and a factor associated with
$Y$, and a factor associated with $Z$, and the factor associated with $Y$
and the factor associated with $Z$ \emph{each} contain at least on factor
of $(M-1)$.  (Indeed, $\mathbf{C}(M)$ for any such diagram is equal to 
$\frac{1}{M}$, times the chromatic polynomial for the diagram obtained 
from the given one by removing all the 45-paths that separate the windows
in $Y$ from one another and from the window in $X$, times the chromatic
polynomial for the diagram obtained from the given one by removing all the
45-paths that separate the windows in $Z$ from one another and from the 
window in $X$, and each of these last two chromatic polynomials includes
at least one factor of $(M-1)$.)

\emph{Hence for any diagram, such that the set} $U$ \emph{of all the 
windows of that diagram, can be partitioned into three nonempty parts}
$X$, $Y$, \emph{and} $Z$, \emph{such that} $X$ \emph{contains exactly one
window, and no window in} $Y$ \emph{shares any 45-path with any window in}
$Z$, \emph{the coefficient} $\left.\frac{d}{dM}\mathbf{C}(M)\right|_{M=1}$
vanishes, \emph{so that that diagram makes no contribution at all to the 
right-hand side of the relevant group-variation equation.}

This is a crucial result of this paper, since it immediately eliminates
vast classes of diagrams from the right-hand sides of the 
group-variation equations.

We now define, in the context of our diagrams, an ``island'' to be, 
precisely, a vacuum bubble.  Thus an island is a connected component of a
diagram, that does \emph{not} involve any of the $n$ original Wilson 
loops.  In the above examples, the diagrams with islands are:
(\ref{E135}) - (\ref{E157}) inclusive, (\ref{E163})-(\ref{E167})
inclusive, (\ref{E171}), (\ref{E172}), and (\ref{E180}) - (\ref{E183})
inclusive.  And among these, (\ref{E139}) has three islands, (\ref{E138}),
(\ref{E164}), (\ref{E166}), and (\ref{E167}) each have two islands, and
the remainer all have one island.

We now see that it follows immediately from the above results, that if a 
diagram has an island, then it must have \emph{no} 45-paths other than
those that form part of that island, if it is to give a nonvanishing
contribution.  Indeed, the chromatic polynomial of a diagram with an
island, is equal to $\frac{1}{M}$, times the chromatic polynomial of the 
diagram obtained from the given diagram by removing all 45-paths that
do \emph{not} form part of the island, times the chromatic polynomial of 
the diagram obtained from the given diagram by removing the island, and if
there are any 45-paths that do \emph{not} form part of the island, then
\emph{both} these last two factors will include a factor $(M-1)$.
\emph{Hence we immediately conclude, in particular, that if a diagram is
to give a nonvanishing contribution, it must contain at most one island,
and, if it} does \emph{have an island, it must have} no \emph{45-paths 
that do} not \emph{form part of that island.}

We call a diagram an ``island diagram'' if it has an island, and a
``non-island diagram'' if it has no island.

We next define, in the context of our diagrams, a ``band'' to be a 
connected component of what remains of a diagram after we remove all the 
islands and all the ``original paths'', (i.e. all the paths that form
wholes or parts of the $n$ original Wilson loops).  In the above examples,
the diagrams with \emph{one} band are (\ref{E112}), (\ref{E115}), 
(\ref{E116}), (\ref{E118}), (\ref{E120}) - (\ref{E134}) inclusive,
(\ref{E155}), (\ref{E156}), (\ref{E168}), (\ref{E169}), (\ref{E171}), and
(\ref{E175}), the diagrams with \emph{two} bands are (\ref{E113}),
(\ref{E117}), (\ref{E119}), (\ref{E157}), (\ref{E159}), (\ref{E173}),
(\ref{E176}), and (\ref{E179}), the diagrams with \emph{three} bands are
(\ref{E114}), (\ref{E160}), (\ref{E170}), (\ref{E172}), and (\ref{E178}),
the diagrams with \emph{four} bands are (\ref{E161}) and (\ref{E177}),
and the diagram with \emph{five} bands is (\ref{E162}).  We note that each
band has at least two ``free ends'', (i.e. where a 45-path ends at an
original path).

We now find another ``selection rule'': if a diagram has a band, all of 
whose free ends are on a \emph{single} one of the $n$ original closed
paths, then that diagram must have \emph{no} 45-paths, apart from those in
that band, if it is to give a nonzero contribution.  This implies, in
particular, that a diagram that gives a nonvanishing contribution to the
right-hadn side of the group-variation equation for $f_0(W_1,g^2)$, has at
most one band.

There is, however, for $n\geq2$, no such limit on the number of bands,
each of which has its free ends on two or more \emph{different} original
closed paths.  Indeed, the generalization of examples (\ref{E159}) -
(\ref{E162}) to the ``$b$-spoked wheel'', ($b\geq2$), contributes, for 
$n=2$, with the coefficient $(-1)^b$.  (To see this, we pick one of the 
$b$ windows, and classify the allowed colourings by the total number, 
$c\geq1$, of the $b$ windows that are coloured the same colour as the 
chosen window.  The contribution to $\mathbf{C}(M)$ from colourings with a
total of $c$ windows coloured the same as the chosen window, is given by
$M(M-1)^c(M-2)^{b-2c}$ times an integer coefficient that counts the number
of ways of choosing the $(c-1)$ windows that are to be coloured the same
as the chosen window.  Only $c=1$ contributes to the derivative at $M=1$,
and the number of ways in this case is 1.)

We note that, for each diagram that gives a nonvanishing contribution, 
apart from the diagrams that have \emph{no} 45-paths, the chromatic
polynomial $\mathbf{C}(M)$ includes precisely one factor of $(M-1)$.  Thus
evaluating the derivative at $M=1$ reduces to removing the factor of 
$(M-1)$, and evaluating what remains at $M=1$.

\section{The Group-Variation Equation for the vacuum expectation value of one Wilson loop}

The explicit form of the group-variation equation for $f_0(W_1,g^2)$ is:
	\unitlength 1.00pt
\linethickness{0.61pt}
	\[f_0(W_1,g^2)+g^2\frac{d}{dg^2}f_0(W_1,g^2)=
	\left\{\raisebox{-42pt}{\rule{0pt}{86pt}}\right.
	\left(\raisebox{-30pt}{\rule{0pt}{62pt}}\right.
\raisebox{-16pt}{

}+\textrm{non-island diagrams with }l\geq3
	\left.\raisebox{-30pt}{\rule{0pt}{62pt}}\right)+
\]
	\unitlength 1.00pt
\linethickness{0.61pt}
	\[+
	\left(\raisebox{-38pt}{\rule{0pt}{78pt}}\right.
	\raisebox{-38pt}{
%
}+\:\textrm{island diagrams with }l\geq2
	\left.\raisebox{-38pt}{\rule{0pt}{78pt}}\right)
	\left.\raisebox{-42pt}{\rule{0pt}{86pt}}\right\}
\end{equation}

Here $l$ denotes, as usual, the number of new paths, (i.e. 45-paths), 
minus the number of action vertices, and we continue to use the 
convention, as stated before (\ref{E111}), that where a diagram has a
``symmetry factor'' associated with a rotational symmetry, (for example, a
factor $\frac{1}{2}$ for the diagram (\ref{E112}), and a factor 
$\frac{1}{3}$ for the diagram (\ref{E115})), this factor is considered to 
be an intrinsic part of the mathematical expression corresponding to the
diagram, and is thus not displayed explicitly.

The left-hand side of (\ref{E184}) is as given by (\ref{E91}), with $r=0$
and $n=1$.  Now the first term in the right-hand side of (\ref{E184}),
namely the diagram:
	\unitlength 1.00pt
\linethickness{0.61pt}
\begin{equation}
	\label{E185}
	\raisebox{-16pt}{
\begin{picture}(34.00,34.00)
\put(34.00,24.00){\line(-1,1){10.00}}
\put(24.00,34.00){\line(-1,0){13.00}}
\put(11.00,34.00){\line(-1,-1){10.00}}
\put(1.00,24.00){\line(0,-1){13.00}}
\put(1.00,11.00){\line(1,-1){10.00}}
\put(11.00,1.00){\line(1,0){13.00}}
\put(24.00,1.00){\line(1,1){10.00}}
\put(34.00,11.00){\line(0,1){13.00}}
\put(31.00,8.00){\line(-1,0){4.00}}
\put(31.00,8.00){\line(0,-1){4.00}}
\end{picture}
}
\end{equation}
is simply the diagrammatic form of $f_0(W_1,g^2)$, so we could of course
cancel this term between the two sides of (\ref{E184}).  However, each 
side of equation (\ref{E184}) has a very simple form in terms of
\emph{Feynman} diagrams, as follows: each side of equation (\ref{E184}) is
equal to the sum of all the leading large-$N$ \emph{Feynman} diagrams that
contribute to the vacuum expectation value of the Wilson loop $W_1$, with
each \emph{Feynman} diagram being simply multiplied by its number of 
windows, (i.e. by its number of closed loops of Kronecker deltas).

In fact, for each $n\geq1$, each side of the group-variation equation for
\\ $f_0(W_1,\ldots,W_n,g^2)$, when written with the left-hand side being 
given by (\ref{E91}), with the appropriate $n$, and $r=0$, (and of course,
the overall power of $N$ cancelled out), is equal to the sum of all the
\emph{Feynman} diagrams contributing to $f_0(W_1,\ldots,W_n,g^2)$, with
each \emph{Feynman} diagram being simply multiplied by its number of 
windows.

Indeed, as follows immediately from the discussion preceding (\ref{E54}),
each Feynman diagram contributing to $f_0(W_1,\ldots,W_n,g^2)$ with $t$
powers of $g^2$, has precisely $t+2-n$ windows, i.e. $t+2-n$ closed loops
of Kronecker deltas.  Thus each Feynman diagram with $u$ windows that 
contributes to $f_0(W_1,\ldots,W_n,g^2)$, has precisely $u+n-2$ powers of
$g^2$.  Hence $g^2\frac{d}{dg^2}$ simply has the effect of multiplying 
each $u$-windowed Feynman diagram by a factor $(u+n-2)$, which immediately
gives the result stated.

We therefore choose the definitive form of each group-variation equation,
(for purposes of referring to its ``left-hand side'' or its ``right-hand
side''), to be that where the left-hand side is given by the appropriate
case of (\ref{E91}), (with the overall power of $N$ cancelled out), in
order to retain this simple interpretation, in terms of Feynman diagrams,
for $r=0$.

We can now sketch a very rough \emph{direct} proof that the right-hand 
side of the group-variation equation for each $f_0$ is indeed equal to 
the same sum of Feynman diagrams, with the same coefficients, as the
left-hand side.  Consider any \emph{Feynman} diagram that contributes to
$f_0(W_1,\ldots,W_n,g^2)$, and let $u$ be its number of windows.  Now the
number of ways of colouring each of the $u$ windows of this \emph{Feynman}
diagram \emph{independently} in any of $M$ colours is simply $M^u$.
Taking the derivative of this with respect to $M$ at $M=1$ gives
$\left(uM^{u-1}\right)_{M=1}=u$, which is the coefficient of that Feynman
diagram in the left-hand side of the group-variation equation for 
$f_0(W_1,\ldots,W_n,g^2)$.  And, given any colouring of the $u$ windows of
this Feynman diagram independently in any of $M$ colours, we erase from
that diagram all propagators that have windows of the same colour on each
side.  The result is, roughly speaking, a valid colouring of one of our
right-hand side group-variation equation diagrams, and we immediately see,
that out of all the $M^u$ independent colourings of the $u$ windows of 
that Feynman diagram with $M$ colours, the number of times we get that
particular group-variation diagram is precisely given by the chromatic
polynomial, $\mathbf{C}(M)$, for that diagram, so that, taking the 
derivative with respect to $M$ at $M=1$, the result is precisely
$\left.\frac{d}{dM}\mathbf{C}(M)\right|_{M=1}$, in agreement with the 
group-variation equations.  This argument is of course incomplete since
we have totally ignored the kinematic and Lorentz-index structure of the 
vertices.  We also have to consider all the distinct ways that the cubic
and quartic vertices arise from the sums over paths given by the 
group-variation equation gauge-field and Fadeev-Popov propagators, for 
example four-gluon vertices arise firstly from $\sigma\to0$ contributions
to (\ref{E27}) where one or more of the individual short straight 
segments have \emph{two} $A_{\mu a}$'s each on them, (which gives the
$AA$ terms in (\ref{E28})), secondly from $F_{\mu\nu}$ insertions as given
by (\ref{E18}), and thirdly from $\frac{1}{E}$ as given by (\ref{E16}),
(\ref{E20}), and (\ref{E30}).  (The $\sigma\to0$ limit of (\ref{E27}) is
of course dominated by contributions where most of the straight segments
have \emph{no} $A_{\mu a}$'s on them, some have one $A_{\mu a}$ on them
and some have two $A_{\mu a}$'s on them.)  Finally we use Mills-type
\cite{Mills} propagator gauge invariance identities ot transform the sums
of Feynman diagrams we get in the right-hand sides of the group-variation
equations, which involve extra vertices corresponding to our choice
(\ref{E6}) of gauge-fixing action, and the corresponding Fadeev-Popov 
action (\ref{E7}), to sums of ordinary Feynman diagrams.  For full details
of all this see the next paper in this series.

\section{The Group-Variation Equation for the correlation function of two Wilson loops}

The explicit form of the group-variation equation for $f_0(W_1,W_2,g^2)$
is:
	\unitlength 1.00pt
\linethickness{0.61pt}
	\[g^2\frac{d}{dg^2}f_0(W_1,W_2,g^2)=
	\left\{\raisebox{-58pt}{\rule{0pt}{118pt}}\right.
	\left(\raisebox{-51pt}{\rule{0pt}{104pt}}\right.
	\raisebox{-51pt}{

}+
\]
	\unitlength 1.00pt
\linethickness{0.61pt}
	\[+\textrm{ non-island diagrams with }l\geq3
	\left.\raisebox{-51pt}{\rule{0pt}{104pt}}\right)
	+
	\left(\raisebox{-51pt}{\rule{0pt}{104pt}}\right.
	\raisebox{-51pt}{
%
}+\textrm{ island diagrams with }l\geq1
	\left.\raisebox{-51pt}{\rule{0pt}{104pt}}\right)
	\left.\raisebox{-58pt}{\rule{0pt}{118pt}}\right\}
\end{equation}

\section{Effective mass for the 45-paths, from the window weights in the path integrals}

We now outline a demonstration that the solutions of the group-variation
equations have the correct qualitative behaviour, namely the Wilson area
law \cite{Wilson} for the vacuum expectation value of one Wilson loop, and 
massive glueball saturation of the correlation functions of two or more
Wilson loops.

Let us \emph{suppose} that the vacuum expectation values of large Wilson
loops have the qualitative behaviour $e^{-\mu^2A}$, where $\mu$ is a fixed
mass and $A$ is the area of the minimal-area orientable spanning surface 
of the loop \cite{Courant}.  What consequences will this have for the sums 
over 45-paths bordered by simply-connected windows in the right-hand sides
of the group-variation equations?  It is clear that the sums over paths
will be suppressed in comparison to the ``free-path'' case (i.e. with
$\frac{-1}{\partial^2}$ given by the $\sigma\to0$ limit of (\ref{E27}) with
$\left(W_{xz_1z_2\ldots z_ny}\right)_{AB}$ removed), since the 
``free-path'' weight-factor of every path is now multiplied by an
additional factor whose magnitude is $\leq1$, and whose magnitude gets
smaller and smaller, as the area of the minimal-area spanning surface of 
the paths around the edge of any simply-connected window, gets larger and
larger.

\subsection{Quantitative estimate of the effective mass}

To get a first quantitative estimate of this effect we consider a 45-path
whose ends are well separated, and which has on each side, a
simply-connected window, such that all the vertices of those two windows
are well separated from one another, and all lie roughly in a single
two-dimensional plane.  Let the ends of the 45-path of interest be at $x$
and $y$, and let $B$ denote the 2-plane in which our two windows roughly
lie.  We consider paths from $x$ to $y$ whose projections into the 2-plane
$B$ follow the straight line from $x$ to $y$.  We also require that in each
of the $4-2=2$ dimensions perpendicular to the 2-plane $B$, the components
of our paths in these two ``transverse'' dimensions, considered as 
functions of distance along the straight line from $x$ to $y$, remain small
near $x$ and $y$, but may become larger, but not too large, between $x$ and
$y$, but well away from both.  We choose a coordinate system that has its
first axis along the straight line from $x$ to $y$, and its second axis
perpendicular to this line in the 2-plane $B$, so that the third and fourth
axes are perpendicular to the 2-plane $B$.  Then under the conditions 
stated, and in the approximation of retaining only zeroth, first, and
second order terms in the transverse components of our paths, the area of
the minimal-area spanning surface of either of our two windows, is equal
to the area of the projection into the 2-plane $B$, (which, under our
assumptions, is fixed), plus, for each transversed dimension $z$, a
contribution:
\begin{equation}
	\label{E187}
	\int_{-\infty}^{\infty}\int_{-\infty}^{\infty}\mathrm{d}s\mathrm{d}t
	\frac{(z(s)-z(t))^2}{4\pi(s-t)^2}
\end{equation}

Here $s$ and $t$ represent distance along the straight line from $x$ to 
$y$, and we are able to extend the limits to $\pm\infty$ due to our
requirement that $z$ be negligibly small except in the region between $x$
and $y$ and well away from both of them.  ((\ref{E187}) is given by the 
solution of Laplace's equation in a half-plane with the boundary function
$z$.)

Our first estimate of the window-weighted path integrals is then based on
the following approximations:

(i)  We neglect all effects on the window areas of deviations from
straightness of the projection of our path into the 2-plane $B$.  This
will have \emph{no effect} on the sum of the ``planar components'' of the
two areas, provided the projection of our path has no self-intersections,
since whatever is lost by one window is gained by the other.  However we
may assume that we will \emph{underestimate} the suppression of paths
whose projections into the 2-plane $B$ have self-intersections.

(ii)  We use the formula (\ref{E187}) for the transverse components of the
window areas, with the further approximation that the parameters $s$ and
$t$ are simply proportional to the number of segments along the path.  
Thus in term $n$ in (\ref{E27}) we remove the factor $\left(W_{xz_1z_2
\ldots z_ny}\right)_{AB}$ and include instead, for each of our two windows
and each of the two dimensions perpendicular to the 2-plane $B$, a factor
given by the exponential of:
\begin{equation}
	\label{E188}
	-\frac{\mu^2}{4\pi}\sum_{\begin{array}{c}1\leq s\leq n \\ 1\leq t\leq n
	\\ s\neq t \end{array}}\frac{\left(z_s-z_t\right)^2}{(s-t)^2}
\end{equation}

This is actually not as bad as it may seem, since we know from Douglas's
solution of Plateau's problem \cite{Courant} that the area of the
minimal-area orientable spanning surface of \emph{any} simple closed path
may be expressed in the quadratic form (\emph{E187}), summed over 
\emph{all} $d$ dimensions, (in our case, 4), in which the path exists,
provided that the \emph{parametrization} of the path is the special
parametrization that satisfies Douglas's variational condition
\cite{Courant}.  This implies, in particular, that for any \emph{planar}
simple closed path without self-intersections, there is a parametrization
such that the area enclosed by the path is given by (\ref{E187}), summed
now over the two dimensions \emph{in} the plane of the path.  This
parametrization must be the \emph{same} as that for which (\ref{E187}),
summed over the $4-2=2$ dimensions \emph{perpendicular} to the base-plane
of the path, correctly gives the transverse contributions to the area of
the minimal-area spanning surface, for small perturbations of the path
perpendicular to its base plane.  (Thus for a simple planar closed path,
the correct parametrization is given by the conformal transformation that
maps the real axis onto the path, for example the Schwarz-Christoffel
transformation fi the path cnsists of a finite number of straight 
segments.)  It follows that (\ref{E188}) is only wrong to the extent that
the parametrization given by simply counting the number of straight
segments along the path from $x$, is not the solution of Douglas's
variational condition.  And in the central region of the itegration domain
of the two components of the $z_i$'s in the 2-plane $B$, where these two
components of the $z_i$'s are distributed uniformly and in order along the
straight line from $x$ to $y$, our parametrization does coincide roughly
with the solution of Douglas's variational condition.  (Douglas's
variational condition is simply the requirement that the parametrization
of the path be such that the sum of (\ref{E187}) over \emph{all} $d$
dimensions, (i.e. 4 dimensions in our case), in which the path exists, has
the minimum possible value among all parametrizations of the path.  The
real axis may of course be conformally mapped to the unit circle if 
required by $s\to\frac{s-i}{s+i}$.)

(iii)  Our third approximation is that we treat the effects of the ends of
our path in the simplest possible way.

Now we readily find, by induction on $n$, that for all $n\geq0$:
\begin{equation}
	\label{E189}
	\int\mathrm{d}z_1\ldots\int\mathrm{d}z_n\frac{e^{-\frac{\left(x-z_1
	\right)^2}{4\sigma}}}{\sqrt{4\pi\sigma}}\frac{e^{-\frac{\left(z_1-z_2
	\right)^2}{4\sigma}}}{\sqrt{4\pi\sigma}}\ldots\frac{e^{-\frac{\left(z_n-
	y\right)^2}{4\sigma}}}{\sqrt{4\pi\sigma}}=\frac{e^{-\frac{\left(x-y
	\right)^2}{4\sigma(n+1)}}}{\sqrt{4\pi\sigma(n+1)}}
\end{equation}
Now if we set $x=y=0$ in this formula, as is appropriate, in our chosen
coordinate system, for the two dimensions perpendicular to the 2-plane 
$B$, we may obtain the same result by finding, for $n\geq1$, the 
eigenvalues of the quadratic form in the $z_s$'s, $1\leq s\leq n$, that 
occurs in the exponent in the left-hand side.

Indeed, the normalized eigenstates of this quadratic form are given by:
\begin{equation}
	\label{E190}
	z_s=\sqrt{\frac{2}{(n+1)}}\sin qs\qquad\qquad\qquad\qquad1\leq s\leq n
\end{equation}
where
\begin{equation}
	\label{E191}
	q=\frac{m\pi}{n+1}\qquad\qquad\qquad\qquad1\leq m\leq n
\end{equation}
The corresponding eigenvalues are given by:
\begin{equation}
	\label{E192}
	\frac{(1-\cos q)}{2\sigma}=\frac{1}{\sigma}\sin^2\left(\frac{q}{2}
	\right)
\end{equation}
We do indeed find that the product, over all the eigenvalues $\lambda$, of
$\sqrt{\frac{\pi}{\lambda}}$, is equal to $\frac{(4\pi\sigma)^{\frac{n}
{2}}}{\sqrt{n+1}}$, obtaining incidentally, for all integers $n\geq1$, the
identities:
\begin{equation}
	\label{E193}
	\prod_{m=1}^n\sin\left(\frac{m\pi}{2(n+1)}\right)=2^{-n}\sqrt{n+1}
\end{equation}

We shall estimate the effect of including in the integrand of 
(\ref{E189}), a factor given by the exponential of (\ref{E188}) for the 
window on each side of our path, (or in other words, a net factor given by
the exponential of twice (\ref{E188})), by calculating, in first order
perturbation theory, the modifications to the eigenvalues (\ref{E192})
produced by including twice (\ref{E188}) in the exponent, and we readily
find, in the approximation of neglecting the effects of the ends of the
path, that (\ref{E192}) is modified by the addition of:
\begin{equation}
	\label{E194}
	\frac{2\mu^2}{\pi}\left((1-\cos q)+\frac{1}{4}(1-\cos 2q)+\frac{1}{9}
	(1-\cos 3q)+\ldots\right)
\end{equation}
(In fact, in this approximation, the eigenstates (\ref{E190}),
(\ref{E191}) are also eigenstates of (\ref{E188}).)

Now (\ref{E194}) is equal to:
\begin{equation}
	\label{E195}
	\mu^2\left(\left|q\right|-\frac{q^2}{2\pi}\right)
\end{equation}
for all $q$ such that $-2\pi\leq q\leq2\pi$.  However the convergence is
very slow for small $\left|q\right|$, (and for $q$ near $-2\pi$ and near
$2\pi$), hence, bearing in mind that the relevant values (\ref{E191}) of
$q$ all lie between 0 and $\pi$, we may expect a problem, associated with
our neglect of the details at the ends of the path, for $q\to0$.

Now adding (\ref{E195}) to (\ref{E192}) is equivalent to multiplying
(\ref{E192}) by:
\begin{equation}
	\label{E196}
	\left(1+\frac{\mu^2\sigma\left(\left|q\right|-\frac{q^2}{2\pi}\right)}
	{\sin^2\left(\frac{q}{2}\right)}\right)
\end{equation}

Now in the notation of equations (\ref{E25}) and (\ref{E26}), $(n+1)
\sigma=s$, where $s$ is the integration variable in equation (\ref{E24}).
We are interested in the limit $\sigma\to0$ with $s$ fixed, so the 
product of (\ref{E196}) over all the values of $q$, as given by 
(\ref{E191}), exponentiates.  Thus multiplication of the integrand of 
(\ref{E189}), which refers to \emph{one} transverse dimension, by the 
exponential of (\ref{E188}), results in multiplying the ``free path''
factor
	\[\left(e^{s\partial^2}\right)_{x,y}=\frac{e^{-\frac{(x-y)^2}{4s}}}
	{(4\pi s)^2}
\]
in the integrand of (\ref{E24}) by:
	\[\exp\left(-\frac{\mu^2s}{2\pi}\sum_{m=1}^n\left(\frac{\pi}{n+1}
	\right)\left(
	\frac{\left(\left|q\right|-\frac{q^2}{2\pi}\right)}{\sin^2\left(
	\frac{q}{2}\right)}\right)\right)\qquad\qquad\qquad\qquad\to
\]
\begin{equation}
	\label{E197}
	\to\qquad\qquad\qquad\qquad \exp\left(-\frac{\mu^2s}{2\pi}\int_0^\pi
	\mathrm{d}q\left(\frac{\left(\left|q\right|-\frac{q^2}{2\pi}\right)}
	{\sin^2\left(\frac{q}{2}\right)}\right)\right)
\end{equation}

Now we have $4-2=2$ transverse dimensions, hence within our approximations
the effect of our window weights on our sum over paths is simply to 
multiply the free path factor in the integrand of equation (\ref{E24}) by
the square of (\ref{E197}).  This means that the integrand of the $s$
integral in (\ref{E24}) has now become precisely that for a
\emph{massive}, \emph{free} scalar particle, with mass squared given by:
\begin{equation}
	\label{E198}
	\frac{\mu^2}{\pi}\int_0^\pi\mathrm{d}q\left(\frac{\left(\left|q\right|-
	\frac{q^2}{2\pi}\right)}{\sin^2\left(\frac{q}{2}\right)}\right)
\end{equation}

Now as we anticipated, there is a problem for $q\to0$: the integral 
diverges logarithmically.  This is because the sum in (\ref{E194}) should
have been cut off after about $n$ terms due to the ends of the path.
This does not matter for $q$ not close to 0, since the sum converges well
for $q$ not close to 0, (or an integer multiple of $2\pi$), but the sum
converges very slowly for $q$ close to 0.  (This is a direct consequence
of the discontinuity of the derivative of (\ref{E195}) at $q=0$ - the 
Gibbs phenomenon.)  Therefore the integrand in (\ref{E198}) should be
smoothly cut off as $q\to0$.  Now the integrand in (\ref{E198}) is well
approximated for all $0<q<2\pi$ by:
\begin{equation}
	\label{E199}
	\frac{4}{q}+\frac{4}{2\pi-q}-\left(\frac{8}{\pi}-\frac{\pi}{2}\right)
\end{equation}

If we replace the term $\frac{4}{q}$ by $\frac{4}{\epsilon}$ for $0\leq q
\leq\epsilon$ then the integral in (\ref{E198}) becomes:
\begin{equation}
	\label{E200}
	4+4\ln\left(\frac{\pi}{\epsilon}\right)+4\ln2-\left(8-\frac{\pi^2}{2}
	\right)=4\ln\left(\frac{\pi}{\epsilon}\right)+3.7
\end{equation}
For example, $\epsilon=1$ gives 8.3, hence an effective mass $1.6\mu$, 
while $\epsilon=\frac{1}{4}$ gives 13.8, hence an effective mass $2.1\mu$.

For $q$ away from 0, the integrand in (\ref{E198}) is not sensitive to the
effects of the ends of the path, hence since the integrand in (\ref{E198})
decreases monotonically from $q=0$ to $q=\pi$, with the derivative being
zero at $q=\pi$, and the value $\frac{\pi}{2}$ of the integrand at $q=\pi$
being the absolute minimum value of the integrand function for all
$-2\pi\leq q\leq2\pi$, a conservative estimate of the effective mass is
given by replacing the integrand in (\ref{E198}) by its minimum value
$\frac{\pi}{2}$ for all $0\leq q\leq\pi$, which gives the value $\mu
\sqrt{\frac{\pi}{w}}=1.3\mu$.

The important point is that the predominant effect, on the right-hand
sides of the group-variation equations, of a qualitative behaviour
$e^{-\mu^2A}$ of the vacuum expectation values of large Wilson loops, 
where $A$ is the area of the minimal-area orientable spanning surface of
the loop, is that for each 45-path such that at least one of the two
windows beside it, is simply connected, the sum over paths is
approximately the free propagator for a \emph{massive} particle, with
mass approximately $\mu$, between those two vertices, \emph{and thus is
suppressed exponentially, by a factor} $e^{-\mu r}$, \emph{for} $r\geq
\mu$, \emph{where} $r$ \emph{is the distance between the ends of the 
path.}

\section{Consequences of the effective mass, for different types of diagrams}

Now in the group-variation equation (\ref{E184}) for $f_0(W_1,g^2)$,
\emph{every} 45-path has at least one simply-connected window beside it,
hence every path brings in a factor $e^{-\mu r}$, where $r$ is the
distance between the ends of the path.  For the one-loop islands we may
fix two or three points on the path, which corresponds to representing
$e^{s\bar{D}^2}$ in (\ref{E32}) as $e^{\frac{s}{3}\bar{D}^2}
e^{\frac{s}{3}\bar{D}^2}e^{\frac{s}{3}\bar{D}^2}$ (for three points 
fixed), and we again obtain, for each pair of consecutive fixed points
around the path, an exponential suppression factor $e^{-\mu r}$, where
$r$ is the distance between those two fixed points.  Thus when the size of
the loop $W_1$ is larger than $\frac{1}{\mu}$, the predominant 
contributions to the right-hand side of the group-variation equation
(\ref{E184}) for $f_0(W_1,g^2)$ come from, firstly, non-island diagrams,
whose one band is roughly of size $\frac{1}{\mu}$, and thus creeps like a
blob along the loop $W_1$, and secondly, from \emph{island} diagrams, with
the size of the island being roughly $\mu$.  (Of course, as we have 
already noted, the first term in the right-hand side, which has no band
or island, is simply $f_0(W_1,g^2)$, amd cancels the $f_0(W_1,g^2)$ term
in the left-hand side.)

Furthermore, looking at the right-hand side of the group-variation 
equation (\ref{E186}) for $f_0(W_1,W_2,g^2)$, we see that apart from the
first term, which is simply $f_0(W_1,W_2,g^2)$ itself, there are, firstly,
non-island diagrams with exactly one band, with all the ends of that band
being on \emph{one} of the two loops $W_1$ and $W_2$.  Every 45-path in
such a band has at least one simply-connected window beside it, and thus
the predominant contributions of such diagrams come from a band of size
$\frac{1}{\mu}$, which thus, for two large and well-separated loops, 
creeps like a blob along the loop to which it is attached.  Secondly, 
there are non-island diagrams with one or more bands, each of which has at
least one end on \emph{both} the loops $W_1$ and $W_2$.  (The simplest
such diagram with exactly one band has $l=3$.)  In such diagrams
\emph{every} window is simply-connected, hence again the predominant
contributions of such diagrams come from bands of size $\frac{1}{\mu}$.
Thus if we define $r$ to be the smallest distance between any point on
$W_1$, and any point on $W_2$, and $r$ is greater than $\frac{1}{\mu}$, 
then the total contributions of such diagrams are suppressed by an 
exponential factor $e^{-\mu r}$.  Thirdly there are island diagrams where,
just as in the corresponding diagrams in the group-variation equation for
$f_0(W_1,g^2)$, among all the windows beside paths of the island, exactly
\emph{one} of those windows is \emph{not} simply-connected.  We call
island diagrams of this type \emph{Type-1 island diagrams.}  Every 45-path
of such an island has at leat one simply-connnected window beside it, 
hence the predominant contributions of such island diagrams, when at least
one of $W_1$ and $W_2$ and $r$, (the shortest distance between $W_1$ and
$W_2$), is largee, come from islands of size $\mu$.  And finally there are
island diagrams of a new sort, the simplest being example (\ref{E165}),
where among all the windows beside paths of the island, two or more of
those windows are \emph{not} simply-connected.  We call all island 
diagrams of this type \emph{Type-2 island diagrams.}

We note that for all $n\geq1$, it is precisely the Type-1 island diagrams
that result in $f_0(W_1,\ldots,W_{n+1},g^2)$ occurring in the right-hand
side of the group-variation equation for $f_0(W_1,\ldots,W_n,g^2)$.
Indeed, in a Type-1 island diagram in the right-hand side of the 
group-variation equation for $f_0(W_1,\ldots,W_n,g^2)$, the one window
beside paths of the island that is \emph{not} simply connected, is
topologically a sphere with $n+1$ holes, (or a disk with $n$ holes), and
thus corresponds to $f_0(W_1\ldots,W_{n+1},g^2)$.  And in every diagram in
the right-hand side of the group-variation equation for
$f_0(W_1,\ldots,W_n,g^2)$ that is \emph{not} a Type-1 island diagram,
every window is topologically a disk with at most $n-1$ holes, and thus
corresponds to $f_0(W_1,\ldots,W_m,g^2)$, where $1\leq m\leq n$.  Thus
although the group-variation equations for the $f_0$'s do indeed couple
together the various $f_0$'s, the mixing is of a very simple form: in the
right-hand side of the group-variation equation for $f_0(W_1,\ldots,W_n,
g^2)$, the \emph{only} $f_0(W_1,\ldots,W_m,g^2)$ with $m>n$ that occurs is
$f_0(W_1,\ldots,W_{n+1},g^2)$, and this \emph{only} occurs in the Type-1
island diagrams.  Furthermore, the dependence of the right-hand side of 
the group-variation equation for $f_0(W_1,\ldots,W_n,g^2)$ on 
$f_0(W_1,\ldots,W_{n+1},g^2)$, is \emph{linear}: in terms of 
$f_0(W_1,\ldots,W_{n+1},g^2)$, the right-hand side of the group-variation
equation for $f_0(W_1,\ldots,W_n,g^2)$, is equal to a term independent of
$f_0(W_1,\ldots,W_{n+1},g^2)$, plus a term linear in $f_0(W_1,\ldots,W_
{n+1},g^2)$.

\section{The u and d quark static masses}

We note in passing that the quantitative value of the effective mass 
produced by the window weights with just \emph{one} window beside the path
is in fact of great interest, since it determines the effective quark mass
to be used in static quark calculations, (e.g. Ref 
\cite{Perkins 160-163}).  The effective mass squared for $4-2=2$ 
transverse dimensions and \emph{one} window beside the path is half
(\ref{E198}), hence if, as suggested, we replace the integrand of 
(\ref{E198}) by its minimum value of $\frac{\pi}{2}$ for all $0\leq q\leq
\pi$, we obtain $\mu\sqrt{\frac{\pi}{4}}=0.89\mu$.  Experimentally
\cite{Perkins 299}, $\mu$ is equal to $\frac{1}{\sqrt{2\pi\alpha'}}=0.41
\ \textrm{GeV}$, where $\alpha'=0.93\ \textrm{GeV}^{-2}$ is the universal
Regge slope, from which we obtain $\mu\sqrt{\frac{\pi}{4}}=0.37
\ \textrm{GeV}$, in agreement with the observed $u$ and $d$ static quark
masses \cite{Perkins 160-163}.

\section{Qualitative behaviour of the correlation functions}

We now develop a hypothesis for the qualitative behaviour
of $f_0(W_1,\ldots,W_n,g^2)$ for $n\geq2$, in preparation
for stating, in the next chapter, our ansatz for the
correlation functions,
 which we will show, in Chapter 7, is 
consistent with the group-variation equations, once
it has been modified, by the inclusion of appropriate
pre-exponential factors, for the terms in the
correlation functions.

\subsection{Minimal-area spanning surfaces of higher topology}

Let the sizes and separations of $W_1,\ldots,W_n$ all be $\geq\frac{1}
{\mu}$.

Let $S$ be the minimal-area orientable spanning surface of 
$W_1,\ldots,W_n$.  Formally, $S$ is defined to be the 2-dimensional
manifold which, among all the measurable orientable 2-dimensional
manifolds with boundary $W_1,\ldots,W_n$, is the one with the smallest
possible area.  (The orientations of the manifolds considered are required
to agree with the given orientations of $W_1,\ldots,W_n$.)  $S$ is
\emph{not} required to be connected.  Courant \cite{Courant} has shown 
that an $S$ realizing the minimum possible area always exists.  Usually
$S$ is unique, but in certain cases there may be two or more different
surfaces $S$ which have the minimum possible area.  Our loops
$W_1,\ldots,W_n$ are always defined by a finite number of parameters, (in
practice they are always formed from a finite number of straight segments,
but we may consider circles in examples).  The degenerate cases occur 
when $W_1,\ldots,W_n$ has two or more different \emph{locally} minimal
spanning surfaces, in the sense of having everywhere vanishing mean 
curvature, and as the parameters defining $W_1,\ldots,W_n$ are varied
continuously, a transition takes place between which of the different
\emph{locally} minimal spanning surfaces, is the one which gives the 
absolute minimum value of the area.  Thus the degenerate cases occur at a
set of points of measure zero in the finite-dimensional space of the
parameters defining $W_1,\ldots,W_n$, and we may for practical purposes
assume that $S$ is unique.

For example, suppose that $n=2$ and that we have two circles of equal
radius $r$, lying in a common 3-dimensional subspace of our 4-dimensional
Euclidean space, such that both circles are perpendicular to the straight
line between their centres.  Let the two circles have \emph{opposite}
orientations, (so that, if view as wheels, they would rotate in opposite
directions about their common axis), and let $d$ be the distance between
their centres.  Then one locally minimal oriented spanning surface of 
$W_1$ and $W_2$, which exists for all $d>0$, is simply the union of the
flat disks which fill the two circles.  It has two connected components, 
and its area is $2\pi r^2$.  And for $d\leq1.3255 r$ there is also a
\emph{locally} minimal oriented spanning surface which has the topology of
a cylinder, (i.e. a sphere with two holes).  This surface is a surface of
revolution about the common axis of the two ``wheels''.  It is described
by:
\begin{equation}
	\label{E201}
	s=\frac{1}{\alpha}\cosh\alpha t
\end{equation}
where $t$ denotes the distance along the common axis of the two ``wheels''
from the point midway between them, $s$ denotes the radial distance
perpendicular to this axis, and $\alpha$ is determined by:
\begin{equation}
	\label{E202}
	\alpha r=\cosh\left(\alpha\frac{d}{2}\right)
\end{equation}

The solution of (\ref{E202}) may be found graphically by superimposing, on
the graph of the curve $y=\cosh x$, the straight line through the origin
with slope $\frac{2r}{d}$.  If the straight line cuts the curve at
$(x,y)$, $x>0$, then a solution of (\ref{E202}) is given by $\alpha=
\frac{2x}{d}$.  We see that if $\frac{2r}{d}\geq1.5089$, (where $1.5089=
\frac{\cosh1.1997}{1.1997}$, and 1.1997 is the positive solution of
$x\sinh x=\cosh x$), then (\ref{E202}) in fact has \emph{two} solutions,
each of which gives a \emph{locally} minimal orientable spanning surface
of $W_1$ and $W_2$.  However for $\frac{2r}{d}>1.5089$ the solution with 
the larger value of $x$, and hence the larger value of $\alpha$, always
has larger area than the other one, and hence may be discarded.  For
$\frac{2r}{d}=1.5089$ the two solutions coincide, while for $\frac{2r}{d}
<1.5089$ there are \emph{no} locally minimal orientable spanning surfaces
with cylinder topology: the straight line misses the curve $y=\cosh x$
altogether.

Furthermore, even when the locally minimal surface of cylinder topology
does exist, its area is not necessarily smaller than the area of the two
disks.  Its area is:
\begin{equation}
	\label{E203}
	\int_{-\frac{d}{2}}^{\frac{d}{2}}2\pi s\sqrt{1+\left(\frac{\mathrm{d}s}
	{\mathrm{d}t}\right)^2}\mathrm{d}t=\frac{\pi}{\alpha}\left(d+\frac{
	\sinh\alpha d}{\alpha}\right)
\end{equation}
where $\alpha$ is determined, for $\frac{2r}{d}\geq1.5089$, or in other
words, for $d\leq1.3255 r$, as the smaller of the two positive solutions 
of (\ref{E202}).  For small $\frac{d}{r}$, the area is approximately equal
to $2\pi rd$, and for fixed $r$, it is a monotonically increasing function
of $d$.  It becomes equal to $2\pi r^2$ for $d=1.0554r$, (where $1.0554=
\frac{2\times0.6392}{\cosh0.6392}$, and 0.6392 is the solution of
$\cosh^2x=x+\frac{1}{2}\sinh2x$).

Thus in the present example, our surface $S$, which by definition is the
\emph{absolute} minimal-area orientable spanning surface of the two 
circles, is equal to the cylinder-topology locally minimal spanning 
surface, as just constructed, for $d<1.0554r$, and equal to the union of 
the flat disks which fill the two circles, for $d>1.0554r$.  The 
degenerate case, where $S$ could be either the cylinder-topology surface
or the union of the two flat disks, occurs only for $d=1.0554r$ which
defines a measure-zero subspace of the two-dimensional space of the 
parameters $d$ and $r$.

We note that if we had chosen the two circles to have the \emph{same}
orientation, (so that, viewed as wheels, they would rotate in the 
\emph{same} direction about their common axis), then $S$, the absolute
minimal area orientable spanning surface of the two circles, would have
been the union of the two flat disks for \emph{all} $d>0$.

\subsection{Minimal-length spanning trees}

Returning now to the general case of $W_1,\ldots,W_n$, let 
$S_1,\ldots,S_m$, where $1\leq m\leq n$, denote the separate connected
components of our absolute minimal-area orientable spanning surface $S$.
We now require to find the ``tree'' of straight line segments, of the 
smallest possible total length, such that the ends of the straight line
segments are either at junctions of the tree or on connected components of
$S$, and such that the union of $S$ and all the straight line segments is
\emph{connected}.

For example, if $m=1$, so that $S$ is itself connected, we do not need any
straight line segments at all.  If $m=2$, we have exactly one straight
line, which has one end on $S_1$ and one end on $S_2$.  It is the shortest
possible straight line segment that has one end on $S_1$ and one end on
$S_2$.  (There may, once again, in some cases be a degeneracy, even a
continuous degeneracy, among different possible such shortest line
segments.  This occurs, for example, in the above example of two circles
in any axisymmetric configuration within a three-dimensional subspace of
our four-dimensional space, when the absolute minimal-area orientable
spanning surface of the two circles is the union of the two flat disks.
Such cases, once again, occur only in domains of measure zero in the 
finite-dimensional spaces of the parameters of our loops.  Furthermore,
our ansatz will depend, firstly, on the total length of all the straight
line segments, and secondly, on the lengths of the individual straight 
line segments, but not on the precise positions of the individual line
segments.)  If $m=3$, we may either have two straight line segments, each
of which has its ends on two different members of $\{S_1,S_2,S_3\}$, (in
which case one of $S_1$, $S_2$, and $S_3$ has two straight lines ending on
it, and the other two each have one straight line ending on them), or else
we may have three straight lines, each of which has one end on one of
$S_1$, $S_2$, and $S_3$, and its other end at a junction where all three
straight lines meet.

Such ``minimal-length spanning trees'' have, as is well known, the 
following properties:\\
\noindent(i) At most three straight lines meet at any junction of
straight lines.\\
\noindent(ii) At any junction where three straight lines meet, all three
straight lines lie within a single 2-plane, and the angle between each
pair of lines is $\frac{2\pi}{3}$.

Indeed, let us consider any junction where exactly three straight lines
meet.  In general, the position of the junction, plus the positions of the
other ends of each of the three straight lines, defines a
three-dimensional space.  Within this three-dimensional space, we project
the junction perpendicularly onto the two-dimensional plane defined by the
outer ends of the three straight lines, and if the junction was not in 
this two-dimensional plane to start with, we thereby obtain a new position
of the junction giving a strictly smaller total length of the three 
straight lines.  Thus we may assume that the junction lies within the 
two-dimensional plane defined by the outer ends of the three straight 
lines.  Let the angles made at the junction by the second and third
straight lines, measured in the same sense from the first straight line,
be $\phi$ and $\psi$ respectively.  Let us now consider an alternative
position of the junction, obtained by moving the old junction a distance
$\epsilon$ along the straight line in this two-plane starting at the old
junction, and makin an angle $\theta$ with the first straight line, 
measured in the same sense as before.  Then if the lengths of the original
straight lines were $p$, $q$, and $r$, respectively, the lengths of the
new straight lines are $\sqrt{p^2+\epsilon^2-2p\epsilon\cos\theta}$,
$\sqrt{q^2+\epsilon^2-2q\epsilon\cos(\theta-\phi)}$, and
$\sqrt{r^2+\epsilon^2-2r\epsilon\cos(\theta-\psi)}$, respectively.  Thus
by expanding in powers of $\epsilon$, and dropping terms of second and
higher order in $\epsilon$, we find that the increase in length, for small
$\epsilon$, is:
	\[-\epsilon(\cos\theta+\cos(\theta-\phi)+\cos(\theta-\psi))=
\]
\begin{equation}
	\label{E204}
	=-\epsilon((1+\cos\phi+\cos\psi)\cos\theta+(\sin\phi+\sin\psi)\sin
	\theta)
\end{equation}
This has the form:
\begin{equation}
	\label{E205}
	-\epsilon\xi\cos(\theta-\zeta)
\end{equation}
where
\begin{equation}
	\label{E206}
	\xi^2=(1+\cos\phi+\cos\psi)^2+(\sin\phi+\sin\psi)^2
\end{equation}
and
\begin{equation}
	\label{E207}
	\tan\zeta=\frac{\sin\phi+\sin\psi}{1+\cos\phi+\cos\psi}
\end{equation}

Thus we can always find a $\theta$ which results in a reduction in the
total length, unless $\xi=0$.  Now $\xi=0$ implies:
\begin{equation}
	\label{E208}
	1+\cos\phi+\cos\psi=\sin\phi+\sin\psi=0
\end{equation}
And $\sin\phi+\sin\psi=0$ implies either $\psi=-\phi$ or $\psi=\pi+\phi$,
and $\psi=\pi+\phi$ implies $\cos\psi=-\cos\phi$, hence cannot be the 
solution.  Hence $\psi=-\phi$, hence $\cos\psi=\cos\phi=-\frac{1}{2}$,
hence $\phi=\pm\frac{2\pi}{3}$.

We now consider three points $a$, $b$, and $c$, and identify their 
minimal-length spanning tree.  We define:
	\[\alpha\equiv(b-c)^2,\ \beta\equiv(c-a)^2,\ \gamma\equiv(a-b)^2,
	\ \eta\equiv\frac{1}{2}(\alpha+\beta+\gamma),
\]
	\[A\equiv\frac{1}{4}\sqrt{2(\beta\gamma+\gamma\alpha+\alpha\beta)-
	(\alpha^2+\beta^2+\gamma^2)}
\]
	\[=\textrm{area of triangle abc},
\]
	\[\mathcal{X}\equiv\eta+2\sqrt{3}A,
\]
	\[\mathcal{A}\equiv\sqrt{3}(\eta-\alpha)+2A=\sqrt{3}(\mathcal{X}-
	\alpha)-4A,
\]
	\[\mathcal{B}\equiv\sqrt{3}(\eta-\beta)+2A=\sqrt{3}(\mathcal{X}-
	\beta)-4A,
\]
	\[\mathcal{C}\equiv\sqrt{3}(\eta-\gamma)+2A=\sqrt{3}(\mathcal{X}-
	\gamma)-4A,
\]
\begin{equation}
	\label{E209}
\mathcal{Z}\equiv\frac{\mathcal{ABC}}{4A\sqrt{3}\mathcal{X}},
	\ s\equiv\frac{\mathcal{Z}}{\mathcal{A}},\ t\equiv\frac{\mathcal{Z}}
	{\mathcal{B}},\ u\equiv\frac{\mathcal{Z}}{\mathcal{C}}
\end{equation}
and note the identities:
	\[(\eta-\beta)+(\eta-\gamma)=\alpha,\ \textrm{etc.,}
\]
	\[A=\frac{1}{4}\sqrt{2(\eta-\alpha)\alpha+2(\eta-\beta)\beta+2(\eta-
	\gamma)\gamma}
\]
	\[=\frac{1}{2}\sqrt{(\eta-\beta)(\eta-\gamma)+(\eta-\gamma)(\eta-\alpha)
	+(\eta-\alpha)(\eta-\beta)}
\]
	\[=\frac{1}{2}\sqrt{\beta\gamma-(\eta-\alpha)^2},\ \textrm{etc.,}
\]
	\[\mathcal{A}+\mathcal{B}+\mathcal{C}=\sqrt{3}\mathcal{X},
\]
	\[\mathcal{BC}+\mathcal{CA}+\mathcal{AB}=4A\sqrt{3}\mathcal{X},
	\qquad\qquad s+t+u=1,
\]
\begin{equation}
	\label{E210}
	\mathcal{A}^2+\mathcal{B}^2+\mathcal{C}^2=3\mathcal{X}^2-8A\sqrt{3}
	\mathcal{X}
\end{equation}

We note that it immediately follows from $(\eta-\beta)+(\eta-\gamma)=
\alpha$, etc., that the sum of any two of $\mathcal{A}$, $\mathcal{B}$, 
and $\mathcal{C}$, is $\geq0$, hence that at most one of $\mathcal{A}$,
$\mathcal{B}$, and $\mathcal{C}$, can be $<0$.  Furthermore, if
$\widehat{cab}$ denotes the angle between the lines $\overline{ab}$ and
$\overline{ac}$, then $\cos\widehat{cab}$ is greater than or less than
$-\frac{1}{2}$, according as $(\beta+\gamma-\alpha)$ is greater than or
less than $-\sqrt{\beta\gamma}$,hence according as $2(\alpha-\eta)$ is 
less than or greater than $\sqrt{\beta\gamma}$, hence according as
$4(\alpha-\eta)^2$ is less than or greater than $\beta\gamma$, hence 
according as $2A$ is greater than or less than $\sqrt{3}(\alpha-\eta)$,
hence according as $\mathcal{A}$ is greater than or less than 0.  Hence
$\mathcal{A}$ is greater than or less than 0 according as $\widehat{cab}$
is less than or greater than $\frac{2\pi}{3}$.  And similarly, 
$\mathcal{B}$ is greater than or less than 0 according as $\widehat{abc}$
is less than or greater than $\frac{2\pi}{3}$, and $\mathcal{C}$ is
greater than or less than 0 according as $\widehat{bca}$ is less than or
greater than $\frac{2\pi}{3}$.  (of course, at most one of the three 
angles of the triangle $abc$ can be greater than $\frac{2\pi}{3}$.)  We
assume first that the angles $\widehat{cab}$, $\widehat{abc}$, and
$\widehat{bca}$ are all greater than zero, so that $\mathcal{A}$, 
$\mathcal{B}$, and $\mathcal{C}$ are all greater than zero.  Then the 
position $x$ of the junction is given by:
\begin{equation}
	\label{E211}
	x=sa+tb+uc
\end{equation}
where $s$, $t$, and $u$ are as defined in (\ref{E209}).  The identity
$s+t+u=1$ in (\ref{E210}) means that $s$, $t$, and $u$ are the barycentric
coordinates of $x$ with reference to $a$, $b$, and $c$, and the assumption
that $\mathcal{A}$, $\mathcal{B}$, and $\mathcal{C}$ are all greater than
zero implies that $s$, $t$, and $u$ are all greater than zero, hence that
the junction lies withing the triangle $abc$.  And using (\ref{E209}) and
(\ref{E210}), we readily calculate that:
\begin{equation}
	\label{E212}
	(x-a)^2=\frac{\mathcal{A}^2}{3\mathcal{X}},\qquad(x-b)^2=\frac{
	\mathcal{B}^2}{3\mathcal{X}},\qquad(x-c)^2=\frac{\mathcal{C}^2}
	{3\mathcal{X}}
\end{equation}
Hence, bearing in mind the assumption that $\mathcal{B}$ and $\mathcal{C}$
are both greater than zero, we find that:
\begin{equation}
	\label{E213}
	\cos\widehat{bxc}=\frac{(x-b)^2+(x-c)^2-(b-c)^2}{2\left|x-b\right|
	\left|x-c\right|}=\frac{\mathcal{B}^2+\mathcal{C}^2-3\mathcal{X}\alpha}
	{2\mathcal{BC}}
\end{equation}
which is readily confirmed, using (\ref{E209}) and (\ref{E210}), to be
equal to $-\frac{1}{2}$.  And similarly, we find $\cos\widehat{cxa}=
\cos\widehat{axb}=-\frac{1}{2}$.  We may also check readily that:
\begin{equation}
	\label{E214}
	\frac{\partial}{\partial x_\mu}(\left|x-a\right|+\left|x-b\right|+\left|
	x-c\right|)=\frac{(x-a)_\mu}{\left|x-a\right|}+\frac{(x-b)_\mu}{\left|x
	-b\right|}+\frac{(x-c)_\mu}{\left|x-c\right|}
\end{equation}
does indeed vanish when $x$ takes the value (\ref{E211}).  We also find 
from (\ref{E210}), that when $x$ takes the value (\ref{E211}), the total
length of the three straight line segments is given by:
\begin{equation}
	\label{E215}
	\left|x-a\right|+\left|x-b\right|+\left|x-c\right|=\sqrt{\mathcal{X}}
\end{equation}

We may confirm that whenever all three angles $\widehat{cab}$,
$\widehat{abc}$, and $\widehat{bca}$ are \emph{strictly} less than
$\frac{2\pi}{3}$, $\sqrt{\mathcal{X}}$ is \emph{strictly} less than the
sum of the lengths of any two edges of the triangle $abc$, as follows.  To
show that $\sqrt{\mathcal{X}}$ is strictly less than $\sqrt{\beta}+
\sqrt{\gamma}$, or in other words, that $\mathcal{X}$ is strictly less
than $\beta+\gamma+2\sqrt{\beta\gamma}$, we note that the assumption that 
$\widehat{cab}$ is strictly less than $\frac{2\pi}{3}$ impliess that 
$\sqrt{\beta\gamma}$ is strictly greater than $\alpha-\beta-\gamma=2
(\alpha-\eta)$.  Hence
\begin{equation}
	\label{E216}
	0<(\sqrt{\beta\gamma}+2(\eta-\alpha))^2
\end{equation}
hence
\begin{equation}
	\label{E217}
	3(\beta\gamma-(\eta-\alpha)^2)<(2\sqrt{\beta\gamma}+(\eta-\alpha))^2
\end{equation}
hence
\begin{equation}
	\label{E217a}
	2\sqrt{3}A<\left|2\sqrt{\beta\gamma}+(\eta-\alpha)\right|
\end{equation}

Now by assumption $\frac{1}{2}\sqrt{\beta\gamma}+(\eta-\alpha)$ is greater
than zero, hence $2\sqrt{\beta\gamma}+(\eta-\alpha)$ is certainly greater
than zero, hence (\ref{E217a}) implies:
\begin{equation}
	\label{E218}
	2\sqrt{3}A<2\sqrt{\beta\gamma}+(\eta-\alpha)
\end{equation}
or in other words, by (\ref{E209}):
\begin{equation}
	\label{E219}
	\mathcal{X}<\beta+\gamma+2\sqrt{\beta\gamma}
\end{equation}

Thus whenever all three angles $\widehat{cab}$, $\widehat{abc}$, and 
$\widehat{bca}$ are all strictly less than $\frac{2\pi}{3}$, the 
minimal-length spanning tree of $a$, $b$, and $c$ has a junction strictly
in the interior of the triangle $abc$.

Now if $\widehat{cab}$ is equal to $\frac{2\pi}{3}$, $\mathcal{A}$ is
equal to 0, which immediately imples that $s=1$, $t=u=0$, hence that the
``junction'' $x$ is at $a$.  Thus we see immediately that if any of the
angles of the triangle $abc$ is equal to $\frac{2\pi}{3}$, then the 
minimal-length spanning tree of $a$, $b$, and $c$ is equal to the two
edges of the triangle $abc$ that meet at the vertex where the angle is
$\frac{2\pi}{3}$.

And finally, if any of the angles of the triangle $abc$ is strictly
\emph{greater} than $\frac{2\pi}{3}$, then there is \emph{no} point $x$ in
the two-plane defined by $a$, $b$, and $c$, such that the angles between
the three lines $\overline{xa}$, $\overline{xb}$, and $\overline{xc}$, are
all equal to $\frac{2\pi}{3}$.  For suppose $\theta\equiv\overline{cab}$
is strictly greater than $\frac{2\pi}{3}$.  We choose a coordinate system
in the two-plane defined by $a$, $b$, and $c$, such that $a=(d,e)$,
$b=(0,0$, and $c=(f,0)$.  Then $\alpha=f^2$, $\beta=(d-f)^2+e^2$, and 
$\gamma=d^2+e^2$.  We consider, within this fixed two-plane, with $b$ and
$c$ fixed, the locus of all possible positions of $a$ such that $\theta$
has the given value.  It is determined by:
\begin{equation}
	\label{E220}
	\alpha=\beta+\gamma-2\sqrt{\beta\gamma}\cos\theta
\end{equation}
which implies:
\begin{equation}
	\label{E221}
	4\beta\gamma\cos^2\theta=(\beta+\gamma-\alpha)^2
\end{equation}
or in other words:
\begin{equation}
	\label{E222}
	(e^2+d(d-f))^2\sin^2\theta-e^2f^2\cos^2\theta=0
\end{equation}
or in other words:
\begin{equation}
	\label{E223}
	((e^2+d(d-f))\sin\theta-ef\cos\theta)((e^2+d(d-f))\sin\theta+ef\cos
	\theta)=0
\end{equation}
which means that the possible positions of $a=(d,e)$ lie on two circles,
each of radius $\frac{f}{2\sin\theta}$, one centred at $\frac{f}{2}(1,
\cot\theta)$, and the other centred at $\frac{f}{2}(1,-\cot\theta)$.  Now
we lost the sign of $\cos\theta$ in going from (\ref{E220}) to 
(\ref{E221}), and we readily verify that, since $\theta>\frac{\pi}{2}$,
the correct arcs of these two circles are the two \emph{shorter} arcs
between $b=(0,0$, and $c=(f,0)$.  (These are the two arcs which lie inside
the compond figure formed by the two intersecting circles.)  Thus the 
possible positions of $a=(d,e)$ lie on these two arcs, which form a convex
shape, roughly like a lemon, with endpoints at $b=(0,0)$ and $c=(f,0)$.
We assume $\beta$ and $\gamma$ are both strictly greater than zero, hence
the endpoints are excluded.  The maximum possible value of $e$ is
$\frac{f(1+\cos\theta)}{2\sin\theta}=\frac{f\sin\theta}{2(1-\cos\theta)}$,
which for $\frac{\pi}{2}<\theta<\pi$ is a strictly decreasing function of
$\theta$.  And in exactly the same way we find that the locus of the
points $x$ in this two-plane such that $\widehat{bxc}=\frac{2\pi}{3}$,
consists of two circular arcs which are obtained from those which give the
possible positions of $a$, by replacing $\theta$ by $\frac{2\pi}{3}$.
These two arcs form the boundary of a convex lemon-shaped domain which
contains all possible positions of $a$ strictly in its interior, (since by
assuption $a$ is not equal to either endpoint).  It immediately follows
from this that there is no point $x$ on these two arcs such that all three
of the angles, at $x$, between the lines $\overline{xa}$, $\overline{xb}$,
and $\overline{xc}$, are equal to $\frac{2\pi}{3}$, for when $x$ is not
equal to either endpoint, all three of these line segments point into the 
convex domain bounded by the two arcs, hence the angle between one pair of
them is greater than $\pi$, while if $x$ is equal to one of the endpoints,
then the angle between the line $\overline{xa}$ and the straight line from
$x$ to the other endpoint, is less than $\pi-\theta$, hence less than
$\frac{\pi}{3}$.  Hence, as stated, if any of the angles of the triangle
$abc$ is strictly greater than $\frac{2\pi}{3}$, then there is no point 
$x$ in the two-plane defined by $a$, $b$, and $c$, such that the angles
between the three lines $\overline{xa}$, $\overline{xb}$, and
$\overline{xc}$, are all equal to $\frac{2\pi}{3}$.  Hence, in this case,
the minimal-length spanning tree of $a$, $b$, and $c$ cannot have a 
junction where three lines meet, hence it must consist of two of the edges
of the triange $abc$, and the two shortest edges are the two that meet at
the vertex where the angle is greater than $\frac{2\pi}{3}$.

We can now show that at most three straight line segments can meet at any
junction in a minimal-length spanning tree.  For suppose a minimal-length
spanning tree has a junction at which four or more straight line segments
meet.  Then there must be at least one pair of segments meeting at the 
junction such that the angle between them at the junction is strictly less
than $\frac{2\pi}{3}$.  For suppose the angle between every pair of 
segments at the junction is greater than or equal to $\frac{2\pi}{3}$.  We
choose a coordinate system with the origin at the junction, and for each
$i$, $1\leq i\leq n$, where $n$ is the number of line segments meeting at
the junction, we define $x_i$ to be the point at unit distance out from
the junction along the $\textrm{i}^\textrm{th}$ line segment, (extended if
necessary).  Then by assumption:
\begin{equation}
	\label{E224}
	x_1^2=x_2^2=\ldots=x_n^2=1
\end{equation}
and, for all $1\leq i<j\leq n$:
\begin{equation}
	\label{E225}
	x_i.x_j\leq-\frac{1}{2}
\end{equation}

Hence
\begin{equation}
	\label{E226}
	(x_1+x_2+\ldots+x_n)^2=n+2\sum_{1\leq i<j\leq n}x_i.x_j\leq\frac{n(3-n)}
	{2}
\end{equation}
which is impossible for $n>3$.  Hence, as stated, there must be at least
one pair of line segments meeting at the junction such that the angle
between them at the junction is strictly less than $\frac{2\pi}{3}$. 
Given such a pair of segments, we form an isosceles triangle, with apex at
the junction, by going out from the junction an equal distance, strictly
greater than zero, along each of them, without going past the outer end of
either segment.  The odd angle of this isosceles triangle is strictly
smaller than $\frac{2\pi}{3}$ by assumption, hence all its angles are
strictly less than $\frac{2\pi}{3}$.  Hence, by the foregoing, the
minimal-length spanning tree of the vertices of this triangle has a 
junction strictly inside this triangle, and the total length of the three
straight line segments that form this spanning tree, is strictly less than
the sum of the lengths of any two edges of this triangle.  Hence by
replacing the two edges of this triangle that meet at the original 
junction, by the minimal-length spanning tree of this triangle, we obtain
a spanning tree of strictly smaller total line length than the given 
spanning tree, and the number of line segments meeting at the given
junction has been reduced by one.

\chapter{Ansatz For The Vacuum Expectation Values And Correlation Functions, And The Island Diagram Mechanism}

\section{Ansatz for the vacuum expectation values and correlation functions}

We can now state our ansatz for the behaviour of $f_0(W_1,\ldots,W_n,g^2)$
for $n\geq2$, when the sizes and separations of $W_1,\ldots,W_n$ are all
$\geq\frac{1}{\mu}$.

\subsection{The massive scalar propagator}

We first note that for $\left|x-y\right|$ large compared to $\frac{1}{m}$,
the free massive scalar propagator:
\begin{equation}
	\label{E227}
	\int_0^\infty\mathrm{d}s\frac{e^{-\frac{(x-y)^2}{4s}}}{(4\pi s)^2}
	e^{-m^2s}
\end{equation}
in our four-dimensional Euclidean space, is found, by a standard steepest
descents approximation to the $s$ integral about the peak of the 
exponential factor at $s=\frac{\left|x-y\right|}{2m}$, to approach the
asymptotic form:
\begin{equation}
	\label{E228}
	\sqrt{\frac{m}{32\pi^3\left|x-y\right|^3}}e^{-m\left|x-y\right|}
\end{equation}

As a check on this, we note that (\ref{E228}) immediately gives the 
Yukawa potential:
\begin{equation}
	\label{E229}
	\frac{e^{-mr}}{4\pi r}
\end{equation}
for the static interaction between two heavy particles, due to exchange of
scalar particles of mass $m$.  Indeed, we consider two parallel straight
lines separated by a distance $r$.  We put $x$ at a fixed position on one
of the two lines, and integrate $y$ along the other line, expanding
$\left|x-y\right|$ in the exponent as $\sqrt{r^2+z^2}=r+\frac{z^2}{2r}$
plus higher order terms which we neglect, where $z$ is the distance along
the second line from the point on it closest to $x$.  The resulting
Gaussian integral with respect to $z$ immediately gives (\ref{E229}).  (Of
course, calculating the effect of scalar exchange in this way with the
exact propagator (\ref{E227}) gives the Yukawa potential (\ref{E229})
without any approximations at all.)

\subsection{Factors in the ansatz}

Our ansatz for the behaviour of $f_0(W_1,\ldots,W_n,g^2)$, for $n\geq2$,
when the sizes and separations of $W_1,\ldots,W_n$ are all $\geq\frac{1}
{\mu}$, is the product of the following factors:\\
\noindent(i) a factor $e^{-\mu^2A}=e^{-\mu^2(A_1+\ldots+A_p)}$, where $A$
is the total area of our absolute minimal-area orientable spanning surface
$S$ of $W_1,\ldots,W_n$, and $A_1,\ldots,A_p$, $1\leq p\leq n$, are the
areas of the separate connected components $S_1,\ldots,S_p$ of $S$.\\
\noindent(ii) for each separate straight line in our minimal-length 
spanning tree of $S_1,\ldots,S_p$, a factor:
\begin{equation}
	\label{E230}
	\sqrt{\frac{m}{32\pi^3L^3}}e^{-mL}
\end{equation}
where $L$ is the length of that straight line, and $m>0$ is the mass of
the lightest glueball.\\
\noindent(iii) for each point where a straight line of our minimal-length
spanning tree ends on one of the connected components of $s$, a factor
$f$, where $f$ represents the coupling of the lightest glueball to a 
minimal-area orientable spanning surface.\\
\noindent(iv) for each junction where three straight lines of our 
minimal-length spanning tree meet, a factor $h$, where $h$ represents the
three-glueball coupling constant.

Thus our ansatz depends on precisely four parameters, namely $\mu$, $m$,
$f$, and $h$.  In this paper we will give the first approximation to the
ratio $\frac{m}{\mu}$, leaving the first calculations of $f$, $h$, and
the ratio $\frac{\mu}{\Lambda}$, where $\Lambda$, (in the range 0.1 GeV
to 0.5 GeV), is the standard QCD running coupling parameter 
\cite{Perkins 307}, to our next paper.

\subsection{Short-distance factors}

Of course in practice we have to divide $f_0(W_1,\ldots,W_n,g^2)$ by a
short-distance factor for each $W_i$, which takes the form of the sum of 
all the Feynman diagrams contributing to $f_0(W_i,g^2)$, but with all the
long-distance effects removed, (most simply by cutting off all the 
propagators smoothly at long distances), and which cancels the divergences
which occur when a subdiagram of a Feynman diagram contributing to 
$f_0(W_1,\ldots,W_n,g^2)$, which is attached to $W_i$ but is not connected
by any \emph{propagators} to any other part of the Feynman diagram, 
shrinks to a very small size on $W_i$ \cite{Polyakov}.  We then have to
rewrite the group-variation equations in terms of these ratios, and our
ansatz applies to these ratios.  The ratios contain a new parameter, 
namely the length $X$ associated with the long-distance cutoff of the 
propagators in the short-distance factors, but the possible dependence on
this parameter is constrained by the fact that it must cancel out of all
physical quantities.  In fact, if we calculate the ratios for one value,
$X_1$, of this parameter, and if we also calculate the (finite) ratio of
the \emph{short}-distance factors for $X_1$ and for another value, $X_2$,
and multiply by this ratio of \emph{short}-distance factors, then the
dependence on $X_1$ must cancel out, to be replaced by the equivalent
dependence on $X_2$.

However it will not be necessary to divide by the short-distance factors
in this paper, and we thus, for this paper, apply our ansatz directly to
$f_0(W_1,\ldots,W_n,g^2)$.  We note that the ansatz encompasses the Wilson
area law for $n=1$, and we thus apply it for all $n\geq1$.

\section{Substituting in the ansatz}

We will see that the group-variation equations force us to make one small,
but crucial, change to the ansatz, which will \emph{not} spoil the Wilson
area law or massive glueball saturation, and that the modified ansatz will
then give a consistent solution of the group-variation equations when the
sizes and separations of $W_1,\ldots,W_n$ are all $\geq\frac{1}{\mu}$. 
But first let us consider the qualitative results of substituting our
ansatz into the right-hand sides of the group-variation equations.

\subsection{\emph{Every} 45-path now gets the effective mass}

We first note that, by our ansatz, \emph{every} 45-path in a right-hand
side group-variation equation diagram, irrespective of whether the windows
beside it are simply-connected or multiply-connected, now gets the same
effective mass that we derived before.  The most conservative, (i.e. the
smallest), likely value of this effective mass is, as explained after
equation (\ref{E200}), $\mu\sqrt{\frac{\pi}{2}}=1.3\mu$.  We note here 
that another possible cause of the logarithmic divergence in (\ref{E198})
for small $q$, in addition to our neglect of the effects of the ends of 
the paths, may be that our use, in (\ref{E188}), of a parameter along the
path that is simply proportional to the number of straight line segments
counted along the path from one end, rather than the parameter that truly
minimizes Douglas's functional, may be all right ``locally'', i.e. for 
$q$ not too small, but give a systematic over-estimate of the suppression,
(i.e. over-estimate of the contribution to the effective mass), for the 
long-wavelength modes of small $q$.  This would support our suggestion 
that the integrand in (\ref{E198}) be replaced, for all $0\leq q\leq\pi$,
by the value $\frac{\pi}{2}$ it takes at $q=\pi$.

\section{The island-diagram mechanism}

Let us now consider the result of substituting our ansatz into an
\emph{island} diagram in the right-hand side of the group-variation 
equation (\ref{E184}) for $f_0(W_1,g^2)$.  Let us suppose that the loop
$W_1$ is roughly planar, that it has no self-intersections, and that its
size is large compared to $\frac{1}{\mu}$.  Now the contributions of this
diagram will be dominated by the contributions of islands of size roughly
$\frac{1}{\mu}$, due to the effective mass that the 45-paths get, and this
will be approximately true no matter how large the loop $W_1$ is.
Furthermore, when the island is roughly in the plane of $W_1$, or more
precisely, roughly ``within'' the minimal-area orientable spanning surface
of $W_1$, (since we don't assum $W_1$ is \emph{exactly} planar), and when
the island is also roughly configured, in position space, such that the
orientation of its ``outer boundary'' is roughly consistent with the 
orientation of the minimal-area orientable spanning surface of $W_1$, then
by our ansatz, the window weight associated with the window that 
\emph{surrounds} the island in the island diagram, namely 
$f_0(W_1,W_2,g^2)$, where $W_2$ is the closed loop in position space
defined by the outer boundary of the island, will be roughly equal to 
$e^{-\mu^2A_{12}}$, where $A_{12}$ is the area of the minimal-area
orientable spanning surface of $W_1$ and $W_2$.  Now under the conditions
stated, $A_{12}$ will be approximately equal to $A$, where $A$ is the area
of the minimal-area orientable spanning surface of $W_1$, minus the area
of the minimal-area orientable spanning surface of $W_2$.  But the area of
the minimal-area orientable spanning surface of $W_2$ is proportional to 
$\frac{1}{\mu^2}$, and by assumption, $A$ is large compared to $\frac{1}
{\mu^2}$, hence $A_{12}$ will be approximately equal to $A$, hence the 
factor $e^{-\mu^2A_{12}}$ associated with the window that surrounds the 
island, is approximately equal to $e^{-\mu^2A}$.  And this will be true
under the conditions stated, no matter where the island, of size roughly
$\frac{1}{\mu}$, lies within the minimal-area orientable spanning surface
of the much larger loop $W_1$.  Hence, provided the contributions from 
configurations where the centre of the island does \emph{not} lie within
the minimal-area orientable spanning surface of $W_1$, fall off rapidly
enough as the distance betweent he centre of the island, and the nearest
point on the minimal-area orientable spanning surface of $W_1$, increases,
we may expecte that the contribution of this island diagram to the 
right-hand side of the group-variation equation (\ref{E184}) for 
$f_0(W_1,g^2)$, is equal to a constant, $Y$, times $A$, the area of the 
minimal-area orientable spanning surface $S$ of $W_1$, times 
$e^{-\mu^2A}$, or in other words, is equal to $YAf_0(W_1,g^2)$, where $Y$
is equal to the integral over all configurations of the island, subject to
the projection onto $S$ of the position of its centre, being fixed.  For
large $W_1$, and for the projection of the centre of the island onto $S$
not being too close to the edge of $S$, (i.e. to $W_1$), we may expect
$Y$ to be roughly independent of the position of the projection of the 
centre of the island onto $S$.  Furthermore, we may expect this to be the
predominant behaviour of \emph{every} island diagram, with the only
difference between different island diagrams, being different values of
the constant $Y$.

We now define $L$ to be the ``diameter'' of $W_1$, or, in other words, $L$
to be the largest distance between any two points of $W_1$.  We consider
a family of loops, each identical in shape to $W_1$, but differing in
size, i.e. having a different value of $L$.  We denote the member of this
family whose diameter is $L$ by $W_{1L}$, and we denote the area $A_L$ of
the minimal-area orientable spanning surface $S_L$ of $W_{1L}$ by
$a_1L^2$, where $a_1$ is of course independent of $L$.  Then we
immediately see that for large $L$, the contribution of each
\emph{island} diagram to the right-hadn side of the group-variation 
equation for $f_0(W_{1L},g^2)$, is equal to a constant times
$a_1L^2f_0(W_{1L},g^2)$, while the contribution of each \emph{non}-island
diagram is not greater than a constant times $Lf_0(W_{1L},g^2)$, (since
the first non-island diagram is of course simply $f_0(W_{1L},g^2)$, while
every other non-island diagram has exactly one band, which moves like a
blob of size $\frac{1}{\mu}$ along the path $W_{1L}$).  Thus for large
$L$, the dominant contributions to the right-hand side of the 
group-variation equation (\ref{E184}) for $f_0(W_{1L},g^2)$ come from the
\emph{island} diagrams.

Let us now consider the left-hand side of the group-variation equation
(\ref{E184}) for $f_0(W_{1L},g^2)$.  Now within the scope of this paper,
where we do not divide the $f_0$'s by short-distance factors, $f_0(W_{1L},
g^2)$ will satisfy the simple renormalization group equation (\ref{E102}).
(The short-distance factors will give small additional terms in 
(\ref{E102}) which will not alter our general results, and which we 
neglect in this paper.)  Now
\begin{equation}
	\label{E231}
	\beta(g)\frac{\partial}{\partial g}=2g\beta(g)\frac{\partial}{\partial
	g^2}
\end{equation}
hence, multiplying both sides of the group-variation equation (\ref{E184})
for $f_0(W_{1L},g^2)$ by:
\begin{equation}
	\label{E232}
	-\frac{2g\beta(g)}{g^2}
\end{equation}
we see that the term $g^2\frac{\mathrm{d}}{\mathrm{d}g^2}f_0(W_{1L},g^2)$
in the left-hand side of (\ref{E184}) gives simply:
\begin{equation}
	\label{E233}
	L\frac{\partial}{\partial L}f_0(W_{1L},g^2)
\end{equation}
But by our ansatz, $f_0(W_{1L},g^2)$ is equal to $e^{-\mu^2a_1L^2}$, hence
(\ref{E233}) is equal to:
\begin{equation}
	\label{E234}
	L\frac{\partial}{\partial L}e^{-\mu^2a_1L^2}=-2\mu^2a_1L^2f_0(W_{1L},g^2
	)
\end{equation}

Hence this is the leading term in the left-hand side of (\ref{E184}) for
large $L$, with the term $f_0(W_{1L},g^2)$ being smaller by two powers of 
$L$.  And comparing with our previous estimate of the right-hand side of
(\ref{E184}) for large $L$, which has now become, due to multiplying by
(\ref{E232}), the sum, over all the island diagrams, of:
\begin{equation}
	\label{E235}
	-\frac{2g\beta(g)}{g^2}Y_ia_1L^2f_0(W_{1L},g^2)
\end{equation}
where the subscript $i$ on $Y_i$ identifies the island diagram concerned,
we see that, within our estimate, our ansatz satisfies the 
group-variation equation (\ref{E184}) for $f_0(W_{1L},g^2)$ for large
$L$, \emph{provided the sign comes out right}.  And furthermore, with this
proviso, $\mu^2$ is given by:
\begin{equation}
	\label{E236}
	\mu^2=\frac{\beta(g)}{g}\sum_iY_i
\end{equation}
where the sum on $i$ runs over all the island diagrams.

Now before we consider the crucial question of the sign, we note, firstly,
that our estimate has in fact not been quite right.  When we come to
estimate more precisely the contribution of an island diagram to the 
right-hand side of ({E184}) for large $L$, we find, indeed, a factor of 
the area, as expected, but we also find another, unwanted, factor equal to
$\ln(\mu^2A)$.  This extra factor is unacceptable.  It arises because the
\emph{rate} at which $f_0(W_{1L},W_2,g^2)$ falls off, according to our
ansatz, as the small loop $W_2$, of size roughly $\frac{1}{\mu}$, and 
oriented consistently with the minimal-area orientable spanning surface
$S_{1L}$ of $W_{1L}$, moves out of the two dimensions of $S_{1L}$,
\emph{decreases,} as $L$ increases by a factor of the reciprocal of 
$\ln(\mu^2A)$, rather than being independent of $A$.  This will force us
to make a small change to our ansatz, after which the Wilson area law will
satisfy the group-variation equations in exactly the manner just 
described.

We also note here, secondly, that an island diagram mechanism, very
similar to what we have just described for the area law, also results in
the massive glueball saturation of the correlation functions, as the
\emph{separations} of the loops $W_1,\ldots,W_n$ increase while their
\emph{sizes} remain fixed, giving a consistent solution of the 
group-variation equations.  In this case we will find that our ansatz
works perfectly without any alterations at all.

\section{The signs of the island diagrams, and the critical value of $g^2$}

But before giving the details of these calculations, we now consider the
crucial question of the sign of the island diagrams.  We first note that
the relative sign of the two terms in (\ref{E102}) is indeed correct, and
in accord with the conventional form of the renormalization group
equations and the conventional defintion of the $\beta$ function
\cite{DJGLH 192}.  The renormalization group equations are normally 
expressed in terms of a normalizaton mass $\mu$.  This normalization mass
$\mu$ is \emph{not} the same as the mass $\mu$ that occurs in the Wilson
area law, notwithstanding the unfortunate use of the same symbol.  Rather
the \emph{normalization} mass $\mu$ is an \emph{input} to the QCD
calculation, along with an \emph{input} value of the coupling constant
$g$.

\subsection{BPHZ renormalization, and restoration of Ward identities by finite counterterms}

In our BPHZ approach \cite{BPHZ 19}, the role of the normalization
mass $\mu$ is played by $\frac{1}{R}$, where $R$ is the length that 
characterizes the smooth \emph{long}-distance cutoff that we have to 
impose on gluon and Fadeev-Popov propagators in counterterms, in order to
avoid \emph{long}-distance divergences in counterterms.  Ward identities
are restored by the addition of simple finite counterterms \cite{Grassi Hurth Steinhauser}.
For example,
if $\sigma\equiv(x-y)^2$, and $b(\sigma)$ and $c(\sigma)$ satisfy
\begin{equation}
	\label{E237}
	2\frac{\mathrm{d}}{\mathrm{d}\sigma}(b(\sigma)+\sigma c(\sigma))+3c(
	\sigma)=0
\end{equation}
for all $\sigma>0$, and $b(\sigma)$ is bounded by a constant times
$\sigma^{-3}$ as $\sigma\to0$ and $c(\sigma)$ is bounded by a constant
times $\sigma^{-4}$ as $\sigma\to0$, then the following renormalized
expression:
	\[\Bigg\{\bigg\{\int\int\mathrm{d}^4x\mathrm{d}^4y\left(\delta_{\mu\nu}
	b(\sigma)+(x-y)_\mu(x-y)_\nu c(\sigma)\right)\quad\times
\]
	\[\times A\raisebox{-4pt}{$\stackrel{\displaystyle{(x)}}
 	{\scriptstyle{\mu a}}$}\left(
 	A\raisebox{-4pt}{$\stackrel{\displaystyle{(y)}}{\scriptstyle{\nu a}}$}
 	-\theta(R^2-\sigma)\left(
 	A\raisebox{-4pt}{$\stackrel{\displaystyle{(x)}}{\scriptstyle{\nu a}}$}+
 	(y-x)_\alpha\partial_\alpha
 	A\raisebox{-4pt}{$\stackrel{\displaystyle{(x)}}{\scriptstyle{\nu a}}$}+
 	\frac{1}{2}(y-x)_\alpha(y-x)_\beta\partial_\alpha\partial_\beta
 	A\raisebox{-4pt}{$\stackrel{\displaystyle{(x)}}{\scriptstyle{\nu a}}$}
 	\right)\right)\bigg\}
\]
\begin{equation}
	\label{E238}
	+\quad\frac{\pi^2 R^4}{2}\left(b(R^2)+R^2c(R^2)\right)\int\mathrm{d}^4
	x\left(
	A\raisebox{-4pt}{$\stackrel{\displaystyle{(x)}}{\scriptstyle{\mu a}}$}
	A\raisebox{-4pt}{$\stackrel{\displaystyle{(x)}}{\scriptstyle{\mu a}}$}+
	\frac{R^2}{12}
	A\raisebox{-4pt}{$\stackrel{\displaystyle{(x)}}{\scriptstyle{\mu a}}$}
	\partial^2
	A\raisebox{-4pt}{$\stackrel{\displaystyle{(x)}}{\scriptstyle{\mu a}}$}
	\right)\Bigg\}
\end{equation}
(where $\theta(s)$ is the step function, $\theta(s)=1$ for $s\geq0$,
$\theta(s)=0$ for $s<0$), is finite, and is also exactly gauge-invariant
under the linear gauge transformation $A_{\mu a}\to A_{\mu a}+\partial_\mu
\epsilon_a$.  The linear gauge variation of the additional finite
counterterm, (the third line of (\ref{E238})), exactly cancels the linear
gauge variation of the standard $BPHZ$ renormalized form, (the first two
lines of (\ref{E238})).  Furthermore, since (\ref{E238}) is finite and
linearly gauge-invariant for \emph{all} $R>0$, and the $R$-dependent terms
in (\ref{E238}) are all \emph{local} functionals of $A_{\mu a}$, 
(involving at most two derivatives), the derivative of (\ref{E238}) with
respect to $R^2$ is a finite, linearly gauge-invariant, \emph{local}
functional of $A_{\mu a}$.  In fact, the derivative of (\ref{E238}) with
respect to $R^2$, is equal to the manifestly linearly gauge-invariant
expression:
\begin{equation}
	\label{E239}
	\frac{\pi^2R^6c(R^2)}{24}\int\mathrm{d}^4x
	A\raisebox{-4pt}{$\stackrel{\displaystyle{(x)}}{\scriptstyle{\mu a}}$}
	\left(\partial^2
	A\raisebox{-4pt}{$\stackrel{\displaystyle{(x)}}{\scriptstyle{\mu a}}$}-
	\partial_\mu\partial_\nu
	A\raisebox{-4pt}{$\stackrel{\displaystyle{(x)}}{\scriptstyle{\nu a}}$}
	\right)
\end{equation}

This is an example of the fact that, although we do indeed have, in our
BPHZ approach, some power-counting renormalizable counterterms of 
non-gauge-invariant structure, these non-gauge-invariant counterterms have
no physical effects at all, because they are completely independent of
$R$.  This is in complete contrast to the gauge-invariant counterterms,
which depend on $R$, and have finite derivatives with respect to $R^2$.
In our approach, the renormalization group arises from the fact that we 
can exactly compensate for changes of $R$ from one finite value, strictly
greater than zero, to another finite value, strictly greater than zero, by
appropriate finite rescalings of $g^2$ and $A_{\mu a}$.  We take the view
that the BPHZ-renormalized perturbation expansion is generated by a 
``seed'' action which, for the ordinary Feynman diagram expansion, (as
opposed to the group-variation equations), is the integral of the standard
action density:
\begin{equation}
	\label{E240}
	\frac{N}{4g^2}F_{\mu\nu a}F_{\mu\nu a}+\frac{N}{g^2}\left(iB_a(\partial
	_\mu A_{\mu a})+\frac{\alpha}{2}B_aB_a+\psi_a(\partial_\mu(\mathrm{D}_
	\mu\phi)_a)\right)
\end{equation}
where $F_{\mu\nu a}=\partial_\mu A_{\nu a}-\partial_\nu A_{\mu a}+f_{abc}
A_{\mu b}A_{\nu c}$ and $(\mathrm{D}_\mu\phi)_a=\partial_\mu\phi_a+A_{\mu
b}f_{abc}\phi_c$.  The full action is equal to the seed action plus the 
counterterms.  The seed action depends on one physical parameter, namely
$g^2$, (since dependence on $\alpha$ cancels out of physical quantities), 
and the counterterms depend on two physical parameters, namely $g^2$ and
$R$.  We define a canonical procedure, given in detail in our next paper,
where at each loop order we have the canonical BPHZ counterterms, as
defined in our previous paper, \cite{BPHZ}, \cite{BPHZ 10}, 
\cite{BPHZ 19}, plus precisely-defined Ward-identity-restoring finite
counterterms, of which (\ref{E238}) is the simplest example.  The 
canonical BPHZ counterterms depend on $R$ through the smooth 
long-distance cutoffs imposed on the propagators in the counterterms.  For
example, if $t$ is a fixed real number strictly greater than $1$, and 
$f(s)$ is a member of $\mathbf{R}^{\mathbf{R}}$ that is infinitely
differentiable for all $s\in\mathbf{R}$, equal to 1 for all $s\leq1$, and
equal to 1 for all $s\geq t$, we may multiply each propagator in a 
counterterm by $f\left(\frac{(x-y)^2}{R^2}\right)$, where $x$ and $y$ are
the positions of the ends of the propagator.  A possible form for $f(s)$
for $1\leq s\leq t$ is $f(s)=\frac{1}{1+e^{\frac{-1}{s-1}}e^{\frac{1}{t-s}
}}$.  (The expression (\ref{E238}) is easily generalized to deal with a
proper smooth long-distance cutoff in the counterterms rather than the 
simple step-function cutoff as in (\ref{E238}).)

The Ward identity is expressed in terms of the effective action $\Gamma$,
(the generating functional of the proper vertices), in a standard manner
\cite{IZ page 597}.  Indicating position arguments by subscripts $x$, 
$y$, $\ldots$ , and introducing sources $J_{\mu a x}$ and $K_{ax}$ for 
$A_{\mu ax}$ and $B_{ax}$, respectively, and anticommuting sources
$\xi_{ax}$ and $\zeta_{ax}$ for $\phi_{ax}$ and $\psi_{ax}$, respectively,
we define $W(J,K,\xi,\zeta)$, the generating functional of the connected
Green functions, (i.e. the correlation functions), by:
\begin{equation}
	\label{E241}
	e^W=e^{W(J,K,\xi,\zeta)}=\left\langle e^{JA+KB+\xi\phi+\zeta\psi}
	\right\rangle
\end{equation}
where the angular brackets indicate the standard functional average over
the $A$, $B$, $\phi$, and $\psi$ fields, with the weight given by the 
exponential of the negative of the integral of the action density
(\ref{E240}).  We indicate the derivative with respet to a quantity by
putting the $\hat{\ }$ above that quantity, for example $\hat{J}_{\mu ax}$
means $\frac{\delta}{\delta J_{\mu ax}}$.  Then we define the ``classical
fields'' in the standard way by:
\begin{equation}
	\label{E242}
	A_{\mu ax}=\hat{J}_{\mu ax}W,\qquad B_{ax}=\hat{K}_{ax}W,\qquad\phi_{ax}
	=\hat{\xi}_{ax}W,\qquad\psi_{ax}=\hat{\zeta}_{ax}W
\end{equation}
(where these ``classical fields'' of course have no relation to the 
integration variables in the functional integrals in (\ref{E241}),
notwithstanding the use of the same symbols), and we define the effective
action $\Gamma$ by:
\begin{equation}
	\label{E243}
	\Gamma=JA+KB+\xi\phi+\zeta\psi-W
\end{equation}

The independent variables $J$, $K$, $\xi$, and $\zeta$ are to be expressed
in terms of $A$, $B$, $\phi$, and $\psi$, which are then taken as the
independent variables of $\Gamma$.  Hence:
\begin{equation}
	\label{E244}
	J_{\mu ax}=\hat{A}_{\mu ax}\Gamma,\qquad K_{ax}=\hat{B}_{ax}\Gamma,
	\qquad\xi_{ax}=-\hat{\phi}_{ax}\Gamma,\qquad\zeta_{ax}=-\hat{\psi}_{ax}
	\Gamma
\end{equation}

We define the matrix:
\begin{equation}
	\label{E245}
	M=\left(\begin{array}{cccc}\hat{A}_{\mu ax}\hat{A}_{\nu by}\Gamma &
	\hat{A}_{\mu ax}\hat{B}_{by}\Gamma & \hat{A}_{\mu ax}\hat{\phi}_{by}
	\Gamma & \hat{A}_{\mu ax}\hat{\psi}_{by}\Gamma \\ \hat{B}_{ax}\hat{A}
	_{\nu by}\Gamma & \hat{B}_{ax}\hat{B}_{by}\Gamma & \hat{B}_{ax}\hat{\phi}
	_{by}\Gamma & \hat{B}_{ax}\hat{\psi}_{by}\Gamma \\ \hat{\phi}_{ax}
	\hat{A}_{\nu by}\Gamma & \hat{\phi}_{ax}\hat{B}_{by}\Gamma & \hat{\phi}
	_{ax}\hat{\phi}_{by}\Gamma & \hat{\phi}_{ax}\hat{\psi}_{by}\Gamma \\
	\hat{\psi}_{ax}\hat{A}_{\nu by}\Gamma & \hat{\psi}_{ax}\hat{B}_{by}
	\Gamma & \hat{\psi}_{ax}\hat{\phi}_{by}\Gamma & \hat{\psi}_{ax}
	\hat{\psi}_{by}\Gamma \end{array}\right)
\end{equation}
and its right inverse:
\begin{equation}
	\label{E246}
	N=\left(\begin{array}{cccc} \hat{J}_{\mu ax}\hat{J}_{\nu by}W & \hat{J}_
	{\mu ax}\hat{K}_{by}W & \hat{J}_{\mu ax}\hat{\xi}_{by}W & \hat{J}_{\mu
	ax}\hat{\zeta}_{by}W \\ \hat{K}_{ax}\hat{J}_{\nu by}W & \hat{K}_{ax}
	\hat{K}_{by}W & \hat{K}_{ax}\hat{\xi}_{by}W & \hat{K}_{ax}\hat{\zeta}_
	{by}W \\ -\hat{\xi}_{ax}\hat{J}_{\nu by}W & -\hat{\xi}_{ax}\hat{K}_{by}W
	& -\hat{\xi}_{ax}\hat{\xi}_{by}W & -\hat{\xi}_{ax}\hat{\zeta}_{by}W \\
	-\hat{\zeta}_{ax}\hat{J}_{\nu by}W & -\hat{\zeta}_{ax}\hat{K}_{by}W &
	-\hat{\zeta}_{ax}\hat{\xi}_{by}W & -\hat{\zeta}_{ax}\hat{\zeta}_{by}W
	\end{array}\right)
\end{equation}
where we note that:
\begin{equation}
	\label{E247}
	MN=1
\end{equation}
follows immediately from (\ref{E242}) and (\ref{E244}).  (The summation 
convention is applied to \emph{all} repeated indices, Lorentz, group, and
position, in the matrix multiplication.)  The Ward identity to be 
satisfied by $\Gamma$ may then be written:
	\[0=\int\mathrm{d}^4x\bigg\{\left(\hat{A}_{\mu ax}\Gamma\right)\left(
	\hat{x}_\mu\phi_{ax}+A_{\mu bx}f_{abc}\phi_{cx}+f_{abc}N_{A_{\mu bx}
	\phi_{cx}}\right)
\]
\begin{equation}
	\label{E248}
	-\frac{1}{2}f_{abc}\left(\hat{\phi}_{ax}\Gamma\right)\left(\phi_{bx}
	\phi_{cx}-N_{\phi_{bx}\phi_{cx}}\right)+i\left(\hat{\psi}_{ax}\Gamma
	\right)B_{ax}\bigg\}
\end{equation}

It expresses the invariance of $\Gamma$ under a modified BRST variation,
which differs from the standard BRST variation, under which the seed
action (\ref{E242}) is invariant, by the addition of the term:
\begin{equation}
	\label{E249}
	f_{abc}N_{A_{\mu bx}\phi_{cx}}=f_{abc}\hat{J}_{\mu bx}\hat{\xi}_{cx}W
\end{equation}
to the variation of $A_{\mu ax}$, and the addition of the term:
\begin{equation}
	\label{E250}
	\frac{1}{2}f_{abc}N_{\phi_{bx}\phi_{cx}}=-\frac{1}{2}f_{abc}\hat{\xi}_
	{bx}\hat{\xi}_{cx}W
\end{equation}
to the variation of $\phi_{ax}$.  These additional terms are themselves
expressed in terms of $\Gamma$ by (\ref{E245}) and (\ref{E247}).

The matrix $N$ may be developed in a standard loop expansion, which, with
our conventions, is equivalent to an expansion in powers of $g^2$.  In a
schematic notation, where we represent all fields by $\Phi$, ignore
Bose-Fermi distinctions, and indices $i$, $j$, $\ldots$ , run over all
field degrees of freedom, (i.e. over field type, and also over Lorentz,
group, and position, as appropriate to each field type), and where $A$
denotes the seed action, the first term in $N$, (i.e. the zero-loop term),
which is of order $g^2$, is the matrix inverse of $\frac{\delta^2A(\Phi)}
{\delta\Phi_i\delta\Phi_j}$.  In fact, if we denote $\left.\frac{\delta^2
A(\Phi)}{\delta\Phi_i\delta\Phi_j}\right|_{\Phi=0}$ by $m_{ij}$, and 
define $n_{ij}$ by $m_{ik}n_{kj}=\delta_{ij}$, then the zero-loop term in
$N$ is:
\begin{equation}
	\label{E251}
	N_{1ij}=n_{ij}-n_{ik}\!\!\;
	\left(\frac{\delta^2A(\Phi)}{\delta\Phi_k\delta\Phi_l}-m_{kl}\right)
	n_{lj}+n_{ik}\!\!\;
	\left(\frac{\delta^2A(\Phi)}{\delta\Phi_k\delta\Phi_l}-m_{kl}\right)
	\!\!\;n_{lp}\!\!\;
	\left(\frac{\delta^2A(\Phi)}{\delta\Phi_p\delta\Phi_q}-m_{pq}\right)
	\!\!\;n_{qj}-\ldots
\end{equation}
where the subscript 1 of $N_{1ij}$ refers to the power of $g^2$ that 
$N_{1ij}$ includes, not the number of loops.  We note that $n_{ij}$ is the
matrix of free propagators, and that $N_{1ij}$ is the corresponding matrix
of propagators in the presence of the ``background fields'' $\Phi$.  The
one-loop and higher-loop terms in $N$ all have the form of 
Feynman-diagram propagator correction with the appropriate number of 
loops, where each line in the Feynman diagram is interpreted as $N_{1ij}$,
each cubic vertex is interpreted as $-\frac{\delta^3A(\Phi)}{\delta\Phi_k
\delta\Phi_p\delta\Phi_q}$, and each quartic vertex is interpreted as
$-\frac{\delta^4A(\Phi)}{\delta\Phi_k\delta\Phi_l\delta\Phi_p\delta\Phi
_q}$.  For example, the one-loop term in $N$, which we denote by $N_{2ij}$
because it includes two powers of $g^2$, is given by:
	\[N_{2ij}=\bigg\{\frac{1}{2}N_{1ik}\left(\frac{\delta^3A(\Phi)}{\delta
	\Phi_k\delta\Phi_p\delta\Phi_q}\right)N_{1pr}N_{1qs}\left(\frac{\delta^3
	A(\Phi)}{\delta\Phi_l\delta\Phi_r\delta\Phi_s}\right)N_{1lj}
\]
\begin{equation}
\label{E252}
-\frac{1}{2}N_{1ik}\left(\frac{\delta^4A(\Phi)}{\delta\Phi_k\delta\Phi_l
\delta\Phi_p\delta\Phi_q}\right)N_{1pq}N_{1lj}\bigg\}
\end{equation}

The counterterms are determined, for the chosen value of $R$, inductively
in the number of loops, so that (\ref{E24}) is satisfied.  At each loop
order, we begin by including the canonical BPHZ counterterms, as defined
in our previous paper \cite{BPHZ}, \cite{BPHZ 10}, \cite{BPHZ 19}, as
generated by the seed action plus all the Ward-identity-restoring finite
counterterms found at lower loop orders.  These canonical BPHZ 
counterterms depend on $R$, which is by definition the largest value of
$\left|x-y\right|$ for which a propagator in a counterterm, with ends at
$x$ and $y$, is equal to the corresponding unmodified propagator.  (In
other words, $R$ is the value of $\left|x-y\right|$ where the smooth
long-distance cutoffs imposed on the propagators in the counterterms,
begins to set in.  The fact that the propagators in our counterterms 
differ from the corresponding propagators in the ``direct'' terms for
$\left|x-y\right|>R$ was of course allowed for in all the proofs in our
previous paper \cite{BPHZ}, \cite{BPHZ 10}, \cite{BPHZ 19}.  In Theorems 1
and 2 in that paper, we only required the existence of a finite real
number $S>0$ such that the propagators in the counterterms are equal to 
the corresponding propagators in the ``direct'' terms for all $\left|x-y
\right|\leq S$.)  The inclusion of the canonical BPHZ counterterms makes
$\Gamma$ a finite functional of the ``classical fields'' $A_{\mu ax}$,
$B_{ax}$, $\phi_{ax}$, and $\psi_{ax}$, up to and including the current
loop order, but (\ref{E248}) is not in general satisfied, (just as the
first two lines of (\ref{E238}) give a finite functional of 
$A\raisebox{-4pt}{$\stackrel{\displaystyle{(x)}}{\scriptstyle{\mu a}}$}$,
which is not, however, linearly gauge-invariant).  We then add precisely
defined, $R$-dependent, finite counterterms, (analogous to the third line
of (\ref{E238})), after which (\ref{E248}) is exactly satisfied up to and
including the current loop order.  (For details we refer to our next 
paper.)

We then find, as we mentioned before, that if we \emph{change} $R$ to 
another finite value, say $R_2$, also strictly greater than zero, then
this is precisely equivalent to leaving $R$ \emph{unaltered}, and making
instead appropriate finite rescalings of $g^2$, $\alpha$, and the fields.
In fact, if we denote the seed action (\ref{E240}) by $S(A,B,\phi,\psi,
g^2,\alpha)$, and the sum of all the counterterms, generated from
$S(A,B,\phi,\psi,g^2,\alpha)$, by the procedure just described, by $C(A,B,
\phi,\psi,g^2,\alpha,R)$, (so that the full action is equal to $S(A,B,
\phi,\psi,g^2,\alpha)+C(A,B,\phi,\psi,g^2,\alpha,R)$), then we find the
identity:
	\[S(A,B,\phi,\psi,g^2\alpha)+C(A,B,\phi,\psi,g^2,\alpha,R_2)=
\]
\begin{equation}
	\label{E253}
	=S\left(\frac{Z_1}{Z_3}A,\frac{Z_1}{Z_3^2}B,\phi,\tilde{Z}_3\psi,
	\frac{Z_1^2}{Z_3^3}g^2,Z_3\alpha\right)+C\left(\frac{Z_1}{Z_3}A,
	\frac{Z_1}{Z_3^2}B,\phi,\tilde{Z}_3\psi,\frac{Z_1^2}{Z_3^3}g^2,
	Z_3\alpha,R\right)
\end{equation}
where $Z_3$, $Z_1$, and $\tilde{Z}_3$ are \emph{finite} functions of 
$g^2$, $\alpha$, and $\frac{R_2}{R}$, given through order $g^2$, (i.e.
through one-loop order), by:
	\[Z_3=1+\frac{g^2}{48\pi^2}(26-6\alpha)\ln\left(\frac{R_2}{R}\right)
\]
	\[Z_1=1+\frac{g^2}{48\pi^2}(17-9\alpha)\ln\left(\frac{R_2}{R}\right)
\]
\begin{equation}
	\label{E254}
	\tilde{Z}_3=1+\frac{g^2}{48\pi^2}(9-3\alpha)\ln\left(\frac{R_2}{R}
	\right)
\end{equation}

(Only the renormalization of the product $\psi\phi$ is determined - the 
manner in which the factor $\tilde{Z}_3$ is divided between $\phi$ and 
$\psi$ is arbitrary.)

We note that, since the seed action is of order $\frac{1}{g^2}$, while the
leading counterterms, (i.e. the one-loop counterterms), are independent of
$g^2$, we may, correct through the terms of orders $(g^2)^{-1}$ and
$(g^2)^0$ in (\ref{E253}), set all the $Z$ factors equal to 1 in the 
\emph{counterterm} term in the right-hand side of (\ref{E253}).  Hence
correct through this order, $S\left(\frac{Z_1}{Z_3}A,\frac{Z_1}{Z_3^2}B,
\phi,\tilde{Z}_3\psi,\frac{Z_1^2}{Z_3^3}g^2,Z_3\alpha\right)-S(A,B,\phi,
\psi,g^2,\alpha)$ is simply equal to \\$C(A,B,\phi,\psi,g^2,\alpha,R_2)-
C(A,B,\phi,\psi,g^2,\alpha,R)$, which is how (\ref{E254}) is obtained.

\subsection{The critical value of $g^2$}

Now as we mentioned above in the discussion following the statement of
our ansatz, we in practice have to divide $f_0(W_1,\ldots,W_n,g^2)$ by a
short-distance factor for each $W_i$, which takes the form of the sum of
all the Feynman diagrams contributing to $f_0(W_i,g^2)$, but with all the
propagators cut off smoothly at long distances, in the same manner as in
the counterterms, but with the onset of the long-distance cutoffs possibly
occurring at a different value of $\left|x-y\right|$, say at $\left|x-y
\right|=T$, (where $T$ is finite and strictly greater than zero), from the
onset value $R$ of the long-distance cutoffs on the propagators in the 
counterterms.  We allow $T$ to differ from $R$ because, as we will see, 
the real number $\frac{1}{\mu R}$, where $\mu^2$ is the coefficient of the
area in the Wilson area law in our ansatz, is computable from the 
group-variation equations in terms of the ``input'' value of $g^2$, as the
value of $\frac{L}{R}$ where the running coupling $\bar{g}^2\left(\frac{L}
{R}\right)$, with the initial value $\bar{g}^2(1)=g^2$, reaches a critical
value that is computable from the group-variation equations.  The critical
value is not a fixed point: it is the point where the magnitude of
$\frac{\beta(g)}{g}$ reaches a sufficiently \emph{large} value, as can be
seen from (\ref{E236}).

\subsection{The signs of the island diagrams}

The values of $Y_i$ for the three one-loop island
diagrams have no explicit dependence on $g^2$.  Obviously, $\displaystyle
\sum_iY_i$ has got to be \emph{negative}, and, indeed, the net
contribution to $\displaystyle\sum_iY_i$ from the three one-loop island
diagrams has surely also got to be \emph{negative}.  We will return to 
this crucial question shortly.

\subsection{$g^2$ never evolves to a value larger than the critical value}

We note here that, once $g^2$ reaches the critical value, the 
renormalization group has done its job.  The right-hand side of the
group-variaton equation for $f_0(W_1,\ldots,W_n,g^2)$, whenever any of the
sizes or separations of the $W_i$'s is larger than $\frac{1}{\mu}$, is
dominated by the contributions of island diagrams of size $\frac{1}{\mu}$.
Thus no matter how large the sizes and separations of the $W_i$'s, the
running coupling never evolves to a scale larger than $\frac{1}{\mu}$, and
never reaches a size larger than the critical value.\footnote{Note added:
Of course, this result depends crucially on the fact that an island
diagram contains \emph{no} propagators that are not part of the island.}
This is in agreement with experiment: there is no concrete experimental
evidence that $\alpha_s$ ever reaches a value significantly larger than 
the value 0.3 found for charmonium \cite{Perkins 179}.  Note added: the
largest value of $\alpha_s$ for which there is currently experimental
evidence, is the value $\alpha_s(m_\tau)=\alpha_s(1748\textrm{ MeV})=
0.35$, observed in $\tau$ decay \cite{beta in MS bar 1}.  As mentioned in
the Introduction, it is interesting to note that in Figure 9.2, on page 
19, of Chapter 9, \textit{Quantum Chromodynamics}, of reference
\cite{beta in MS bar 1}, the experimental value of $\alpha_s(m_\tau)$
lies approximately 1.6 standard deviations above the best-fit curve to
all measurements of $\alpha_s$, which indicates that the curve of 
$\alpha_s(\mu)$ may be curving upwards towards a vertical slope here, as
expected as $\alpha_s$ approaches the critical value.  In fact, as
discussed in the Introduction, and in Subsection \ref{Subsection 4LB},
the large-$N_c$ limit, of the general form of the four-loop 
$\beta$-function given in reference \cite{beta in MS bar 2}, indicates
that the series for our $\frac{\beta(g)}{g}$ probably diverges for 
$\frac{g^2}{4\pi}$ somewhere in the range 0.53 to 1.05, and most likely 
near the lower end of this range.  Since this $\frac{g^2}{4\pi}$ is equal
to $\frac{3}{2}$ times the value $\alpha_s$ would have in the absence of
quarks, which is about 
$\frac{3}{2}\times\frac{1}{1.22}\times\alpha_s=1.23\alpha_s$ at 1748 MeV,
(where the factor 1.22 corresponds to three quarks lighter than
1748 MeV), we expect the critical value of $\alpha_s$ to lie somewhere in
the range 0.43 to 0.85, and most likely near the lower end of this range.
Conversely, the observation of $\alpha_s(1748\textrm{ MeV})=0.35$
indicates that the critical value of our $\frac{g^2}{4\pi}$ is larger
than, and probably quite close to, 0.43.  We should also note that
$\frac{\beta(g)}{g}$ is expected to \emph{diverge} for our 
$\frac{g^2}{4\pi}$ somewhere in the range 0.53 to 1.05, and the critical
value of $\frac{g^2}{4\pi}$ is expected to be strictly \emph{less} than
the value of $\frac{g^2}{4\pi}$ where the series diverges, with 
a corresponding reduction in the expected critical value of $\alpha_s$.

\subsection{Normalization of $g^2$}
\label{g^2}

Our $\frac{g^2}
{4\pi}$ is larger than $\alpha_s$ as conventionally defined by a factor
$\frac{3}{2}$.  The factor of 3 is due ot the explicit factor of $N$ in
(\ref{E50}) and in (\ref{E240}), while the factor of $\frac{1}{2}$ is 
because we normalize our fundmental representation generators by 
(\ref{E43}), whereas the generators traditionally used in extracting
$\alpha_s$ from experiments are defined with an extra factor of $\frac{1}
{2}$ in the right-hand side of (\ref{E43}) \cite{DJGLH 5.20}, 
\cite{Quigg}, \cite{Perkins 178}, \cite{Perkins 398}.  Including an extra
factor of $\frac{1}{2}$ in the right-hand side of (\ref{E43}) has the
effect of reducing structure constants by a factor $\frac{1}{\sqrt{2}}$,
hence of increasing the value of $g$, as extracted from any experimental
result, by a factor $\sqrt{2}$, since only the product $gf_{abc}$ is
actually measured.

\subsection{The renormalization group, and the critical value of $g^2$}

Now in our BPHZ approach, there is no ``bare'' coupling constant, and we
never do any infinite rescalings of the fields.  Infinite rescalings of 
the fields would be completely pointless, because the counterterms do
\emph{not} have a gauge-invariant or BRST-invariant structure.  Only the
seed action has a BRST-invariant structure, and the property of the 
counterterms that ``corresponds'' to BRST-invariance is expressed by
(\ref{E253}): leaving the fields, $g^2$, and $\alpha$ unaltered, and 
replacing $R$ by $R_2$, is equivalent to leaving $R$ unaltered, and making
finite rescalings of the fields, $g^2$, and $\alpha$, as in the right-hand
side of (\ref{E253}).  The seed action in the right-hand side of 
(\ref{E253}) also has a BRST invariance, which differs from that of the
seed in the left-hand side by the rescalings of the fields.

Consequently we must consider carefully the definition and significance of
$\beta(g)$.  Now in conventional formulations of renormalization group
equations \cite{RGE}, the \emph{normalization} mass $\mu$, which is an
\emph{input}, and corresponds to $\frac{1}{R}$ in our approach, and
$\beta(g)$ always occur in the combination:
\begin{equation}
	\label{E255}
	\left(\mu\frac{\partial}{\partial\mu}+\beta(g)\frac{\partial}{\partial
	g}+\textrm{ other terms }\right)f=0
\end{equation}
Here the ``other terms'' depend on the identity of $f$, and when $f$ is
$f_0(W_1,\ldots,W_n,g^2)$, the only ``other terms'' are those associated
with the short-distance factors, which we omit in this paper.  We note 
that if $W_1,\ldots,W_n$ is a set of loops whose ``diameter'', i.e. the
largest distance between any two points on the loops, is equal to 
$\frac{1}{\mu}$, and $W_{1L},\ldots,W_{nL}$ denotes the set of loops 
obtained from $W_1,\ldots,W_n$ by uniformly rescaling all their sizes and
separations by a factor $\mu L$, so that their diameter becomes $L$, then
$f_0(W_{1L},\ldots,W_{nL},g^2)$ can only depend on $\mu$ and $L$ through
the combination $\mu L$, hence the relative sign of the two terms in 
(\ref{E102}) is indeed correct.

Now, as we mentioned, the conventional ``normalization mass'' $\mu$ 
corresponds to our $\frac{1}{R}$.  Therefore our version of (\ref{E255}) 
is:
\begin{equation}
	\label{E256}
	\left(-R\frac{\partial}{\partial R}+\beta(g)\frac{\partial}{\partial g}+
	\textrm{ other terms }\right)f=0
\end{equation}

Now when the ``other terms'' are absent, the meaning of (\ref{E256}) is 
that it gives the curves of constant $f$ in the $(R,g)$ plane.  Indeed, 
under variations of our input parameters $R$ and $g$, and assuming that
$f$ is a physical quantitiy, and hence independent of $\alpha$:
\begin{equation}
	\label{E257}
	\mathrm{d}f=\frac{\partial f}{\partial R}\mathrm{d}R+\frac{\partial f}
	{\partial g}\mathrm{d}g
\end{equation}

Hence when the ``other terms'' are absent, the meaning of (\ref{E256}) is
that the curves of constant $f$, i.e. $\mathrm{d}f=0$, in the $(R,g)$
plane, are given by:
\begin{equation}
	\label{E256a}
	R\frac{\mathrm{d}g}{\mathrm{d}R}=-\beta(g)
\end{equation}
This is the physical significance of the renormalization group from our
point of view.  Now from equation (\ref{E253}) we see that, up to the 
effects of rescaling the fields and $\alpha$, (which are what can give the
``other terms'' in (\ref{E256})), calculated quantities $f$ will be left
unaltered if we \emph{simultaneously} replace $R$ by $R_2$ and $g^2$ by
$g_2^2$, where:
\begin{equation}
	\label{E257a}
	g_2^2=\frac{Z_3^3}{Z_1^2}g^2
\end{equation}
Hence from (\ref{E256a}) we see that $\beta(g)$ is given by:
\begin{equation}
	\label{E258}
	\beta(g)=-R_2\frac{\mathrm{d}g_2}{\mathrm{d}R_2}
\end{equation}
where the derivative is to be evaluated at constant $R$ and $g$, after
which $R_2$ is to be set equal to $R$, and $g_2$ set equal to $g$.  Thus
from (\ref{E254}) and (\ref{E257a}) we find:
\begin{equation}
	\label{E259}
	2g\beta(g)=-\frac{11g^4}{12\pi^2}+\textrm{ terms of order }g^6
\end{equation}

Thus for small $g^2$ the ``curves of constant $f$'' in the $(R,g)$ plane
have the form:
\begin{equation}
	\label{E260}
	\frac{g^2}{4\pi}=\frac{1}{\frac{11}{6\pi}\ln\left(
	\frac{\textrm{constant}}{R^2}\right)}
\end{equation}

The ``curves of constant $f$'' in the $(R,g)$ plane look like:
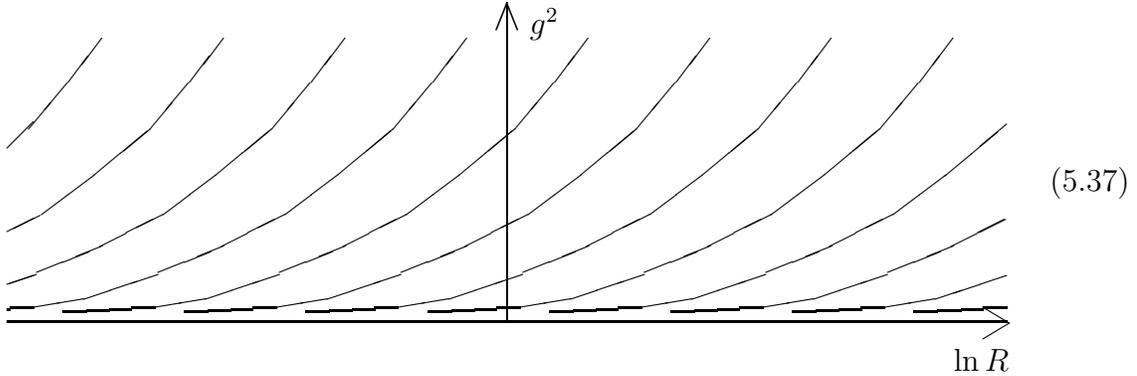
\unitlength 1.00pt
\linethickness{0.61pt}
\begin{equation}
	\label{E261}
	\raisebox{-69pt}{
\begin{picture}(380.00,140.34)
\put(11.00,25.00){\line(6,1){19.00}}
\put(57.00,25.00){\line(6,1){19.00}}
\put(103.00,25.00){\line(6,1){19.00}}
\put(149.00,25.00){\line(6,1){19.00}}
\put(195.00,25.00){\line(6,1){19.00}}
\put(241.00,25.00){\line(6,1){19.00}}
\put(287.00,25.00){\line(6,1){19.00}}
\put(333.00,25.00){\line(6,1){19.00}}
\put(30.00,28.00){\line(3,1){28.00}}
\put(76.00,28.00){\line(3,1){28.00}}
\put(122.00,28.00){\line(3,1){28.00}}
\put(168.00,28.00){\line(3,1){28.00}}
\put(214.00,28.00){\line(3,1){28.00}}
\put(260.00,28.00){\line(3,1){28.00}}
\put(306.00,28.00){\line(3,1){28.00}}
\put(12.00,38.00){\line(5,2){25.00}}
\put(58.00,38.00){\line(5,2){25.00}}
\put(104.00,38.00){\line(5,2){25.00}}
\put(150.00,38.00){\line(5,2){25.00}}
\put(196.00,38.00){\line(5,2){25.00}}
\put(242.00,38.00){\line(5,2){25.00}}
\put(288.00,38.00){\line(5,2){25.00}}
\put(334.00,38.00){\line(5,2){25.00}}
\put(36.00,48.00){\line(2,1){24.00}}
\put(82.00,48.00){\line(2,1){24.00}}
\put(128.00,48.00){\line(2,1){24.00}}
\put(174.00,48.00){\line(2,1){24.00}}
\put(220.00,48.00){\line(2,1){24.00}}
\put(266.00,48.00){\line(2,1){24.00}}
\put(312.00,48.00){\line(2,1){24.00}}
\put(14.00,60.00){\line(4,3){20.00}}
\put(60.00,60.00){\line(4,3){20.00}}
\put(106.00,60.00){\line(4,3){20.00}}
\put(152.00,60.00){\line(4,3){20.00}}
\put(198.00,60.00){\line(4,3){20.00}}
\put(244.00,60.00){\line(4,3){20.00}}
\put(290.00,60.00){\line(4,3){20.00}}
\put(336.00,60.00){\line(4,3){20.00}}
\put(11.00,24.67){\line(-1,0){9.00}}
\put(57.00,24.67){\line(-1,0){9.00}}
\put(103.00,24.67){\line(-1,0){9.00}}
\put(149.00,24.67){\line(-1,0){9.00}}
\put(195.00,24.67){\line(-1,0){9.00}}
\put(241.00,24.67){\line(-1,0){9.00}}
\put(287.00,24.67){\line(-1,0){9.00}}
\put(333.00,24.67){\line(-1,0){9.00}}
\put(379.00,24.67){\line(-1,0){9.00}}
\put(34.00,75.00){\line(6,5){21.00}}
\put(80.00,75.00){\line(6,5){21.00}}
\put(126.00,75.00){\line(6,5){21.00}}
\put(172.00,75.00){\line(6,5){21.00}}
\put(218.00,75.00){\line(6,5){21.00}}
\put(264.00,75.00){\line(6,5){21.00}}
\put(310.00,75.00){\line(6,5){21.00}}
\put(9.00,92.50){\line(5,6){14.58}}
\put(55.00,92.50){\line(5,6){14.58}}
\put(101.00,92.50){\line(5,6){14.58}}
\put(147.00,92.50){\line(5,6){14.58}}
\put(193.00,92.50){\line(5,6){14.58}}
\put(239.00,92.50){\line(5,6){14.58}}
\put(285.00,92.50){\line(5,6){14.58}}
\put(331.00,92.50){\line(5,6){14.58}}
\put(24.00,110.00){\line(3,4){12.75}}
\put(70.00,110.00){\line(3,4){12.75}}
\put(116.00,110.00){\line(3,4){12.75}}
\put(162.00,110.00){\line(3,4){12.75}}
\put(208.00,110.00){\line(3,4){12.75}}
\put(254.00,110.00){\line(3,4){12.75}}
\put(300.00,110.00){\line(3,4){12.75}}
\put(346.00,110.00){\line(3,4){12.75}}
\put(48.00,24.00){\line(-1,0){8.33}}
\put(94.00,24.00){\line(-1,0){8.33}}
\put(140.00,24.00){\line(-1,0){8.33}}
\put(186.00,24.00){\line(-1,0){8.33}}
\put(232.00,24.00){\line(-1,0){8.33}}
\put(278.00,24.00){\line(-1,0){8.33}}
\put(324.00,24.00){\line(-1,0){8.33}}
\put(370.00,24.00){\line(-1,0){8.33}}
\put(39.67,23.67){\line(-1,0){8.67}}
\put(85.67,23.67){\line(-1,0){8.67}}
\put(131.67,23.67){\line(-1,0){8.67}}
\put(177.67,23.67){\line(-1,0){8.67}}
\put(223.67,23.67){\line(-1,0){8.67}}
\put(269.67,23.67){\line(-1,0){8.67}}
\put(315.67,23.67){\line(-1,0){8.67}}
\put(361.67,23.67){\line(-1,0){8.67}}
\put(31.00,23.33){\line(-1,0){9.00}}
\put(77.00,23.33){\line(-1,0){9.00}}
\put(123.00,23.33){\line(-1,0){9.00}}
\put(169.00,23.33){\line(-1,0){9.00}}
\put(215.00,23.33){\line(-1,0){9.00}}
\put(261.00,23.33){\line(-1,0){9.00}}
\put(307.00,23.33){\line(-1,0){9.00}}
\put(353.00,23.33){\line(-1,0){9.00}}
\put(13.33,59.67){\line(-2,-1){12.33}}
\put(12.00,37.33){\line(-3,-1){11.00}}
\put(2.00,24.00){\line(-1,0){1.00}}
\put(1.00,85.00){\line(1,1){10.00}}
\put(1.00,19.33){\line(1,0){122.00}}
\put(122.33,19.33){\line(1,0){130.00}}
\put(251.00,19.33){\line(1,0){117.67}}
\put(368.33,19.33){\line(1,0){10.67}}
\put(355.67,74.67){\line(6,5){23.33}}
\put(358.00,48.00){\line(2,1){20.67}}
\put(379.00,37.00){\line(-3,-1){27.00}}
\put(190.00,19.67){\line(0,1){120.67}}
\put(190.00,140.00){\line(-2,-5){4.00}}
\put(190.00,140.00){\line(2,-5){4.00}}
\put(380.00,19.00){\line(-5,3){10.00}}
\put(380.00,19.00){\line(-5,-3){10.00}}
\put(204.00,138.00){\makebox(0,0)[ct]{$g^2$}}
\put(369.33,1.00){\makebox(0,0)[cb]{$\ln R$}}
\end{picture}
}
\end{equation}

Each of these curves corresponds to a different choice of the constant in
(\ref{E260}), and each of these curves gives a possible behaviour of the
``running coupling'' $\bar{g}^2(R)$.

We note that, for the reasons given above, for each value of the logarithm
in (\ref{E260}), our $\frac{g^2}{4\pi}$ is larger than $\alpha_s$ for the
corresponding value of the logarithm \cite{Perkins 307}, \cite{Quigg 223}
by a factor $\frac{3}{2}$.

Now the critical value of $g^2$ is certainly calculable from the
group-variation equations.  In brief, each $Y_i$ in the right-hand side of
(\ref{E236}) will be equal, on dimensional grounds, to $\mu^2$, tims a
calculable \emph{numerical} factor, times a power of $g^2$ that can be 
read off the corresponding island diagram.  For the one-loop islands, this
power of $g^2$ is \emph{zero}.  Thus $\mu^2$ cancels out, and we are left
with an equation of the form $1=\frac{\beta(g)}{g}$ times a function of
$g^2$ that begins with a term independent of $g^2$.  Hence (\ref{E236})
fixes $g^2$, and the condition is roughly that $\frac{\beta(g)}{g}$ be
equal to the reciprocal of the sum of the numerical factors in the $Y_i$
corresponding to the three one-loop island diagrams.  (This of course 
means that, as we noted, the sum of these three $Y_i$ must be
\emph{negative}.)  Now (\ref{E236}) of course depends on our ansatz, and 
we thus see that, as we noted above, throughout the domain described by
our ansatz, (i.e. when the sizes and separations of the loops are all
larger than $\frac{1}{\mu}$), $g^2$ remains equal to the critical value:
$g^2$ never goes above the critical value.  Now at smaller scales, the
behaviour described by our ansatz has not yet been achieved, and $g^2$ can
be smaller.  What this means in practice is that we can do our 
calculations with an input value of $g^2$ as small as we like: the only
restrictions on the input value of $g^2$ are that it must be strictly
greater than zero, and not greater than the critical value.  However, the
input value of $g^2$ chosen fixes $\frac{1}{\mu R}$, (where $\mu^2$ is the
coefficient of the area in the Wilson area law in our ansatz, so that 
$\frac{1}{\mu}$ is the typical island size), according to the 
renormalization group.  If we choose the input value of $g^2$ equal to the
critical value, then $\frac{1}{\mu R}$ is a calculable number of order 1.
And as we reduce the input value of $g^2$ below the critical value, 
$\frac{1}{\mu R}$ increases very rapidly, in a manner given precisely by
the renormalization group.  For example, (referring to experimental data),
the experimental value of $\mu$ is 0.41 GeV, as we noted before.  The
experimental value of $\alpha_s$ at about 4 GeV is approximately 0.3,
\cite{Perkins 179}, and the experimental value of $\alpha_s$ at about 40
GeV is approximately 0.16, \cite{Perkins 300}.  Now our $\frac{g^2}{4\pi}$
is $\frac{3}{2}$ times the value $\alpha_s$ would have in the absence of 
quarks, and the presence of the quarks rougly increases $\alpha_s$, in
comparison to the value it would have in the absence of the quarks, by a
factor 1.3 at 4 GeV, and by a factor 1.4 at 40 GeV.  Hence we infer that 
the critical value of our $\frac{g^2}{4\pi}$ is greater than 0.35, and 
that if we choose our input value of $\frac{g^2}{4\pi}$ equal to 0.35, we
will find that $\frac{1}{\mu R}$ is approximately equal to 10, (which 
would mean that our input parameter $R$ is approximately equal to $0.25
\textrm{ GeV}^{-1}$), and that if we choose our input value of $\frac{g^2}
{4\pi}$ equal to 0.17, we will find that $\frac{1}{\mu R}$ is
approximately equal to 100, (which would mean that our input parameter $R$
is approximately equal to $0.025\textrm{ GeV}^{-1}$).

\section{The result of dividing by the short-distance factors}

The reason for this is as follows.  Our $f_0(W_1,\ldots,W_n,g^2)$ satisfy
a version of (\ref{E256}) where the ``other terms'' are associated 
exclusively with the short-distance factors which we have to divide out.
These short-distance factors are characterized by a parameter $T$, which,
as we explained above, is the value of $\left|x-y\right|$ where the 
long-distance cutoffs imposed on the propagators in the short-distance
factors, begins.  $T$ is finite and strictly greater than zero.  The value
of $T$ cancels out of all physical quantities, such as $\mu$, and the mass
$m$ of the lightest glueball.  We allow $T$ to differ from $R$ because, as
we just noted, if we choose our input value of $g^2$ equal to the critical
value, we will find that $\frac{1}{\mu R}$ is equal to a calculable number
of order 1.  Thus if we choose our input value of $g^2$ equal to the
critical value, our input parameter $R$ is approximately equal to the 
typical island size.  Hence if we required $T$ to be equal to $R$, we 
would have practical problems when we explicitly restore the 
short-distance factors in the right-hand sides of the group-variation
equations, as we have to do, because the short-distance factors would then
become subject to strong-coupling effects.  Hence we allow $T$ to differ
from $R$, so that if we choose our input value of $g^2$ close to the 
critical value, we are free to choose $T$ to be substantially smaller than
$R$, to ensure that the short-distance factors are not subject to 
strong-coupling effects.  When we express the group-variation equations in
terms of the $f_0$'s with the short-distance factors divided out, we have
to restore the short-distance factors explicitly in each window of each
right-hand side group-variation equation diagram.  This results in partly
recovering the \emph{Feynman} diagram expansions of the left-hand sides of
the group-variation equations, but in such a way that all the 
long-distance information is contained in the ``long-distance factors'',
(i.e. the $f_0$'s of the right-hand side windows with their 
short-distance factors divided out), and the propagators in the ``Feynman
diagrams'' are all cut off smoothly at long distances, with the onset of
the cutoffs occurring at $\left|x-y\right|=T$.  We also divide each
right-hand side group-variation equation diagram by the short-distance
factor associated with the left-hand side $f_0$, which for the island
diagrams trivially cancels the restoration of the short-distance factors
associated with the left-hand side loops $W_1,\ldots,W_n$, and for 
non-island diagrams cancels the main part of the restoration of the 
short-distance factors for the paths that form part of the left-hand side
loops, with some complications occurring at the junctions where a 45-path
ends on a left-hand side loop.

The sum of ``Feynman diagrams'' we obtain when we restore the 
short-distance factors in the windows of the right-hand side 
group-variation equation diagrams also differs from the Feynman diagram
expansion of the left-hand side of the group-variation equation, (which,
as we mentioned after equation (\ref{E184}), is simply the sum of the 
Feynman diagrams contributing to the left-hand side $f_0$, with each of 
these Feynman diagrams being multiplied by it number of windows, for the
following reason:

We define a ``band'' of a Feynman diagram contributing to an expectation
value or correlation function of Wilson loops, to be a connected component
of what remains of the diagram whenall the Wilson loops are removed.  When
we substitute the short-distance factor, as defined above, into a 
\emph{non-simply-connected} window of a right-hand side group-variation
diagram, we recover not the perturbative expansion, with modified
propagators, of the correlation function of Wilson loops that border that
window, but rather the perturbative expansion, with modified propagators,
of the product of the vacuum expectation values of the Wilson loops that
border that window.  (All factors of $N$ have of course been removed - 
what we have is $\displaystyle\prod_if_0(W_i,g^2)$, where the $W_i$'s are
the Wilson loops that border that window.)  The Feynman diagrams
contributing to this product of vacuum expectation values have no bands 
with ends on two or more different $W_i$'s, in complete contrast to the
Feynman diagrams contributing to the correlation function of the $W_i$'s.
What this means in practice is that it is our ``long-distance factors'',
(i.e. our $f_0$'s divided by their short-distance factors, as defined 
above), that become singular when two \emph{different} $W_i$'s intersect
one another.  This is not a serious problem, because configurations where
two different Wilson loops intersect one another are not very important in
four dimensions, but it does mean that, for the full recovery of the 
renormalized perturbation expansion of the left-hand sides of the 
group-variation equations, our ansatz for the long-distance factors must
be refined to include this feature.

The perturbative expansions, (i.e. in powers of $g^2$), of the 
long-distance factors, (i.e. the $f_0$'s divided by their short-distance
factors), have a simple form: they can be expressed by the same sum of
Feynman diagrams that contributes to $f_0$, provided we modify the 
mathematical expression that corresponds to each diagram as follows, and
also partly drop the requirement of planarity, as follows.  We call the
bands, (as defined above), of a Feynman diagram contributing to 
$f_0(W_1,\ldots,W_n)$, ``mono-bands'', ``duo-bands'', ``trio-bands'', and
so on, according to the number of different $W_i$'s they have ends on. 
Thus a mono-band has all its ends on a single $W_i$, a duo-band has ends
on precisely two different $W_i$'s, a trio-band has ends on precisely
three different $W_i$'s, and so on.  We then relax the requirement of
planarity as follows:  every band is still required to be planar ``within
itself'', and the relations among all bands other than mono-bands are
still required to be planar, as before.  However, the requirement of
planarity in the relations between mono-bands and other bands is dropped:
it is as if, to each mono-band, every other band is invisible.  We define 
the contribution to $f_0(W_1,\ldots,W_n)$ of each band configuration that 
is now allowed, but which was not allowed before, to be zero.  Now our
long-distance factor is, by definition:
\begin{equation}
	\label{E262}
	\frac{f_0(W_1,\ldots,W_n,g^2)}{\displaystyle\prod_{i=1}^n\tilde{f}_0
	(W_i,g^2)}
\end{equation}
where $\tilde{f}_0(W_i,g^2)$ is defined by the same sum of Feynman
diagrams as $f_0(W_i,g^2)$, but with all propagators cut off smoothly at
long distances, with the onsets of the long-distance cutoffs occurring at
$\left|x-y\right|=T$.  We expand each factor $\frac{1}{\tilde{f}_0(W_i,
g^2)}$ as:
\begin{equation}
	\label{E263}
	\frac{1}{\tilde{f}_0(W_i,g^2)}=1-\left(\tilde{f}_0(W_i,g^2)-1\right)+
	\left(\tilde{f}_0(W_i,g^2)-1\right)^2-\ldots
\end{equation}

Let us now consider one of our generalized diagrams contributing to \\
$f_0(W_1,\ldots,W_i,g^2)$, (i.e. where we have dropped the requirement of
planarity with reference to the relations between different mono-bands,
and between mono-bands and other bands, and defined the contributions to
$f_0(W_1,\ldots,W_n,g^2)$ from the new configurations that were previously
forbidden, to be zero).  Let us consider a specific configuration, in
position space, of the vertices of this diagram, and let us consider
\emph{all} the contributions to (\ref{E262}) of the precise set of bands
and configuration of their vertices that we have here, when we expand all
the denominators in (\ref{E262}) as in (\ref{E263}).  Let $A$ represent 
all the bands \emph{except} the mono-bands, i.e. $A$ contains all the 
duo-bands, all the trio-bands, etc.  We treat $A$ as a single indivisible
unit, since no band in $A$ can come from anywhere except $f_0(W_1,\ldots,
W_n,g^2)$, (i.e. no band in $A$ can come from any of the
$\left(\tilde{f}_0(W_i,g^2)-1\right)$'s).  Let the mono-bands of this
generalized diagram ``contributing'' to $f_0(W_1,\ldots,W_n,g^2)$ be
$B_1,B_2,\ldots,B_r$.  We are considering a specific configuration in
position space of the vertices of all the bands, hence we may assum that 
all $B_i$ are different, since even if two of the $B_i$'s look identical
\emph{as diagrams}, configurations where all their corresponding vertices
have identical positions in configuration space have measure zero, and 
may hence be ignored.  Now each mono-band $B_j$ is attached to a specific
one of the $W_i$'s, and if $B_j$ is attached to $W_i$, then $B_j$ can also
occur in $\tilde{f}_0(W_i,g^2)$, in which case we denote it by $\tilde{B}_
j$, to signify that all its propagators have long-distance cutoffs, 
commencing at $\left|x-y\right|=T$.

Now let us suppose, for example, that there are just two mono-bands, $B_1$
and $B_2$.  We then write:
	\[f_0(W_1,\ldots,W_n,g^2)=\Big\{1+\ldots+\left\langle A\right\rangle+
	\ldots+\left\langle B_1\right\rangle+\ldots+\left\langle B_2
	\right\rangle+\ldots+\left\langle AB_1\right\rangle+
\]
\begin{equation}
\label{E264}
	+\ldots+\left\langle AB_2\right\rangle+\ldots+\left\langle B_1B_2
	\right\rangle+\ldots+\left\langle AB_1B_2\right\rangle+\ldots\Big\}
\end{equation}
where the bracket notation displays all the bands in a diagram, $A$ is an
abbreviation of $D_1\ldots D_dT_1\ldots T_t\ldots$, where the $D_i$'s are
the duo-bands, the $T_i$'s are the trio-bands, and so on, and each $B_i$,
$D_j$, $T_k$, etc., indicates not only a specific band \emph{as a 
diagram}, but also a specific position-space configuration of the vertices
of that band, and we explicitly display all the contributions to
$f_0(W_1,\ldots,W_n,g^2)$ that can contribute to the contribution to our
long-distance factor (\ref{E262}) that involves \emph{precisely} the bands
in $A$, and also all the $B_i$'s, where each $B_i$ may or may not have a
``twiddle''.

And in the same bracket notation, if the mono-bands attached to $W_i$, for
example, are precisely $B_j$ and $B_k$, we write:
\begin{equation}
	\label{E265}
	\tilde{f}_0(W_i,g^2)=1+\ldots+\left\langle \tilde{B}_j\right\rangle+
	\ldots+\left\langle \tilde{B}_k\right\rangle+\ldots+\left\langle 
	\tilde{B}_j\tilde{B}_k\right\rangle+\ldots
\end{equation}

We then expand each $\frac{1}{\tilde{f}_0(W_i,g^2)}$ as in (\ref{E263}),
and extract all the terms that involve precisely $A$, and also each $B_i$
precisely once, where each $B_i$ may occur either without a ``twiddle'' or
with a ``twiddle'', (but not both ways in the same term).

The general result is that we have a sum over all the partitions of
$\left\langle AB_1B_2\ldots B_r\right\rangle$ into nonempty parts, (with
$A$ being treated as an indivisible unit), such that every $B_i$ in the
\emph{same} part as $A$, occurs \emph{without} a ``twiddle'', and every
$B_i$ \emph{not} in the same part as $A$, occurs with a ``twiddle'', (i.e.
is $\tilde{B_i}$), and the partitions are restricted by the requirement
that no two $\tilde{B}_j$'s attached to two \emph{different} $W_i$'s, may
occur in the same part, (which means that every part that does \emph{not}
contain $A$, is associated with a \emph{specific} $W_i$).  And if, for
each $1\leq i\leq n$, the number of parts of the partition associated with
$W_i$, (in the sense that all the $\tilde{B}_i$'s in that part, are 
attached to $W_i$), is $m_i$, then the coefficient of that partition is:
\begin{equation}
\label{E266}
	(-1)^{m_1+m_2+\ldots+m_n}m_1!m_2!\ldots m_n!
\end{equation}

For example, if the mono-bands are $B_1$ and $B_2$, and $B_1$ and $B_2$ 
are both attached to the \emph{same} $W_i$, we have:
\begin{equation}
	\label{E267}
	\left\langle AB_1B_2\right\rangle-\left\langle AB_1\right\rangle
	\left\langle \tilde{B}_2\right\rangle-\left\langle AB_2\right\rangle
	\left\langle \tilde{B}_1\right\rangle-\left\langle A\right\rangle
	\left\langle \tilde{B}_1\tilde{B}_2\right\rangle+2\left\langle A
	\right\rangle\left\langle \tilde{B}_1\right\rangle\left\langle 
	\tilde{B}_2\right\rangle
\end{equation}
while if the mono-bands are $B_1$ and $B_2$, and $B_1$ and $B_2$ are
attached to two \emph{different} $W_i$'s, then we have:
\begin{equation}
	\label{E268}
	\left\langle AB_1B_2\right\rangle-\left\langle AB_1\right\rangle
	\left\langle \tilde{B}_2\right\rangle-\left\langle AB_2\right\rangle
	\left\langle \tilde{B}_1\right\rangle+\left\langle A\right\rangle
	\left\langle \tilde{B}_1\right\rangle\left\langle \tilde{B}_2
	\right\rangle
\end{equation}

In these formulae, and in the general result, if there are \emph{no} bands
in $A$, (i.e. if all the bands are mono-bands), we simply set $A=1$.  This
is the form of the result that applies for $n=1$.  (of course, for 
$n\geq2$, there is always at least one band in $A$, since every Feynman
diagram contributing to $f_0(W_1,\ldots,W_n,g^2)$, is connected.)

We note that if \emph{all} the mono-bands $B_1,B_2,\ldots,B_r$ are 
attached to the \emph{same} $W_i$, then we simply have the ``correlation
function'' of $A$ and all the $B_i$'s, with each $B_i$ that is \emph{not}
in the same part as $A$, becoming a $\tilde{B}_i$.  We also note that our
general result here has nothing to do with planar diagrams or the 
large-$N$ limit.  In fact, the general result is more simply stated for
the general-$N$ case.  The interpretation of the bracket notation is 
simply that for each ``$\tilde{B}$ part'' that is associated with a loop
$W_i$, we
have a separate ``duplicate'' of the loop $W_i$, ``superimposed'' on the 
loop $W_i$, such that the fundamental representation matrices $t_a$ at the
ends of the mono-bands in that ``$\tilde{B}$ part'', have their own matrix
product and trace around that ``duplicate'' of the loop $W_i$, (which
comes from the expansion of one of the denominator factors), rather than
``mixing in'' with the matrix products around the ``original'' loop $W_i$,
which is included in the part that contains $A$, and comes from the 
numerator in (\ref{E262}).

We can now see how dividing out the short-distance factors has cancelled 
all the linear short-distance divergences along the loops.  These occur
when a mono-band, or more generally, a group, or cluster, of monobands, 
all on the same $W_i$, which may be nested, (in the large-$N$ case, or, in
the case of unrestricted $N$, generally ``entangled'' with one another,
all shrink togethr to a point on the loop $W_i$, such that the freedom of
movement of this point along the loop $W_i$ is not restricted by this
group of bands being entangled with any other band.  Now when such a group
of mono-bands is not entangled with any other band, (by which we mean, 
that the matrices $t_a$ at the vertices where gluon propagators of these
bands end on the loop $W_i$, occur in a single sequence, unbroken by the 
presence of any $t_a$ belonging to any band \emph{not} in this group), the
structure constants $f_{abc}$ at the action vertices in these bands, and 
the matrices $(t_a)_{jk}$ at the vertices where gluon propagators of these
bands end on the loop $W_i$, are contracted into an $\textrm{SU}(N)$-invariant
tensor $X_{jk}$, which has precisely one ``quark representation index'',
$j$, one ``anti-quark representation index'', $k$, and \emph{no} adjoint
representation indices, and hence satisfies, for any infinitesimal
parameters $\epsilon_a$, $a\in \textrm{SU}(N)$:
\begin{equation}
	\label{E269}
	\left(\delta_{jp}+\left(\epsilon_a t_a\right)_{jp}\right)\left(\delta_
	{kq}+\epsilon_b\left(\left(t_b\right)_{kq}\right)^\ast\right)X_{pq}=
	X_{jk}
\end{equation}
which implies immediately that, for all $t_a$:
\begin{equation}
	\label{E270}
	\left(t_a\right)_{jp}X_{pk}-X_{jq}\left(t_a\right)_{qk}=0
\end{equation}
or in other words that, considered as a matrix, $X_{jk}$ commutes with 
all the $t_a$'s \cite{Polyakov}.  (Equation (\ref{E269}) follows 
immediately from the invariant tensor properties of $f_{abc}$ and $\left(
t_a\right)_{jk}$.)  Now the $t_a$'s are the generators of the fundamental
representation of $\textrm{SU}(N)$, which is irreducible, hence the fact that 
$X_{jk}$ commutes with all the $t_a$'s implies immediately that $X_{jk}$ 
is a multiple of the unit matrix, $X_{jk}=X\delta_{jk}$, say.

Let us now number our bands such that the $s$ mono-bands which are going
to shrink to a point on one of the $W_i$'s, are $B_1$, $B_2$, $\ldots$, 
$B_s$.  We assume $s\geq1$.  Now all the $s$ ``shrinking'' mono-bands are
of course attached to a \emph{single} one of the $W_i$'s: let the $W_i$ to
which the $s$ ``shrinking'' bands are attached, be $W_1$.

Let us now consider all the terms in our long-distance factor, as derived
above, which involve precisely a specified set, $A$, of duo-bands, 
trio-bands, and so on, precisely the $s$ ``shrinking'' mono-bands 
$B_1,B_2,\ldots,B_s$, and a precise set $B_{s+1},\ldots,B_t$ of additional
monobands.  These terms correspond, as shown above, to all partitions of
$\left\langle AB_1B_2\ldots B_sB_{s+1}\ldots B_t\right\rangle$, such that
in every part, (i.e. every bracket), that does \emph{not} contain $A$, all
the $B_j$'s in that part are attached to the \emph{same} $W_i$.  Now
bearing in mind the origins of the brackets, we call the one part of each
such partition that contains $A$, the ``numerator part'', and all the 
other parts of each such partition, ``denominator parts''.  Thus each
denominator part is associated with a specific $W_i$, in the sense that 
all the mono-bands in that part, are attached to that $W_i$.  Furthermore,
all the $B$'s in the numerator part are \emph{without} ``twiddles'', while
all the $B$'s in every denominator part, have ``twiddles''.  And if, for
each $1\leq i\leq n$, the number of denominator parts associated with
$W_i$, is $m_i$, then the coefficient of that partition is given by
(\ref{E266}).

Now the ``shrinking'' mono-bands, $B_1$, $B_2$, $\ldots$, $B_s$, are by
assumption not ``entangled'' with any other band, in the sense defined
above.  Hence by the result just given, we have, for each denominator part
associated to $W_1$, in any of our partitions, that if, for example, that
denominator part is $\left\langle \tilde{B}_2\tilde{B}_5\tilde{B}_7
\tilde{B}_9\tilde{B}_{15}\right\rangle$, where $\tilde{B}_2$, $\tilde{B}
_5$, and $\tilde{B}_7$ are ``shrinking'' mono-bands, and $\tilde{B}_9$ and
$\tilde{B}_{15}$ are ``non-shrinking'' mono-bands attached to $W_1$, the
following identity, in our bracket notation:
\begin{equation}
	\label{E271}
	\left\langle \tilde{B}_2\tilde{B}_5\tilde{B}_7\tilde{B}_9\tilde{B}_{15}
	\right\rangle=\left\langle \tilde{B}_2\tilde{B}_5\tilde{B}_7
	\right\rangle\left\langle \tilde{B}_9\tilde{B}_{15}\right\rangle
\end{equation}
In other words, each denominator part associated to $W_1$, factorizes
exactly into a product of two factors, one of which is the bracket 
containing all the ``shrinking'' mono-bands in that denominator part, and 
the other of which is the bracket containing all the ``non-shrinking''
mono-bands in that denominator part.  (Any bracket that contains \emph{no}
bands, is by definition equal to 1.  This applies, in particular, for 
$n=1$, to the numerator bracket, when the numerator bracket contains just
$A$, since for $n=1$, there are \emph{no} duo-bands, trio-bands, etc.)

And we also have a corresponding result for the numerator bracket: if, for
example, the numerator bracket is $\left\langle AB_3B_6B_{10}B_{12}B_{19}
\right\rangle$, where $B_3$ and $B_6$ are ``shrinking'' mono-bands, and
$B_{10}$, $B_{12}$, and $B_{19}$ are ``non-shrinking'' mono-bands, then we
have the identity:
\begin{equation}
	\label{E272}
	\left\langle AB_3B_6B_{10}B_{12}B_{19}\right\rangle=\left\langle AB_{10}
	B_{12}B_{19}\right\rangle\left\langle B_3B_6\right\rangle
\end{equation}
In other words, the numerator bracket exactly factors into a product of
two factors, one of which is the numerator bracket containing just $A$ and
all the \emph{non-shrinking} mono-bands in the original numerator bracket,
and the other of which is a bracket containing all the \emph{shrinking}
mono-bands in the original numerator bracket:this factor is like a
denominator bracket, but without the ``twiddles''.

Let us now suppose that the position-space configuration of the vertices
of the ``shrinking'' mono-bands is sufficiently ``shrunk'' that , for 
every propagator in each of these bands, the distance between the ends of
that propagator is less than or equal to $T$, where $T$ is the onset point
of the smooth long-distance cutoffs imposed on the propagators in the
$\tilde{B}$'s.  then the smooth long-distance cutoffs imposed on the 
``non-counterterm'' propagators in the $\tilde{B}$'s have \emph{no 
effect}, hence for $1\leq j\leq s$, (where $s$ is the number of 
``shrinking'' mono-bands), we may set $\tilde{B}_j=B_j$.  (We note that,
in the \emph{counterterms} in the $\tilde{B}$'s, we must use the same
smooth long-distance cutoffs as in the counterterms \emph{in the}
$B$\emph{'s}, i.e. with the onset of the smooth long-distance cutoffs
occurring at $R$, \emph{not} at $T$.)

We thus see that, due firstly to the factorization identities (\ref{E271})
and (\ref{E272}), and secondly to the smallness, in configuration space,
of the shrinking mono-bands, our original set of partitions of 
$\left\langle AB_1B_2\ldots B_sB_{s+1}\ldots B_t\right\rangle$, has 
collapsed to a strict subset of this original set of partitions, whose
members may be characterized as follows: a member of our ``final set'' of
partitions of $\left\langle AB_1B_2\ldots B_sB_{s+1}\ldots B_t
\right\rangle$, consists of an allowed partition of $\left\langle AB_{s+1}
\ldots B_t\right\rangle$, together with an arbitrary partition of 
$\left\langle B_1B_2\ldots B_s\right\rangle$.  We have to calculate the
total coefficient with which we obtain each member of this ``final set''
of partitions.  Let us consider a member of this ``final set'' of 
partitions, which has, as before, for each $1\leq i\leq n$, a total of
$m_i$ denominator parts associated with $W_i$.  (By a ``denominator 
part'', we still mean any part that does not contain $A$.)  For 
convenience, we also define $m=m_1$.  Let the number of parts of our 
partition of $\left\langle B_1B_2\ldots B_s\right\rangle$ be $u$.  Thus
we have $1\leq u\leq m$.  We have to identify, among all the original
partitions, all the possible ``sources'' of this final partition.  These
are given by all possible mergings of $k$ parts of our partition of
$\left\langle B_1B_2\ldots B_s\right\rangle$, where $0\leq k\leq
\mathrm{min}(u,m+1-u)$, with either denominator parts, \emph{associated
with} $W_1$, of our partition of $\left\langle AB_{s+1}\ldots B_t\right
\rangle$, or with the numerator part of our partition of $\left\langle A
B_{s+1}\ldots B_t\right\rangle$, subject to the restriction that at most
one part of our partition of $\left\langle B_1B_2\ldots B_s\right\rangle$,
can merge with any given part of our partition of $\left\langle AB_{s+1}
\ldots B_t\right\rangle$.  Now for each possible such merging of $k$ parts
of our partition of $\left\langle B_1B_2\ldots B_s\right\rangle$, into 
parts of our partition of $\left\langle AB_{s+1}\ldots B_t\right\rangle$,
subject to the restrictions stated, the coefficient of the original 
partition obtained by that merging, is obtained from the coefficient of
our ``final'' partition, by multiplying by:
\begin{equation}
	\label{E273}
	(-1)^k\frac{(m-k)!}{m!}
\end{equation}
since the total number of denominator parts associated with $W_1$ has been
reduced by $k$.  And the number of distinct possible such mergings, 
involoving $k$ parts of our partition of $\left\langle B_1B_2\ldots B_s
\right\rangle$, is:
\begin{equation}
	\label{E274}
	\frac{(m+1-u)!}{(m+1-u-k)!k!}\frac{u!}{(u-k)!k!}k!
\end{equation}
Hence the total coefficient of our final partition is:
\begin{equation}
	\label{E275}
	\frac{(m+1-u)!u!}{m!}\sum_{k=0}^{\mathrm{min}(u,m+1-u)}(-1)^k
	\frac{(m-k)!}{(m+1-u-k)!(u-k)!k!}
\end{equation}
times the original coefficient of our final partition.  Now, defining
$v=m+1-u$, so that we have both $u\geq1$ and $v\geq1$, the sum in 
(\ref{E275}) is equal to:
\begin{equation}
	\label{E276}
	F(u,v)=\sum_{k=0}^{\mathrm{min}(u,v)}(-1)^k\frac{(u+v-k-1)!}
	{(u-k)!(v-k)!k!}
\end{equation}
This expression $F(u,v)$ is symmetric in $u$ and $v$, and is well-known to
vanish for all integer $u$ and $v$ such that $u\geq1$ and $v\geq1$ are
both true.  Indeed, we see immediately that $F(u,1)=0$ for all integers
$u\geq1$.  And for all integers $u$, $v$, such that $u\geq v\geq2$, we
have:
	\[F(u,v)=\sum_{k=0}^v(-1)^k\frac{(u+v-k-1)!}{(u-k)!}\frac{1}{v}\left(
	\frac{1}{(v-k-1)!k!}+\frac{1}{(v-k)!(k-1)!}\right)=
\]
	\[=\frac{1}{v}\left(\sum_{k=0}^{v-1}(-1)^k\frac{(u+v-k-1)!}{(u-k)!
	(v-k-1)!k!}+\sum_{k=0}^{v-1}(-1)^{k+1}\frac{(u+v-k-2)!}{(u-k-1)!
	(v-k-1)!k!}\right)=
\]
\begin{equation}
	\label{E277}
	=-\left(\frac{v-1}{v}\right)F(u,v-1)
\end{equation}
Hence by induction on $v$, $F(u,v)=0$ for all integers $u$ and $v$ such 
that $u\geq v\geq1$, hence, by symmetry, $F(u,v)=0$ for all integers $u$
and $v$ such that $u\geq1$ and $v\geq1$ are both true.  Hence our 
long-distance factor (\ref{E262}) has no contributions at all from any
configurations where a group of mono-bands, all on the same $W_i$, and not
entangled with any other band, in the sense defined above, are 
sufficiently small in configuration space, that all their propagators have
$\left|x-y\right|\leq T$.  Now by power-counting, these are the only
configurations which can produce linear divergences along the loops $W_i$,
hence our long-distance factor (\ref{E262}) is completely free from such
linear divergences.

Now these linear divergences, which we have just verified all cancel out 
of (\ref{E262}), are all associated with subdiagrams, of Feynman diagrams
contributing to \\
$f_0(W_1,\ldots,W_n,g^2)$, that have precisely two ``path
legs'', and \emph{no} gluon legs.  Our long-distance factors (\ref{E262})
\emph{do} still have residual divergences on the $W_i$'s associated with
subdiagrams, of Feynman diagrams contributing to 
$f_0(W_1,\ldots,W_n,g^2)$, that have precisely two ``path legs'', and 
\emph{one} gluon leg.  These subdiagrams look like ``vertex corrections''
to a vertex obtained by contracting the subdiagram to a point, so that the
gluon leg ends directly on the path.  We find by power-counting that these
divergences are at worst logarithmic, and that they are \emph{not} 
affected by the presence of corners in the path.  (The logarithmic 
``corner divergences'' discussed in \cite{Polyakov} are associated with
subdiagrams with \emph{no} gluon legs, and, being entirely due to 
configurations where a group of mono-bands on a single $W_i$, not 
entangled with any other band, are such that all their propagators are
``shorter'' than $T$, they are totally absent from (\ref{E262}) by the 
proof just given.)  These ``vertex'' divergences, due to subdiagrams with
one gluon leg and two ``path legs'', simply give a fixed infinite
renormalization of the gluon field
$A\raisebox{-4pt}{$\stackrel{\displaystyle{(x)}}{\scriptstyle{\mu a}}$}$
at the vertex, in the diagram obtained from the given diagram by 
contracting the divergent subdiagram to a point, where the gluon leg meets
the path.  However, in our BPHZ approach, there are no infinite
renormalizations of the fields, and we therefore have to introduce
counterterms for these divergent subdiagrams by hand.  The counterterm for
each such divergent subdiagram is defined once only: for an infinite 
straight path.  This one counterterm cancels the divergence even when the 
divergent subdiagram is ``going round a corner'' - this follows directly
from power-counting, since when we calculate the change in the counterterm
that would be required near a corner, the ``contraction point'' of the
subdiagram, (e.g. the mean position of the vertices on the path, or the
position of one of the vertices on the path), is no longer fixed, hence we
get one extra integral along the path, which makes the ``corner 
discrepancy'' finite.  These extra counterterms are defined in our 
standard manner \cite{BPHZ 13 219}.  We have to choose a set of 
contraction weights for the subdiagram - the fact that the divergence is
only logarithmic means that the counterterm is independent of the choice
of the set of contraction weights.  (In general, for a
translation-invariant theory, changing the set of contraction weights
changes the counterterm by a total divergence, but when the degree of 
divergence is zero, changing the set of contraction weights has no effect
on the counterterm.)  The propagators in these extra counterterms are cut
off smoothly at long distances in exactly the same manner as the 
propagators in the usual counterterms, with the onsets of the cutoffs
occurring at $\left|x-y\right|=R$.  We also have to add \emph{finite}
counterterms, which are necessary to ensure that, when we substitute the
\emph{perturbative} expansions of the long-distance factors of the windows
into the right-hand sides of the group-variation equations, we obtain the
correct renormalized perturbative expansions of the left-hand sides of the
group-variation equations.  These extra finite counterterms may be
calculated in perturbation theory.  They make their appearance in the
solution of the group-variation equations via the \emph{initial} valules
of the $f_0$'s, at small $g^2$, (or equivalently, at small $L$), which are
also calculated in perturbation theory.  They also have to be included in
some non-island diagrams.

Now these counterterms, which we require for these subdiagrams with one
gluon leg and two ``path legs'', do \emph{not} arise from any counterterms
in the action.  They are special counterterms which, in our BPHZ approach,
form an integral part of the definition of the long-distance factors of
Wilson-loop vacuum expectation values and correlation functions. 
Nevertheless, we find that, if we change $R$ to another finite value, 
$R_2$, also strictly greater than zero, the effect on these counterterms 
is exactly equivalent to leaving $R$ \emph{unaltered}, and multiplying by
$A\raisebox{-4pt}{$\stackrel{\displaystyle{(x)}}{\scriptstyle{\mu a}}$}$,
in the path-ordered phase factor, by exactly the same finite factor
$\frac{Z_1}{Z_3}$ as occurs in (\ref{E253}).  (The finite change in the
counterterm for a given subdiagram with two ``path legs'' and one gluon
leg, contributes to the $\frac{Z_1}{Z_3}$ factor for the 
$A\raisebox{-4pt}{$\stackrel{\displaystyle{(x)}}{\scriptstyle{\mu a}}$}$
at the vertex formed by contracting that subdiagram to a point.)  Hence we
can completely cancel the effects of the rescaling of
$A\raisebox{-4pt}{$\stackrel{\displaystyle{(x)}}{\scriptstyle{\mu a}}$}$
in (\ref{E253}) and in the path-ordered phase factor, by simply rescaling
the integration variable 
$A\raisebox{-4pt}{$\stackrel{\displaystyle{(x)}}{\scriptstyle{\mu a}}$}$
in the functional integral.  This is the reason, in our approach, why 
there are no ``extra terms'' associated with field rescalings in the
renormalization group equations (\ref{E102}) and (\ref{E256}) for our
$f_0$'s: the only ``other terms'' are those associated with the 
short-distance factors.

We note here one further crucial effect of dividing out the short-distance
factors.  We have to give a convergent limiting procedure for calculating
the path integrals in the right-hand sides of the group-variation 
equations, (i.e. the sums over the 45-paths), in such a way that if, 
having restored the \emph{short}-distance factors in the right-hand side
windows as described above, we then substitute for each right-hand side
window's \emph{long}-distance factor (\ref{E262}), the \emph{perturbative}
expansion of that long-distance factor, we exactly recover, in the limit
of our convergent procedure for calculating the path integrals, the 
renormalized perturbative expansion of the left-hand side of that
group-variation equation.  (This is, of course, the essential test of the
validity of our limiting procedure for calculating the path integrals.)
The property of the long-distance factors (\ref{E262}) that makes this
possible, is that the dependence of the long-distance factors on the 
details of the path is \emph{soft}: the long-distance factors do not 
depend sensitively on the details of the path, even when that path is 
quite jagged.  (All our paths are formed from finite numbers of straight
segments.)  This soft dependence of the long-distance factors on the
details of the path is due to the $\frac{\mathrm{d}x_{\mu_i}(s_i)}
{\mathrm{d}s_i}$ factors in the definition (\ref{E1}) of the path-ordered
phase factor: these factors tend to average out the details of the path,
provided the dependence, on $x(s_1),\ldots,x(s_n)$, of whatever 
expectation value 
$A\raisebox{-4pt}{$\stackrel{\displaystyle{(x(s_1))}}
{\scriptstyle{\mu_1a_1}}$}\ldots
A\raisebox{-4pt}{$\stackrel{\displaystyle{(x(s_n))}}
{\scriptstyle{\mu_na_n}}$}$ is involved in, is not too singular.

\section{Limiting procedure for calculating the path integrals}

Our limiting procedure for calculating the path integrals consists, in
brief, of the following.  We first restore the short-distance factors, as
sums of renormalized Feynman diagrams, with their propagators cut off
smoothly at long distances, with the onsetof the long-distance cutoffs
occurring at $\left|x-y\right|=T$, in the windows of the right-hand side
group-variation equation diagrams, and also divide by the short-distance
factors for the loops of the left-hand side $f_0$.  We then express all 
the path sums for the \emph{gluon} 45-paths in terms of $\frac{1}
{\bar{D}^2}$ by means of (\ref{E16}), (\ref{E18}), (\ref{E20}), and 
(\ref{E30}).  (The path sums for the Fadeev-Popov 45-paths are given
directly by $\frac{1}{\bar{D}^2}$, as we see from the first term in
(\ref{E23}).)  We then make a sequence of approximations to $\frac{1}
{\bar{D}^2}$, as follows.  At \emph{every} stage in our sequence of 
approximations, we now go to the limit of zero segment lenght, or in other
words, zero width of each individual Gaussian.  \emph{But}, we do not
``tie'' the \emph{long}-distance factor to the path all the way along its
length: we only tie the \emph{short}-distance factor precisely to the 
path all along its length.  Our approach is the same as before, (equations
(\ref{E24}) - (\ref{E26})), except that the parameter $\sigma$ that 
characterizes each approximation now represents no the ``maximum 
tolerable'' Gaussian width of an individual segment, but rather the
``maximum tolerable'' \emph{total} of all the infinitesimal Gaussian 
widths between successive points where we ``tie'' the long-distance factor
to the path.  (By ``Gaussian width'' we mean the actual coefficient of
$\bar{D}^2$ in an exponent, \emph{not} the square root of that 
coefficient.)  What this means in practice is that, for the 
$\sigma$-approximation to $\frac{1}{\bar{D}^2}$, we break up the 
$s$-integral in equation (\ref{E24}) as in equation (\ref{E25}) as before.
then for $n\geq1$, and for $n\sigma\leq s\leq(n+1)\sigma$, we break up
$\left(e^{s\bar{D}^2}\right)_{Ax,By}$ into a product of $(n+1)$ factors,
exactly as before in equation (\ref{E26}).  Now equation (\ref{E26}) is
exact.  What we now do is, for each of the factors $\left(e^{\frac{s
\bar{D}^2}{n+1}}\right)_{C_iz_i,C_{i+1}z_{i+1}}$ in (\ref{E26}), (where
$C_0z_0\equiv Ax$ and $C_{n+1}z_{n+1}\equiv By$), to represent this factor
again by a path integral, but, bearing in mind that the Gaussian width
$\frac{s}{n+1}$ of this factor is $\leq\sigma$, we approximate the
\emph{long}-distance factor's path, for \emph{every} path in this path
integral, by the straight line from $z_i$ to $z_{i+1}$.  The 
\emph{short}-distance factor follows the true path exactly.  The result is
that, for insertion into (\ref{E26}), we approximate $\left(e^{\frac{s
\bar{D}^2}{n+1}}\right)_{C_iz_i,C_{i+1},z_{i+1}}$ by the straight line
path from $z_i$ to $z_{i+1}$ for the \emph{long}-distance factor, times:
	\[\sum_{m=0}^\infty\int_0^\infty\ldots\int_0^\infty\mathrm{d}t_1\ldots
	\mathrm{d}t_m\delta\left(\frac{s}{n+1}-(t_1+\ldots+t_m)\right)\quad
	\times
\]
\begin{equation}
	\label{E278}
	\times\left(e^{t_m\partial^2}(\partial A+A\partial+AA)e^{t_{m-1}
	\partial^2}(\partial A+A\partial+AA)\ldots(\partial A+A\partial+AA)
	e^{t_1\partial^2}\right)_{C_iz_i,C_{i+1}z_{i+1}}
\end{equation}
where $A\equiv
A\raisebox{-4pt}{$\stackrel{\displaystyle{(x)}}{\scriptstyle{\mu a}}$}
\left(t_a\right)_{BC}$ here is an effective field in terms of which we
represent the short-distance factor.  Thus at \emph{every} stage of our
sequence of approximations to the path integral, we treat the 
\emph{short}-distance factor exactly: it is only the \emph{long}-distance
factor whose path is approximated by $(n+1)$ straight segments in 
(\ref{E26}).  Thus at every stage in our sequence of approximations, we
retain the correct short-distance behaviour of the underlying renormalized
Feynman diagrams.  Our procedure converges due to the soft dependence of 
the long-distance factors on the details of the paths.  We do have to make
special allowance, however, for cases where two \emph{different} loops in
the border of one of our non-simply connected windows approach one
another closely, for in such a case, as we noted before, the 
short-distance divergence is in the long-distance factor, \emph{not} the
short-distance factor.  However, due to the fact that we never have more
than one island, and that if we do have an island, there are no 45-paths
that do \emph{not} form part of that island, we \emph{never}, in the 
right-hand sides of the group-variation equations, have to do a 
path-integral for more than one of the connected borders of a non-simply
connected window.  Thus the only configurations that need special
attention in this regard are those where a border of an island intersects,
or comes close to, one of the loops in the left-hand side $f_0$.  For 
details and renormalization, we refer to our next paper.

\chapter{Convergence Of The Sums In The Right-Hand Sides, Renormalization In Position Space, And The Signs Of The Island Diagrams}

\section{Renormalization group equations for the ansatz parameters}

We now return to the discussion, begun before, of the dependence of $\mu$,
where $\mu^2$ is the coefficient of the area in the Wilson area law in our
ansatz, on our input parameters $g^2$ and $R$.  Our long-distance factors
(\ref{E262}) satisfy, as we stated before, a renormalization group 
equation of the form (\ref{E256}), where the ``other terms'' are 
associated exclusively with the short-distance factors which we divide
$f_0(W_1,\ldots,W_n,g^2)$ by in (\ref{E262}).  Now these ``extra terms''
involve the partial derivative $\frac{\partial}{\partial T}$, where $T$ is
the onset value $\left|x-y\right|$ of the long-distance cutoffs imposed on
the propagators in the $\tilde{f}_0$'s in the denominator of (\ref{E262}).
Our ansatz does not contain $T$ at all, because it only refers to the 
essential features of the long-distance behaviour of our $f_0$'s.  The
$T$-dependence of our long-distance factors does not interfere at all with
these essential features: as we know, the $T$-dependence of our
long-distance factors (\ref{E262}) is constrained by the fact that if we
multiply these long-distance factors by the (finite) ratio of the
short-distance factors for $T$ divided by the short-distance factors for 
another value, say $T_2$, the dependence of this product on $T$ must
completely cancel out, to be replaced by the equivalent dependence on
$T_2$.  It follows that the factors in our ansatz obey (\ref{E256}) 
exactly with \emph{no} ``other terms'' at all.  In particular, for a 
single Wilson loop, we have:
\begin{equation}
	\label{E279}
	\left(-R\frac{\partial}{\partial R}+\beta(g)\frac{\partial}{\partial g}
	\right)e^{-\mu^2A}=0
\end{equation}
Now $A$, the area of the minimal-area spanning surface of our loop, is of
course completely independent of $R$ and $g$, hence (\ref{E279}) implies:
\begin{equation}
	\label{E280}
	-R\frac{\partial\mu}{\partial R}+\beta(g)\frac{\partial\mu}{\partial g}
	=0
\end{equation}
And on dimensional grounds, the dependence of $\mu$ on $R$ is simply by an
overall factor of $\frac{1}{R}$, (so that $\mu R$ is a function only of
$g^2$), hence (\ref{E280}) implies:
\begin{equation}
	\label{E281}
	\mu R+\beta(g)\frac{\mathrm{d}(\mu R)}{\mathrm{d}g}=0
\end{equation}
Hence, comparing with equation (\ref{E256}), we see that the functional
dependence of $\mu R$ on $g^2$ is identical in form to a ``curve of 
constant $f$'' in the $(R,g)$ plane, which in other words, means that the
functional relationship between $g^2$ and $\mu R$ is identical to a 
possible behaviour of the ``running coupling'' $\bar{g}^2(R)$.  From 
equation (\ref{E260}) and diagram (\ref{E261}) we therefore see that for
small $g^2$, $\mu R$ decreases very rapidly as $g^2$ decreases towards
zero, hence, as stated, $\frac{1}{\mu R}$, which is the typical island
size measured in units of $R$, \emph{increases} very rapidly as the input
value of $g^2$ decreases towards zero.

\section{Convergence of the sums in the right-hand sides of the 
Group-Variation Equations}

We note here that, although, as we saw above, the experimental value
$\alpha_s=0.3$ for charmonium implies that the critical value of 
$\frac{g^2}{4\pi}$, as given by equation (\ref{E236}), must be greater 
than 0.35, we cannot get a good indication of the actual critical value
from experimental results, since the experimental value of $\mu$, which is
approximately  where the running coupling reaches the critical value, is
0.41 GeV, while quoted values of $\Lambda_s$ range from 0.1 GeV to 
0.5 GeV \cite{Perkins 308}.  \footnote{More recent experimental data show
$\Lambda_s$ stabilizing around 0.2 GeV \cite{beta in MS bar 1}.} 
Furthermore, the
critical value will be renormalization-scheme dependent.  (That is not a
serious problem, however, because it is easy to specify a simple and
natural renormalization scheme.  For example, we can require that our
additional Ward-identity-restoring finite counterterms contain no
$A\raisebox{-4pt}{$\stackrel{\displaystyle{(x)}}{\scriptstyle{\mu a}}$}
\partial_\mu\partial_\nu
A\raisebox{-4pt}{$\stackrel{\displaystyle{(x)}}{\scriptstyle{\nu a}}$}$
terms, which is the choice made in (\ref{E238}).)  The actual calculation
of the critical value of $g^2$ from (\ref{E236}) requires the full
renormalization of the group-variation equations, which we treat in our
next paper, and we therefore turn now to the following consideration.

The order $g^6$ term in $2g\beta(g)$ is known \cite{Caswell}, 
\cite{Jones}, \cite{DJGLH 196},
\cite{Quigg 8.3}, and like the order $g^4$ term, it is 
renormalization-scheme-independent \cite{DJGLH 178}.  From the quoted
results, we find that our $\beta(g)$ is given by:
\begin{equation}
	\label{E282}
	2g\beta(g)=-\frac{11g^4}{12\pi^2}-\frac{17g^6}{48\pi^4}+
	\textrm{ terms of order }g^8
\end{equation}
Now of course, it is possible to find a renormalization scheme where, by
adding suitable \emph{extra} finite counterterms at each order in the loop
expansion of the effective action $\Gamma$, every term in $\beta(g)$
beyond the two displayed above, vanishes.  Such a renormalization scheme
would be, from the point of view of the group-variation equations, 
extremely unnatural and undesirable.  it seems much more likely that, in
a natural renormalization scheme, such as that suggested above, where we
fix the Ward-identity-restoring finite counterterms by requiring them to 
include no term
$A\raisebox{-4pt}{$\stackrel{\displaystyle{(x)}}{\scriptstyle{\mu a}}$}
\partial_\mu\partial_\nu
A\raisebox{-4pt}{$\stackrel{\displaystyle{(x)}}{\scriptstyle{\nu a}}$}$,
allow no other finite counterterms, and at each order in the loop
expansion of $\Gamma$, choose the BPHZ counterterms to have the canonical
form given in our previous paper \cite{BPHZ}, as generated by the seed
action plus all the lower-order finite counterterms, the expansion
coefficients in $2g\beta(g)$ will have roughly the same behaviour as the
expansion coefficients in the expansions of other physical quantities in
powers of $g^2$.  Let us assume a weak form of this hypothesis, namely 
that, in a natural renormalization scheme, such as that suggested above, 
the expansion coefficients in $2g\beta(g)$ do \emph{not} behave
\emph{better}, (with regard to convergence), than the expansion 
coefficients in the expansions of other physical quantities, in the
\emph{explicit} powers of $g^2$ that multiply the diagrams in the
right-hand sides of the group-variation equations, (namely, for each
diagram, a power of $g^2$ equal to its number of 45-paths minus its number
of action vertices), and also the explicit powers of $g^2$ in the 
short-distance factors, when we restore them in the windows of the 
right-hand side group-variation equation diagrams.  Now of course, these
expansion coefficients depend on the ansatz for the long-distance factors
which we substitute into the right-hand side of the group-variation
equations, so to make the hypothesis precise, we assume it applies with 
the ansatz for the long-distance factors we discussed above, when that
ansatz is modified to include the correct dependence on the onset length
$T$ of the smooth long-distance cutoffs impose on the propagators in the
short-distance factors we divide by in (\ref{E262}).  (Note added: of 
course, we also assume it is true when the actual solution of the 
group-variation equations is substituted into the right-hand sides of the
group-variation equations.)

We next note that, for planar diagram field theories, such as large-$N_c$
QCD, in four Euclidean dimensions, 't Hooft \cite{'t Hooft a}
\cite{'t Hooft b} has shown
that, firstly, if there are no divergent subdiagrams, then the 
perturbation-theory expansion coefficients grow at worst geometrically,
and there exists a circle of convergence in the $g^2$-plane, of radius
strictly greater than zero, inside which the expansions, in powers of 
$g^2$, of \emph{all} physical quantities converge, and, secondly, if the 
theory is asymptotically free when the divergent subdiagrams are included,
the input value of $g^2$ is sufficiently small, that there exists a mass
to terminate the growth of the running coupling at large distances, then
there exists an absolutely convergent procedure, based on skeleton 
expansions, difference equations in momentum space, and the 
renormalization group, for calculating \emph{all} physical quantities.

Now the lower bounds on the radii of the circles of convergence obtained
by 't Hooft are several orders of magnitude smaller than  the 
experimental lower bound, 0.35, on the critical value of 
$\frac{g^2}{4\pi}$, but let us, nevertheless, suppose that the general
behaviour of the expansion coefficients, in the expansions of physical
quantities in the ``explicit'' powers of $g^2$, (as defined above), is not
worse than geometric.  Then by our hypothesis, the behaviour of the 
expansion coefficients in $2g\beta(g)$ is \emph{at best} geometric, and,
if the expansions of other physical quantities in the ``explicit'' powers
of $g^2$ have a finite radius of convergence, then the expansion of 
$2g\beta(g)$ in powers of $g^2$ also has a finite radius of convergence,
which is not greater than the radius of convergence for other physical
quantities.  Now from (\ref{E282}) we see that the first two coefficients
in the expansion of $2g\beta(g)$ have the \emph{same} sign.  Let us 
suppose that this trend continues, and that in a natural renormalization
scheme, \emph{all} the expansion coefficients in $2g\beta(g)$ have the
\emph{same} sign.  Then if the expansion of $2g\beta(g)$ in powers of 
$g^2$ has a finite radius of convergence, $2g\beta(g)$ will tend to 
$-\infty$ as $g^2$ approaches the radius of convergence from below along
the positive real axis.  This means that the ``curves of constant $f$'' in
diagram (\ref{E261}) will curl upwards to a vertical slope at the finite
value of $g^2$ given by the radius of convergence.  That does not matter,
because from equation (\ref{E236}) we see that, \emph{assuming} that
$\displaystyle\sum_iY_i$ is negative, (we return to this point shortly),
if $\frac{\beta(g)}{g}$ goes to $-\infty$ for a finite value of $g^2$,
strictly greater than zero, then the critical value of $g^2$, as 
determined by (\ref{E236}), will be strictly smaller than the value of
$g^2$ where $\frac{\beta(g)}{g}$ goes to $-\infty$.  This is so because
the right-hand side of (\ref{E236}) is on dimensional grounds equal to
$\mu^2$, times a function of $g^2$, hence if we divide both sides of
(\ref{E236}) by $\mu^2$, we have in the right-hand side a function of 
$g^2$ which sweeps out all real values from 0 to $+\infty$ as $g^2$
increases from 0 to the radius of convergence.  (The leading terms in
$\displaystyle\sum_iY_i$ come from the one-loop island diagrams and are
independent of $g^2$.  \footnote{Note added: that is, they have no 
explicit dependence on $g^2$.}  There is certainly no reason to expect
$\displaystyle\sum_iY_i$ to vanish as $2g\beta(g)$ goes to $-\infty$.)

Now by assumption, the expansion coefficients in $2g\beta(g)$ do 
\emph{not} behave \emph{better} than the expansion coefficients in the 
expansions of other physical quantities in the ``explicit'' powers of 
$g^2$, as defined above, hence it follows immediately that if, in a 
natural renormalization scheme, the trend in (\ref{E282}) continues, and
all the expansion coefficients in $2g\beta(g)$ have the same sign, then at
the critical value of $g^2$, as determined by (\ref{E236}), the expansions
of physical quantities, (such as the right-hand sides of the
group-variation equations, for a reasonable ansatz), in the ``explicit''
powers of $g^2$, will \emph{converge}, and the convergence rate will be
at least geometrical.  In other words, if all the expansion coefficients
in $2g\beta(g)$ have the same sign, then the singularity in 
$\frac{\beta(g)}{g}$ will shield all other physical quantities from 
having divergent expansions, by ensuring that the critical value of $g^2$
is strictly smaller than the radius of convergence of the expansion of any
physical quantity in the ``explicit'' powers of $g^2$.

As a check we note that, if we assume that the two terms given in
(\ref{E282}) are the first two terms of a geometric series, then the 
radius of convergence is given by $\frac{g^2}{4\pi}=\frac{11\pi}{17}=
2.03$, which is safely larger than the experimental lower bound of 0.35 on
the critical value of $\frac{g^2}{4\pi}$.

We note here that, with a simple derivative in the left-hand sides, and
integrals in the right-hand sides, the group-variation equations have the
ideal form for solving by repeated substitution of the left-hand sides,
after integrating with respect to $L$ or $g^2$, from initial conditions
given for small $L$ or small $g^2$ by perturbation theory, into the 
right-hand sides.  If a metric can be found in the  function space of our
long-distance factors (\ref{E262}), such that if when we substitute two
different ans$\ddot{\textrm{a}}$tze into the right-hand sides of the 
group-variation equations, the distance in function space between the two
``outputs'' is strictly less than the distance in function space between
the two ``inputs'', then this procedure will converge to a unique solution
of the group-variation equations.

The considerations just given indicate that it is likely that, if a
reasonable ansatz is substituted into the right-hand sides of the 
group-variation equations, and the input value of $g^2$ is not larger than
the critical value, then the sums over the diagrams in the right-hand 
sides of the group-variation equations will converge.

We may, in practice, want to consider iterating the group-variation
equations with truncated versions of their right-hand sides, and then
repeating the process with less truncated versions of their right-hand
sides, and so on, and it is likely that such a procedure would also 
converge.

\subsection{Application of the four-loop $\beta$-function}
\label{Subsection 4LB}

From the general form of the four-loop $\beta$-function in
$\overline{\mathrm{MS}}$, given in reference \cite{beta in MS bar 2}, we
find that our $\frac{\beta(g)}{g}$ is given by:
	\[
\frac{\beta(g)}{g}=-\left(\frac{11}{3}\frac{g^2}
{8\pi^2}+\frac{34}{3}\left(\frac{g^2}{8\pi^2}\right)^2+\frac{2857}{54}
\left(\frac{g^2}{8\pi^2}\right)^3+315.49\left(\frac{g^2}{8\pi^2}\right)^4
+\textrm{order}\left(g^{10}\right)\right)=
\]
\begin{equation}
\label{FLB1}
=-\left(0.5836\frac{g^2}{4\pi}+0.2871\left(\frac{g^2}{4\pi}\right)^2+
0.2133\left(\frac{g^2}{4\pi}\right)^3+0.2024\left(\frac{g^2}{4\pi}\right)
^4+\textrm{order}\left(g^{10}\right)\right)
\end{equation}
(We recall that, as explained in Subsection \ref{g^2}, our 
$\frac{g^2}{4\pi}$
is equal to $\frac{3}{2}$ times the value $\alpha_s$ would have in the
absence of quarks.)  The coefficients are decreasing, but that is an
artifact of our choice of expansion parameter.  The ratios of successive pairs of coefficients in the expansion are increasing, but at a decreasing rate.  If we assume that the difference, between 
two successive pairs of ratios, changes by a fixed factor, for each
successive pair of ratios, then the ratio, of two successive 
coefficients, tends asymptotically to 1.89, which means that the series 
will diverge when $\frac{g^2}{4\pi}=0.53$.  And if we assume no more than
that the ratios of successive pairs of coefficients will go on 
increasing, then we can conclude that the series will diverge for
$\frac{g^2}{4\pi}\leq 1.05$.  The four-loop term is essential to
reach these conclusions.

\section{BPHZ renormalization in position space}

We think of our BPHZ counterterms \cite{BP}, \cite{BS}, \cite{Hepp},
\cite{Zimmerman}, \cite{BPHZ}, as enormously deep reservoirs, whose
actual depth, (at small distances), is unknown, and of no interest.  We
only see the surface of the reservoir, (at large distances), whose height
corresponds to $\ln(R)$.  And the fact that, way down in its depths, our
reservoir may have irregularities and caverns that do not correspond to 
the simple BRST-invariance properties of its surface, is of no concern to
us at all.  Now in correspondence with Theorem 2 of our previous paper
\cite{BPHZ 157}, we do indeed assume that the depths of our reservoirs,
though enormous, are finite, and that the amount of irregularity is
limited, (the propagators in all our counterterms and ``direct terms'' are
\emph{smoothly} regularized at extremely short distances).  We also assume
that the deep structures of our ``positive'' and ``negative'' reservoirs,
(the counterterms and the direct terms), although unknown, exactly match.
We then find that the calculated values of observable quantities, above or
close to the surfaces of the reservoirs, have finite limits as the depths
of the reservoirs tend to $\infty$.  In our BPHZ approach, we \emph{never}
have to refer to the actual depths of the reservoirs, because we always
add the contributions of a positive reservoir and a negative reservoir 
with the same deep structure before calculating a volume.  The reason we 
have to assume that the amount of irregularity at great depths is limited,
(propagators are \emph{smoothly} regularized at extremely small 
distances), is that when we work in position space, certain reservoirs,
namely those associated with one-line \emph{reducible} subdiagrams with
two, three, or four legs, can appear superficially, (i.e. by 
power-counting), to have infinite volumes, when in fact their volumes are
finite.  The calculation of the volumes of these reservoirs in position
space is only conditionally convergent, but there is only one natural way
to approach the limit: namely from propagators \emph{smoothly} regularized
at exremely small distances.  \emph{Any} smooth short-distance
regularization satisfying the general conditions of Theorem 2 of our
previous paper \cite{BPHZ 157} will give the same result for the volume.
The reason for this is that, as noted on page 236 of that paper, the 
actual uses made of the regularization are extremely limited: they amount
to allowing us to sweep derivatives past the ``key'' propagators of 
one-line-reducible subdiagrams by integrations by parts and translation
invariance, without picking up any ``short distance surface terms'' due to
having to cut a small sphere centred at one end of such a propagator, out
of the integration domain of the other end of such a propagator.  Once 
these manipulations are done, absolute convergence is attained, and the
smooth regularization is no longer necessary.  Examples of smooth
regularizations with the required properties are constructed on pages 235
and 236 of that reference: we never need to use their detailed forms, it
is sufficient that they exist.

We call these enormously deep reservoirs, whether they occur in the 
counterterms or in the ``direct'' parts of our calcultions, and whose
depths are unknown but must match between different parts of our
calculations, ``counters'', (analogous to ``differentials'').

We in fact did \emph{not} assume translation invariance in the above
reference: only a much weaker assumption, called ``translation 
smoothness'', was made.  Defined on pages 224 and 225 of that paper, and
included in a general form in the conditions for Theorem 2, translation
smoothness essentially requires the power-counting behaviour of a 
propagator as $y$ tends to $x$ to be independent of $x$, or at least not
worse than the standard translation-invariant power-counting behaviour,
while allowing the \emph{coefficient} of the singularity as $y$ tends to 
$x$, to depend smoothly on $x$.  (Examples of translation smooth
propagators are given by ordinary propagators, multiplied by completely
smooth functions of $x$ and $y$.)  Thus the question of whether 
translation invariance is necessary for BPHZ finiteness has been answered:
translation invariance is \emph{not} necessary, translation smoothness is
sufficient.  We note, however, that while for translation invariant 
theories, changing the choice of the contraction weights 
\cite{BPHZ 13 219} used to define a counterterm, (i.e. the weights
assigned to the vertices of the ``direct term'', in the linear combination
of the positions of those vertices, that defines the position of the 
contraction point of the counterterm), results in changing the counterterm
only by a total derivative, and in fact leaves the counterterm unaltered 
if the degree of divergence is zero, we do not find a corresponding result
in non-translation-invariant theories.  Thus the choice of the contraction
weights becomes a dynamical issue in non-translation-invariant situations,
for example in the presence of background gravitational fields.

When one forms the \emph{position-space} integrand for a particular 
Feynman diagram and a particular BPHZ forest for that Feynman diagram, one
has to detach the inner ends of the legs, of each subdiagram that is a 
member of that forest, from the Feynman diagram vertices they are 
initially attached to, and move them to the position of the contraction
point of that subdiagram.  The contraction point of a subdiagram is a
linear combination of the positions of the vertices of the subdiagram, for
example their mean position.  Other choices are more convenient in 
practice, for example one can number all the vertices of the Feynman 
diagram, and define the contraction point of each subdiagram to be the 
highest-numbered vertex of that subdiagram.  The choice of the contraction
points of members of the forest must satisfy the following constraint
\cite{BPHZ 13 219}: the contraction point of each member of the forest
must be expressible as a linear combination of the contraction points of
the largest members of the forest that are strict subsets of that member
of the forest, plus the positions of any vertices of that subdiagram that
are not vertices of any strictly smaller member of the forest.  Both the
``mean position'' choice and the ``highest-numbered vertex'' choice that
we just mentioned, satisfy this criterion.  (The reason we define
``greenwoods'' in \cite{CCT} and ``woods'' in \cite{BPHZ}, which are in
one-to-one correspondence with the usual forests, is to avoid having to
make frequent separate references to ``the vertices of that subdiagram 
that are not vertices of any strictly smaller member of that forest''.)

It is very helpful in position-space BPHZ to have a separate index or 
label for every propagator end.  We can then represent each vertex of the
Feynman diagram by the set of all the propagator ends at that vertex.  The
set $V$ of all the vertices of the Feynman diagram then becomes a 
partition of the set of all the propagator ends.  In \cite{BPHZ} we
represent the set $V$ of all the vertices of a Feynman diagram by a 
partition of a set that includes all the propagator ends, but may have
other members as well.  The advantage of this is that not only all
subdiagrams, but also all diagrams obtained from the given diagram by
\emph{contracting} some of its subdiagrams, can be represented in exactly
the same way.  In fact, diagrams obtained from the given diagram by
contracting some of its subdiagrams, can be represented by partitions of
exactly the same set as the given diagram.

The operation of moving the inner ends of the legs of the subdiagrams that
are members of the forest, to the contraction points of those subdiagrams,
is performed in a sequence in which smaller members of the forest precede
any larger members of the forest that contain them.  The final result is
that each propagator end is moved to the contraction point of the largest
member of the forest that has that propagator end as a member, but does
not have the other end of that propagator as a member.  If a propagator
end is not a membmer of any member of the forest, it is left in its
original position.  (Again, working with woods enables us to avoid having
to state this last case separately.)

These movements of propagator ends, which are an essential part of BPHZ in
position space, cause the following practical problem: to ensure that our
BPHZ-renormalized feynman diagrams are convergent in position space, we
want to use the Cluster Convergence Theorem \cite{CCT}, which involves 
cutting up a position space integrand into a finite number of sectors,
characterized by the way the vertices are clustered in position space.  If
$V$ is a finite set, $x$ is a map from $V$ into position space, and 
$\sigma$ is a real number such that $0<\sigma<1$, then a $\sigma$-cluster
of $x$ is a nonempty subset $S$ of $V$ such that for all $i\in S$, 
$j\in S$, $k\in S$, and $l\in V$ such that $l$ is \emph{not} a member of
$S$, $\left|x_i-x_j\right|<\sigma\left|x_k-x_l\right|$.  The set of all
the $\sigma$-clusters of $x$ is a forest of $V$.  For reasons of 
convenience, we also explicitly specify that \emph{all} one-member subsets
of $V$ are $\sigma$-clusters of $x$.  The set of all the $\sigma$-clusters
of $x$ is then a forest of $V$ that includes all the one-member subsets of
$V$.  This is called a greenwood of $V$ in \cite{CCT}.  The Cluster
Convergence Theorem then guarantees absolute convergence if for every
greenwood $G$ of $V$, the absolute value of the integrand can be bounded,
throughout the secto of configuration space where the set of all the
$\sigma$-clusters of $x$ is equal to $G$, by a constant times a product of
powers of the distances between the vertices, that satisfies a standard
power-counting condition for every member of $G$ that has two or more
members.  The problem caused by the movements of propagator ends, as
described above, for the different BPHZ forest, is that for a given
position-space configuration of the vertices of the original Feynman
diagram, the clusters defined by the positions of the contraction points
of the members of a forest, may be completely different for different
forests, and completely different from the clusters defined by the 
positions of the vertices in the original Feynman diagram.  For example, 
if we have a triangle subdiagram that has a separate vertex insertion in
each of its three corners:
	\unitlength 1.00pt
\linethickness{0.61pt}
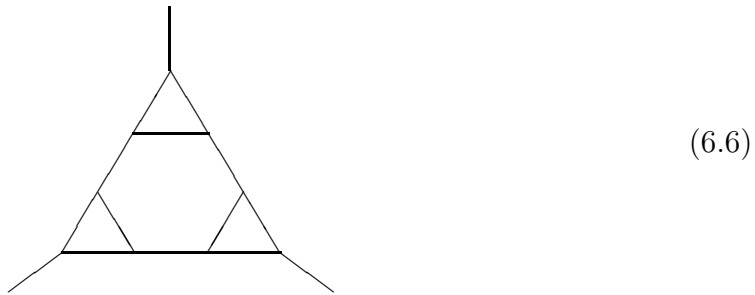
\begin{equation}
	\label{E283}
	\raisebox{-54pt}{
\begin{picture}(124.00,109.00)
\put(21.00,16.00){\line(1,0){83.00}}
\put(21.00,16.00){\line(3,5){41.40}}
\put(62.40,85.00){\line(3,-5){41.40}}
\put(62.00,85.00){\line(0,1){24.00}}
\put(104.00,16.00){\line(4,-3){20.00}}
\put(48.00,61.00){\line(1,0){29.00}}
\put(35.00,39.00){\line(3,-5){13.80}}
\put(90.00,39.00){\line(-3,-5){13.80}}
\put(21.00,16.00){\line(-4,-3){20.00}}
\end{picture}
}
\end{equation}
then the mean positions of the vertices of each of the three ``corner''
triangles can be close to one another in position space even if 
\emph{none} of the nine vertices of the original subdiagram are close to
one another in position space.  This means that, if we are using the
``mean position'' choice of contraction points, and we consider a BPHZ
forest that includes the three corner triangles as members, we will find,
after moving, for each corner triangle, the ``inner ends'' of the legs of
that triangle to the contraction point of that triangle, as represented
symbolically here:
	\unitlength 1.00pt
\linethickness{0.61pt}
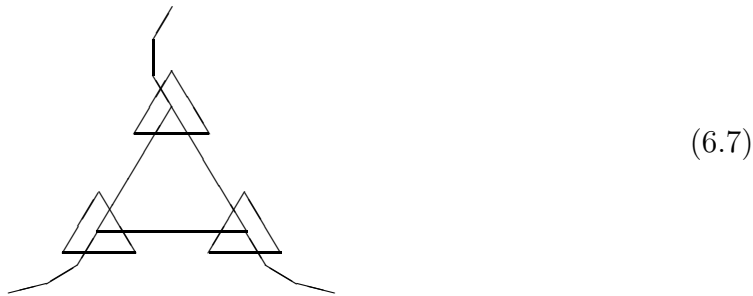
\begin{equation}
	\label{E284}
	\raisebox{-54pt}{
\begin{picture}(124.00,109.00)
\put(35.00,39.00){\line(3,-5){13.80}}
\put(90.00,39.00){\line(-3,-5){13.80}}
\put(48.67,16.00){\line(-1,0){27.67}}
\put(21.00,16.00){\line(3,5){13.60}}
\put(48.00,61.00){\line(3,5){14.40}}
\put(62.40,85.00){\line(3,-5){14.40}}
\put(76.67,16.00){\line(1,0){27.00}}
\put(103.67,16.00){\line(-3,5){13.60}}
\put(90.80,24.00){\line(-1,0){56.80}}
\put(34.00,24.00){\line(3,5){28.40}}
\put(62.33,71.33){\line(3,-5){28.40}}
\put(90.33,24.00){\line(3,-5){7.60}}
\put(97.93,11.33){\line(5,-3){11.73}}
\put(109.67,4.29){\line(4,-1){14.33}}
\put(34.33,24.00){\line(-3,-5){7.60}}
\put(26.73,11.33){\line(-5,-3){11.73}}
\put(15.00,4.29){\line(-4,-1){14.33}}
\put(62.67,109.00){\line(-3,-5){7.40}}
\put(55.27,96.67){\line(0,-1){13.67}}
\put(55.27,83.00){\line(3,-5){7.20}}
\put(48.33,61.00){\line(1,0){28.00}}
\end{picture}
}
\end{equation}
that we now have a ``cluster'' configuration in position space for this
subdiagram, due to the assumed closeness, in position space, of the mean
positions of these three corner triangles to one another, even though
\emph{none} of the nine vertices of the original triangle are close to 
one another in position space.  This means that to decide, for each
position-space configuration of the vertices of the original Feynman
diagram, which BPHZ forests' contributions need to be combined together to
obtain integrands which satisfy the conditions for applying the Cluster
Convergence Theorem, it is \emph{not} good enough just to look at the 
position-space clusters in the originial Feynman diagram: we have to look
separately at the position-space clusters for each individual BPHZ forest,
since the precise set of contraction points is different for each BPHZ
forest.  But on the other hand, for each possible greenwood of 
position-space clusters of the vertices, we have to combing the BPHZ
forest into groups, called ``good sets of woods'', such that for each
group, the sum of the integrands of the members of that group, satisfies a
bound that lets us apply the Cluster Convergence Theorem.  The appropriate
groupings are defined by pairs $(P,Q)$ of forests or woods, such that 
$P\subseteq Q$, and the members of the grouping defined by the pair
$(P,Q)$ are all the forests, or woods, $F$, such that $P\subseteq F
\subseteq Q$.  But, given a configuration $x$ of the vertices of the 
original Feynman diagram, (i.e. a map from $V$ into position space), and
a BPHZ forest $F$, how should we decide to which pair $(P,Q)$ the forest
$F$ should be assigned?  The improvement of short-distance behaviour
obtained by adding together the integrands of all the BPHZ forests in the
group defined by the pair $(P,Q)$, occurs for the members of $Q$ that are
not members of $P$.  This is because the group of BPHZ forests defined by
the pair $(P,Q)$ includes precisely all the BPHZ forests that include 
every member of $P$, and, independently for each member of $Q$ that is
\emph{not} a member of $P$, may or may not include that member of $Q$.  We
get an extra minus sign for each extra member of $Q$ included, and the
resulting cancellations have the effect of shifting enough denominator
posers of intervals out from the members of $Q$ that are not members of
$P$, onto the legs of those members of $Q$, to enable the members of $Q$
that are not members of $P$, to satisfy the conditions for applying the
Cluster Convergence Theorem.  Thus to obtain manifest convergence we need
to ensure that every position-space cluster that is power-counting 
divergent in the  original Feynman diagram, becomes a member of $Q$ that 
is not a member of $P$, and also that no member of $P$ is small in 
position space.  Here ``cluster'' and ``small in position space'' refer,
for each subdiagram that is a member of $F$, to what we have after
contracting all members of $F$ that are strict subsets of that member of
$F$.  Let us define, for any BPHZ forest $F$, subdiagram $A$, and 
configuration $x$ of the vertices of the original Feynman diagram,
$\mathbf{L}(F,A,x)$ to be the largest distance between any two vertices of
$A$, \emph{after} all members of $F$ that are strict subsets of $A$, have
been contracted.  (So in other words, $\mathbf{L}(F,A,x)$ is the
``diameter'' of $A$, \emph{after} all members of $F$ that are strict
subsets of $A$, have been contracted.)  Now the example above shows that
$\mathbf{L}(F,A,x)$ can be very small even when \emph{none} of the 
distances between pairs of vertices of $A$ are small in the original
Feynman diagram.  There is another possibility that can also affect the
identification of the position-space clusters.  A $\sigma$-cluster is a
subset of the vertices whose ``diameter'' , in a particular 
position-space configuration, is less than $\sigma$ times the smalles 
distance between any member of the $\sigma$-cluster and any vertex that is
\emph{not} a member of the $\sigma$-cluster.  It is possible for the
contraction of a subdiagram to put a vertex, namely the contraction point
of that subdiagram, near a set of vertices that previously had no other
vertex near them.  Hence to identify the forests $P$ and $Q$, which define
the ``good grouping of forests'' to which we should assign the BPHZ forest
$F$, for a given configuration $x$ of the vertices of the original
Feynman diagram, we have to take account, for each subdiagram $A$, not 
only of the contractions of all members of $F$ that are strict subsets of
$A$, (and hence use $\mathbf{L}(F,A,x)$ rather than the diameter of $A$ in
the original Feynman diagram), but also to take account of the 
contractions of members of $F$ that are \emph{disjoint} from $A$.  To do
this we note that, as mentioned above, it is very helpful to have a 
separate index or label for every propagator end.  We can then represent
each propagator by the two-member set whose members are the two ends of 
that propagator.  Now as we noted before, after doing all the contractions
for a particular BPHZ forest, each propagator end has moved to the 
contraction point of the largest member of the forest that has that 
propagator end as a member, but does not have the other end of that
propagator as a member.  We can express this by defining, first, the set
$H$ to be the set of all the propagators, and, secondly, for each BPHZ
forest $F$ and propagator end $i$, $\mathcal{Z}(F,H,i)$ to be the largest
member of $F$ that has $i$ as a member but does \emph{not} have the other
end of that propagator as a member.  We also define, for every subdiagram
$A$, $x_A$ to be the contraction point of $A$.

Thus the simplest reasonable generalization of the $\sigma$-cluster
criterion to the case where we have to take into account the contractions
associated with the members of the BPHZ forest $F$, would be to require
that for every propagator end $i$ that is a member of $A$, and every
propagator end $j$ that is \emph{not} a member of $A$, we have the 
following inequality:
\begin{equation}
	\label{E285}
	\mathbf{L}(F,A,x)<\sigma\left|x_{\mathcal{Z}(F,H,i)}-
	x_{\mathcal{Z}(F,H,j)}\right|
\end{equation}
However this is over-stringent: there is no point in applying such a
restriction when the propagator end $j$ is in some part of the Feynman
diagram that has no direct connection at all to the subdiagram $A$.  Thus
we only require (\ref{E285}) ot be satisfied when $\{i,j\}$ is a 
propagator such that $i$ is a member of $A$ and $j$ is \emph{not} a member
of $A$.

Now we are here considering the case of Theorem 1 of \cite{BPHZ}, where we
allow counterterms for \emph{all} power-counting divergent connecte
subdiagrams, both one-line reducible and one-line irreducible.  (The
additional considerations necessary when only one-line \emph{irreducible}
counterterms are allowed, are given in Theorem 2 of \cite{BPHZ}.)

We therefore try the following identification of the pair $P$ and $Q$ of
forests that defines the ``good grouping of forests'' to which we should
assign $F$, for the given configuration $x$ of the vertices of the 
original Feynman diagram:
\begin{equation}
	\label{E286}
	\left.\begin{array}{l}\textrm{$P$ is the set of all the members $A$ of
	$F$ such that there} \\ \textrm{exists a propagator $\{i,j\}$ such that
	$i$ is a member of $A$,} \\ \textrm{$j$ is \emph{not} a member of $A$,
	and (\ref{E285}) is \emph{not} true.}\end{array}\right\}
\end{equation}
\begin{equation}
	\label{E287}
	\left.\begin{array}{l}\textrm{$Q$ is the union of $F$, and the set of 
	all the power-counting} \\ \textrm{divergent connected subdiagrams $A$
	that overlap no member} \\ \textrm{$F$, such that (\ref{E285}) is true
	for \emph{all} propagators $\{i,j\}$ such} \\ \textrm{that $i$ is a
	member of $A$ and $j$ is \emph{not} a member of $A$.}\end{array}\right\}
\end{equation}

Now (\ref{E286}) and (\ref{E287}) certainly give a pair $(P,Q)$ such that
$P\subseteq F\subseteq Q$ is true, and we see that, roughly speaking, the
members of $P$ are the members of $F$ that are \emph{not} in
``position-space cluster configurations'' after the contractions for the
members of $F$ have been done, while the members of $Q$ that are 
\emph{not} members of $P$, (and will hence get their short-distance 
behaviour improved when we combine the contributions of all the BPHZ 
forests in the ``good set of forests defined by the pair $(P,Q)$), are,
roughly speaking, all the power-counting divergent connected subdiagrams
that do \emph{not} overlap any member of $F$, and which \emph{are} in
``position-space cluster configurations'' after the contractions for the 
members of $F$ have been done.  But we now have to ask the following
questions:

\noindent 1) Is $Q$ a forest?  In other words, do we have a guarantee that
none of the members of $Q$ overlap one another?

\noindent 2) Suppose $Q$ \emph{is} a forest.  Do we have a guarantee that
for \emph{every} member $G$ of the ``good set of forests'' defined by the
pair $(P,Q)$, or in other words, for \emph{every} forest $G$ such that
$P\subseteq G\subseteq Q$, if we calculate a forest $P'$ and a set $Q'$
from (\ref{E286}) and (\ref{E287}) with $F$ replaced by $G$, we will find
that $P'=P$ and $Q'=Q$?

Now our purpose is to make the convergence, in position space, of our
BPHZ-renormalized Feynman diagram manifest in the following way.  We cut
up the set of all ordered pairs $(F,x)$ of a BPHZ forest $F$, and a
configuration $x$ of the vertices of the original Feynman diagram, into a
finite number of sectors, each characterized by a pair $(P,Q)$ of BPHZ
forests such that $P\subseteq Q$.  The sector associated with the pair
$(P,Q)$ of BPHZ forests, is to be equal to the Cartesian product of the 
set of all BPHZ forests $F$ such that $P\subseteq F\subseteq Q$, and an
appropriate sector of the configuration space of the vertices of the 
original Feynman diagram.  Now when we add the integrands for all the BPHZ
forests $F$ such that $P\subseteq F\subseteq Q$, we get an improvement in
the short-distance behaviour of all the  subdiagrams that are members of 
$Q$ but are \emph{not} members of $P$.  Thus the appropriate sector of the
configuration space of the vertices of the pair $(P,Q)$, is a sector 
where all the members of $P$, after performing the contractions for the
members of $P$, are \emph{not} ``clusters'', while the members of $Q$ that
are \emph{not} members of $P$, after performing the contractions for the
members of $P$, or maybe for the members of $Q$, \emph{are} ``clusters''.
There is an inherent ambiguity with regard to whether we should contract
\emph{all} the members of $Q$ before identifying the clusters, or whether
we should just contract the members of $P$ before identifying the 
clusters.  In other words, should we replace $F$ in (\ref{E285}) by $P$ or
by $Q$?  Now if we just want to identify the clusters ``roughly'', then
this ambiguity doesn't matter, because since the members of $Q$ that are
\emph{not} members of $P$ are required to be small in position space
whichever criterion we use, contracting them does not move any propagator
end by a large distance.

But for our demonstration of manifest convergence to be valid, our 
sectors, associated with the pairs $(P,Q)$ of BPHZ forests, must 
\emph{exactly} tesselate the Cartesian product of the set of all the BPHZ
forests and the set of all configurations of the vertices of the original
Feynman diagram.  There must be no gaps or overlaps at all.  (In the
present case, where we allow counterterms for one-line-reducible
subdiagrams as well as for one-line-irreducible subdiagrams, we obtain
absolute convergence, and no integrations by parts are done, so provided
the tesselation is exact, there is no problem with cutting up the 
integration domain into a finite number of sectors.)

The tesselation problem is solved by making the following requirement: we
must give a precise rule by which, given any BPHZ forest $F$ and any 
configuration $x$ of the vertices of the original Feynman diagram, we can
calculate uniquely the pair $(P,Q)$ of BPHZ forests to whose sector the
pair $(F,x)$ belongs.  Our first attempt at guessing such a rule is given
above in (\ref{E286}) and (\ref{E287}).

Now of course the set $Q$ we calculate by such a rule has got to be a BPHZ
forest, hence for any acceptable such rule, the answer to question 1) 
above has got to be, ``Yes''.

Furthermore, we require that the sector associated with the pair $(P,Q)$ 
of BPHZ forests, is to be the \emph{Cartesian product} of the set of all
forests $F$ such that $P\subseteq F\subseteq Q$, and a certain domain of 
the configuration space of the vertices of the original Feynman diagram.

In other words, for each pair $(P,Q)$ of BPHZ forests such that 
$P\subseteq Q$, and for each point $x$ of the configuration space of the
vertices of the original Feynman diagram, we require that either the pair
$(F,x)$ belongs to the sector defined by $(P,Q)$ for \emph{all} the 
forests $F$ such that $P\subseteq F\subseteq Q$, or else the pair $(F,x)$
belongs to the sector defined by $(P,Q)$ for \emph{none} of the forests
$F$ such that $P\subseteq F\subseteq Q$.  It immediately follows from 
this, that for any acceptable rule for calculating $P$ and $Q$, given $F$
and $x$, the answer to question 2) above has also got to be, ``Yes''.

Now in fact, when $Q$ is given by (\ref{E287}), the anwer to question 1)
above is, ``Yes''.  However, when $P$ and $Q$ are given by (\ref{E286})
and (\ref{E287}), the answer to question 2) above is, ``No''.  Indeed, one
can construct examples where, starting from the empty forest as $F$, one
has two non-intersecting subdiagrams, both of which are members of $Q$ by
(\ref{E287}).  However contracting one of the two subdiagrams moves the 
outer end of a leg of the second subdiagram towards the second subdiagram,
and prevents the second subdiagram from being a member of $Q$ according to
(\ref{E287}), if we start from the forest whose only member is the first
subdiagram, as $F$.

Instead of attempting to guess an acceptable improved version of 
(\ref{E286}) and (\ref{E287}), we now give a construction for which a 
``Yes'' answer to question 2) above is automatic.  For each configuration
$x$ of the vertices of the original Feynman diagram, we define a set
$\Omega$, whose members are \emph{all} the pairs $(P,Q)$ of BPHZ forests
such that $P\subseteq Q$ and, for every member $A$ of $Q$ that is not a
member of $P$, there \emph{exists} a forest $F$ such that $P\subseteq F
\subseteq Q$ is true, and (\ref{E285}) is true for all propagators 
$\{i,j\}$ such that $i$ is a member of $A$ and $j$ is not a member of 
$A$.  We note that the pair $(F,F)$ is a member of $\Omega$ for \emph{all}
BPHZ forests $F$.

Now the members of $\Omega$ are, in a sense, all the pairs $(P,Q)$ of 
BPHZ forests which might reasonably try to ``claim'' $x$ as part of the
configuration space domain in which they define a good set of forests.
Two members $(P,Q)$, and $(P',Q')$ of $\Omega$ are in direct competition
for $x$ if $P\cup P'\subseteq Q\cap Q'$ is true, for in that case there
exist BPHZ forests $F$, for example $P\cup P'$, such that both
$P\subseteq F\subseteq Q$ and $P'\subseteq F\subseteq Q'$ are true.  This
competition is resolved in the following way.  It turns out that if the
fixed real number $\sigma$, which enters the definition of $\Omega$ by
(\ref{E285}), satisfies $0<\sigma\leq\frac{1}{8}$, then for any members
$(P,Q)$, and $(P',Q')$ of $\Omega$ such that $P\cup P'\subseteq Q\cap Q'$
is true, the set $Q\cup Q'$ is a forest, (i.e. no member of $Q$ overlaps
any member of $Q'$), and the pair $(P\cap P',Q\cup Q')$ is also a member
of $\Omega$.  This is proved in Lemmas 6 and 7 in \cite{BPHZ}.  It 
immediately follows from this result, as shown in Lemma 8 of \cite{BPHZ},
that if $X$ is any nonempty subset of $\Omega$, such that for any 
partition of $X$ into two non-empty parts $Y$ and $Z$, there exists a 
member $(P,Q)$ of $Y$ and a member $(P',Q')$ of $Z$ such that $P\cup P'
\subseteq Q\cap Q'$ is true, then the union of the $Q$'s of \emph{all} the
members of $X$ is a forest, and the set $(\tilde{P},\tilde{Q})$, where
$\tilde{P}$ is the intersection of the $P$'s of \emph{all} the members of
$X$, and $\tilde{Q}$ is the union of the $Q$'s of \emph{all} the members
of $X$, is a member of $\Omega$.  Now the \emph{maximal} such subsets $X$
of $\Omega$ do not intersect one another, and the set of all of them forms
a partition of $\Omega$.  Furthermore, as we noted above, for \emph{every}
BPHZ forest $F$, the pair $(F,F)$ is a member of $\Omega$.  Now since the
maximal such subsets $X$ of $\Omega$ form a partition of $\Omega$, for 
each BPHZ forest $F$, the pair $(F,F)$ is a member of precisely on such
maximal subset $X$ of $\Omega$.  We then, bearing in mind that the
definition of $\Omega$ refers to a particular position-space configuration
$x$ of the vertices of the original Feynman diagram, define, for each
BPHZ forest $F$, the pair $(P,Q)$ to whose sector the pair $(F,x)$ 
belongs, to be the pair $(\tilde{P},\tilde{Q})$, where $\tilde{P}$ is the
intersection of the $P$'s of all the members of the maximal such subset
$X$ of $\Omega$ of which $(F,F)$ is a member, and $\tilde{Q}$ is the 
union of the $Q$'s of all the members of the maximal such subset $X$ of 
$\Omega$ of which $(F,F)$ is a member.  This gives a solution of the 
tesselation problem for which the answer to both questions 1) and 2) above
is, ``Yes''.  In practice, for any BPHZ forest $F$, the pair $(P,Q)$ to
whose sector the pair $(F,x)$ belongs, is given by the smallest subset $P$
of $F$, and the largest ``superset'' $Q$ of $F$, such that $(P,Q)$ is a 
member of $\Omega$: our construction guarantees, (provided $\sigma$ is 
chosen such that $0<\sigma\leq\frac{1}{8}$), that this determines $P$ and
$Q$ uniquely, as $\tilde{P}$ and $\tilde{Q}$ as defined above.  The
characteristic funciton, in configuration space, of the sector defined by
the pair $(P,Q)$ of BPHZ forests, is constructed on pages 46 and 47 of
\cite{BPHZ}.

In practice, two modifications have to be made to the definition of 
$\Omega$ given above.  The first modification is simply to impose an
absolute upper limit on the diameter $\mathbf{L}(F,A,x)$ of any cluster.
This is due to the smooth long-distance cutoffs we impose on the 
propagators in our counterterms, and ensures that within any cluster, the
propagators in the counterterm are identical to the original propagators.
(The onset of the smooth long-distance cutoffs imposed on the propagators
in the counterterms occurs at a value of $\left|x-y\right|$ strictly
greater than zero.)  The second modification involves restricting the 
propagators $\{i,j\}$ to which we apply (\ref{E285}) in the definition of
$\Omega$.  For a given member $A$ of $Q$ that is not a member of $P$, 
instead of applying (\ref{E285}) for \emph{all} propagators $\{i,j\}$ such
that $i$ is a member of $A$ and $j$ is not a member of $A$, (or in other
words, for \emph{all} legs of $A$), we apply (\ref{E285}) only for all 
propagators $\{i,j\}$ such that $i$ is a member of $A$, and $j$ is not a
member of $A$, but \emph{is} a member of every member of $P$ that contains
$A$ as a subset.  In other words, we apply (\ref{E285}) only to those legs
of $A$ that are \emph{internal} lines of every member of $P$ that contains
$A$ as a subset.  The reason for this is that, in the course of the
convergence proof, for the sector defined by the pair $(P,Q)$ of BPHZ
forests, we eventually bound the sum of the integrands for all the BPHZ
forests $F$ such that $P\subseteq F\subseteq Q$, by an expression that
\emph{factors} over the members of $P$, and has a form such that we can 
apply the Cluster Convergence Theorem \cite{CCT} separately for the 
integral associated with each member of $P$.  However with the original
form of the definition of $\Omega$, this factorization is spoilt by the
occurrence of inequalities between the sizes of clusters in the factor 
associated with one member of $P$, and distances between vertices in the 
factors associated with larger members of $P$.  The above modification to
the definition of $\Omega$ removes this problem.  However it turns out
that with the modified definition of $\Omega$, the size of the fixed real
number $\sigma$ that occurs in (\ref{E285}) has to be restricted slightly
further to ensure that the tesselation problem is still solved.  A
sufficient restriction is $0<\sigma\leq\frac{3}{25}$.  For details we
refer to \cite{BPHZ}.

The additional considerations necessary to demonstrate manifest BPHZ
convergence in position space when only one-line \emph{irreducible}
counterterms are allowed, are based on an identity given in Lemma 34 of
\cite{BPHZ}, which expresses the sum over BPHZ forests that have only
one-line \emph{irreducible} members, as an alternating sum over quantities
that also include forest that have one-line \emph{reducible} members.  The
extra forests cancel out of the alternating sum, and each individual term
in the alternating sum can be treated by an extension of the techniques
used in the previous case.  Part of the demonstration involves showing 
that if, given any power-counting divergent one-line-\emph{reducible}
subdiagram, we do a separate full BPHZ renormalization for each separate
one-line-\emph{irreducible} component of that subdiagram, (including
forest with both one-line-irreducible and one-line-reducible members
within each of those separate one-line-\emph{irreducible} components), 
then the \emph{overall} counterterm for this expression is finite, 
provided we define it as a limit from propagators that are \emph{smoothly}
regularized at extremely short distances - this is where we have to do the
integrations by parts that we mentioned before.  After sweeping excessive
derivatives on the ``key'' propagators of this expression, ``off the 
edge'' of this overall counterterm by means of integrations by prts and
translation smoothness, we are left with an absolutely convergent
expression.  For details we refer to Theorem 2 of \cite{BPHZ}.

\section{Renormalization group equation for a family of loops}

We now briefly consider again the renormalization group equation 
(\ref{E102}) for a family of loops $W_{1L},\ldots,W_{nL}$, where the loops
$W_{1L},\ldots,W_{nL}$ are obtained from a set of loops $W_1,\ldots,W_n$,
such that the maximu distance between any two points on any of the loops
$W_1,\ldots,W_n$ is $R$, (where $R$ is the onset point of the smooth
long-distance cutoffs we impose on the propagators in our BPHZ 
counterterms), by uniformly rescaling all the sizes and separations of the
loops $W_1,\ldots,W_n$ until the maximum distance between any two points 
on any of them is $L$.  (As usual in this paper we neglect the additional
terms in (\ref{E102}) associated with the short-distance factors we divide
our $f_0$'s by.)

The ``curves of constant $f$'', i.e. of $\mathrm{d}f=0$, in the $(L,g)$
plane, are given by:
\begin{equation}
	\label{E288}
	L\frac{\mathrm{d}g}{\mathrm{d}L}=\beta(g)
\end{equation}
These curves look like:
	\unitlength 1.00pt
\linethickness{0.61pt}
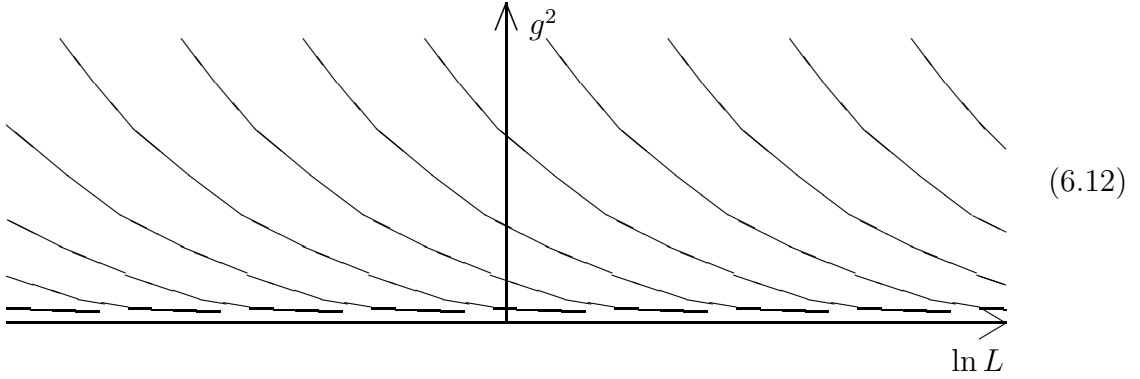
\begin{equation}
	\label{E289}
	\raisebox{-69pt}{
\begin{picture}(379.00,140.34)
\put(369.00,25.00){\line(-6,1){19.00}}
\put(323.00,25.00){\line(-6,1){19.00}}
\put(277.00,25.00){\line(-6,1){19.00}}
\put(231.00,25.00){\line(-6,1){19.00}}
\put(185.00,25.00){\line(-6,1){19.00}}
\put(139.00,25.00){\line(-6,1){19.00}}
\put(93.00,25.00){\line(-6,1){19.00}}
\put(47.00,25.00){\line(-6,1){19.00}}
\put(350.00,28.00){\line(-3,1){28.00}}
\put(304.00,28.00){\line(-3,1){28.00}}
\put(258.00,28.00){\line(-3,1){28.00}}
\put(212.00,28.00){\line(-3,1){28.00}}
\put(166.00,28.00){\line(-3,1){28.00}}
\put(120.00,28.00){\line(-3,1){28.00}}
\put(74.00,28.00){\line(-3,1){28.00}}
\put(368.00,38.00){\line(-5,2){25.00}}
\put(322.00,38.00){\line(-5,2){25.00}}
\put(276.00,38.00){\line(-5,2){25.00}}
\put(230.00,38.00){\line(-5,2){25.00}}
\put(184.00,38.00){\line(-5,2){25.00}}
\put(138.00,38.00){\line(-5,2){25.00}}
\put(92.00,38.00){\line(-5,2){25.00}}
\put(46.00,38.00){\line(-5,2){25.00}}
\put(344.00,48.00){\line(-2,1){24.00}}
\put(298.00,48.00){\line(-2,1){24.00}}
\put(252.00,48.00){\line(-2,1){24.00}}
\put(206.00,48.00){\line(-2,1){24.00}}
\put(160.00,48.00){\line(-2,1){24.00}}
\put(114.00,48.00){\line(-2,1){24.00}}
\put(68.00,48.00){\line(-2,1){24.00}}
\put(366.00,60.00){\line(-4,3){20.00}}
\put(320.00,60.00){\line(-4,3){20.00}}
\put(274.00,60.00){\line(-4,3){20.00}}
\put(228.00,60.00){\line(-4,3){20.00}}
\put(182.00,60.00){\line(-4,3){20.00}}
\put(136.00,60.00){\line(-4,3){20.00}}
\put(90.00,60.00){\line(-4,3){20.00}}
\put(44.00,60.00){\line(-4,3){20.00}}
\put(369.00,24.67){\line(1,0){9.00}}
\put(323.00,24.67){\line(1,0){9.00}}
\put(277.00,24.67){\line(1,0){9.00}}
\put(231.00,24.67){\line(1,0){9.00}}
\put(185.00,24.67){\line(1,0){9.00}}
\put(139.00,24.67){\line(1,0){9.00}}
\put(93.00,24.67){\line(1,0){9.00}}
\put(47.00,24.67){\line(1,0){9.00}}
\put(1.00,24.67){\line(1,0){9.00}}
\put(346.00,75.00){\line(-6,5){21.00}}
\put(300.00,75.00){\line(-6,5){21.00}}
\put(254.00,75.00){\line(-6,5){21.00}}
\put(208.00,75.00){\line(-6,5){21.00}}
\put(162.00,75.00){\line(-6,5){21.00}}
\put(116.00,75.00){\line(-6,5){21.00}}
\put(70.00,75.00){\line(-6,5){21.00}}
\put(371.00,92.50){\line(-5,6){14.58}}
\put(325.00,92.50){\line(-5,6){14.58}}
\put(279.00,92.50){\line(-5,6){14.58}}
\put(233.00,92.50){\line(-5,6){14.58}}
\put(187.00,92.50){\line(-5,6){14.58}}
\put(141.00,92.50){\line(-5,6){14.58}}
\put(95.00,92.50){\line(-5,6){14.58}}
\put(49.00,92.50){\line(-5,6){14.58}}
\put(356.00,110.00){\line(-3,4){12.75}}
\put(310.00,110.00){\line(-3,4){12.75}}
\put(264.00,110.00){\line(-3,4){12.75}}
\put(218.00,110.00){\line(-3,4){12.75}}
\put(172.00,110.00){\line(-3,4){12.75}}
\put(126.00,110.00){\line(-3,4){12.75}}
\put(80.00,110.00){\line(-3,4){12.75}}
\put(34.00,110.00){\line(-3,4){12.75}}
\put(332.00,24.00){\line(1,0){8.33}}
\put(286.00,24.00){\line(1,0){8.33}}
\put(240.00,24.00){\line(1,0){8.33}}
\put(194.00,24.00){\line(1,0){8.33}}
\put(148.00,24.00){\line(1,0){8.33}}
\put(102.00,24.00){\line(1,0){8.33}}
\put(56.00,24.00){\line(1,0){8.33}}
\put(10.00,24.00){\line(1,0){8.33}}
\put(340.33,23.67){\line(1,0){8.67}}
\put(294.33,23.67){\line(1,0){8.67}}
\put(248.33,23.67){\line(1,0){8.67}}
\put(202.33,23.67){\line(1,0){8.67}}
\put(156.33,23.67){\line(1,0){8.67}}
\put(110.33,23.67){\line(1,0){8.67}}
\put(64.33,23.67){\line(1,0){8.67}}
\put(18.33,23.67){\line(1,0){8.67}}
\put(349.00,23.33){\line(1,0){9.00}}
\put(303.00,23.33){\line(1,0){9.00}}
\put(257.00,23.33){\line(1,0){9.00}}
\put(211.00,23.33){\line(1,0){9.00}}
\put(165.00,23.33){\line(1,0){9.00}}
\put(119.00,23.33){\line(1,0){9.00}}
\put(73.00,23.33){\line(1,0){9.00}}
\put(27.00,23.33){\line(1,0){9.00}}
\put(366.67,59.67){\line(2,-1){12.33}}
\put(368.00,37.33){\line(3,-1){11.00}}
\put(378.00,24.00){\line(1,0){1.00}}
\put(379.00,85.00){\line(-1,1){10.00}}
\put(379.00,19.33){\line(-1,0){122.00}}
\put(257.67,19.33){\line(-1,0){130.00}}
\put(129.00,19.33){\line(-1,0){117.67}}
\put(11.67,19.33){\line(-1,0){10.67}}
\put(24.33,74.67){\line(-6,5){23.33}}
\put(22.00,48.00){\line(-2,1){20.67}}
\put(1.00,37.00){\line(3,-1){27.00}}
\put(190.00,19.67){\line(0,1){120.67}}
\put(190.00,140.00){\line(2,-5){4.00}}
\put(190.00,140.00){\line(-2,-5){4.00}}
\put(379.00,19.00){\line(-5,3){10.00}}
\put(379.00,19.00){\line(-5,-3){10.00}}
\put(368.33,1.00){\makebox(0,0)[cb]{$\ln L$}}
\put(204.00,138.00){\makebox(0,0)[ct]{$g^2$}}
\end{picture}
}
\end{equation}

On \emph{these} ``curves of constant $f$'', $g^2$ \emph{decreases} as $L$
increases: the \emph{opposite} behaviour to the ``running coupling''.  But
that is exactly right: to use this diagram, we suppose that $f_0(W_{1L},
\ldots,W_{nL},g^2)$ has been calculated at one particular value of $L$,
say $L_0$, for all $g^2$.  Then to calculate $f_0(W_{1L},\ldots,W_{nL},
g^2)$ for some particular value of $g^2$, say $g_1^2$, at some other value
of $L$, say $L_1$, we identify the particular ``curve of constant $f$''
that passes through the point $(L_1,g_1)$, and trace that curve back until
it intersects the vertical line $L=L_0$.  Then if $g_0$ is the value of 
$g$ where this curve intersects the vertical line $L=L_0$, we have:
\begin{equation}
	\label{E290}
	f_0(W_{1L_1},\ldots,W_{nL_1},g_1^2)=f_0(W_{1L_0},\ldots,W_{nL_0},g_0^2)
\end{equation}
We see immediately from diagram (\ref{E289}) that $g_0^2$ will be greater
than $g_1^2$ if $L_1$ is greater than $L_0$, and $g_0^2$ will be less than
$g_1^2$ is $L_1$ is less than $L_0$.  In fact on dimensional grounds, for
fixed $g_1$, $g_0$ only depends on $L_0$ and $L_1$ through the ratio
$\frac{L_0}{L_1}$.  Hence $L_0\left.\frac{\mathrm{d}g_0}{\mathrm{d}L_0}
\right|_{g_1}=-L_1\left.\frac{\mathrm{d}g_0}{\mathrm{d}L_1}\right|_{g_1}$,
hence by (\ref{E288}):
\begin{equation}
	\label{E291}
	L_1\left.\frac{\mathrm{d}g_0}{\mathrm{d}L_1}\right|_{g_1}=-\beta(g_0)
\end{equation}
Hence, by comparison with (\ref{E256}), we see that the dependence of 
$g_0$ on $L_1$, for fixed $L_0$ and $g_1$, is indeed given precisely by 
the ``running coupling''.

\section{The contribution of the one-loop islands must have the
\emph{opposite} sign to what a scalar boson would give}

We now consider the crucial question of the signs of the island diagrams,
and of the sign of $\displaystyle\sum_iY_i$ in equation (\ref{E236}),
which, as we have noted, must be \emph{negative}.  Now from equation
(\ref{E184}), we see that the coefficients $\left.\frac{\mathrm{d}}
{\mathrm{d}M}\mathbf{C}(M)\right|_{M=1}$ tend to alternate in sign.
However, as we explained in the discussion after equation (\ref{E282}), we
expect that for input values of $g^2$ not greater than the critical value,
and for reasonable ans$\ddot{\textrm{a}}$tze for our long-distance factors
(\ref{E262}), the sums over the \emph{explicit} powers of $g^2$ that occur
in the right-hand sides of the group-variation equations, and in the 
short-distance factors when we restore them in the windows of the 
right-hand side group-variation equation diagrams, will \emph{converge},
and that the convergence rate will be geometric.  It is thus highly
unlikely that the sum over the contributions of all the island diagrams
will have the opposite sign from the net contribution of the leading 
island diagrams, namely the one-loop islands.  Therefore the net 
contribution of the one-loop island diagrams to $\displaystyle\sum_iY_i$ 
in (\ref{E236}) must be \emph{negative}.  (Of course, since \emph{every}
island diagram has \emph{exactly} one island, and \emph{no} 45-paths that
do \emph{not} form part of that island, it makes no difference whether we
say ``one-loop islands'' or ``one-loop island diagrams''.)

Let us consider carefully what determines the signs of the one-loop island
diagrams.  The factor $\left.\frac{\mathrm{d}}{\mathrm{d}M}\mathbf{C}(M)
\right|_{M=1}$ is equal to 1.  Now if, instead of Landau-gauge vector
bosons, and Fadeev-Popov scalar fermions, we had a simple scalar boson,
then the one-loop island would simply be $-\frac{1}{2}\mathrm{tr}\ln\left(
\bar{D}^2\right)$, which we see from (\ref{E32}), the paragraph after
(\ref{E32}), and (\ref{E27}), is simply a sum over closed paths, \emph{all
with positive coefficients}, with each path being weighted, by the 45-term
in (\ref{E67}) and our general rules of right-hand side group-variation
equation diagrams, by a product of two oppositely-directed fundamental
representation Wilson loops, each in a \emph{different} $\textrm{SU}(N)$ group, 
(i.e. with a different set of $\textrm{SU}(N)$ gauge fields).  And if the island
diagram is like(\ref{E135}) or (\ref{E163}), which in the discussion
between equations (\ref{E200}) and (\ref{E201}), we called Type-1 island
diagrams, then one of the two $\textrm{SU}(N)$ groups occurs in no other path, so
we have a Wilson-loop vacuum expectation value, (corresponding to the 
interior window of the island), for that $\textrm{SU}(N)$ Wilson loop, while the 
other $\textrm{SU}(N)$ group also occurs in the other connected components of the
border of the non-simply connected window surrounding the island, so that
we have a Wilson-loop correlation function involving the border of the
island, and the other connected components of the border of that 
non-simply connected window, (which are the loops that occur in the 
left-hand side $f_0$), for that $\textrm{SU}(N)$ group.  And if the island diagram
is a Type-2 island diagram, for example (\ref{E165}), (\ref{E181}), or
(\ref{E182}), then each of the two $\textrm{SU}(N)$ fundamental representation
Wilson loops weighting the island's 45-path is involved in a correlation
function with one or more left-hand side loops.

Now these vacuum expectation values and correlation functions may all be
expected to be predominantly \emph{positive}, hence there is no doubt that
if, instead of our Landau-gauge vector bosons and Fadeev-Popov scalar
fermions, we had a simple scalar boson, then the contribution of the 
one-loop island would be \emph{positive}: the \emph{opposite} of what we
need for (\ref{E236}) and the group-variation equations in general.  Now
(\ref{E236}) is certainly correct, (subject to the modification which we
explain later, which has no effect on the sign).  \emph{We are forced to
the conclusion that the contribution of the one-loop islands must have 
the} opposite \emph{sign to what we would expect if, instead of our
Landau-gauge vector bosons and Fadeev-Popov scalar fermions, we had a 
simple scalar boson.}  Now (\ref{E236}) applies when our ansatz is valid,
which requires all the sizes and separations of the left-hand side loops 
to be greater than $\frac{1}{\mu}$, but as we will see, the island 
diagrams also play a major role in driving the transition from 
asymptotic-freedom behaviour to the area law behaviour of our ansatz.  in
the \emph{transition} domain, the most important island configurations are
those where the island path closely tracks the left-hand side path, (in
the group-variation equation for $f_0(W_1,g^2)$), and the window 
surrounding the island is a thin ribbon whose edges are the left-hand side
loop and the border of the island.\footnote{Note added: this statement may
be incorrect, but it may still be possible to investigate the transition
region with the non-simply connected window treated in perturbation
theory, thus obtaining an equation for the transition region which closes
among the $f_0(W_1,g^2)$.}  In this domain we treat the \emph{non-simply
connected} window \emph{by perturbation theory}, thus decoupling the 
group-variation equations for the multi-loop correlation functions, and
obtaining, approximately, a simple linear differential equation for
$f_0(W_{1L},g^2)$.  The conclusion with regard to the sign is as before:
the sign of the contribution from the one-loop island diagrams must be
\emph{opposite} to what we would expect if, instead of our Landau-gauge
vector bosons and Fadeev-Popov scalar fermions, we had a simple scalar
boson.  Our conclusion must be that the fact that $2g\beta(g)$ is 
negative, must be the first of a whole family of results, which guarantee
that the net contribution of the one-loop island diagrams to the 
right-hand sides of the group-variation equations in their standard forms
(\ref{E184}), (\ref{E186}), etc., are negative under a rather wide range
of conditions.  Is this possible?

\section{The required sign change occurs in the simplest example}

Now from (\ref{E34}) and (\ref{E41}) we see that the contribution of the
vector boson one-loop island is:
\begin{equation}
	\label{E292}
	-2\mathrm{tr}\ln\left(\bar{D}^2\right)-\frac{1}{2}\mathrm{tr}\ln E+
	\frac{1}{2}\mathrm{tr}\left(G_0F\right)+\frac{1}{4}\mathrm{tr}\left(G_0
	FG_0F\right)+\frac{1}{6}\mathrm{tr}\left(G_0FG_0FG_0F\right)+\ldots
\end{equation}
where $g^2G_0$ denotes the propagator matrix (\ref{E20}), $F$ denotes the
matrix $\left(\begin{array}{cc} 2\bar{F}_{\mu\sigma} & 0 \\ 0 & 0
\end{array}\right)$, the operator $E$ is defined by (\ref{E16}), and 
$\ln E$ is to be expanded in powers of $(E-1)$.

The Fadeev-Popov contribution is $\mathrm{tr}\ln(\bar{D}^2)$, hence the 
total contribution of the one-loop islands is:
\begin{equation}
	\label{E293}
	-\mathrm{tr}\ln\left(\bar{D}^2\right)-\frac{1}{2}\mathrm{tr}\ln E+
	\frac{1}{2}\mathrm{tr}\left(G_0F\right)+\frac{1}{4}\mathrm{tr}\left(G_0
	FG_0F\right)+\frac{1}{6}\mathrm{tr}\left(G_0FG_0FG_0F\right)+\ldots
\end{equation}
This is to be interpreted in terms of window-weighted path integrals in
the same way as we discussed above for a ``scalar boson'' one-loop island.
Now the first term in (\ref{E293}) is simply twice the result for a
``scalar boson'' island, hence is expected to be \emph{positive} for the
same reasons as before.  Hence the remaining terms must reverse the sign.

To consider whether this is possible, we have first to generalize 
(\ref{E67}) to cases where we have $\bar{D}_\mu$ or $\bar{F}_{\mu\nu}$
insertions along the paths, to deal with the $\mathrm{tr}\ln E$ and
$\mathrm{tr}\left(G_0FG_0F\ldots G_0F\right)$ terms in (\ref{E293}).  The
required results may be obtained directly from (\ref{E67}), by making use
of the $\sigma\to0$ limit of (\ref{E29}), and the equation:
\begin{equation}
	\label{E294}
	\bar{F}_{\mu\nu a}f_{AaB}=\left(\bar{D}_\mu\bar{D}_\nu-\bar{D}_\nu
	\bar{D}_\mu\right)_{AB}
\end{equation}
which follows directly from (\ref{E5}), (\ref{E4}), and the definition of
$\bar{F}_{\mu\nu a}$ after equation (\ref{E11}).

Now our path integrals have to be treated by the limiting procedure 
discussed in connection with expression (\ref{E278}) above.  The $\bar{D}
_\mu$ insertions affect both the long-distance factor and the 
short-distance factor.  The effect on the long-distance factor is in
general nonzero and finite.  From (\ref{E29}) we see it essentially 
consists in the insertion of \emph{one} extra segment, of length 
$\epsilon$ say, into the sequence of a finite number of straight segments
which defines the long-distance factor, and evaluating the derivative with
respect to $\epsilon$ and $\epsilon=0$.  Only one straight segment of the
original path need be affected: it is replaced by the new segment of 
length $\epsilon$, plus a segment ``nearly'' the same as the original
segment (as $\epsilon\to0$), making the third side of a triangle with the
original segment and the new segment of length $\epsilon$.  The change to
the area of the minimal-area orientable spanning surface resulting from 
such a change to the path is in general of order $\epsilon$, so we expect
the effect of $\bar{D}_\mu$ insertions to get a finite contribution from
the long-distance factor.  We may expect this result to remain true if we
take the limit in the derivative simultaneously with the limit $\sigma\to
0$, as suggested in (\ref{E29}), rather than taking the limit in the
derivative with $\sigma$ fixed.  However the predominant effect of the
$\bar{D}_\mu$ and $\bar{F}_{\mu\nu}$ insertions comes from the 
\emph{short}-distance factor, which must be treated in the path integrals
as discussed in connection with expression (\ref{E278}).  Thus to 
determine the signs of the island diagrams we need the full
renormalization of the group-variation equations, which is treated in our
next paper.  Here we shall simply show that the $\mathrm{tr}\ln E$ and
$\mathrm{tr}\left(G_0FG_0F\ldots G_0F\right)$ terms in (\ref{E293}) do
indeed produce the required sign reversal in the simplest meaningful
calculation, namely the change in the counterterm for the one-loop islands
when we change the onset point of the smooth long-distance cutoffs imposed
on the propagators in the counterterms, from $R$ to $R_2$.  Thus we 
neglect any effect of the $\bar{D}_\mu$ and $\bar{F}_{\mu\nu}$ insertions
on the long-distance factors.  We work through quadratic order in the 
effective fields introduced as in (\ref{E273}) to represent the 
short-distance factor.  Now in general we have to introduce an independent
set of $\textrm{SU}(N)$ effective gauge fields for each window of our right-hand
side group-variation equation diagrams: these correspond to different
$\textrm{SU}(N)$ subgroups of $\textrm{SU}(NM)$ and $(\textrm{SU}(N))^M$, as in (\ref{E57}).  However
through quadratic order the effects of the effective fields introduced in
different windows decouple from one another: this is because the traces of
all the $\textrm{SU}(N)$ fundamental representation generators are zero, so we need
either zero or two fields in each trace to get a nonzero result.  Thus we
consider explicitly the contributions to (\ref{E293}) quadratic in the
effective fields $A_{\mu ax}$ introduced for one ``side'' of the closed
45-path that defines a one-loop island, and of zeroth order in the 
effective fields, (another set of $\textrm{SU}(N)$ gauge fields, independent of 
$A_{\mu ax}$, introduced for the other ``side'' of that closed 45-path.
We will extract the change in the logarithmically divergent part of the
counterterm when we change the onset point of the smooth long-distance
cutoffs imposed on the propagators in the counterterms, from $R$ to $R_2$.
Now as explained after (\ref{E240}), the long-distance cutoffs on the
propagators in the counterterms are imposed by multiplying them by a
factor $f\left(\frac{(x-y)^2}{R^2}\right)$, where $x$ and $y$ are the
positions of the ends of the propagator, and $f(s)$ is a fixed member of
$\mathbf{R}^{\mathbf{R}}$ that is infinitely differentiable for all $s\in
\mathbf{R}$, equal to 1 for $s\leq1$, and equal to 0 for $s\geq t$, where
$t$ is a fixed finite real number strictly greater than 1.  The 
consequence of this for the present calculation, where all the relevant
``subdiagrams'' have just two vertices, $x$ and $y$, one of which is 
chosen as the contraction point, is that, after the angular integrations
are done, the contribution to the logarithmically divergent part of the
counterterm from the domain $\left|x-y\right|\geq R$ has the form:
\begin{equation}
	\label{E295}
	\int_R^{R\sqrt{t}}\frac{\mathrm{d}u}{u}h\left(\frac{u}{R}\right)
\end{equation}
for some smooth real function $h(s)$, and is thus completely independent
of $R$.  Thus if, for example, we assume $R_2\geq R$, then the change to 
the logarithmically divergent part of the counterterm, when $R$ is 
replaced by $R_2$, is given entirely by the domain $R\leq\left|x-y\right|
\leq R_2$, calculated with the \emph{original} propagators.

We now determine the contributions to this calculation of the separate
terms in (\ref{E293}).  We define the matrix $\Delta$ by:
\begin{equation}
	\label{E296}
	\Delta_{xy}=\left(\frac{-1}{\partial^2}\right)_{xy}=\frac{1}{4\pi^2
	(x-y)^2}
\end{equation}
Now
\begin{equation}
	\label{E297}
	\bar{D}^2=\partial^2+A\partial+\partial A+AA
\end{equation}
hence by (\ref{E33}) we have:
	\[\mathrm{tr}\ln\left(\bar{D}^2\right)=\mathrm{tr}(\ln(\partial^2)+\ln
	(1-\Delta(A\partial+\partial A+AA)))=
\]
\begin{equation}
	\label{E298}
	=\mathrm{tr}\left(\ln(\partial^2)-\Delta(A\partial+\partial A+AA)-
	\frac{1}{2}\Delta(A\partial+\partial A)\Delta(A\partial+\partial A)-
	\ldots\right)
\end{equation}
where the terms not explicitly displayed are of third degree or higher in
$A_{\mu ax}$.  Now the term $\mathrm{tr}\ln(\partial^2)$ is of no 
interest, and the trace on the group indices vanishes for the terms linear
in $A_{\mu ax}$, hence for our calculation we may take
\begin{equation}
	\label{E299}
	-\mathrm{tr}\ln(\bar{D}^2)=\mathrm{tr}(\Delta AA)+\frac{1}{2}\mathrm{tr}
	(\Delta(A\partial+\partial A)\Delta(A\partial+\partial A))
\end{equation}
But the counterterm for $\mathrm{tr}(\Delta AA)$ is completely independent
of $R$, hence the only term in $-\mathrm{tr}\ln(\bar{D}^2)$ that 
contributes to our calculation is the term:
\begin{equation}
	\label{E300}
	\frac{1}{2}\mathrm{tr}(\Delta(A\partial+\partial A)\Delta(A\partial+
	\partial A))
\end{equation}
This term gives
\begin{equation}
	\label{E301}
	-\left(\Delta_{xy}\left(\hat{y}_\mu\hat{y}_\nu\Delta_{xy}\right)-\left(
	\hat{y}_\mu\Delta_{xy}\right)\left(\hat{y}_\nu\Delta_{xy}\right)\right)
	A_{\mu ax}A_{\nu ay}
\end{equation}
where $\hat{y}_\mu$ means $\frac{\partial}{\partial y_\mu}$, etc., and the
overall minus sign comes from (\ref{E43}), and is due to the fact that our
representation matrices $t_a$ are \emph{anti}hermitian.  Now (\ref{E301})
is quadratically divergent, so, choosing $x$ as the contraction point, the
logarithmically divergent part of the counterterm is given by taking the
second-degree term $\frac{1}{2}(y-x)_\alpha(y-x)_\beta\hat{x}_\alpha
\hat{x}_\beta A_{\nu ax}$ in the Taylor expansion of $A_{\nu ay}$ about
the point $y=x$.  Hence, defining $u\equiv y-x$, (\ref{E301}) gives:
	\[-\;\!\!\int_{R\leq\left|u\right|\leq R_2}\;\!\!\mathrm{d}^4u
	\left(\frac{1}{4\pi^2u^2}\left(
	\frac{-2\delta_{\mu\nu}}{4\pi^2\left|u\right|^4}+\frac{8u_\mu u_\nu}{4
	\pi^2\left|u\right|^6}\right)-\frac{-2u_\mu}{4\pi^2\left|u\right|^4}
	\frac{-2u_\nu}{4\pi^2\left|u\right|^4}\right)A_{\mu ax}\frac{1}{2}
	u_\alpha u_\beta\hat{x}_\alpha\hat{x}_\beta A_{\nu ax}
\]
	\[=\frac{-1}{(4\pi^2)^2}\int_{R\leq\left|u\right|\leq R_2}\mathrm{d}^4u
	\left(\frac{-2\delta_{\mu\nu}u_\alpha u_\beta}{\left|u\right|^6}+\frac{
	4u_\mu u_\nu u_\alpha u_\beta}{\left|u\right|^8}\right)\frac{1}{2}A_{\mu
	ax}\hat{x}_\alpha\hat{x}_\beta A_{\nu ax}=
\]
	\[=\frac{-1}{8\pi^2}\ln\left(\frac{R_2}{R}\right)\left(\frac{-2\delta_
	{\mu\nu}\delta_{\alpha\beta}}{4}+\frac{4\left(\delta_{\mu\nu}\delta_
	{\alpha\beta}+\delta_{\mu\alpha}\delta_{\nu\beta}+\delta_{\mu\beta}
	\delta_{\nu\alpha}\right)}{24}\right)\frac{1}{2}A_{\mu ax}\hat{x}_\alpha
	\hat{x}_\beta A_{\nu ax}
\]
\begin{equation}
	\label{E302}
	=\frac{-1}{96\pi^2}\ln\left(\frac{R_2}{R}\right)F_{\mu\nu ax}F_{\mu\nu
	ax}
\end{equation}
where in the last line here we neglected a total derivative, and of course
since we are neglecting terms of third degree and higher in $A_{\mu ax}$,
$F_{\mu\nu ax}F_{\mu\nu ax}$ is equal to \\
$\left(\hat{x}_\mu A_{\nu ax}-
\hat{x}_\nu A_{\mu ax}\right)\left(\hat{x}_\mu A_{\nu ax}-\hat{x}_\nu A_
{\mu ax}\right)$.

Now by (\ref{E16}), $E$ is equal to $\bar{D}_\mu\frac{1}{\bar{D}^2}\bar{D}
_\nu$, and $\mathrm{tr}\ln E$, and the $\frac{1}{E}$ that occurs in $G_0$
by (\ref{E20}), are to be expanded in powers of $(E-1)$.  Now:
\begin{equation}
	\label{E303}
	\mathrm{tr}\ln E=\mathrm{tr}(E-1)-\frac{1}{2}\mathrm{tr}((E-1)^2)+\ldots
\end{equation}
But $\mathrm{tr}E=1$, hence through quadratic order in the fields we have:
\begin{equation}
	\label{E304}
	\mathrm{tr}\ln E=-\frac{1}{2}\mathrm{tr}((E-1)^2)
\end{equation}

Now $(E-1)$ begins with a term linear in the fields, hence we see from
(\ref{E304}) that to calculate $\mathrm{tr}\ln E$ through quadratic order
in the fields we \emph{only} need the term in $(E-1)$ \emph{linear} in the
fields.  Furthermore, to calculate $\mathrm{tr}\left(G_0F\right)$ through 
quadratic order in the fields we again only need the term in $(E-1)$ 
linear in the fields, and to calculate $\mathrm{tr}\left(G_0FG_0F\right)$
through quadratic order in the fields we may set $E=1$.  Thus for our
calculation we only need the term in $(E-1)$ linear in the fields.

Now by (\ref{E294}), $[\bar{D}_\mu,\bar{D}_\nu]=\bar{F}_{\mu\nu}$, hence
$[\bar{D}_\mu,\bar{D}^2]=\bar{F}_{\mu\nu}\bar{D}_\nu+\bar{D}_\nu\bar{F}
_{\mu\nu}$, hence
\begin{equation}
	\label{E305}
	[\bar{D}_\mu,\frac{1}{\bar{D}^2}]=-\frac{1}{\bar{D}^2}[\bar{D}_\mu,
	\bar{D}^2]\frac{1}{\bar{D}^2}=-\frac{1}{\bar{D}^2}\left(\bar{F}_{\mu\nu}
	\bar{D}_\nu+\bar{D}_\nu\bar{F}_{\mu\nu}\right)\frac{1}{\bar{D}^2}
\end{equation}
hence
	\[E-1=[\bar{D}_\mu,\frac{1}{\bar{D}^2}]\bar{D}_\mu=-\frac{1}{\bar{D}^2}
	\left(\bar{F}_{\mu\nu}\bar{D}_\nu+\bar{D}_\nu\bar{F}_{\mu\nu}\right)
	\frac{1}{\bar{D}^2}\bar{D}_\mu=
\]
	\[=-\Delta\left(\bar{F}_{\mu\nu}\partial_\nu+\partial_\nu\bar{F}_{\mu
	\nu}\right)\Delta\partial_\mu+\textrm{ quadratic in the fields}
\]
\begin{equation}
	\label{E306}
	=\partial_\mu\Delta\bar{F}_{\mu\nu}\Delta\partial_\nu+
	\textrm{ quadratic in the fields}
\end{equation}
Now (\ref{E306} is based on (\ref{E294}), and when inserted in a 45-path,
includes contributions from the effective fields introduced for 
\emph{both} ``sides'' of that 45-path.  However, as explained above, the
contributions from the two sides of the closed 45-path that forms a
one-loop island decouple through the quadratic order to which we are
working, and we are calculating the contributions from \emph{one} side of
the 45-path.  (If we were calculating the contributions from both sides of
the path, we would have needed additional terms in (\ref{E297}), involving
a second set of $\textrm{SU}(N)$ gauge fields, and we would have had to display, in
(\ref{E297}), the separate $\textrm{SU}(N)$ fundamental representation indices for 
each side of the path.)  Now the 45-term in (\ref{E67}) shows that, of the
two $\textrm{SU}(N)$ fundamental representation path-ordered phase factors along 
the 45-path, one is directed in the same direction as the 45-path, and the
other is directed in the opposite direction to the 45-path, and we readily
verify, from (\ref{E294}), (\ref{E29}), and (\ref{E67}), that if we 
neglect the effect of $\bar{D}_\mu$ and $\bar{D}_\nu$ on the long-distance
factors, then the insertion of $\bar{F}_{\mu\nu}$ into a 45-path is
equivalent to the insertion of $F_{\mu\nu}$ into the $\textrm{SU}(N)$ fundamental
representation path-ordered phase factor directed in the \emph{same}
direction as the 45-path, \emph{minus} the insertion of $F_{\mu\nu}$ into
the $\textrm{SU}(N)$ fundamental representation path-ordered phase factor directed
in the \emph{opposite} direction to the 45-path, with the gauge fields in
each inserted $F_{\mu\nu}$ being those of the phase factor into which the
insertion occurs.  The same result may be obtained directly from the 
formula for $\bar{D}_\mu$ as inserted into a 45-path:
	\[\left(\bar{D}_\mu\right)_{jk,qp}=\delta_{jq}\delta_{kp}\partial_\mu+
	A_{1\mu a}\left(t_a\right)_{jq}\delta_{kp}+\delta_{jq}A_{2\mu a}\left(
	\left(t_a\right)_{kp}\right)^\ast=
\]
\begin{equation}
	\label{E307}
	=\delta_{jq}\delta_{pk}\partial_\mu+A_{1\mu a}\left(t_a\right)_{jq}
	\delta_{pk}-\delta_{jq}A_{2\mu a}\left(t_a\right)_{pk}
\end{equation}
where the $A_1$'s and $A_2$'s are two completely independent sets of
$\textrm{SU}(N)$ gauge fields, with the $A_1$'s occurring in the $\textrm{SU}(N)$ 
fundamental representation phase factor directed in the \emph{same}
direction as the 45-path, and the $A_2$'s occurring in the $\textrm{SU}(N)$
fundamental representation phase factor directed in the \emph{opposite}
direction to the 45-path.  Indeed, we find immediately from (\ref{E307})
that:
	\[\left(\bar{D}_\mu\right)_{jk,qp}\left(\bar{D}_\nu\right)_{qp,sr}-
	\left(\bar{D}_\nu\right)_{jk,qp}\left(\bar{D}_\mu\right)_{qp,sr}=
\]
	\[=\Big\{\left(\partial_\mu A_{1\nu a}-\partial_\nu A_{1\mu a}+f_{abc}
	A_{1\mu b}A_{1\nu c}\right)\left(t_1\right)_{js}\delta_{rk}
\]
	\[-\delta_{js}\left(\partial_\mu A_{2\nu a}-\partial_\nu A_{2\mu a}+
	f_{abc}A_{2\mu b}A_{2\nu c}\right)\left(t_a\right)_{rk}\Big\}=
\]
\begin{equation}
	\label{E308}
	=F_{1\mu\nu a}\left(t_a\right)_{js}\delta_{rk}-\delta_{js}F_{2\mu\nu a}
	\left(t_a\right)_{rk}
\end{equation}

Hence, bearing in mind that, as explained, the effects of the two sides of
the 45-path decouple through the quadratic order to which we are working,
we will complete this calculation for the ``side'' of the 45-path directed
in the \emph{same} direction as the 45-path itself, and accordingly
replace $\bar{F}_{\mu\nu}$ in (\ref{E306}) by $F_{1\mu\nu}=F_{1\mu\nu a}
\left(t_a\right)_{ij}$.  Hence from (\ref{E304}) and (\ref{E306}) we find
that the contribution to the $-\frac{1}{2}\mathrm{tr}\ln E$ term in 
(\ref{E293}) from ``our'' side of the 45-path is:
\begin{equation}
	\label{E309}
	\frac{1}{4}\mathrm{tr}\left(\partial_\mu\Delta F_{1\mu\nu}\Delta\partial
	_\nu\partial_\sigma\Delta F_{1\sigma\tau}\Delta\partial_\tau\right)
\end{equation}
Now $\Delta_{xy}$ satisfies the identity:
\begin{equation}
	\label{E310}
	\left(\Delta\partial_\mu\partial_\nu\Delta\right)_{xy}=\frac{-\delta_
	{\mu\nu}}{8\pi^2(x-y)^2}+\frac{(x-y)_\mu(x-y)_\nu}{4\pi^2\left|x-y
	\right|^4}
\end{equation}
hence (\ref{E309} is equal to:
\begin{equation}
	\label{E311}
	-\frac{1}{4}\int\int\mathrm{d}^4x\mathrm{d}^4y\left(\frac{-\delta_{\nu
	\sigma}}{8\pi^2u^2}+\frac{u_\nu u_\sigma}{4\pi^2\left|u\right|^4}\right)
	\left(\frac{-\delta_{\tau\mu}}{8\pi^2u^2}+\frac{u_\tau u_\mu}{4\pi^2
	\left|u\right|^4}\right)F_{1\mu\nu ax}F_{1\sigma\tau ay}
\end{equation}
where the overall minus sign again comes from (\ref{E43}), and we have
defined $u\equiv y-x$ as before.  Now (\ref{E311}) is logarithmically
divergent, hence, again choosing $x$ as the contraction point, its
contribution to our calculation is:
	\[\frac{-1}{4(4\pi^2)^2}\int_{R\leq\left|u\right|\leq R_2}\mathrm{d}^4u
	\left(\frac{\delta_{\nu\sigma}\delta_{\tau\mu}}{4\left|u\right|^4}-
	\frac{\delta_{\nu\sigma}u_\tau u_\mu+\delta_{\tau\mu}u_\nu u_\sigma}{2
	\left|u\right|^6}+\frac{u_\nu u_\sigma u_\tau u_\mu}{\left|u\right|^8}
	\right)F_{1\mu\nu ax}F_{1\sigma\tau ax}
\]
	\[=\frac{-1}{32\pi^2}\ln\left(\frac{R_2}{R}\right)\left(\delta_{\nu
	\sigma}\delta_{\tau\mu}\left(\frac{1}{4}-\frac{1}{8}-\frac{1}{8}\right)+
	\frac{\delta_{\nu\sigma}\delta_{\tau\mu}+\delta_{\nu\tau}\delta_{\sigma
	\nu}+\delta_{\nu\mu}\delta_{\sigma\tau}}{24}\right)F_{1\mu\nu ax}F_{1
	\sigma\tau ax}
\]
\begin{equation}
	\label{E312}
	=0
\end{equation}

We next determine the contribution of the term $\frac{1}{2}\mathrm{tr}
\left(G_0F\right)$ in (\ref{E293}).  Now by (\ref{E20}) we have:
	\[\mathrm{tr}\left(G_0F\right)=\mathrm{tr}\left(\left(\frac{-1}{\bar{D}
	^2}\delta_{\mu\nu}+\frac{1}{\bar{D}^2}\bar{D}^\mu\frac{1}{E}\bar{D}_\nu
	\frac{1}{\bar{D}^2}\right)2\bar{F}_{\nu\sigma}\right)=
\]
	\[\mathrm{tr}\left(\frac{1}{\bar{D}^2}\bar{D}_\mu\frac{1}{E}\bar{D}_\nu
	\frac{1}{\bar{D}^2}2\bar{F}_{\nu\mu}\right)=
\]
\begin{equation}
	\label{E313}
	=-\mathrm{tr}\left(\Delta\bar{F}_{\mu\nu}\Delta\bar{F}_{\mu\nu}\right)-
	\mathrm{tr}\left(\Delta\partial_\mu\partial_\sigma\Delta\bar{F}_{\sigma
	\tau}\Delta\partial_\tau\partial_\nu\Delta2\bar{F}_{\mu\nu}\right)+
	\textrm{ cubic}
\end{equation}
where we used (\ref{E294}) and (\ref{E306}).  We again extract the
contribution of \emph{one} side of the 45-path by replacing $\bar{F}_{\mu
\nu}$ by $F_{1\mu\nu}$, and note that the second term in (\ref{E313}) has
the same form as (\ref{E309}), so that by (\ref{E312}) it gives no
contribution to our calculation.  The first term in (\ref{E313}) is
logarithmically divergent, hence with the same conventions as before, the
contribution to our calculation of the $\frac{1}{2}\mathrm{tr}\left(G_0F
\right)$ term in (\ref{E293}) is:
\begin{equation}
	\label{E314}
	\frac{1}{2(4\pi^2)^2}\int_{R\leq\left|u\right|\leq R_2}\frac{\mathrm{d}^4
	u}{\left|u\right|^4}F_{1\mu\nu ax}F_{1\mu\nu ax}=\frac{1}{16\pi^2}\ln
	\left(\frac{R_2}{R}\right)F_{1\mu\nu ax}F_{1\mu\nu ax}
\end{equation}
where the overall sign is again due to (\ref{E43}).

Finally we determine the contribution of the $\frac{1}{4}\mathrm{tr}\left(
G_0FG_0F\right)$ term in (\ref{E293}).  From (\ref{E20}) we find:
\begin{equation}
	\label{E315}
	\mathrm{tr}\left(G_0FG_0F\right)=4\mathrm{tr}\left(\left(\Delta\delta_
	{\mu\nu}+\Delta\partial_\mu\partial_\nu\Delta\right)\bar{F}_{\nu\sigma}
	\left(\Delta\delta_{\sigma\tau}+\Delta\partial_\sigma\partial_\tau
	\Delta\right)\bar{F}_{\tau\mu}\right)+\textrm{ cubic}
\end{equation}
We again extract the contribution of \emph{one} side of the 45-path by
replacing $\bar{F}_{\mu\nu}$ by $F_{1\mu\nu}$.  Expression (\ref{E315}) is
again logarithmically divergent, hence by (\ref{E296}) and (\ref{E310}),
and with the same conventions as before, the contribution to our 
calculation of the $\frac{1}{4}\mathrm{tr}\left(G_0FG_0F\right)$ term in
(\ref{E293}) is:
	\[-\int_{R\leq\left|u\right|\leq R_2}\mathrm{d}^4u\left(\frac{\delta_
	{\mu\nu}}{8\pi^2u^2}+\frac{u_\mu u_\nu}{4\pi^2\left|u\right|^4}\right)
	\left(\frac{\delta_{\sigma\tau}}{8\pi^2u^2}+\frac{u_\sigma u_\tau}{4\pi^2
	\left|u\right|^4}\right)F_{1\nu\sigma ax}F_{1\tau\mu ax}=
\]
	\[=\frac{-1}{8\pi^2}\ln\left(\frac{R_2}{R}\right)\left(\delta_{\mu\nu}
	\delta_{\sigma\tau}\left(\frac{1}{4}+\frac{1}{8}+\frac{1}{8}\right)+
	\frac{\delta_{\mu\nu}\delta_{\sigma\tau}+\delta_{\mu\sigma}\delta_{\nu
	\tau}+\delta_{\mu\tau}\delta_{\nu\sigma}}{24}\right)F_{1\nu\sigma ax}
	F_{1\tau\mu ax}
\]
\begin{equation}
	\label{E316}
	=\frac{1}{16\pi^2}\ln\left(\frac{R_2}{R}\right)F_{1\mu\nu ax}F_{1\mu\nu
	ax}
\end{equation}
where the overall sign is again due to (\ref{E43}).

Hence, bearing in mind that the result (\ref{E302}) is the contribution of
\emph{one} side of the 45-path for the term $-\mathrm{tr}\ln(\bar{D}^2)$
in (\ref{E293}), we find from (\ref{E302}), (\ref{E312}), (\ref{E314}),
and (\ref{E316}), that through quadratic order in the effective fields for
\emph{one} side of the 45-path, the change in the counterterm for the 
one-loop islands, when we change the onset point of the smooth 
long-distance cutoffs imposed on the propagators in the counterterms, from
$R$ to $R_2$, is:
\begin{equation}
	\label{E317}
	(-1+6+6)\frac{1}{96\pi^2}\ln\left(\frac{R_2}{R}\right)F_{1\mu\nu ax}F_
	{1\mu\nu ax}=\frac{11}{96\pi^2}\ln\left(\frac{R_2}{R}\right)F_{1\mu\nu
	ax}F_{1\mu\nu ax}
\end{equation}
Thus the required sign change has indeed occurred.

\section{The sign reversal is a consequence of the sign of the 
$\beta$-function}

Now in fact the calculation we have just done precisely parallels a 
calculation of the leading $\beta$-function coefficient by the
``gauge-covariant background field method'' \cite{GCBFM}.  That method is
based on the general result that the effective action $\Gamma$, (i.e. the
generating functional of the proper vertices), is equal to minus the sum
of all the connected, one-line-irreducible vacuum bubbles, calculated with
an action that is obtained from the usual action $A(\phi)$, (where $\phi$
denotes al the fields we functionally integrate over), by replacing $\phi$
by $\Phi+\phi$, (where $\Phi$ represents all the ``classical'' fields, or
``background'' fields, and is the argument of $\Gamma$), and deleting the
term linear in $\phi$ in the power series expansion of $A(\Phi+\phi)$
about $\phi=0$.  In other words, $\Gamma{\Phi}$ is equal to minus the sum
of all the connected, one-line-irreducible vacuum bubbles, calculated with
the action:
\begin{equation}
	\label{E318}
	A(\Phi+\phi)-\phi_i\frac{\delta A(\Phi)}{\delta\Phi_i}
\end{equation}
We are here using the same schematic notation that we used in (\ref{E251})
and (\ref{E252}).  The ``free'' $\Phi$ propagator, in the presence of the
``background fields'' $\Phi$, is given by the matrix $N_{1ij}$ of 
(\ref{E251}).  The cubic vertex given by the action (\ref{E318}) is
$-\frac{\delta^3A(\Phi)}{\delta\Phi_i\delta\Phi_j\delta\Phi_k}$, the 
quartic vertex is $-\frac{\delta^4A(\Phi)}{\delta\Phi_i\delta\Phi_j\delta
\Phi_k\delta\Phi_l}$, and so on, just as in the development of the matrix
$N$, as illustrated in (\ref{E252}).  For the one-loop effective action,
the only term in (\ref{E318}) needed is the term:
\begin{equation}
	\label{E319}
	\frac{1}{2}\phi_i\phi_j\frac{\delta^2A(\Phi)}{\delta\Phi_i\delta\Phi_j}
\end{equation}
The one-loop effective action is half the trace of the logarithm of the
matrix $\frac{\delta^2A(\Phi)}{\delta\Phi_i\delta\Phi_j}$.  (The effective
action is the \emph{negative} of the sum of the vacuum bubbles due to our
conventions (\ref{E241}) and (\ref{E243}).)

In the ``gauge-covariant background field method'', the above general
results are distorted by modifying the gauge-fixing action so that the 
full quantum action, including the gauge-fixing and Fadeev-Popov terms,
has exactly the same gauge-invariance in terms of the ``classical'' gauge
field, say $\bar{A}_{\mu ax}$, as the original action had in terms of
$A_{\mu ax}$.  In other words, if $\bar{A}_{\mu ax}$ denotes the 
``classical'' gauge field, (that becomes an argument of $\Gamma$), and 
$A_{\mu ax}$ denotes the ``quantum'' gauge field, (that we functionally
integrate over), then where, in the gauge-fixing term, we would take
$\partial_\mu\left(\bar{A}_{\mu ax}+A_{\mu ax}\right)$ by the above
prescription, the ``gauge-covariant background field method'' instead 
takes $\partial_\mu A_{\mu ax}+\bar{A}_{\mu bx}f_{abc}A_{\mu cx}$.  The
result of this is that the ``gauge-covariant background field method''
calculates no the true effective action $\Gamma(\bar{A},\bar{B},
\bar{\phi},\bar{\psi})$, but rather a different quantity $\tilde{\Gamma}
(\bar{A})$.  This quantity $\tilde{\Gamma}(\bar{A})$ has exactly the same
gauge-invariance in terms of $\bar{A}_{\mu ax}$ as the original action has
in terms of $A_{\mu ax}$, whereas the true effective action $\Gamma(
\bar{A},\bar{B},\bar{\phi},\bar{\psi})$ is \emph{not} gauge-invariant, but
instead satisfies the Ward identity (\ref{E248}).  However the fact that
$\tilde{\Gamma}(\bar{A})$ has exactly the same gauge-invariance in terms 
of $\bar{A}_{\mu ax}$ as the original action has in terms of $A_{\mu ax}$,
means that, in our BPHZ approach, and with our conventions as in 
(\ref{E50}) and (\ref{E240}), the change in the one-loop counterterm for
$\tilde{\Gamma}(\bar{A})$, when we change the onset point of the smooth
long-distance cutoffs imposed on the propagators in the counterterm from
$R$ to $R_2$, is equal to $\frac{N}{4}$, times the coefficient of $g^4$ in
$2g\beta(g)$, times $\ln\left(\frac{R_2}{R}\right)$, times $\bar{F}_{\mu
\nu ax}\bar{F}_{\mu\nu ax}$, where
\begin{equation}
	\label{E320}
\bar{F}_{\mu\nu ax}=\partial_\mu
\bar{A}_{\nu ax}-\partial_\nu\bar{A}_{\mu ax}+f_{abc}\bar{A}_{\mu b}
\bar{A}_{\nu c}
\end{equation}
The relevance of this to the group-variation equations is that there is a
precise correspondence between expression (\ref{E12}), which in the 
context of the group-changing equations is the sum of all the terms in
$F_{\mu\nu a}F_{\mu\nu a}+F_{\mu\nu A}F_{\mu\nu A}$ that contain exactly
two $A_\mu$'s with upper-case group indices, and the expression:
\begin{equation}
	\label{E321}
	2\left(\left(\bar{D}_\mu A_\nu\right)_a\left(\bar{D}_\mu A_\nu\right)_a
	-\left(\bar{D}_\nu A_\mu\right)_a\left(\bar{D}_\mu A_\nu\right)_a+
	\bar{F}_{\mu\nu a}f_{abc}A_{\mu b}A_{\nu c}\right)
\end{equation}
where $\left(\bar{D}_\mu A_\nu\right)_a\equiv\partial_\mu A_{\nu a}+
\bar{A}_{\mu b}f_{abc}A_{\nu c}$, and $\bar{F}_{\mu\nu a}$ is defined by
(\ref{E320}), which in the context of the ``gauge-covariant background
field method'' is the sum of all the terms in 
$F\!\raisebox{-4pt}{$\stackrel{\displaystyle{(\bar{A}+A)}}
{\scriptstyle{\mu\nu a\qquad\ \ }}$}
F\!\raisebox{-4pt}{$\stackrel{\displaystyle{(\bar{A}+A)}}
{\scriptstyle{\mu\nu a\qquad\ \ }}$}$ that contain exactly two $A_{\mu a}$'s.
Indeed, if in (\ref{E12}) we first put a bar above all the $A_\mu$'s with
\emph{lower}-case group indices, (which occur in $\bar{D}_\mu$ by
(\ref{E5}) and in $\bar{F}_{\mu\nu a}$ by the definition after
(\ref{E11})), and then replace all upper-case group indices by lower-case
group indices, we precisely obtain (\ref{E321}).  Furthermore, the same
treatment applied to the third and fourth terms in the gauge-fixing action
(\ref{E6}) changes them precisely into the gauge-fixing terms for $A_{\mu
a}$ in the ``gauge-covariant background field method'' when a gauge-fixing
auxiliary field $B_a$ is used in fixing the gauge of $A_{\mu a}$ in that
method.  Furthermore, if we set $\beta=0$ in (\ref{E6}), as we of course
do in the context of the group-changing equations and group-variation
equations, then all the terms in the group-changing Fadeev-Popov action 
that contain upper-case indices and contribute to expectation values and
correlation functions that contain no explicit Fadeev-Popov fields or
gauge-fixing auxiliary fields, may be put in the form (\ref{E23}).  And if
we apply the same procedure as before to the first term in (\ref{E23}), 
namely putting a bar over the $A_{\mu a}$ that occurs in $\bar{D}_\mu$,
then replacing all upper-case indices by lower-case indices, we precisely
obtain the part of the Fadeev-Popov action in the ``gauge-covariant
background field method'' that is quadratic in the ``quantum fields''.

The consequence of this is that, as stated above, our calculation, through
quadratic order in the effective fields for one side of the 45-path, of 
the change in the counterterm for the one-loop islands, when we change the
onset point of the smooth long-distance cutoffs imposed on the propagators
in the counterterms from $R$ to $R_2$, precisely parallels the calculation
of the change in the one-loop counterterm for $\tilde{\Gamma}(\bar{A})$,
when we change the onset point of the smooth long-distance cutoffs imposed
on the propagators in the counterterms from $R$ to $R_2$.  In fact, to
obtain the corresponding change in the one-loop counterterm for 
$\tilde{\Gamma}(\bar{A})$ from (\ref{E317}), we multiply by $-1$, (since
the effective action is the \emph{negative} of the sum of the vacuum
bubbles), replace $F_{1\mu\nu ax}F_{1\mu\nu ax}$ by $\bar{F}_{\mu\nu ax}
\bar{F}_{\mu\nu ax}$, multiply by 2, (due to the two ``sides'' of an
adjoint representation path in terms of fundamental representation paths),
and multiply by $N$, (for the fundamental representation trace on the 
other ``side'' of the adjoint representation path.)  This indeed gives the
coefficient of $g^4$ in $2g\beta(g)$ in (\ref{E259}) by the relation 
stated above.  This corresponds to the fact that, to one-loop order, the
full action $S(A,B,\phi,\psi,g^2,\alpha)+C(A,B,\phi,\psi,g^2,\alpha,R)$,
where $S(A,B,\phi,\psi,g^2,\alpha)$ is the seed action (\ref{E240}), and
$C(A,B,\phi,\psi,g^2,\alpha,R)$ is the sum of all the counterterms, is
left unaltered, if we \emph{simultaneously} replace $R$ by $R_2$ and $g^2$
by $g_2^2$, where:
\begin{equation}
	\label{E322}
	\frac{N}{4g_2^2}+\frac{11N}{48\pi^2}\ln\left(\frac{R_2}{R}\right)=
	\frac{N}{4g^2}
\end{equation}
\emph{and} make the required finite rescalings of the fields, which are
not detected by the ``gauge-covariant background field method''.  We
immediately obtain (\ref{E259}) from (\ref{E322}) by use of (\ref{E258}).

Thus the fact that the required sign change occurred in (\ref{E317}) is
not an accident: it is an immediate corollary of the fact that 
$2g\beta(g)$ is negative.  And we see furthermore that, as noted before
(\ref{E292}), the fact that $2g\beta(g)$ is negative must, in the context
of the group-variation equations, be the first of a whole family of
results, which guarantee that the net contribution of the one-loop island
diagrams to the right-hand sides of the group-variation equations in their
standard forms (\ref{E184}), (\ref{E186}), etc., are negative under a
rather wide range of conditions.

\chapter{Verification And Correction Of The Ansatz, And Zeroth Order Value Of $m_{0^{++}}/\surd\sigma$, If The Pre-Exponential Factor In The Cylinder-Topology Term Is Non-Marginal}

We now leave further investigation of the signs of the island diagrams to
our next paper, and, \emph{assuming} they come out right, (i.e. negative),
consider in more detail the result of substituting our ansatz for 
$f_0(W_1,\ldots,W_n,g^2)$, when the sizes and separations of $W_1,\ldots,
W_n$ are all $\geq\frac{1}{\mu}$, as described after (\ref{E227}) above,
into the right-hand sides of the group-variation equations.  Our purpose 
is to determine whether our ansatz solves the group-variation equations
when the sizes and separations of $W_1,\ldots,W_n$ are all $\geq\frac{1}
{\mu}$.

\section{The ansatz satisfies the Group-Variation Equations for the
correlation functions}

We begin by recalling that, in the discussion between (\ref{E200}) and 
(\ref{E201}), we defined an island diagram to be a Type-1 island diagram
if, among all the windows beside paths of the island, exactly \emph{one}
of those windows is \emph{not} simply connected, and we defined all other
island diagrams to be Type-2 island diagrams.  We note that \emph{all} the
island diagrams in the right-hand side of the group-variation equation
(\ref{E184}) for $f_0(W_1,g^2)$ are Type-1 island diagrams, and we also
note that, among all the windows beside paths of the island in a Type-2
island diagram, at least two of those windows are \emph{not} simply
connected.

We recall from the start of this section that the principal effect of the
window weights on the right-hand side group-variation diagrams is that the
path sum for each 45-path gives approximately the free propagator for a
\emph{massive} scalar particle, and that , by the paragraph after 
(\ref{E200}), a conservative estimate of the effective mass is $\mu
\sqrt{\frac{\pi}{2}}=1.3\mu$, where $\mu^2$ is the coefficient of the area
in the Wilson area law in our ansatz.  We have to note, however, that this
estimate has taken no account of the details of the path sums for our
Landau-gauge vector bosons, specifically the $F_{\mu\nu}$ insertions and 
the $\frac{1}{E}$ term in $G_0$, which, as we know, have to reverse the 
sign of the one-loop island diagrams in comparison to what we would get
with a simple scalar boson, and which indeed do reverse the sign in the
example considered above.  We nevertheless assume that these effects do 
not reduce the effective mass of a 45-path below $1.3\mu$.

We begin by asking whether substituting our ansatz into the right-hand 
sides of the group-variation equations, correctly reproduces the factor
$\sqrt{\frac{m}{32\pi^3L^3}}e^{-mL}$ for each separate straight line of
length $L$ in our minimal-length spanning tree of $S_1,\ldots,S_p$, where
$S_1,\ldots,S_p$ are the separate connected components of our absolute
minimal-area orientable spanning surface of $W_1,\ldots,W_n$, and $m$ is
the mass of the lightest glueball, as in (\ref{E230}).  To investigate 
this, we consider loops $W_1,\ldots,W_n$, $n\geq2$, such that the diameter
of $W_1,\ldots,W_n$, (i.e. the maximum distance between any two points on
$W_1,\ldots,W_n$), is $R$, (where $R$ is the onset point of the smooth
long-distance cutoffs imposed on the propagators in our counterterms), and
such that the \emph{separations} between the loops $W_1,\ldots,W_n$ are
all much larger than any of the \emph{sizes} of the loops $W_1,\ldots,
W_n$.  We define $W_{1L},\ldots,W_{nL}$ to be the loops obtained from
$W_1,\ldots,W_n$ by uniformly scaling all the sizes and relative positions
of $W_1,\ldots,W_n$ by the scale factor $\frac{L}{R}$.  We now, as before,
(in (\ref{E232})), multiply both sides of the group-variation equation for
$f_0(W_{1L}\ldots,W_{nL},g^2)$ by $-\frac{2g\beta(g)}{g^2}$, after which,
by (\ref{E102}), (and, as usual in this paper, neglecting the terms in the
renormalization group equation (\ref{E102}) associated with the 
short-distance factors), the term $g^2\frac{\mathrm{d}}{\mathrm{d}g^2}f_0
(W_{1L},\ldots,W_{nL},g^2)$ in the left-hand side of the group-variation
equation for $f_0(W_{1L},\ldots,W_{nL},g^2)$ becomes simply:
\begin{equation}
\label{E323}
	L\frac{\partial}{\partial L}f_0(W_{1L},\ldots,W_{nL},g^2)
\end{equation}
Now by our assumption that the \emph{separations} between the loops 
$W_1,\ldots,W_n$ are all much larger than their \emph{sizes}, the absolute
minimal-area orientable spanning surface $S_L$ of $W_{1L},\ldots,W_{nL}$
will have a separate connected component $S_{iL}$ for each loop $W_{iL}$,
which wil be the minimal-area orientable spanning surface of $W_{iL}$, of
area $a_iL^2$, say.  Let the lengths of the separate straight lines
forming the minimal-length spanning tree of $S_{1L},\ldots,S_{nL}$ be
$b_1L,\ldots,b_qL$, where $(n-1)\leq q\leq(2n-3)$.  We assume that the
$a_i$'s have been chosen sufficiently small, and the separations between
the $W_i$'s sufficiently large, that there is a large range of $L$ where
our ansatz applies, and $(a_1+\ldots+a_n)L^2$ is small compared to
$(b_1+\ldots+b_q)L$.  Then in this range of $L$, the dominant term in the
left-hand side of the group-variation equation for $f_0(W_{1L},\ldots,
W_{nL},g^2)$, after multiplication by $-\frac{2\beta(g)}{g}$, comes from
(\ref{E322}) acting on the exponents $e^{-mb_jL}$ in (\ref{E230}), and 
gives:
\begin{equation}
	\label{E324}
	-m\left(b_1+\ldots+b_q\right)Lf_0(W_{1L},\ldots,W_{nL},g^2)
\end{equation}
Now, what diagrams will give the predominant contributions to the 
right-hand side of the group-variation equation for $f_0(W_{1L},\ldots,
W_{nL},g^2)$ in this situation?  We see immediately that the answer
depends on the actual value of the ratio $\frac{m}{\mu}$.  Indeed, if $m$
is less than twice the effective mass $1.3\mu$ of the 45-paths, then only
island diagrams can contribute significantly, while if $m$ is greater than
twice $1.3\mu$, then diagrams such as (\ref{E159}) and (\ref{E178}) will
give the largest contributions to the right-hand side.  But $m$
\emph{cannot} be strictly larger than twice the effective mass of the
45-paths, because if that were the case, then the left-hand side of the
group-variation equation for $f_0(W_{1L},\ldots,W_{nL},g^2)$ would have a
more rapid exponential fall-off with increasing $L$ than the leading terms
in the right-hand side, (unless the leading terms in the right-hand side
somehow cancelled, which there is no reason to expect, and which we assume
does \emph{not} occur).  Can $m$ be \emph{equal} to twice the effective
mass of the 45-paths, or, more precisely, equal to the lowest-mass state
of a cylinder with two 45-paths running along it?  If this were the case,
we would have to consider diagrams such as (\ref{E159}) and (\ref{E178}),
but we readily see that, if the spectrum of the cylinder with two 45-paths
along it is discrete, (as it must be), with lowest mass equal to $m$, then
these diagrams must have the same $L$-dependence as our ansatz, and hence
be smaller, by one power of $L$, than the leading term in the left-hand
sides of the relevant group-variation equations, as given by (\ref{E324}).
Thus, \emph{if} our ansatz satisfies the group-variation equations in this
region, then the leading terms in the right-hand sides must come from the
island diagrams, irrespective of whether $m$ is strictly less than, or 
equal to, the lowest-mass state of a cylinder with two 45-paths along it.
However, these two cases differ with respect to the particular types of
island configurations that will give the leading contributions: if $m$ is
strictly less than the lowest-mass state of a cylinder with two 45-paths
along it, then the dominant configurations will come from islands of size
$\frac{1}{\mu}$, with the contributions of larger islands being suppressed
exponentially, while if $m$ is equal to the lowest-mass state of a 
cylinder with two 45-paths along it, then important contributions will
also come from islands ``stretched'' along the spanning tree.

Let us first assume that $m$ is strictly less than the lowest-mass state
of a cylinder with two 45-paths along it, so that the dominant 
contributions will come from islands of $L$-independent size $\frac{1}
{\mu}$.  Which island diagrams will give the main contributions?  Let us
first consider the Type-1 island diagrams.  Now in a Type-1 island diagram
in the right-hand side of the group-variation equation for $f_0(W_{1L},
\ldots, W_{nL},g^2)$, there is \emph{one} window whose boundary has
$(n+1)$ connected components, namely $W_{1L},\ldots,W_{nL}$, and the outer
boundary of the island, and all the other windows are simply-connected,
and lie ``inside'' the island.  Let $\tilde{W}$ denote the outer boundary
of the island.  Then the window weight for the non-simply connected window
is $f_0(W_{1L},\ldots,W_{nL},\tilde{W},g^2)$.  Now, since we are assuming
that $m$ is strictly less than the lowest mass state of the cylinder with
two 45-paths along it, the predominant contributions will come from
$\tilde{W}$ with the $L$-independent size $\frac{1}{\mu}$.  Thus, provided
$\tilde{W}$ is not right next to one of the $W_{iL}$'s, the absolute
minimal-area orientable spanning surface $\tilde{S}$ of $W_{1L},\ldots,
W_{nL},\tilde{W}$, will have $(n+1)$ connected components, namely the 
absolute minimal-area orientable spanning surfaces of the individual loops
$W_{iL}$ and $\tilde{W}$.  Now clearly the minimum possible total length
of the straight line segments forming the minimal-length spanning tree of 
the $(n+1)$ connected components of $\tilde{S}$ will be attained if
$\tilde{W}$ is situated such that one of the straight line segments of the
minimal-length spanning tree of $S_{1L},\ldots,S_{nL}$, where $S_{iL}$ is
the absolute minimal-area orientable spanning surface of $W_{iL}$, passes
through the absolute minimal-area orientable spanning surface of 
$\tilde{W}$.  And clearly the main contributions to this island diagram
will come from configurations where $\tilde{W}$, with $L$-independent size
$\frac{1}{\mu}$, which is small compared to any of the lengths $b_jL$ of 
the straight-line segments of the minimal-length spanning tree $T_L$ of
$S_{1L},\ldots,S_{nL}$, is close to one of the straight-line segments of
$T_L$.  Now in any such configuration, if $z$ denotes any point of the
absolute minimal-area orientable spanning surface of $\tilde{W}$, and $x$
and $y$ denote the ends of the straight-line segment of $T_L$ that 
$\tilde{W}$ is close to, then the angle $\widehat{xzy}$ will be much 
larger than $\frac{2\pi}{3}$, and in fact approaching $\pi$, hence by the
discussion after (\ref{E219}), the minimal-length spanning tree of $x$, 
$y$, and $z$, will consist of the two straight lines $\overline{xz}$ and
$\overline{yz}$.  Thus the minimal-length spanning tree $\tilde{T}$ of the
$(n+1)$ connected components of $\tilde{S}$ will consist of all the 
straight segments of the minimal-length spanning tree $T_L$ of $S_{1L},
\ldots,S_{nL}$ \emph{except} the segment $\overline{xy}$, plus a straight
line from $x$ to some point $z_1$ on the absolute minimal-area orientable
spanning surface of $\tilde{W}$, plus a straight line from $y$ to some
point $z_2$ on the absolute minimal-area orientable spanning surface of
$\tilde{W}$.

Let us now, for simplicity, assume that $m$ is sufficiently smaller than
the lowest-mass state $M$ of the cylinder with two 45-paths along it,
(which we estimate roughly as twice the effective mass $1.3\mu$ of the
45-paths), that there is no significant tendency for the island to 
``stretch'' along the segment $\overline{xy}$ of the minimal-length
spanning tree of $S_{1L},\ldots,S_{nL}$ that it is close to.  Then we may,
to a good enough approximation, set $z_1=z_2=z$, where $z$ is the centre
of the island, i.e. the mean position of the vertices of the island.  (If,
on the other hand, $M-m$ was small compared to $\mu$, we would have to
allow the island to ``stretch'' to the $L$-independent length $\frac{1}
{M-m}$ along the segment $\overline{xy}$, with $z_1$ and the end nearer
$x$, and $z_2$ at the end nearer $y$, but for simplicity we assume that 
this is not the case.)  We now see that, according to our ansatz, as
described after (\ref{E229}), the contribution of this configuration to
this island diagram, is equal to our ansatz for $f_0(W_{1L},\ldots,W_{nL},
g^2)$, with the following changes: firstly, the factor $\sqrt{\frac{m}{32
\pi^3\left|x-y\right|^3}}e^{-m\left|x-y\right|}$ associated with the
straight line segment $\overline{xy}$ by point (ii) in our ansatz is 
replaced by the factor:
\begin{equation}
	\label{E325}
	\sqrt{\frac{m}{32\pi^3\left|x-z\right|^3}}e^{-m\left|x-z\right|}
	\sqrt{\frac{m}{32\pi^3\left|y-z\right|^3}}e^{-m\left|y-z\right|}
\end{equation}
secondly, by point (i) in our ansatz as applied to $f_0(W_{1L},\ldots,
W_{nL},g^2)$, we get a factor $e^{-\mu^2\tilde{\tilde{A}}}$, where
$\tilde{\tilde{A}}$ is the area of the absolute minimal-area orientable
spanning surface $\tilde{\tilde{S}}$ of $\tilde{W}$, thirdly, by point (i)
or our ansatz as applied, for each ``internal'' window $U_i$ of the
island, to $f_0(U_i,g^2)$, a factor $e^{-\mu^2B_i}$, where $B_i$ is the 
area of the absolute minimal-area orientable spanning surface of $U_i$, 
and fourthly, from point (iii) in our ansatz applied to the ends on
$\tilde{\tilde{S}}$ of the line segments $\overline{xz}$ and 
$\overline{yz}$, a factor $f^2$, where $f$ represents the coupling of the
lightest glueball to a minimal-area orientable spanning surface.

We now see that, due to our assumption, which we made for simplicity, that
$m$ is sufficiently smaller than the mass $M$ of the lowest-mass state of
the cylinder with two 45-paths along it, that there is no significant
tendency for the island to ``stretch'' along the segment $\overline{xy}$,
(in consequence of which we may, in this preliminary analysis, set $z_1=
z_2=z$, where $z$ is the mean position of the vertices of the island), if
we do the path integrals for all the 45-paths in the island, subject to 
the mean position $z$ of the vertices in all the 45-paths being fixed, 
then by the translation-invariance of the extra factors associated with
$\tilde{\tilde{S}}$ and the internal windows of the island, the 
\emph{only} $z$-dependence of the result is through the factor 
(\ref{E325}).

Now every Type-1 island is obtained from some connected, one-line 
irreducible vacuum bubble formed of 45-paths, that may be drawn on the
surface of the 2-sphere without any 45-paths crossing one another, by
cutting $n$ holes in \emph{one} of the windows of that vacuum bubble, and
stretching that window to form the non-simply connected window that
``surrounds'' the island.  Each such vacuum bubble gives rise to as many
Type-1 island diagrams as it has windows.  If the vacuum bubble has some
symmetries, it will have associated with it a symmetry factor given by the
reciprocal of the number of elements in the finite symmetry group under
which the bubble is invariant, and we will also find that different 
choices of the window we put the $n$ holes in lead to the same island 
diagram.  In that case, the factor given by the number of windows that give
the same island diagram, partly cancels the symmetry factor of the bubble,
and any symmetry factor that remains is associated with a rotational
symmetry of the island, as discussed in the paragraphs after (\ref{E104}).
For example, the one-loop vector boson vacuum bubble has a symmetry factor
$\frac{1}{2}$, due to the two-fold symmetry that rotates the two windows
of the bubble into one another.  This cancels the factor of 2 that arises
because we get the \emph{same} Type-1 island diagram irrespective of which
of the two windows we put the $n$ holes into.  (We did not include this
factor of 2 in our study above of the counterterm for the one-loop 
islands.  The Fadeev-Popov one-loop vacuum bubble has no symmetry factor,
and the two possible choices of which of the two windows we put the $n$
holes in, give two different diagrams, e.g. (\ref{E136}) and (\ref{E137}).
However the mathematical expressions corresponding to these two diagrams
are the same.  Thus the Fadeev-Popv contribution also needs to be 
multiplied by 2, hence the result (\ref{E317}) needs to be multiplied by
2.  This doesn't affect the correspondence with the ``covariant background
field method'', except that the extra factor of 2 due to the two 
fundamental representation ``sides'' of an adjoint representation phase
factor, now occurs in (\ref{E317}) as well.  Hence the counterterm-action
for the separate set of $\textrm{SU}(N)$ gauge fields and Fadeev-Popov fields we
introduce for each window, is the same as the standard $\textrm{SU}(N)$ BPHZ
counterterm action, as we should of course expect.)

We now define, for each such vacuum bubble $b$ drawn on the 2-sphere, 
$X_b$ to be the result of doing the path integrals over all the 45-paths 
of $b$, with a window weight $e^{-\mu^2B_i}$ for each window $i$ of $b$,
(where $B_i$ is the area of the absolute minimal-area orientable spanning
surface of the boundary of window $i$ of $b$), subject to the mean 
position of all the vertics in all the 45-paths of $b$, having the fixed
value $z$.  Of course, by translation invariance, $X_b$ is independent of
$z$.  $X_b$ \emph{includes} any symmetry factor for $b$, (for example, for
the vector-boson one-loop vacuum bubble, $X_b$ includes the symmetry
factor $\frac{1}{2}$.)  $X_b$ also \emph{includes} the factor 
$\left.\frac{\mathrm{d}}{\mathrm{d}M}\mathbf{C}(M)\right|_{M=1}$, where
$\mathbf{C}(M)$, the chromatic polynomial of the bubble $b$, is by 
definition the number of distinct ways of colouring the windows of $b$ 
with $M$ colours available, subject to the constraint that if two windows
share a common ``boundary'', (i.e. a common 45-path), they are to be
coloured in two \emph{different} colours, and also to the special rules,
discussed in connection with (\ref{E106}) - (\ref{E110}), for colouring
windows that meet at a quartic vertex that has two Fadeev-Popov ``legs''.
Now this $\mathbf{C}(M)$ is completely unaffected by making holes in the
windows of $b$, hence this factor $\left.\frac{\mathrm{d}}{\mathrm{d}M}
\mathbf{C}(M)\right|_{M=1}$ is identical to the corresponding factor
$\left.\frac{\mathrm{d}}{\mathrm{d}M}\mathbf{C}(M)\right|_{M=1}$ for
\emph{every} island diagram obtained from $b$ by making holes in its 
windows.

We also define $n_b$ to be the number of windows of $b$.

We thus find immediately that, within the approximation that we can set
$z_1=z_2=z$, the contribution of all the Type-1 island diagrams, when we
substitute our ansatz into the right-hand side of the group-variation
equation for $f_0(W_{1L},\ldots,W_{nL},g^2)$, is equal to
	\[f^2\left(\sum_bn_bX_b\right)f_0(W_{1L},\ldots,W_{nL},g^2)
\]
times the sum, over all the straight-line segments in the minimal-length
spanning tree $T_L$ of $S_{1L},\ldots,S_{nL}$, of the integral over $z$ of
$(\ref{E325})$, where $x$ and $y$ are interpreted as the ends of the 
straight-line segment of $T_L$ concerned, divided by the factor 
$\sqrt{\frac{m}{32\pi^3\left|x-y\right|^3}}e^{-m\left|x-y\right|}$ that 
our ansatz for $f_0(W_{1L},\ldots,W_{nL},g^2)$ includes for the segment
$\overline{xy}$.  Now the integral over $z$ of (\ref{E325}) is dominated
by the zone where $z$ is close to the straight line $\overline{xy}$.  We
choose a coordinate system with origin at $x$, and one axis pointing from
$x$ to $y$, and define $u$ to be the component of $z$ in this direction, 
and $v$ to be the three-vector comprising the other three components of 
$z$.  Then in the exponent we may expand:
\begin{equation}
	\label{E326}
	m\left|x-z\right|+m\left|y-z\right|=m\left|x-y\right|+\frac{\left|x-y
	\right|v^2}{2u(\left|x-y\right|-u)}+\textrm{ order }(\left|v\right|^4)
\end{equation}
while in the pre-exponential factors we may approximate $\left|x-z\right|$
as $u$ and $\left|y-z\right|$ as $(\left|x-y\right|-u)$.  The Gaussian
integral over $v$ then cancels all $u$-dependence in the pre-exponential
factors, so the $u$-integral then simply gives the factor 
$\left|x-y\right|$.  Thus for the $z$-integral of (\ref{E325}) we get:
\begin{equation}
	\label{E327}
	\frac{\left|x-y\right|}{2m}\sqrt{\frac{m}{32\pi^3\left|x-y\right|^3}}
	e^{-m\left|x-y\right|}
\end{equation}
which is simply $\frac{\left|x-y\right|}{2m}$ times the factor that our
ansatz for $f_0(W_{1L},\ldots,W_{nL},g^2)$ includes for the segment
$\overline{xy}$.  (As a check, we note that, if we were calculating the
$z$-integral of (\ref{E325}), times a mass counterterm $-2m(\delta m)$,
then by (\ref{E327}) we would find correctly, that for large
$\left|x-y\right|$, and to first order in $\delta m$, we get the 
correction to the large $\left|x-y\right|$ asymptotic form (\ref{E228}) of
the free massive scalar propagator, for the mass shift $m\to m+(\delta 
m)$.)

We thus see that, remembering that we defined the lengths of the separate
straight lines forming the minimal-length spanning tree $T_L$ of
$S_{1L},\ldots,S_{nL}$ to be $b_{1L},\ldots,b_{qL}$, where $(n-1)\leq q
\leq(2n-3)$, and within the approximation that we can set $z_1=z_2=z$, the
contribution of all the Type-1 island diagrams, when we substitute our
ansatz into the right-hand side of the group-variation equation for
$f_0(W_{1L},\ldots,W_{nL},g^2)$, and multiply both sides of the group
variation equation by $-\frac{2\beta(g)}{g}$, is:
\begin{equation}
	\label{E328}
	-\frac{2\beta{g}}{g}\frac{\left(b_1+\ldots+b_q\right)L}{2m}f^2\left(
	\sum_bn_bX_b\right)f_0(W_{1L},\ldots,W_{nL},g^2)
\end{equation}
which has exactly the same functional form as (\ref{E324}).

Now what about the contributions of the other island diagrams?  We now
\emph{change} our terminology for non-Type-1 island diagrams.  Whereas
before, we called \emph{all} non-Type-1 island diagrams, ``Type-2'', we
now define, for all integers $c\geq1$, a ``Type-$c$ island diagram'' to
be,
precisely, an island diagram obtained from one of our connected, one-line
irreducible vacuum bubbles, formed of 45-paths, and drawn on the 2-sphere
with no lines crossing, as above, by making one or more holes in each of
$c$ different windows of the vacuum bubble.  Thus Type-1 island diagrams
are as before, but in general, for each $c\geq1$, a Type-$c$ island 
diagram is an island diagram where precisely $c$ of the windows of the
diagram are \emph{not} simply connected.  (We recall that, as discussed
after examples (\ref{E111}) - (\ref{E183}), no contributing island 
diagram, i.e. no island diagram for which $\left.\frac{\mathrm{d}}
{\mathrm{d}M}\mathbf{C}(M)\right|_{M=1}$ is nonzero, has any 45-paths that
do \emph{not} form part of the island.)  The examples of 
\emph{contributing} Type-2 island diagrams are now (\ref{E165}), 
(\ref{E181}), and (\ref{E182}), and the simplest example of a Type-3
island diagram is now (\ref{E183}).  We note that a Type-$c$ island 
diagram requires making at least $c$ holes in the windows of the vacuum
bubble, hence a Type-$c$ island diagram can only occur for $f_0(W_{1L},
\ldots,W_{nL},g^2)$ such that $n\geq c$, and furthermore, that a Type-$c$
island diagram of course has at least $c$ windows.

Now of course we are at present considering $f_0(W_{1L},\ldots,W_{nL},
g^2)$ such that $n\geq2$.  Let us first consider the contributions of 
Type-2 island diagrams to the right-hand side of the group-variation
equation for $f_0(W_{1L},\ldots,W_{nL},g^2)$.  Now a Type-2 island diagram
has holes in precisely two of the windows of the corresponding vacuum
bubble, and thus partitions the $W_{iL}$'s into two nonempty sets, 
corresponding to which of the $W_{iL}$'s occur in which of the two 
relevant windows of the vacuum bubble.  Now a partition of the external
states of a process into two nonempty sets defines a channel for that
process, and if that process is described by a particular tree diagram,
(or, in our case, our minimal-length spanning tree $T_L$), each given 
channel may or may not correspond to a propagator of that tree diagram.
Now, with the same assumptions as we made in studying the contributions of
the Type-1 islands, namely, that $m$ is strictly smaller than the mass $M$
of the lowest-mass state of the cylinder with two 45-paths along it, and
furthermore, (for simplicity), that $m$ is sufficiently smaller than $M$
that there is no significant tendency for islands to elongate along
segments of $T_L$, we readily see that, since all the individual
straight-line segments of $T_L$ are, by assumption, much longer than the
predominant island size $\frac{1}{\mu}$, if the partition of $\{W_{1L},
\ldots,W_{nL}\}$ defined by a given Type-2 island diagram corresponds to 
one of the straight-line segments of $T_L$, (in the sense that cutting
that segment of $T_L$ partitions $\{W_{1L},\ldots,W_{nL}\}$ into the same
two nonempty parts as defined by that Type-2 island diagram), then that
Type-2 island diagram \emph{does} give a significant contribution, namely
from configurations where the island ``slides up and down'' that 
particular straight line segment of $T_L$, while if the partition of
$\{W_{1L},\ldots,W_{nL}\}$ defined by that Type-2 island diagram does
\emph{not} correspond to any of the straight-line segments of $T_L$ in
that sense, then that Type-2 island diagram does \emph{not} give a
significant contribution, because no matter what configuration the island
is in, either the island is stretched to a size large compared to 
$\frac{1}{\mu}$, or else the relevant minimal-length spanning trees have
a total length (of all their segments) significantly greater than the 
total length of $T_L$.  Thus we can associate each Type-2 island diagram
that gives a significant contribution, with a particular segment of $T_L$,
namely the segment of $T_L$ that defines the same partition of $\{W_{1L},
\ldots,W_{nL}\}$ as that Type-2 island diagram.

Now, as we know, each Type-2 island diagram is obtained from a 
corresponding vacuum bubble by making the appropriate holes in the 
appropriate two windows of that vacuum bubble.  We readily see, in a
manner precisely analogous to what we found for the Type-1 islands, that 
if a Type-2 island diagram defines a partition of $\{W_{1L},\ldots,W_{nL}
\}$ that matches that defined by a segment of $T_L$ of length $b_jL$, then
after multiplying by $-\frac{2\beta(g)}{g}$, its contribution to the
right-hand side of the group-variation equation for $f_0(W_{1L},\ldots,
W_{nL},g^2)$ is:
\begin{equation}
	\label{E329}
	-\frac{2\beta(g)}{g}\frac{b_jL}{2m}f^2X_bf_0(W_{1L},\ldots,W{nL},g^2)
\end{equation}
where the definition of $X_b$ is exactly as before.  We note, however,
that if, due to symmetries of the vacuum bubble $b$, there are several
distinct ways of obtaining the given Type-2 island diagram by putting the
appropriate holes in two distinct windows of $b$, then (\ref{E329}) just
gives the contribution of \emph{one} of those ways to the contribution of
that Type-2 island diagram.  Now for a given straight-line segment of
$T_L$, (and the corresponding partition of $\{W_{1L},\ldots,W_{nL}\}$ into
two nonempty parts), and a given one of our vacuum bubbles $b$, there is
a total of $n_b(n_b-1)$ such contributions (\ref{E329}) to the right-hand
side of the group-variation equation for $f_0(W_{1L},\ldots,W_{nL},g^2)$,
corresponding to the $n_b(n_b-1)$ different ways of assigning the two 
parts into which $\{W_{1L},\ldots,W_{nL}\}$ is partitioned, to two 
distinct windows of $b$, (where $n_b$ denotes, as before, the number of 
windowsw of $b$).  We note that (\ref{E329}) has no dependence on the 
particular segment of $T_L$ concerned, other than through the length
$b_jL$ of that segment.  In particular, (\ref{E329}) does \emph{not}
depend on the particular partition of $\{W_{1L},\ldots,W_{nL}\}$ defined
by that segment.  Thus we see that the total effect of adding, to the
contributions (\ref{E328}) of all the Type-1 island diagrams, the 
contributions of all the Type-2 island diagrams, is simply to replace
$n_b$ by $n_b^2$ in (\ref{E328}).  In other words, the total contribution
of all the Type-1 island diagrams and all the Type-2 island diagrams, to
the right-hand side of the group-variation equation for $f_0(W_{1L},
\ldots,W_{nL},g^2)$, is:
\begin{equation}
	\label{E330}
	-\frac{2\beta(g)}{g}\frac{\left(b_1+\ldots+b_q\right)L}{2m}f^2\left(
	\sum_bn_b^2X_b\right)f_0(W_{1L},\ldots,W_{nL},g^2)
\end{equation}

Furthermore, we readily see that \emph{no} island diagram of Type-3, or 
any higher type, makes any significant contribution to the right-hand side
of the group-variation equation for $f_0(W_{1L},\ldots,W_{nL},g^2)$, for 
no matter what configuration the isand is in, either the island is 
stretched to a size large compared to $\frac{1}{\mu}$, or the 
minimal-length spanning trees concerned have a total length significantly
larger than the total length of $T_L$, (i.e. the total length of the
segments of $T_L$), or else the island is ``anchored'' at one of the
vertices of $T_L$ where three straight segments of $T_L$ meet.  (This 
third possibility arises for a Type-3 island diagram when the partition of
$\{W_{1L},\ldots,W_{nL}\}$ into three nonempty parts, as defined by that
Type-3 island diagram, matches the partition of $\{W_{1L},\ldots, W_{nL}
\}$ into three nonempty parts defined by removing that vertex from $T_L$.
In this case, the island can be small, and the total length of the 
spanning trees involved be no longer than the total length of $T_L$, 
provided the island is close to that vertex.  However, since, as shown in
connection with (\ref{E204}) - (\ref{E208}) above, the angles between the
three segments meeting at that vertex will all be $\frac{2\pi}{3}$, it is
impossible for the island to move away from that vertex without either
stretching to a size significantly greater than $\frac{1}{\mu}$, or the
total length of the spanning trees involved becoming significantly greater
than the total length of $T_L$.  Thus these cases cannot produce the extra
factor of $L$ that occurs in (\ref{E324}) and (\ref{E330}).)\footnote{Note
added: example (\ref{E183}) is a simple case in which this situation
occurs.}

Thus, since the non-island diagrams also do not make any significant
contributions to the right-hand side of the group-variation equation for
$f_0(W_{1L},\ldots,W_{nL},g^2)$ in the region we are investigating, (i.e.
no contributions with the extra power of $L$ that occurs in (\ref{E324})
and (\ref{E330})), we wee that, subject to ur assumption that the mass
$m$ of the lightest glueball is strictly less than the mass $M$ of the
lightest state of the cylinder with two 45-paths along it, and our further
assumption, made for simplicity, that $m$ is sufficiently smaller than
$M$ that there is no significant tendency for islands to elongate along
the straight-line segments of $T_L$, the result of substituting our ansatz
into the right-hand side of the group-variation equation for $f_0(W_{1L},
\ldots,W_{nL},g^2)$, when the separations of $W_{1L},\ldots,W_{nL}$ are
all large compared to $\frac{1}{\mu}$, and $m$ times the total length of 
all the straight-line segments in $T_L$ is large compared to $\mu^2$ times
the sum of the areas of the absolute minimal-area spanning surfaces 
$S_{1L},\ldots,S_{nL}$, of $W_{1L},\ldots,W_{nL}$, is given by 
(\ref{E330}).  And comparing (\ref{E330}) with (\ref{E324}), we see that
our ansatz does indeed give a solution of the group-variation equations in
this domain, and that, furthermore, the mass $m$ of the lightest glueball
is given by:
\begin{equation}
	\label{E331}
	m^2=\frac{\beta(g)}{g}f^2\left(\sum_bn_b^2X_b\right)
\end{equation}
Here $g^2$ is of course to be set equal to the critical value that it 
takes throughout the domain where our ansatz applies, as determined by
(\ref{E236}), or more precisely, by the corrected version of (\ref{E236})
to be given below, and as discussed after (\ref{E261}).  And of course,
since $\frac{\beta(g)}{g}$ is negative, $\displaystyle\sum_bn_b^2X_b$ must
be negative, and, indeed, there is no doubt that the sum $X_b$ for the
vector-boson and Fadeev-Popov one-loop vacuum bubbles, (which have $n_b
=2$), must be negative.

We next note that if, instead of assuming that the mass $m$ of the 
lightest glueball is strictly less than the mass $M$ of the lowest-mass
state of the cylinder with two 45-paths along it, we had assumed that $m$
is \emph{equal} to $M$, we would have found that our ansatz did 
\emph{not} solve the group-variation equations.  This can be seen in the
example of the group-variation equation for $f_0(W_{1L},W_{2L},g^2)$, in
the domain where the separation, approximately $L$, between $W_{1L}$ and
$W_{2L}$ is large compared to $\frac{1}{\mu}$, and $mL$ is large compared
to $\mu^2$ times the sum of the areas of the absolute minimal-area
spanning surfaces of $W_{1L}$ and $W_{2L}$.  In this domain the assumption
that $m=M$ means that there is no exponential suppression of the island 
size along the straight line between $W_{1L}$ and $W_{2L}$: we have to
consider configurations of the one-loop islands (\ref{E163}) and 
(\ref{E165}) where the islands are ``stretched'' to lengths of order $L$
along the straight line between $W_{1L}$ and $W_{2L}$.  And for the
higher-loop islands we have to consider configurations where the vertices
of the island are divided into two nonempty groups, with the two groups
being connected to one another by just two 45-paths, and stretching the
island such that one of these two groups moves towards $W_{1L}$ and the 
other moves towards $W_{2L}$.  And we also have to consider possible
``stretchings'' of the higher-loop islands in two or more ``two-path
connected'' regions.  But such ``stretched'' islands produce \emph{too
many} extra powers of $L$ in the right-hand side of the group-variation
equation for $f_0(W_{1L},W_{2L},g^2)$ to match the single extra power of
$L$ in the left-hand side, as given by (\ref{E324}).  Indeed, considering
just the one-loop islands (\ref{E163}) and (\ref{E165}), (it doesn't 
matter which), if we consider summing over all 45-paths of a stretched
island configuration such that the two ``ends'' of that stretched island
are fixed at $z_1$ and $z_2$, where $z_1$ and $z_2$ are close to the
straight line between $W_{1L}$ and $W_{2L}$, but may be separated from one
another by a distance of order $L$, then the assumption that $M=m$ implies
that the result of summing over those 45-paths, subject to $z_1$ and 
$z_2$ fixed, is a constant numerical multiple of $\sqrt{\frac{m}{32\pi^3
\left|z_1-z_2\right|^3}}e^{-m\left|z_1-z_2\right|}$.  And if $z_1$ is
the end of the island nearer $W_{1L}$, $z_2$ is the end of the island
nearer $W_{2L}$, and $x$ and $y$ are the ends, on the absolute 
minimal-area orientable spanning surfaces of $W_{1L}$ and $W_{2L}$
respectively, of the shortest straight-line segments between those two
spanning surfaces, then our ansatz gives a factor $\sqrt{\frac{m}{32\pi^3
\left|x-z_1\right|^3}}e^{-m\left|x-z_1\right|}$ for the domain between
$W_{1L}$ and $z_1$, and a factor $\sqrt{\frac{m}{32\pi^3\left|z_2-y
\right|^3}}e^{-m\left|z_2-y\right|}$ for the domain between $z_2$ and
$W_{2L}$.  We then find, doing the $z_1$ and $z_2$ integrals in exactly
the same way as before, (i.e. to arrive at (\ref{E327})), we get
\emph{two} extra powers of $L$, (essentially one extra power of $L$ for
each end of the island).  And when we consider higher-loop island diagrams
with several ``stretching'' regions, we get even more extra powers of $L$.
This does \emph{not} match (\ref{E324}), and there is certainly no reason
to expect that all the trms with the extra powers of $L$ would exactly
cancel one another.  We conclude that the mass $m$ of the lightest 
glueball must be \emph{strictly} less than the mass $M$ of the lowest-mass
state of the cylinder with two 45-paths along it.

\section{The area-law domain}

Having now seen that our ansatz does indeed satisfy the group-variation
equations in the ``glueball'' domain of widely separated loops, we return
to the consideration of the ``area law'' domain.  We considered this
domain qualitatively before, obtaining equation (\ref{E236}), but, as we
mentioned in the paragraph after (\ref{E236}), that qualitative estimate
was in fact not quite right.  We consider a loop $W_1$, such that the
largest distance between any two points of $W_1$, (i.e. the ``diameter''
of $W_1$), is $R$, where $R$ is the onset point of the smooth 
long-distance cutoffs we impose on the propagators in our BPHZ 
counterterms, and define $W_{1L}$ to be the loop obtained from $W_1$ by
scaling $W_1$ by the factor $\frac{L}{R}$.  Thus $W_{1L}$ is identical in
shape to $W_1$, but differs in size, having the diameter $L$.  We define 
the area of the absolute minimal-area orientable spanning surface of $W_1$
to be $a_1R^2$, so that the area of the absolute minimal-area orientable
spanning surface of $W_{1L}$ is $a_1L^2$.  we consider the 
group-variation equation for $f_0(W_{1L},g^2)$, and note that, exactly as
before, (equation (\ref{E234})), when we substitute our ansatz into the
left-hand side of the group-variation equation for $f_0(W_{1L},g^2)$, and
multiply by $-\frac{2\beta(g)}{g}$, we get $-2\mu^2a_1L^2f_0(W_{1L},g^2)$.
Furthermore, exactly as before, (in the discussion before (\ref{E231})),
we may expect, for large $L$, the contribution of each \emph{island}
diagram to the right-hand side of the group-variation equation for $f_0(
W_{1L},g^2)$, to behave \emph{roughly} as $L^2f_0(W_{1L},g^2)$, while the
contribution of each \emph{non}-island diagram will behave only as $Lf_0(
W_{1L},g^2)$.  Thus, for large $L$, we only need to consider the 
contributions of the \emph{island} diagrams.  (Of course, only
\emph{Type-1} island diagrams occur in the right-hand side of the 
group-variation equation for $f_0(W_{1L},g^2)$.)

We consider an island diagram contributing to the right-hand side of the
group-variation equation for $f_0(W_{1L},g^2)$, and define $W_2$ to be the
outer boundary of the island.  Then the window weight for the non-simply
connected window that ``surrounds'' the island is given by $f_0(W_{1L},
W_2,g^2)$.  Now due to the effective mass $1.3\mu$ of the 45-paths, the
predominant contributions for this island diagram will be given, for large
$L$, by islands of size $\frac{1}{\mu}$, hence the predominant 
contributions will come from configurations where the diameter of $W_2$ is
roughly $\frac{1}{\mu}$.  Now according to our ansatz, as stated after
(\ref{E229}), the functional form of $f_0(W_{1L},W_2,g^2)$ depends on
whether the absolute minimal-area orientable spanning surface $S$ of 
$W_{1L}$ and $W_2$ has cylinder topology, (i.e. a sphere with two holes
in), or consists of the separate absolute minimal-area orientable spanning
surfaces $S_{1L}$ and $S_2$ of $W_{1L}$ and $W_2$.  Now clearly the 
configurations of $W_2$ which give the largest values of $f_0(W_{1L},W_2,
g^2)$, are those where $W_2$ is a simple loop lying within the 
minimal-area orientable spanning surface $S_{1L}$ of $W_{1L}$, and 
oriented consistently with the orientation of $S_{1L}$ defined by 
$W_{1L}$, (i.e. such that if $W_{1L}$ were shrunk until it coincided with
$W_2$, then the arrows on the two loops would point in \emph{opposite}
directions), for the area $A$ of $S$ is then equal to $A_{1L}-A_2$, where
$A_{1L}$ is the area of $S_{1L}$, and $A_2$ is the area of $S_2$.  Thus 
for a first quantitative estimate we shall assume that the predominant
contributions come from $W_2$ such that $S$ has cylinder topology, and 
then check the validity of this assumption after the calculation.

To examine a specific example, we choose 4-dimensional Cartesian 
coordinates $(w,x,y,z)$, and assume that $W_{1L}$ is the circle of 
diameter $L$ given parametrically by $\frac{L}{2}(\cos\theta,\sin\theta,
0,0)$, $0\leq\theta\leq2\pi$, and $W_2$ is the circle of diameter 
$\frac{1}{\mu}$ given parametrically by $(w,x,y,z)+\frac{1}{2\mu}(\cos
\theta,-\sin\theta,0,0)$, $0\leq\theta\leq2\pi$.  We assume that 
$\sqrt{w^2+x^2}+\frac{1}{2\mu}$ is less than $\frac{L}{2}$, and that $y$
and $z$ are sufficiently small that the area $A$ of the absolute 
minimal-area orientable spanning surface $S$ of $W_{1L}$ and $W_2$ is
given to a good approximation by the appropriate solution of Laplace's
equation in the circle bordered by $W_{1L}$.  (We will also check the 
validity of this assumption afterwards.)  We then find immediately, from
the appropriate solution of Laplace's equation, that in the limit of large
$\mu L$, and without any restriction on $w$ and $x$, other than that given
above, the area $A$ of $S$ is given by:
\begin{equation}
	\label{E332}
	A=\frac{\pi}{4}\left(L^2-\frac{1}{\mu^2}\right)+\frac{\pi\left(y^2+z^2
	\right)}{\ln\left(\mu L\left(1-\frac{4\left(w^2+x^2\right)}{L^2}\right)
	\right)}
\end{equation}
Now, with $W_{1L}$ still being our circle of diameter $L$, but $W_2$ now
unrestricted, we define $J(w,x,y,z)$ to be the contribution to our island
diagram from all configurations of the island such that the mean position
of all the vertices of the 45-paths of the island is equal to $(w,x,y,z)$.
Thus the contribution from this island diagram to the right-hand side of 
the group-variation equation for $f_0(W_{1L},g^2)$ is $\int\int\int\int
\mathrm{d}w\mathrm{d}x\mathrm{d}y\mathrm{d}zJ(w,x,y,z)$.  Now of course
$J(w,x,y,z)$ is largest when $(w,x,y,z)$ lies in, or close to, $S_{1L}$,
i.e. the circular disc of diameter $L$ that fills $W_{1L}$.  Let us 
suppose that $(w,x,y,z)$ lies in, or close to, $S_{1L}$.  Then 
$J(w,x,y,z)$ will certainly get contributions from island whose outer
boundary $W_2$ is such that the absolute minimal-area orientable spanning
surface $S$ of $W_{1L}$ and $W_2$ does \emph{not} have cylinder topology,
but consists instead of $S_{1L}$ and the absolute minimal-area orientable
spanning surface $S_2$ of $W_2$.  For example, $W_2$ could be the same
circle we considered before, but with its arrow pointing in the other
direction, i.e. with its parameter $\theta$ replaced by $-\theta$.
However the contributions of such island configurations will be 
suppressed, due to $S$ having larger area, relative to the contributions
of island configurations such that the projection onto $S_{1L}$ of the
outer boundary $W_2$ of the island, forms a simple loop in $S_{1L}$, (i.e.
a loop without any self-intersections), oriented consistently with 
$S_{1L}$, and such that, for \emph{all} points on $W_2$, the perpendicular
distance from that point to $S_{1L}$ is small.  In fact, the predominant
contributions to $J(w,x,y,z)$ will come from island configurations such 
that the outer boundary $W_2$ of the island has the properties just 
described, and furthermore the projections onto $S_{1L}$ of all the 
remaining 45-paths of the island, (if the island has two or more loops),
all lie within the simple loop formed by the projection of $W_2$ onto
$S_{1L}$, and intersect neither one another nor themselves, and 
furthermore, for \emph{every} point on \emph{every} 45-path of the island,
the perpendicular distance from that point to $S_{1L}$ is small.  Now for
such island configurations, the absolute minimal-area orientable spanning
surfaces of \emph{every} window of the diagram, are given to a good
approximation by the appropriate solutions of Laplace's equation, in the
zones of $S_{1L}$ delineated by the projections onto $S_{1L}$ of the 
45-paths of the island.  Thus the total of the areas of the absolute
minimal-area orientable spanning surfaces of the windows of the diagram,
is equal to $\frac{\pi L^2}{4}$, plus corrections quadratic in the 
perpendicular distances of the points of the 45-paths of the island from
$S_{1L}$.  Furthermore, due to the effective mass $1.3\mu$ of the 
45-paths, the predominant contributions come from islands of size roughly
$\frac{1}{\mu}$, and by assumption, $L$ is large compared to $\frac{1}
{\mu}$.  Furthermore, due to the fact that we are seeking the contribution
of this island diagram to the right-hand side of the group-variation
equation for $f_0(W_{1L},g^2)$ at large $L$, and this contribution behaves
\emph{roughly} as $L^2f_0(W_{1L},g^2)$ at large $L$, we may exclude the
contributions of islands whose projection onto $S_{1L}$ lies within a band
of fixed width $B$ at the edge of $S_{1L}$, where $B$ is large compared to
$\frac{1}{\mu}$, but independent of $L$ as $L$ becomes large, since the
contribution of the excluded island configurations behaves only as
$Lf_0(W_{1L},g^2)$ as $L$ becomes large.  Then, due to the facts that our
islands are small, and their projections onto $S_{1L}$ are not too close
to the edge of $S_{1L}$, the only dependence, on the coordinates $w$ and
$x$ of the mean position $(w,x,y,z)$ of the vertices of the 45-paths of
the island, of the coefficients of the terms in the window areas quadratic
in the perpendicular distances of the points of the 45-paths of the island
from $S_{1L}$, is in the coefficients of the terms quadratic in $y$ and
$z$ in the area of the absolute minimal-area orientable spanning surface
$S$ of the non-simply connected window, bordered by $W_{1L}$ and $W_2$,
that ``surrounds'' the island.  In fact, the \emph{only} dependence on
$w$, $x$, $y$, and $z$, of the window areas, is to a good approximation
given by the terms quadratic in $y$ and $z$ in the area (\ref{E332}) of
$A$.  Indeed, the areas of the minimal-area spanning surfaces of the
\emph{internal} windows of the island, depend only on the 45-paths 
bordering the windows concerned, and are completely independent of $w$,
$x$, $y$, and $z$.  For the non-simply connected window, we note that 
Laplace's equation is linear, and that if we choose polar coordinates
$(r,\theta)$ in the plane of $W_{1L}$, with origin at some point
\emph{strictly} in the interior of the projection of the island onto
$S_{1L}$, (so the origin of the polar coordinates will \emph{not} in 
general coincide with the origin of the Cartesian coordinates $(w,x)$),
then the general solution of Laplace's equation in the zone of $S_{1L}$
outside the projection of the island onto $S_{1L}$, is given by:
\begin{equation}
	\label{E333}
	a+b\ln r+\sum_{n=1}^\infty c_nr^n\cos(n(\theta-\gamma_n))
	+\sum_{n=1}^\infty d_n
	r^{-n}\cos(n(\theta-\delta_n))
\end{equation}
where $a$, $b$, the $c_n$, $(n\geq1)$, and the $d_n$, $(n\geq1)$, are 
constant coefficients, and the $\gamma_n$ and the $\delta_n$, $(n\geq1)$,
are constant angles, chosen to satisfy the boundary conditions.  (Each of
the constant coefficients is actually a 2-vector corresponding to the
$4-2=2$ dimensions perpendicular to $S_{1L}$, and each of the constant
angles also has an undisplayed index indicating which of the two
dimensions
perpendicular to $S_{1L}$ it refers to, but we can treat each of the
two dimensions perpendicular to $S_{1L}$ independently of the other one.)
We define the $\theta=0$ direction for our polar coordinates to be the
direction opposite to the direction given by an arrow pointing from the
origin of our polar coordinates to the centre of $W_{1L}$.  We assume that
the origin of our polar coordinates is at the projection onto $S_{1L}$ of
the mean position of the vertices of the island, and thus has the 
cartesian coordinates $(w,x)$, and we define $s\equiv\sqrt{w^2+x^2}$.
Then if, for example, $W_2$ is the circle specified before (\ref{E332}),
then in the limit of large $\mu L$, and with no restrictions on $w$ and
$x$ other than that $s+\frac{1}{2\mu}$ be less than $\frac{L}{2}$, which
ensures that the projection of $W_2$ onto the plane of $W_{1L}$ lies
within $S_{1L}$, we find that, (remembering that the coefficients are
2-vectors, and that the constant angles also have an undisplayed index):
	\[b=\frac{-(y,z)}{\ln\left(\mu L\left(1-\frac{4s^2}{L^2}\right)\right)},
	\qquad\qquad a=-b\ln\left(\frac{L}{2}\left(1-\frac{4s^2}{L^2}\right)
	\right)
\]
	\[c_n=\frac{b}{n}\left(\frac{4s}{L^2-4s^2}\right)^n,\ (n\geq1),\qquad
	\qquad d_n=0,\ (n\geq1)
\]
\begin{equation}
	\label{E334}
	\gamma_n=\delta_n=0,\ (n\geq1)
\end{equation}
and (\ref{E333}) then becomes:
\begin{equation}
	\label{E335}
	b\left(\ln\left(\frac{2r}{L}\right)-\frac{1}{2}\ln\left(\left(1-\frac{4
	s^2}{L^2}\right)^2+\left(\frac{4rs}{L^2}\right)^2-\frac{8rs}{L^2}\left(
	1-\frac{4s^2}{L^2}\right)\cos\theta\right)\right)
\end{equation}
We readily verify that (\ref{E335}) vanishes exactly on $W_{1L}$, i.e.
when $\sqrt{r^2+s^2+2rs\cos\theta}=\frac{L}{2}$, or in other words, when
$\left(1-\frac{4s^2}{L^2}\right)=\frac{4r^2+8rs\cos\theta}{L^2}$, and on
the circle $r=\frac{1}{2\mu}$, becomes equal to $(y,z)\left(1+
\textrm{ order}\left(\frac{1}{\mu L}\right)\right)$.

Now with $\partial_\mu$ representing the 2-vector $\left(\frac{\partial}
{\partial w},\frac{\partial}{\partial x}\right)$, and denoting 
(\ref{E333}) by $(\tilde{y},\tilde{z})$, the contributions to the area
$A$ of the absolute minimal-area orientable spanning surface $S$ of the
non-simply connected window, from the components of the island 
configuration perpendicular to the plane of $S_{1L}$, are given, through
quadratic order in those components, by the Laplace action:
\begin{equation}
	\label{E336}
	\frac{1}{2}\int\int\mathrm{d}w\mathrm{d}x\left(\left(\partial_\mu
	\tilde{y}\right)\left(\partial_\mu\tilde{y}\right)+\left(\partial_\mu
	\tilde{z}\right)\left(\partial_\mu\tilde{z}\right)\right)=\frac{1}{2}
	\int\mathrm{d}t\left(\tilde{y}\left(n.\partial\tilde{y}\right)+
	\tilde{z}\left(n.\partial\tilde{z}\right)\right)
\end{equation}
where $\mathrm{d}t$ represents the boundary element of the projection onto
$S_{1L}$ of the non-simply connected window, $n$ denotes the outward unit
normal, in the plane of $S_{1L}$, at each point of the boundary of the
projection onto $S_{1L}$ of the non-simply connected window, and the 
integral in the right-hand side of (\ref{E336}) goes over both the outer
boundary, (i.e. $W_{1L}$), of the projection onto $S_{1L}$ of the 
non-simply connected window, and the inner boundary of that domain, (i.e.
the projection onto $S_{1L}$ of $W_2$), and Laplace's equation was of 
course used in obtaining the right-hand side of (\ref{E336}).

We readily verify that substituting (\ref{E334}) and (\ref{E335}) into
(\ref{E336}), gives the terms in (\ref{E332}) quadratic in $y$ and $z$,
plus terms of order $\frac{\left(y^2+z^2\right)}{\mu L}$.  (Only the inner
boundary contributes, since (\ref{E335}) vanishes exactly on the outer
boundary.)

\noindent$\quad$Now 
for more general island configurations, subject to the requirement
that Laplace's equation be approximately applicable to the calculation of 
the ``out-of-plane'' contributions to the area, (which, as we noted above,
will be true for the most important island configurations), the 
``out-of-plane'' components $\tilde{y}$ and $\tilde{z}$ of the absolute
minimal-area orientable spanning surface $S$ of the non-simply connected
windows, will still be given by (\ref{E333}), with $a$, $b$, and the 
$c_n$, $(n\geq1)$, as given by (\ref{E334}), but with the differences that
the 2-vector coefficients $d_n$, $(n\geq1)$, will in general now be
non-zero and of order $\mu^{-(n+1)}$, (since both the ``in-plane
components'' and the ``out-of-plane components'' of the \emph{internal}
coordinates of the island are of size roughly $\frac{1}{\mu}$), the angles
$\gamma_n$ and $\delta_n$, $(n\geq1)$, may be non-zero, and that the $w$
and $x$ that occur in (\ref{E334}) via $s=\sqrt{w^2+x^2}$, and the $y$ and
$z$ that occur in (\ref{E334}), may differ from the components $(w,x,y,z)$
of the mean position of the vertices of the island, by amounts of order
$\frac{1}{\mu}$.  Now we readily verify that, if the projection of the
outer boundary $W_2$ of the island onto $S_{1L}$ is circular, and $(w,x)$
are the ``in-plane'' components of the centre of this circle, and moreover
$y$ and $z$ are equal to the mean ``out-of-plane'' components of $W_2$,
calculated with equal weights assigned to equal elements of the circular
projection of $W_2$ onto $S_{1L}$, then when we substitute (\ref{E333})
into (\ref{E336}), all cross terms between $a$, $b$, and the $c_n$'s on
the one hand, (which are proportional to $y$ and $z$), and the $d_n$'s on
the other hand, (which are proportional to the ``out-of-plane'' components
of the \emph{internal} coordinates of the island), are suppressed by at
least one factor of $\frac{1}{\mu L}$, which means that for large $\mu L$
the entire dependence of the area $A$ on $y$ and $z$ is given by the terms
in (\ref{E332}) quadratic in $y$ and $z$, as modified by multiplying $\mu$
inside the logarithm by a factor of order 1, to get the correct radius of
the circle.  Moreover, since the coefficients $d_n$ do not depend on $w$
and $x$, and they are all suppressed by at least on factor of $\frac{1}
{\mu L}$ on the outer boundary $W_{1L}$, the entire dependence of the area
$A$ on $w$ and $x$ is also given, for large $\mu L$, by the terms in
(\ref{E332}) quadratic in $y$ and $z$, (with other terms dependent on $w$
and $x$ all being suppressed by at least one factor of $\frac{1}{\mu L}$).
Now of course the projection of the outer boundary $W_2$ of the island 
onto $S_{1L}$ will never be exactly circular in practice, and the 
weighting of the vertices of the 45-paths of the island in the calculation
of $y$ and $z$ is different from the weighting just described, but 
nevertheless there will in general be, within the convex hull of the 
45-paths of the island, a point $(w,x,y,z)$, such that, for large $\mu L$,
the \emph{only} dependence of the area $A$ on $w$, $x$, $y$, and $z$, is
through the term in (\ref{E332}) quadratic in $y$ and $z$, with $\mu$
inside the logarithm being multiplied by a factor of order 1, (with all
other terms dependent on any of $w$, $x$, $y$, and $z$, all being 
suppressed by at least one factor of $\frac{1}{\mu L}$).  Here $(w,x)$,
and the factor of order 1 by which we multiply $\mu$ inside the 
logarithm, are determined respectively by the centre and the diameter of a
circle, roughly coincident with the projection onto $S_{1L}$ of the
island, and roughly of diameter $\frac{1}{\mu}$, onto which we transform
the projection onto $S_{1L}$ of the outer boundary of the island, by a
suitable conformal transformation, leaving $W_{1L}$ invariant, of the 
projection onto $S_{1L}$ of the non-simply connected window.  (We could,
of course, by another conformal transformation, move the circle 
representing the island to the centre of $S_{1L}$, but the radius of that
circle then gets divided by precisely the $w$ and $x$ dependent factor in
the logarithm in (\ref{E332}), so that we get the same result again.)  And
$y$ and $z$ are determined by the mean values of the appropriate
``out-of-plane'' components of the points of $W_2$, with the weights in 
the mean being determined by the above conformal transformation, such that
the elements of $W_2$ whose projections onto $S_{1L}$ are transformed to
equal-sized elements of the circle by the conformal transformation, get
equal net weights.

Thus we see that, for large $\mu L$, the entire dependence of 
$J(w,x,y,z)$, (which, as defined after (\ref{E332}), is the contribution
to our island diagram from all configurations of the island such that the
mean position of the vertices of the 45-paths of the island is equal to
$(w,x,y,z)$), on $w$, $x$, $y$, and $z$, is to a good approximation given
by a factor equal to the exponential of $-\mu^2$ times the terms in
(\ref{E332}) quadratic in $y$ and $z$, with corrections being suppressed
by at least one factor of $\frac{1}{\mu L}$.  In fact, for large $\mu L$,
$J(w,x,y,z)$ is to a good approximation equal to $\tilde{J}\equiv J(0,0,
0,0)$, times the exponential of $-\mu^2$ times the terms in (\ref{E332})
quadratic in $y$ and $z$, with the corrections all being suppressed by at
least one power of $\frac{1}{\mu L}$.  Furthermore, the entire dependence
of $\tilde{J}$ on $L$ is through the factor $e^{-\mu^2\frac{\pi}{4}L^2}$,
or in other words, the factor $f_0(W_{1L},g^2)$, which also arises from
the exponentiation of $-\mu^2$ times (\ref{E332}).  Indeed, for large
$\mu L$, the sum over the 45-paths of the island, subject to their mean
position $(w,x,y,z)$ being held fixed, is to a good approximation 
completely independent of $L$: the weights associated with the internal
windows of the island are manifestly independent of $L$, while if $w$,
$x$, $y$, and $z$ are defined as described above, to avoid any 
cross-terms, at leading order in $\frac{1}{\mu L}$, between $a$, $b$, and
the $c_n$'s on the one hand, and the $d_n$'s on the other hand, occurring
in $A$, then the only effect of the non-simply connected window on the 
sums over the 45-paths of the island, subject to $(w,x,y,z)$ being held
fixed, is through the coefficients $d_n$, which are independent of $L$ at
leading order in $\frac{1}{\mu L}$, and the angles $\gamma_n$ and 
$\delta_n$.  And if $w$, $x$, $y$, and $z$ are defined simply as the 
coordinates of the mean positon of the vertices of the island, they will
differ by amounts of no more than $\frac{1}{\mu}$ from the values they
would have to take to avoid the cross terms.  Hence at leading order in
$\frac{1}{\mu L}$, all dependence on $L$ of the sum over the 45-paths of 
the island, subject to their mean position $(w,x,y,z)$ being held fixed,
may be absorbed by adjustments of $w$, $x$, $y$, and $z$, by amounts of at
most $\frac{1}{\mu}$.

Thus if we define $K$ by:
\begin{equation}
	\label{E337}
	\tilde{J}=ke^{-\mu^2\frac{\pi}{4}L^2}=Kf_0(W_{1L},g^2)
\end{equation}
at one particular value of $L$, large compared to $\frac{1}{\mu}$, so that
$K$ is completely independent of $L$, and of $w$, $x$, $y$, and $z$, then
at leading order in $\frac{1}{\mu L}$, the entire dependence of $J(w,x,y,
z)$, on $L$, and on $w$, $x$, $y$, and $z$, is to a good approximation
expressed by the formula:
\begin{equation}
	\label{E338}
	J(w,x,y,z)=K\;f_0(W_{1L},g^2)\;\exp\left(\frac{-\mu^2\pi\left(y^2+z^2
	\right)}{\ln\left(v\mu L\left(1-\frac{4\left(w^2+x^2\right)}{L^2}\right)
	\right)}\right)
\end{equation}
Here $v$ is an absolutely fixed numerical constant of order 1, that 
represents an effective mean value of the factor of order 1, by which we
multiply $\mu$ inside the logarithm, as determined, for each important
configuration of the outer boundary $W_2$ of the island, by the diameter
of the circle, roughly of diameter $\frac{1}{\mu}$, and roughly coincident
with the projection onto $S_{1L}$ of the island, onto which we transform
the projection onto $S_{1L}$ of the outer boundary of the island, by a 
conformal transformation as above.  We note that (\ref{E338}) is valid to
a good approximation due to the fact that, at leading order in $\frac{1}
{\mu L}$, the actual slight additional dependencies of $J(w,x,y,z)$ on
$L$, $w$, $x$, $y$, and $z$, not included in (\ref{E338}), may all be
reproduced by small adjustments of $w$, $x$, $y$, and $z$ in (\ref{E338}),
\emph{by amounts of at most} $\frac{1}{\mu}$, with the actual values of
these small adjustments of $w$, $x$, $y$, and $z$ in (\ref{E338}), 
themselves being dependent on $L$, $w$, $x$, $y$, and $z$.

We can now immediately determine, from (\ref{E338}), the dependence on 
$L$, at large $\mu L$, of the contribution of our island diagram to the
right-hand side of the group-variation equation for $f_0(W_{1L},g^2)$.
Indeed, by the definition, (after (\ref{E332})), of $J(w,x,y,z)$, the
contribution of this island diagram to the right-hand side of the 
group-variation equation for $f_0(W_{1L},g^2)$, is simply the integral,
over $(w,x,y,z)$, of $J(w,x,y,z)$.  And from (\ref{E338}) we find 
immediately:
\begin{equation}
	\label{E339}
	\int\int\mathrm{d}y\mathrm{d}zJ(w,x,y,z)=\frac{K}{\mu^2}f_0(W_{1L},g^2)
	\ln\left(v\mu L\left(1-\frac{4\left(w^2+x^2\right)}{L^2}\right)\right)
\end{equation}
We note that (\ref{E339}) is valid for $(w,x)$ within $S_{1L}$ and not too
close to the edge of $S_{1L}$, (i.e. not too close to $W_{1L}$), and that,
as noted in the discussion before (\ref{E333}), we may neglect, at large
$\mu L$, the contributions of the domains where (\ref{E339}) is not valid,
since their contributions will behave only as $Lf_0(W_{1L},g^2)$ at large
$L$, while the contribution of the domain where (\ref{E339}) is valid
behaves \emph{roughly} as $L^2f_0(W_{1L},g^2)$ at large $L$.

Finally we have to integrate (\ref{E339}) over the domain of $(w,x)$ where
it is valid, say over the domain $\sqrt{w^2+x^2}\leq\frac{L}{2}-B$, where
$B$ is a fixed value, large compared to $\frac{1}{\mu}$, but independent
of $L$ as $L$ becomes large.  We readily find that the integral of
(\ref{E339}) over this domain is exactly:
\begin{equation}
	\label{E340}
	\frac{\pi K}{\mu^2}f_0(W_{1L},g^2)\left(\left(\frac{L}{2}-B\right)^2
	\ln\left(\frac{v\mu L}{e}\right)-B(L-B)\ln\left(\frac{4B(L-B)}{L^2}
	\right)\right)
\end{equation}
where of course $e$ denotes the base of natural logarithms.  The leading
term in (\ref{E340}) at large $L$ is:
\begin{equation}
	\label{E341}
	\frac{K}{\mu^2}f_0(W_{1L},g^2)\frac{\pi L^2}{4}\ln\left(\frac{v\mu L}
	{e}\right)
\end{equation}
which is of course completely independent of $B$.  Now the preceding
calculation applies to \emph{every} island diagram contributing to the
right-hand side of the group-variation equation for $f_0(W_{1L},g^2)$.
The only differences between different island diagrams are in the values
of the constant $K$, and the value of the numerical constant $v$ of order
1.  Thus (\ref{E341}) also gives the leading term at large $L$ in the sum
of the contributions of \emph{all} the island diagrams to the right-hand
side of the group-variation equation for $f_0(W_{1L},g^2)$, with $K$ now
interpreted as the sum of the $K$'s of the individual island diagrams, and
$v$ interpreted as an effective mean value of the $v$'s of the individual
island diagrams.  Hence, in view of the discussion preceding (\ref{E332}),
(\ref{E341}) also gives the leading term at large $L$ in the sum of 
\emph{all} the diagrams in the right-hand side of the group-variation
equation for $f_0(W_{1L},g^2)$.

\section{The need to cancel the extra factor of $\ln(\sigma A)$}

We see that (\ref{E341}) differs from the result we expected from our
qualitative consideration before (\ref{E231}), which resulted in 
(\ref{E236}), by the factor $\ln\left(\frac{v\mu L}{e}\right)$.
Furthermore, due to this extra factor, (\ref{E341}), after multiplication
by $-\frac{2\beta(g)}{g}$, (with $g^2$ of course set equal to the critical
value), does \emph{not} match the behaviour $L\frac{\partial}{\partial L}
f_0(W_{1L},g^2)=-2\mu^2\frac{\pi}{4}L^2f_0(W_{1L},g^2)$ of the leading 
term at large $L$ in the left-hand side of the group-variation equation
for $f_0(W_{1L},g^2)$, after multiplication by $-\frac{2\beta(g)}{g}$.

However, before concluding that our ansatz is wrong, we have to check that
our calculation has been correct.  We have assumed, firstly, that the
absolute minimal-area orientable spanning surface of the non-simply
connected window has \emph{cylinder} topology, (as opposed to the separate
absolute minimal-area orientable spanning surfaces $S_{1L}$ and $S_2$ of
$W_{1L}$ and $W_2$), and secondly, that this cylinder-topology absolute
minimal-area orientable spanning surface of the non-simply connected
window, is given to a good approximation by the appropriate solution of
Laplace's equation, in the projection onto $S_{1L}$ of the non-simply
connected window.  Clearly, for our calculation to be correct, these
assumptions must be valid for all $(w,x,y,z)$ which give significant
contributions to the integral over $(w,x,y,z)$ of $J(w,x,y,z)$.  Now of 
course, (\ref{E341}) only gets contributions from $(w,x)$ in $S_{1L}$ and
not too close to the edge of $S_{1L}$, so we have to check that, for 
$(w,x)$ in this domain, our assumptions are valid for all $(y,z)$ that
give significant contributions to (\ref{E339}).

We note, however, that we do \emph{not} have to check that our assumptions
are valid for \emph{all} island configurations centred at $(w,x,y,z)$:
that is manifestly false, even for $(w,z,y,z)=(0,0,0,0)$, and we did not
assume that.  What we have to check is that, for each $(w,x,y,z)$ in the
domain concerned, our assumptions are valid for \emph{some} island
configurations centred at $(w,x,y,z)$.  Thus it is sufficient to check,
for each $(w,x,y,z)$ in the domain concerned, that our assumptions are
valid when $W_2$ is the circle specified before (\ref{E332}).

Now from (\ref{E338}) we immediately see that the main contributions to 
(\ref{E339}) come from $(y,z)$ such that:
\begin{equation}
	\label{E342}
	\left(y^2+z^2\right)\leq\frac{1}{\mu^2\pi}\ln\left(\mu L\left(1-\frac{4
	\left(w^2+x^2\right)}{L^2}\right)\right)
\end{equation}
where we set $v=1$ because its presence does not have any significant
effect on the magnitude of the right-hand side of (\ref{E342}).

And from (\ref{E333}) - (\ref{E335}), we see that when $W_2$ is the circle
specified before (\ref{E332}), the maximum value of $\left(\partial_\mu
\tilde{y}\right)\left(\partial_\mu\tilde{y}\right)+\left(\partial_\mu
\tilde{z}\right)\left(\partial_\mu\tilde{z}\right)$, (in the notation of
(\ref{E336})), which is realized by the radial derivatives on $W_2$, is
given by $\left(\frac{b}{r}\right)^2$ at $r=\frac{1}{2\mu}$, or in other
words, (remembering that $b$ is a 2-vector), by:
\begin{equation}
	\label{E343}
	\frac{4\mu^2\left(y^2+z^2\right)}{\left(\ln\left(\mu L\left(1-\frac{4
	\left(w^2+x^2\right)}{L^2}\right)\right)\right)^2}
\end{equation}
which by (\ref{E342}) is less than or equal to:
\begin{equation}
	\label{E344}
	\frac{4}{\pi\ln\left(\mu L\left(1-\frac{4\left(w^2+x^2\right)}{L^2}
	\right)\right)}
\end{equation}
for all $(y,z)$ which give the main contributions to (\ref{E339}).  Now
(\ref{E344}) is small compared to 1 for large $\mu L$, hence, since the 
corrections to the Laplace approximation (\ref{E336}) to the 
``out-of-plane'' contributions to the area are of order $\left(\left(
\partial_\mu\tilde{y}\right)\left(\partial_\mu\tilde{y}\right)+\left(
\partial_\mu\tilde{z}\right)\left(\partial_\mu\tilde{z}\right)\right)^2$,
(\ref{E336}) \emph{does} give a good approximation to the ``out-of-plane''
contributions to the area, hence (\ref{E332}) does give a good 
approximation to the area of the ``Laplace surface'', for all $(y,z)$
which give the main contributions to (\ref{E339}).  Now by Courant's 
results \cite{Courant}, the absolute minimal-area orientable spanning
surface will have cylinder topology if there \emph{exists} an orientable
spanning surface with cylinder topology whose area is less than the sum
$\frac{\pi}{4}\left(L^2+\frac{1}{\mu^2}\right)$ of the areas of the
separate absolute minimal-area spanning surfaces of $W_{1L}$ and $W_2$, so
it comes down to whether (\ref{E332}) is less than or greater than 
$\frac{\pi}{4}\left(L^2+\frac{1}{\mu^2}\right)$, and we see immediately
from (\ref{E342}) that (\ref{E332}) is \emph{less} than $\frac{\pi}{4}
\left(L^2+\frac{1}{\mu^2}\right)$, for all $(y,z)$ that give the main
contributions to (\ref{E339}).  Thus for all $(y,z)$ that give the main
contributions to (\ref{E339}), there \emph{do} exist island configurations
of size $\frac{1}{\mu}$, centred at $(y,z)$, such that the absolute
minimal-area orientable spanning surface of the non-simply connected
window has cylinder topology, and is moreover given, to a good 
approximation, by the appropriate solution of Laplace's equation, in the
projection onto $S_{1L}$ of the non-simply connected window.

Thus there is no escape: our assumptions have been valid, and our
calculation has been correct.  Formula (\ref{E341}) \emph{does} give the
leading term in the right-hand side of the group-variation equation for 
$f_0(W_{1L},g^2)$ at large $L$, and, due to the logarithmic factor in
(\ref{E341}), our ansatz does \emph{not} satisfy the group-variation
equation for $f_0(W_{1L},g^2)$ in the area-law domain.  What is the 
significance of this?

Let us first ask whether the behaviour suggested by (\ref{E341}) gives an
acceptable modification to the Wilson area law.  Multiplying (\ref{E341})
by $-\frac{2\beta(g)}{g}$, and assuming that $K$ is negative, we see, by
comparison with (\ref{E233}), that (\ref{E341}) gives an ``output
behaviour'' of $f_0(W_{1L},g^2)$ at large $L$, of the form:
\begin{equation}
	\label{E345}
	a\;e^{-bL^2\ln(cL)}
\end{equation}
where $a$, $b$, and $c$ are constants.  This does \emph{not} match the 
``input behaviour'' $e^{-\mu^2\frac{\pi}{4}L^2}$ of $f_0(W_{1L},g^2)$ at
large $L$.  Does (\ref{E345}) give an acceptable modification of the
Wilson area law?  The answer is \emph{no}: (\ref{E345}) violates the
Seiler bound \cite{Seiler}, \cite{Simon Yaffe}, which says that for simple
planar loops, $f_0(W,g^2)$ cannot fall off faster than $e^{-\mu^2A}$ for
some \emph{fixed} $\mu$, where $A$ is the area of the spanning surface of
$W$.  Thus the area law for $f_0(W,g^2)$ must stand, and our ansatz must
be modified in some other domain.

Now in fact it is obvious what the problem is: for $(w,x)$ in $S_{1L}$
and not too close to the edge of $S_{1L}$, the rate at which $J(w,x,y,z)$,
as defined after (\ref{E332}), falls off for increasing $y$ and $z$, as
given by (\ref{E338}), \emph{decreases} logarithmically with increasing
$L$, due to the logarithmic factor in the denominator in the exponent of
(\ref{E338}), which in turn is due to the logarithmic factor in the 
denominator of the term in (\ref{E332}) quadratic in $y$ and $z$.  In
other words, for island configurations of size $\frac{1}{\mu}$, centred at
$(w,x,y,z)$, such that the absolute minimal-area orientable spanning
surface of the non-simply connected window has cylinder topology, the 
``tightness'' with which that cylinder-topology surface is able to ``draw
back'' the island towards $S_{1L}$, gets smaller and smaller as $L$ gets
larger: for purposes of restricting the freedom of movement of the island
in the directions $y$ and $z$ perpendicular to $S_{1L}$, the
cylinder-topology surface effectively gets ``slacker'' as $L$ gets larger.

\section{Pre-exponential factors for the higher-topology terms in the
ansatz}

Now it is obvious, and confirmed explicitly below, that if, in point (i)
of our ansatz, as presented after (\ref{E229}), we had confined attention
totally to the \emph{separate} absolute minimal-area orientable spanning
surfaces $S_{1L},\ldots,S_{nL}$ of $W_{1L},\ldots,W_{nL}$, and allowed no
consideration at all of the higher-topology surfaces, (i.e. with two or
more ``holes'' per connected component), then our qualitative argument, as
given before (\ref{E231}), would have been exactly right, and our ansatz
\emph{would} have satisfied the group-variation equations in the area-law
domain as well as in the glueball saturation domain.  Indeed, when we
substitute this modified ansatz into the right-hand side of the
group-variation equation for $f_0(W_{1L},\ldots,W_{nL},g^2)$, it is then
the exponential fall-off of the glueball propagator that ``pulls back''
the island to the separate absolute minimal-area spanning surfaces
$S_{1L},\ldots,S_{nL}$, and the ``tightness'' of this is completely
independent of the sizes of the $W_{iL}$'s.

Note added: another possibility, overlooked in the first version of this 
paper, is that each exponential factor in the ansatz that involves the 
area of a non-simply connected, connected component of the absolute
minimal-area orientable spanning surface of $W_{1L},\ldots,W_{nL}$, or in
other words, a connected, absolute minimal-area orientable spanning
surface, whose boundary is $q$ of the $W_{iL}$'s, where $q\geq2$, has 
associated with it a \emph{pre-exponential} factor $\frac{1}{\left(\ln
\left(\mu^2A\right)\right)^{q-1}}$, analogous to the pre-exponential
factors associated with the line segments in the absolute minimal-length
spanning tree.  When this modified ansatz is substituted into the 
right-hand side of the group-variation equations, then, since the 
logarithm is slowly varying at large $L$, the leading order calculation,
at large $L$, would be just as before, but with the area $A$ in the
logarithm simply set equal to the area $A$ of the left-hand side Wilson
loop, which would then cancel the unwanted logarithmic factor in 
(\ref{E341}).  As we will see below, in this case, the group-variation
equation result for the Wilson area law coefficient, $\mu^2$, sets
$\mu^2$ equal to the sum of (\ref{E341}) with twice the unwanted 
logarithmic factor divided out, plus the contribution, calculated below,
from the term in the ansatz, (which, as modified, becomes a sum of terms,
over the various possible topologies of the total absolute minimal-area
orientable spanning surface), where, instead of the cylinder topology
absolute minimal-area orientable spanning surface of $W_{1L}$ and $W_2$,
we take the two separate absolute minimal-area orientable spanning 
surfaces of $W_{1L}$ and of $W_2$, with a minimal-length straight-line
segment, corresponding to the lightest glueball propagator, connecing 
them.  In this case, we do not get the zeroth-order result $m=2.38\mu$
quoted in the introduction to the paper, because there is no obvious
relation between the cylinder contribution and the ``lightest glueball
propagator'' contribution.  However, a third possibility is that the
pre-exponential factor associated with the ``higher-topology'' absolute
minimal-area orientable spanning surfaces gives a stronger suppression
than the ``minimal'' possibility $\frac{1}{\left(\ln\left(\mu^2A\right)
\right)^{q-1}}$, for example, higher powers of the logarithm in the 
denominator, or even powers of the area itself in the denominator, 
analogously to the pre-exponential factors for the straight line segments
of the absolute minimal-length spanning tree.  In this case, the leading
contribution to the right hand side of the group-variation equations, at
large $L$, in the ``area-law domain'', would come totally from the term in
the ansatz where the island is connected to an absolute minimal-area
orientable spanning surface of the left-hand side Wilson loops, by a
straight segment, or in other words, a lightest-glueball propagator.  In
this case, we again get the zeroth-order relation $m=2.38\mu$, as given in
the first version of this paper, where ``suppression factors'' for the 
higher-topology terms were assumed, which, in fact, do not give the
suppression required.  Since the requirement of the self-consistency of 
the ansatz at long distances doesn't seem to choose between these two
possibilities, the correct possibility must be determined by integrating
the group-variation equations from short distances, with boundary
conditions at short distances, (i.e. very small $L$), given by 
renormalization-group-improved perturbation theory.  In practice, the
group-variation equations will be re-expressed as equations for the 
long-distance factors (\ref{E262}) of the Wilson loop vacuum expectation
values and correlation functions, which will be subject to the boundary
conditions that the vacuum expectation values $f_0(W_{1L},g^2)$ tend to 1
as $L$ becomes very small, and the correlation functions $f_0(W_{1L},
\ldots,W_{nL},g^2)$, $(n\geq2)$, tend to the values of the correlation
functions, as determined by renormalization-group-improved perturbation
theory, and with the vacuum expectation values of $W_{1L},\ldots,W_{nL}$
divided out.  We now return to the discussion in the first version of this
paper.

Can we then completely abandon consideration of the higher-topology 
spanning surfaces?  The answer would appear to be no: if we consider
$f_0(W_{1L},W_{2L},g^2)$ when $W_{1L}$ and $W_{2L}$ are two large loops
that track one another closely, but have opposite orientations, (i.e. 
their arrows point in opposite directions), then we must surely expect
$f_0(W_{1L},W_{2L},g^2)$ to be determined by the area of the 
cylinder-topology surface, i.e. by the closed loop of ``ribbon'' that runs
between $W_{1L}$ and $W_{2L}$, rather than the much larger areas of the
separate absolute minimal-area orientable spanning surfaces of $W_{1L}$
and $W_{2L}$.  Indeed, when two such loops track one another \emph{very}
closely, such that, if they are composed of straight segments, then
throughout the length of each segment, the separation between the two
paths is small compared to $\frac{1}{\mu}$, and \emph{also} small compared
to the length of that segment, then $f_0(W_{1L},W_{2L},g^2)$ is given to a
good approximation by perturbation theory, as the exponential of
$-\frac{g^2}{4\pi}$ times the integral, over \emph{one} of the two paths,
of the reciprocal of the shortest distance to the other path.  This bears
no relation at all to the areas of the separate absolute minimal-area
spanning surfaces of the two loops, but can certainly make a natural
transition, as the separation between the two loops increases, to a
dependence on the area of the absolute minimal-area, cylinder-topology
spanning surface of the two loops, i.e. the closed loop of ``ribbon'' 
which runs between the two loops.

Such pairs of loops, which track one another closeley, but have opposite
orientations, are important in the \emph{transition} from the asymptotic
freedom domain to the area law domain.\footnote{Note added: or rather, in
the domain where the long-distance factors are just beginning to differ
from their boundary conditions at very small $L$, as described just
above.}  Here we decouple the group-variation equation for $f_0(W_{1L},g^2
)$ from the other group-variation equations, by treating, in the island
diagrams in the right-hand side of the group-variation equation for 
$f_0(W_{1L},g^2)$, the non-simply connected window by perturbation theory,
(i.e. substituting in the perturbative expansion of 
$f_0(W_{1L},W_2,g^2)$), while treating the \emph{internal} windows of the
island, (i.e. the simply connected windows), non-perturbatively in the
usual way.  In this case, when $L$ is approximately equal to $\frac{1}
{\mu}$, island configurations where $W_2$, the outer boundary of the 
island, closely tracks $W_{1L}$, but has the opposite orientation, are
very important.  In fact, if we consider just the one-loop islands, we
obtain, to a good approximation, a simple first-order \emph{ordinary
linear} differential equation for $f_0(W_1L,g^2)$, with only slight
couplings between $f_0(W_{1L},g^2)$ for loops in different ``scaling
families'', and this results, as $L$ increases towards $\frac{1}{\mu}$
from values small compared to $\frac{1}{\mu}$, in gradually exponentiating
the initially weak dependence of $f_0(W_{1L},g^2)$ on the onset point $R$
of the smooth long-distance cutoffs we impose on the propagators in our
counterterms, and this drives the transition to the area-law behaviour.
Note added:
the statement that we obtain, to a good approximation, an ordinary linear
differential equation for $f_0(W_{1L},g^2)$, when $L$ is approximately
equal to $\frac{1}{\mu}$, is probably not correct.  What seems possible,
however, is that, due to the soft dependence of the long-distance factors
on the details of the path, there will be a region, starting at very small
$L$, where the long-distance factors barely differ from their boundary
conditions at small $L$, as described just above, and continuing to larger
$L$, but ending \emph{well before} $L$ reaches a size of about $\frac{1}
{\mu}$, where the long-distance factor for $f_0(W_{1L},g^2)$ will
approximately satisfy an ordinary first-order differential equation with
respect to $L$, which will be linear if we consider only the one-loop
islands.  This is because, due to the soft dependence of the long-distance
factor for the interior of the island on the details of the island
perimeter, $W_2$, we may, in some approximation, in this region, replace
the long-distance factor for the island, in regions where the perimeter
$W_2$ of the island does not wander too far from the left-hand side
Wilson loop $W_{1L}$, by the long-distance factor for the left-hand side
Wilson loop $W_{1L}$, while when the perimeter of the island is not close
to the left-hand side Wilson loop, we would treat the interior of the 
island, as well as the non-simply connected window, by
renormalization-group-improved perturbation theory.  We now return to the
discussion in the original version of this paper.  The ``initial'' value
of $f_0(W_{1L},g^2)$, at very small $L$, is determined, for each ``scaling
family'' of loops $W_{1L}$, by perturbation theory.  And as the area law
starts to set in as $L$ increases to values greater than $\frac{1}{\mu}$,
the typical island size stops at $\frac{1}{\mu}$, and we begin to use the
group-variation equation for $f_0(W_{1L},W_{2L},g^2)$, rather than 
perturbation theory, to determine the window weight for the non-simply
connected window.

In general, if $W_{1L},\ldots,W_{nL}$ are such that $W_{1L}$ is a very
large loop, and $W_{2L},\ldots,W_{nL}$ are smaller loops, but all of them
large compared to $\frac{1}{\mu}$, and $W_{2L},\ldots,W_{nL}$ all lie
within, or nearly within, the absolute minimal-area orientable spanning
surface $S_{1L}$ of $W_{1L}$, and are all oriented consistently with the
orientation of $S_{1L}$ defined by $W_{1L}$, (so that, in other words, if
$S_{1L}$ is drawn on a flat surface, and $W_{2L},\ldots,W_{nL}$ are drawn
on $S_{1L}$, then the handedness, clockwise or anti-clockwise, of the
arrows on $W_{2L},\ldots,W_{nL}$, is opposite to the handedness of the
arrow on the boundary $W_{1L}$ of $W_{1L}$), and the separations between
the loops are all large compared to $\frac{1}{\mu}$, then we would expect
$f_0(W_{1L},\ldots,W_{nL},g^2)$ to be given roughly by $e^{-\mu^2A}$,
where $A$ is the area of the surface $S$ obtained from $S_{1L}$ by cutting
out of it the separate minimal-area spanning surfaces $S_{2L},\ldots,
S_{nL}$ of $W_{2L},\ldots,W_{nL}$, so that $S$ is, indeed, the 
``higher-topology'', \emph{absolute} minimal-area orientable spanning
surface of $W_{1L},\ldots,W_{nL}$, exactly as specified in point (i) of
our ansatz, as presented after (\ref{E229}).  And we cannot simply dismiss
such sets of loops as of no physical importance, since they are likely to 
be important, for example, in calculating the corrections to meson
propagators from quark-antiquark vacuum bubbles.  How can we reconcile
this with the conclusion we arrived at after (\ref{E345}), namely that for
$f_0(W_{1L},W_{2L},g^2)$, if $W_{1L}$ is large compared to 
$\frac{1}{\mu}$, but the size of $W_{2L}$ is roughly equal to $\frac{1}
{\mu}$, then $f_0(W_{1L},W_{2L},g^2)$ is determined by the \emph{separate}
absolute minimal-area orientable spanning surfaces $S_{1L}$ and $S_{2L}$
of $W_{1L}$ and $W_{2L}$, together with the glueball propagator for the
shortest straight line segment between any point on $S_{1L}$ and any point
on $S_{2L}$, completely irrespective of whether or not the \emph{absolute}
minimal-area orientable spanning surface of $W_{1L}$ and $W_{2L}$ actually
has cylinder topology?

Note added: the following two paragraphs discuss an example in QCD-2, and
its relation to the four-dimensional case, which was intended to motivate
the suppression factors for the higher-topology terms in the ansatz that
were proposed in the original version of this paper.  However, those 
suppression factors do not, in fact, give the suppression required, so the
discussion of the QCD-2 example is, strictly speaking, no longer relevant.

A possible compromise to consider is suggested by QCD-2.  Indeed, in
QCD-2, with coupling constant $\mu$, if $W_{1L}$ and $W_{2L}$ are simple
planar loops, with no self-intersections and no mutual intersections, and
$W_{2L}$ lies inside $W_{1L}$, and their orientations are opposite to one
another, (so that if $W_{1L}$ is shrunk until it coincides with $W_{2L}$,
then their arrows point in \emph{opposite} directions), and the area of 
the non-simply connected windows, insied $W_{1L}$ but outside $W_{2L}$, is
$A$, and the area of the simply-connected window, inside $W_{2L}$, is
$B$, (so that the \emph{total} area enclosed by $W_{1L}$ is $A+B$), then
$f_0(W_{1L},W_{2L},\mu^2)$ is given by:
\begin{equation}
	\label{E346}
	f_0(W_{1L},W_{2L},\mu^2)=e^{-\mu^2A}\left(1-\left(1+2\mu^2B\right)e^{-2
	\mu^2B}\right)
\end{equation}
Now $A$ here is the area of the \emph{absolute} minimal-area orientable
spanning surface of $W_{1L}$ and $W_{2L}$, which has cylinder topology,
and for $B$ large compare to $\frac{1}{\mu^2}$, the factor in brackets in
the right-hand side of (\ref{E346}) becomes equal to 1, so that $f_0(W_
{1L},W_{2L},g^2)$ becomes equal to $e^{-\mu^2A}$, exactly as given by the
initial form of our ansatz, as stated after (\ref{E229}).  However, for 
\emph{small} $B$, (\ref{E346}) becomes
\begin{equation}
	\label{E347}
	2\mu^4B^2e^{-\mu^2A}
\end{equation}
and thus vanishes quadratically in $B$ as $B$ tends to zero.  Now
actually, this behaviour is completely expected.  Any Wilson loop whose
absolute minimal-area orientable spanning surface has zero area, is
identically equal to 1: such a Wilson loop can be the trace of a
zero-length path-ordered phase factor, or, more generally, the trace of a
``tree'' formed of hairpin-shaped path-ordered phase factors, which
exactly double back on themselves.  (Such ``trees'' of hairpin-shaped
path-ordered phase factors are not, however, very important, since they
are suppressed by the \emph{kinematic} weights in path integrals, e.g.
the Gaussian factors in (\ref{E27}).  Note added: they might, however, be
important in connection with chiral symmetry breaking, when a light quark
moves along the perimeter of the loop, since in that case, they look like
the emission of pions into the vacuum.)  Thus any correlation function 
involving a Wilson loop whose absolute minimal-area orientable spanning
surface has zero area, vanishes identically.  This explains why
(\ref{E346}) vanishes at zero $B$.  And the reason (\ref{E346}) vanishes
\emph{quadratically} in $B$, rather than linearly in $B$, for small $B$,
is due to the vanishing of the $\textrm{SU}(N)$ group factor, for all $N$, for any
Feynman diagram that can be separated into two disconnected parts, with
$W_{1L}$ in one part, and $W_{2L}$ in the other part, by cutting exactly
one propagator.  (This is due, as explained after (\ref{E52}), to a
cancellation between the two terms in (\ref{E49}) for that one ``key''
propagator.)

Now of course, although this ``formal'' argument for the vanishing of any
correlation function involving any Wilson loop whose absolute
minimal-area orientable spanning surface has zero area, works just as well
in four dimensions as it does in two dimensions, there are, in four
dimensions, competing tendencies for quantities to become singular as the
sizes of Wilson loops tend to zero.  Thus we did not make any allowance
for these competing tendencies in our initial ansatz, as stated after
(\ref{E229}), but rather left it for the group-variation equations to
determine the correct behaviour.  And in view of our result above, namely
that when $W_{1L}$ is large compared to $\frac{1}{\mu}$, but the size of
$W_{2L}$ is roughly equal to $\frac{1}{\mu}$, $f_0(W_{1L},W_{2L},g^2)$
must be determined by the \emph{separate} absolute minimal-area orientable
spanning surfaces $S_{1L}$ and $S_{2L}$ of $W_{1L}$ and $W_{2L}$, together
with the glueball propagator for the shortest straight line segment
between any point of $S_{1L}$ and any point on $S_{2L}$, completely
irrespective of whether or not the \emph{absolute} minimal-area orientable
spanning surface of $W_{1L}$ and $W_{2L}$ actually has cylinder topology,
and our observation that when the sizes of \emph{all} the loops are large
compared to $\frac{1}{\mu}$, we \emph{do} have to consider the possibility
that the absolute minimal-area orientable spanning surface might have
``higher topology'', (i.e. fewer connected components, and more holes per
connected component), and the above example from QCD-2, we now modify our
initial ansatz, as stated after (\ref{E229}), as follows:

\noindent $f_0(W_1,\ldots,W_n,g^2)$ is now expressed as a \emph{sum} of
terms, each associated with a different topology of spanning surface of
$W_1,\ldots,W_n$, or in other words, with a different partition of $\{W_1,
\ldots,W_n\}$ into parts corresponding to the separate connected
components of the spanning surface.  And the term associated with a given
topology of spanning surface, or in other words, with a given partition of
$\{W_1,\ldots,W_n\}$ into parts corresponding t the separate connected
components of the spanning surface, is given by the product, firstly, of
exactly the same factors as before, (i.e. as stated after (\ref{E229})),
with the proviso that the area $A$ in point (i) is now the area of the
absolute minimal-area orientable spanning surface \emph{with the given
topology}, if it exists, and secondly, of possible additional factors, as
follows.  If \emph{no} absolute minimal-area orientable spanning surface
with the given topology exists, then the additional factor is zero.  And
if an absolute minimal-area orientable spanning surface with the given
topology does exist, (the additional factors specified in the following,
differ from those proposed in the original version of this paper), then
the additional factors consist of the product, over all parts $j$ of the
given partition of $\{W_1,\ldots,W_n\}$ whose number $q_j$ of members is $\geq2$, of a factor
\begin{equation}
	\label{E348}
	\left(\mathbf{F}(\mu^2A_j)\right)^{q_j-1}
\end{equation}
where $A_j$ is the area of the absolute minimal-area, oriented, \emph{and 
connected}, spanning surface of the $q_j\geq2$ members of the 
$j^{\mathrm{th}}$ part of the given partition of $\{W_1,\ldots,W_n\}$,
and $\mathbf{F}(s)$ is a fixed real function, which is either, in the
``marginal'' case, equal to $\frac{1}{\ln(s)}$, or, in the 
``non-marginal'' cases, equal to a product of powers of $\ln(s)$ and $s$,
which is small compared to $\frac{1}{\ln(s)}$, as $s\to+\infty$.  As we
will see, the requirement of self-consistency of the ansatz at large
distances does not appear to distinguish between the ``marginal'' case,
and the ``non-marginal'' cases, so determining the explicit form of
$\mathbf{F}(s)$, as either the ``marginal case'' possibility, or one of
the ``non-marginal'' possibilities, will require integrating the 
group-variation equations from small $L$, with the boundary conditions at
small $L$ as discussed above.

If it were not for the extra factors (\ref{E348}), the behaviour of our
modified ansatz would essentially be the same as the behaviour of the
original ansatz, as stated after (\ref{E229}), since the original ansatz
essentially consists of picking out whichever of the terms of our modified
ansatz, without the extra factors, is largest.  But the extra factors
(\ref{E348}) perform the crucial function of suppressing the contributions
of terms corresponding to ``higher topology'' spanning surfaces, (i.e.
with fewer connected components, and more holes per connected component),
such that they now either give the correct form of the leading 
contribution to the right-hand sides of the group-variation equations in
the ``area-law'' domain, (in the ``marginal case''), which is the same
form as given by the relevant term in the ansatz where the island is 
connected to a connected component of a minimal-area spanning surface of
the left-hand side Wilson loops, by a straight-line segment, or ``lightest
glueball propagator'', or else give a contribution to the right-hand
sides of the group-variation equations in the ``area-law'' domain, (in the
``non-marginal'' cases), which is small compared to the leading term,
which is now given totally by the relevant term in the ansatz where the 
island is connected to a connected component of a minimal-area spanning
surface of the left-hand side Wilson loops, by a straight-line segment, or
``lightest glueball propagator''.

Thus when we substitute the modified ansatz into the right-hand side of 
the group-variation equation for $f_0(W_{1L},\ldots,W_{nL},g^2)$, 
$n\geq1$, the leading terms at large $L$, which come from island diagrams,
with islands of size roughly $\frac{1}{\mu}$, come, in the
\emph{non-marginal} cases, from terms in our ansatz, where any boundaries
of the island that form parts of the borders of non-simply connected
windows, belong to \emph{one-member} parts of the partitions of the sets
of the connected components of the borders of those non-simply connected
windows, that define those terms in our ansatz for those non-simply
connected windows.  The contributions of terms in our ansatz where
boundaries of the island are involved in ``higher topology'' spanning
surfaces of the boundaries of those non-simply connected windows, are
suppressed, in the \emph{non-marginal} cases, by the extra factors
(\ref{E348}).  In the \emph{marginal} case, the leading terms in the 
right-hand side come from the terms just specified, \emph{plus} the terms
that give the leading contribution for the non-modified ansatz, as 
calculated in (\ref{E341}), which is now modified to have the correct
form, by dividing by twice the logarithmic factor.

We can now re-do, with our modified ansatz, the calculation of the leading
term, at large $L$, in the contribution of any island diagram to the
right-hand side of the group-variation equation for $f_0(W_{1L},g^2)$.  We
consider the \emph{non-marginal} cases, since the result for the marginal
case can be obtained from the result for the \emph{non-marginal} cases,
simply by adding (\ref{E341}), with twice the logarithmic factor divided
out.  It is obvious that, since the leading term in our ansatz for
$f_0(W_{1L},W_2,g^2)$, where $W_2$ is the outer boundary of the island, is
now:
\begin{equation}
	\label{E349}
	f^2e^{-\mu^2A_{1L}}e^{-\mu^2A_2}\sqrt{\frac{m}{32\pi^3\left|x-y\right|^3
	}}e^{-m\left|x-y\right|}
\end{equation}
where $f$ is the glueball to surface coupling constant introduced in point
(iii) of our original ansatz, as stated after (\ref{E229}), $A_{1L}$ is 
the area of the absolute minimal-area orientable spanning surface $S_{1L}$ 
of $W_{1L}$, $A_2$ is the area of the absolute minimal-area orientable
spanning surface $S_2$ of $W_2$, and $x$ and $y$ are the ends, on $S_{1L}$
and $S_2$ respectively, of the shortest straight-line segment between any
point on $S_{1L}$ and any point on $S_2$, we now get exactly the right
dependence on $L$, namely, within a given ``scaling family'' of loops
$W_{1L}$, a dependence of the form:
\begin{equation}
	\label{E350}
	aL^2f_0(W_{1L},g^2)
\end{equation}
where $a$ is a constant.

Indeed, we now recall that, as discussed after (\ref{E325}), every Type-1
island diagram is obtained from some connected, one-line irreducible
vacuum bubble formed of 45-paths, that may be drawn on the surface of the
2-sphere without any 45-paths crossing one another, by cutting $n$ holes,
(where $n=1$ in the present case), in \emph{one} of the windows of that
vacuum bubble, and stretching that window to form the non-simply connected
window that ``surrounds'' the island.  And we also recall that, before
(\ref{E326}), we defined, for each such vacuum bubble $b$ drawn on the
2-sphere, $X_b$ to be the result of doing the path integrals over all the
45-paths of $b$, with a window weight $e^{-\mu^2B_i}$ for each window $i$
of $b$, (where $B_i$ is the area of the absolute minimal-area orientable
spanning surface of the boundary of window $i$ of $b$), subject to the
mean position of all the vertices in all the 45-paths of $b$, having the
fixed value $z$.  We recall that, by translation invariance, $X_b$ is
independent of $z$, and we recall that, by definition, $X_b$
\emph{includes} any symmetry factor for $b$, (such as the symmetry factor
$\frac{1}{2}$ for the vector-boson one-loop vacuum bubble), and that, also
by definition, $X_b$ includes the standard factor $\left.\frac{\mathrm{d}}
{\mathrm{d}M}\mathbf{C}(M)\right|_{M=1}$, which is completely unaffected
by making holes in the windows of $b$.

We also recall that, for each of our vacuum bubbles $b$, we define $n_b$
to be the number of windows of $b$.

Let us now, for simplicity, make the assumption, that the mass $m$ of the
lightest glueball, is sufficiently smaller than the mass $M$ of the 
lowest-mass state of the cylinder with two 45-paths along it, that there
is no significant tendency, for a fixed mean position $z$ of all the 
vertices of the 45-paths of the island, for the glueball propagator
between $x$ and $y$, (where $x$ and $y$ are the ends, on $S_{1L}$ and
$S_2$ respectively, of the shortest straight line segment between any
point of $S_{1L}$ and any point of $S_2$), to ``pull'' part of the island
towards $S_{1L}$.  This assumption is precisely analogous to the
assumption which we made, also for simplicity, in studying the glueball
saturation domain, that $m$ is sufficiently smaller than $M$, that there
is no significant tendency for an island to elongate along the straight
line segment of the minimal-length spanning tree that it is close to.

It follows immediately from this assumption, that in (\ref{E349}), instead
of taking $y$ strictly as the closest point, on $S_2$, to any point of
$S_{1L}$, we may, to a good approximation, set $y$ simply equal to any
chosen vertex of $W_2$, or maybe to the mean position of all the vertices
of $W_2$.  We then find that, for the given values of $x$ and $y$, the
contribution of this island diagram to the right-hand side of the
group-variation equation for $f_0(W_{1L},g^2)$, is equal to:
\begin{equation}
	\label{E351}
	f^2e^{-\mu^2A_{1L}}\sqrt{\frac{m}{32\pi^3\left|x-y\right|^3}}
	e^{-m\left|x-y\right|}
\end{equation}
times the path integrals over all the 45-paths of the island, with a
window weight $e^{-\mu^2B_i}$ for each window $i$ of the island,
(\emph{including} the outer boundary window $W_2$), subject to the chosen
vertex of $W_2$, (or, if preferred, the mean position of all the vertices
of $W_2$), having the given value $y$, times the standard factor
$\left.\frac{\mathrm{d}}{\mathrm{d}M}\mathbf{C}(M)\right|_{M=1}$ for that
island diagram, times a possible symmetry factor, associated with
rotational symmetries of the island.  Here $B_i$ is the area of the 
absolute minimal-area orientable spanning surface of the boundary of the
window $i$ of the island, and the factor $e^{-\mu^2B_i}$ for the outer
boundary $W_2$ of the island, is the factor $e^{-\mu^2A_2}$ in
(\ref{E349}).  But by the translation invariance of the sums over the
45-paths of the island, and of the window weights $B_i$, this is precisely
equal to (\ref{E351}), times $X_B$, where $b$ is the vacuum bubble 
corresponding to that island, times a possible integer factor, associated
with symmetries of $b$, equal to the number of different windows of $b$,
such that making a hole in that window, gives the given island diagram.
(The integer factor, if it occurs, is equal to the symmetry factor
assoicated with any rotational symmetries of the island, divided by the
symmetry factor associated with the vacuum bubble $b$.)  We note that, if the vacuum bubble $b$ has two or more loops, then $X_b$, and also the
corresponding island diagrams, also include an explicit power of $g^2$,
equal to the number of 45-paths minus the number of action vertices of 
that vacuum bubble or island.  This power of $g^2$ is in general one less
than the number of loops of that vacuum bubble or island.  We also note
that the factor $e^{-\mu^2A_{1L}}$ in (\ref{E351}), is equal to 
$f_0(W_{1L},g^2)$.

We thus find immediately that, subject to our assumption, made for
simplicity, that $m$ is sufficiently smaller than $M$, that we may, to a
good approximation, set $y$, in (\ref{E349}), equal to the mean position
of the vertices of $W_2$, the contribution of all the island diagrams,
when we substitute our modified ansatz into the right-hand side of the
group-variation equation for $f_0(W_{1L},g^2)$, is equal to 
$f^2\left(\displaystyle\sum_bn_bX_b\right)f_0(W_{1L},g^2)$, times the 
integral:
\begin{equation}
	\label{E352}
	\int_{S_{1L}}\mathrm{d}^2x\int\mathrm{d}^2y\sqrt{\frac{m}{32\pi^3\left|
	x-y\right|^3}}e^{-m\left|x-y\right|}
\end{equation}
since of course, as $L$ becomes large, and correspondingly, both radii of
curvature of $S_{1L}$ tend to zero at all points of $S_{1L}$, $x$ becomes
simply equal to the perpendicular projection onto $S_{1L}$ of the mean
position $y$ of the vertices of the island, and furthermore, again due to
the radii of curvature of $S_{1L}$ tending to zero as $L$ becomes large,
and also due to the suppression of the integrand in (\ref{E351}) and
(\ref{E352}) when $\left|x-y\right|$ is large compared to $\frac{1}{m}$,
we may, for large $L$, represent the four-dimensional integral over the
mean position $y$ of the vertices of the island, as the two-dimensional
integral, over $S_{1L}$, of the position of the perpendicular projection
$x$ of $y$ onto $S_{1L}$, times the two-dimensional integral, over all $y$
such that the perpendicular projection of $y$ onto $S_{1L}$, (or more
precisely, onto the two-plane tangential to $S_{1L}$ at $x$), is equal to
$x$.  (The two-dimensional integral over $y$ in (\ref{E352}) runs over the
two dimensions perpendicular to the two-plane tangential to $S_{1L}$ at
$x$.)

Now, setting $z\equiv y-x$, the two-dimensional integral over $y$ in
(\ref{E352}) is simply:
\begin{equation}
	\label{E353}
	\int\mathrm{d}^2z\sqrt{\frac{m}{32\pi^3\left|z\right|^3}}e^{-m\left|z
	\right|}=2\pi\sqrt{\frac{m}{32\pi^3}}\int_0^\infty\frac{\mathrm{d}r}
	{\sqrt{r}}e^{-mr}=\frac{1}{2\sqrt{2}}
\end{equation}
and the two-dimensional integral over $x$ gives simply $A_{1L}$.  Hence we
find that, for large $L$, the leading term in the sum over all the island
diagrams in the right-hand side of the group-variation equation for
$f_0(W_{1L},g^2)$, and, indeed, the leading term in the sum over
\emph{all} the diagrams in the right-hand side of the group-variation
equation for $f_0(W_{1L},g^2)$, is equal, for all the possible 
non-marginal forms of the pre-exponential function $\mathbf{F}(s))$ in
(\ref{E348}), to:
\begin{equation}
	\label{E354}
	\frac{1}{2\sqrt{2}}A_{1L}f^2\left(\sum_bn_bX_b\right)f_0(W_{1L},g^2)
\end{equation}
Hence, multiplying by $-\frac{2\beta(g)}{g}$, and comparing with
(\ref{E233}) and (\ref{E234}), and noting that $a_1L^2$, in (\ref{E234}),
is equal to $A_{1L}$ in our present notation, we see that our modified
ansatz does indeed exactly satisfy the group-variation equation for
$f_0(W_{1L},g^2)$ at large $L$, and furthermore, that the Wilson area-law
parameter $\mu^2$ is given, for all the possible 
non-marginal forms of the pre-exponential function $\mathbf{F}(s))$ in
(\ref{E348}), by:
\begin{equation}
	\label{E355}
	\mu^2=\frac{1}{2\sqrt{2}}\frac{\beta(g)}{g}f^2\left(\sum_bn_bX_b\right)
\end{equation}
This is the equation that replaces (\ref{E236}) for our modified ansatz,
for all the possible 
non-marginal forms of the pre-exponential function $\mathbf{F}(s))$ in
(\ref{E348}).

(We of course assume, as usual, that $\left(\displaystyle\sum_bn_bX_b
\right)$ is \emph{negative}, and, indeed, that the sum of $X_b$ for the
one-loop vacuum bubbles, corresponding to the island diagrams
(\ref{E135}), (\ref{E136}), and (\ref{E137}), is \emph{negative}.)  We see
that (\ref{E355}), which we have now derived quantitatively from our
modified ansatz, for all the possible 
non-marginal forms of the pre-exponential function $\mathbf{F}(s))$ in
(\ref{E348}), (subject to our assumption, made for simplicity, that the 
mass $m$ of the lightest glueball, is sufficiently smaller than the mass
$M$ of the lowest-mass state of the cylinder with two 45-paths along it,
that there is no significant tendency, for a fixed mean position of all
the vertices of the island, for the glueball propagator to ``pull'' part
of the island towards $S_{1L}$), is essentially identical in structure to
(\ref{E236}), apart from the presence of the factor $f^2$, where $f$ is 
the glueball to surface coupling constant, introduced in point (iii) of
our original ansatz, as stated after (\ref{E229}).

Considering now the marginal case, where the pre-exponential function
$\mathbf{F}(s))$ in (\ref{E348}) has the form $\frac{1}{\ln s}$, it
is necessary, first, to determine whether the leading term (\ref{E341}),
in the contributions of the higher-topology terms in the ansatz, to the
right-hand sides of the group-variation equations, (before multiplying by
the pre-exponential factor $\mathbf{F}(\mu^2A)$, where $A$ is the area of
the higher-topology minimal-area spanning surface involved, which is
equal, to sufficient accuracy for the pre-exponential factor, in the
leading term, to the area of the minimal-area spanning surface of the
appropriate left-hand side Wilson loops), has the same form, with the same
coefficient, irrespective of the shapes of the left-hand side Wilson
loops, or whether its form, or coefficient, which has so far been derived
only for a circular left-hand side Wilson loop, depends on the shapes of
the left-hand side Wilson loops, because if its form, or coefficient,
\emph{did} depend on the shapes of the left-hand side Wilson loops, then
we could eliminate the marginal possibility for $\mathbf{F}(s)$, as being
inconsistent with the Wilson area law, for large $L$.

However, it appears that, at least for planar left-hand side Wilson loops,
even with self-intersections, the coefficient of $\ln(\mu L)$ in
(\ref{E341}) is in fact completely independent of the shape of the
left-hand side Wilson loop.  Indeed, repeating, for a general planar
left-hand side Wilson loop, the analysis that led to (\ref{E341}), we
would take, as the appropriate solution of Laplace's equation, to
sufficient accuracy:
\begin{equation}
	\label{E355a}
	b\left(\ln r +\Phi(W_{1L},\tilde{w},\tilde{x},w,x)\right)
\end{equation}
where $r\equiv\sqrt{(w-\tilde{w})^2+(x-\tilde{x})^2}$, $(w,x)$ are
coordinates in the region of the two-plane, bordered by parts, or the 
whole, of the left-hand side Wilson loop, in which the projection into the
two-plane of the centre of the circle of radius $\frac{1}{2\mu}$ lies,
$(\tilde{w},\tilde{x})$ are the coordinates of the projection into the 
two-plane of the centre of the circle, and $b$ is a two-vector
coefficient, (in the two directions perpendicular to the two-plane defined
by the left-hand side Wilson loop), to be determined.  $\Phi(W_{1L},
\tilde{w},\tilde{x},w,x)$ is the solution of Laplace's equation, in the
coordinates $(w,x)$, regular in the closed region of the two-plane,
bordered by parts, or the whole, of the left-hand side Wilson loop, that
contains the point $(\tilde{w},\tilde{x})$, and which takes the value
$-\ln r = -\ln(\sqrt{(w-\tilde{w})^2+(x-\tilde{x})^2})$ on the border of
that domain, so that (\ref{E355a}) vanishes exactly on the border of that
domain.

We then require that:
\begin{equation}
	\label{E355b}
	b\left(\ln\left(\frac{1}{2\mu}\right)+\Phi(W_{1L},\tilde{w},\tilde{x},
	\tilde{w},\tilde{x})\right)
	=(y,z)+\textrm{ order}\left(\frac{1}{\mu L}\right)
\end{equation}
where $(y,z)$ are the coordinates of the centre of the circle of radius
$\frac{1}{2\mu}$, in the two directions perpendicular to the two-plane,
which fixes
\begin{equation}
	\label{E355c}
	b=\frac{(y,z)}{\left(\ln\left(\frac{1}{2\mu}\right)+\Phi(W_{1L},
	\tilde{w},\tilde{x},\tilde{w},\tilde{x})\right)}+\textrm{ order}\left(
	\frac{1}{\mu L}\right)
\end{equation}

Now $\Phi(W_{1L},\tilde{w},\tilde{x},w,x)$ has no dependence at all on
$\mu$, hence, since all lengths, describing the shape of $W_{1L}$, can be
expressed as dimensionless multiples of $L$, the only way the argument
of the logarithm, in the denominator of (\ref{E355c}), can become
dimensionless, as it must, is for $\Phi(W_{1L},\tilde{w},\tilde{x},
\tilde{w},\tilde{x})$, to contain a term $-\ln\left(\frac{L}{2}\right)$.
Thus $b$ tends, at large $L$, to the two-vector:
\begin{equation}
	\label{E355d}
	\frac{-(y,z)}{\ln(\mu L)}
\end{equation}
exactly as for the case when $W_{1L}$ is circular.  Furthermore, the
leading contribution to (\ref{E336}), which again comes entirely from the
circle of radius $\frac{1}{2\mu}$, since (\ref{E355a}) vanishes on the
outer boundary, is
\begin{equation}
	\label{E355e}
	\pi b^2\ln(\mu L)\to\pi\frac{y^2+z^2}{\ln(\mu L)}
\end{equation}
exactly as for the case when $W_{1L}$ is circular.

That the $-\ln\left(\frac{L}{2}\right)$ term in $\Phi(W_{1L},\tilde{w},
\tilde{x},\tilde{w},\tilde{x})$ really does occur, as stated, can be
checked explicitly, for example, in the extreme case where the ``loop''
$W_{1L}$ becomes the pair of lines $w=\frac{L}{2}$ and $w=-\frac{L}{2}$,
in which case we find:
	\[\Phi(W_{1L},\tilde{w},\tilde{x},w,x)\quad
	=\quad-\ln\left(\frac{L}{2}\right)+
	\qquad\qquad\qquad\qquad\qquad\qquad\qquad\qquad\qquad\qquad
\]
\begin{equation}
	\label{E355f}
	\qquad
	+\int_0^\infty\frac{\mathrm{d}\alpha}{\alpha}e^{-\frac{\alpha L}{2}}
	\left\{\left(\frac{\cosh(\alpha\tilde{x})\cosh(\alpha x)}{\cosh\left(
	\frac{\alpha L}{2}\right)}+\frac{\sinh(\alpha\tilde{x})\sinh(\alpha x)}
	{\sinh\left(\frac{\alpha L}{2}\right)}\right)\cos(\alpha(w-\tilde{w}))
	-1\right\}
\end{equation}

Furthermore, the arguments regarding the decoupling of the integrals over
the shape of the island, from the integral over the mean position
$(w,x,y,z)$ of the vertices of the island, go through exactly as before,
and we again arrive at (\ref{E341}).

Furthermore, since, by (\ref{E342}), the main contributions to
(\ref{E339}) come from $(y,z)$ of magnitude less than or equal to
\begin{equation}
	\label{E355g}
	\frac{1}{\mu}\sqrt{\frac{\ln(\mu L)}{\pi}}
\end{equation}
while for non-planar $W_{1L}$, the radii of curvature of the minimal-area
orientable spanning surface of $W_{1L}$ will be of order $L$, we expect
that, for non-planar loops, we may, in a zone of thickness (\ref{E355g})
about the minimal-area orientable spanning surface of $W_{1L}$, again
choose coordinates where $(w,x)$ represents the position of the
perpendicular projection of the mean position of the vertices of the
island, onto the minimal-area orientable spanning surface of $W_{1L}$, and
$(y,z)$ represents the position of mean position of the vertices of the
island, in the directions perpendicular to the two-plane tangential, at
$(w,x)$, to the minimal-area orientable spanning surface of $W_{1L}$, and
that we will get the same result, (\ref{E341}), again, for the leading
term.

Thus the marginal possibility for the pre-exponential factor
$\mathbf{F}(s)$ also appears to be compatible with the Wilson area law,
and in this case, formula (\ref{E355}), for the Wilson area-law parameter
$\mu^2$, must be modified, by the addition of (\ref{E341}), with twice the
logarithmic factor divided out.  It will be necessary to integrate the
group-variation equations from small $L$, with boundary conditions, at
small $L$, as given by renormalization-group-improved perturbation theory,
in order to determine whether the marginal case occurs or not.

Now comparison of points (i), (ii), and (iii) of our original ansatz, as
stated after (\ref{E229}), shows immediately that the dimension of $f$ is
\emph{length}, so that the dimension of $\frac{1}{f}$ is \emph{mass}.  It
then follows immediately, by reasoning exactly analogous to that which
gave (\ref{E280}) and (\ref{E281}), that $\frac{1}{f}$ obeys precisely the
same renormalization group equation as the Wilson area law parameter
$\mu$, i.e. (\ref{E281}), with $\mu$ replaced by $\frac{1}{f}$.  The
dependence of $f$ on our input parameter $R$, (the onset point of the 
smooth long-distance cutoffs we impose on the propagators in our 
counterterms), is simply through an overall factor $R$, and the dependence
of $\frac{1}{f}$ on our input parameter $g$, is through exactly the same
factor that gives the dependence of $\mu$ on $g$.  Thus the product $f\mu$
is an absolutely fixed real number, completely independent of $R$ and $g$,
which we may expect to be of order 1, (and, of course, calculable from the
group-variation equations).

Furthermore, on dimensional grounds, each $X_b$ is equal to $\mu^4$, times
the \emph{explicit} power of $g^2$ associated with that vacuum bubble,
(i.e. a power of $g^2$ equal to the number of 45-paths minus the number of
action vertices of that vacuum bubble), times a real number independent of
$R$ and $g^2$.  And of course, the power of $g^2$ is zero for the one-loop
vacuum bubbles.  Hence, for the non-marginal cases, we may divide $\mu^2$
out of (\ref{E355}), to obtain an equation for the critical value of
$g^2$, (which applies throughout the domain where our ansatz applies),
exactly as we obtained before from (\ref{E236}), (as discussed between
(\ref{E254}) and (\ref{E255}), and after (\ref{E261})).  If the marginal
case applies, we would expect an analogous result to hold.  (Note that, in
the marginal case, $\mathbf{F}(s)$ will have the form $c \frac{1}{\ln s}$,
where $c$ is a number that will need to be determined by integrating the
group-variation equations from small $L$, with boundary conditions as
given by renormalization-group-improved perturbation theory.)

And furthermore, bearing in mind that for the one-loop vacuum bubbles $b$,
$X_b$ is completely independent of $g^2$, we may expact that, exactly as
before, the critical value of $g^2$ will essentially be determined by the
point where $\frac{\beta(g)}{g}$ reaches a critical value, as determined
by (\ref{E355}), (or its analogue, if the marginal case applies), with
$\mu^2$ divided out.  Hence, exactly as before, (in the discussion after
(\ref{E282})), we may expect that, if, in a natural renormalization
scheme, (where the behaviour of the expansion coefficients in $2g\beta(g)$
is \emph{no better} than the behaviour of the expansion coefficients in
the expansions of other physical quantities in the \emph{explicit} powers
of $g^2$ that multiply the right-hand side group-variation equation
diagrams), the trend shown by the first two terms in (\ref{E282})
continues, and all the expansion coefficients in $2g\beta(g)$ have the
\emph{same} sign, (i.e. \emph{negative}), then, due to the fact that, in 
that case, $-2g\beta(g)$ will tend to infinity as $g^2$ approaches the 
radius of convergence of the expansion of $-2g\beta(g)$ in powers of 
$g^2$, the critical value of $g^2$, as determined by (\ref{E355}), (or its
analogue, if the marginal case applies), will be \emph{strictly less} than
the radius of convergence of the expansion of $-2g\beta(g)$ in powers of
$g^2$, and consequently, due to the assumption that the renormalization 
scheme is natural, the critical value of $g^2$ will also be \emph{strictly
less} than the radii of convergence of the expansions of other physical
quantities  in the explicit powers of $g^2$ that multiply the right-hand
side group-variation equation diagrams.  Hence, exactly as before, we may
expect that, if all the expansion coefficients in $2g\beta(g)$ are
negative in such a natural renormalization scheme, then the expansions of
all physical quantities in the explicit powers of $g^2$ that multiply the
right-hand side group-variation equation diagrams, will converge
geometrically for all $g^2$ less than or equal to the critical value.

\section{The zeroth-order value of $m_{0^{++}}/\surd\sigma$ when the
pre-exponential factor is non-marginal}

Now comparing (\ref{E355}) with (\ref{E331}), we see that, if the 
pre-exponential factor $\mathbf{F}(s)$ is non-marginal, the ratio
$\frac{m^2}{\mu^2}$ is given by:
\begin{equation}
	\label{E356}
	\frac{m^2}{\mu^2}=\frac{2\sqrt{2}\left(\displaystyle\sum_bn_b^2X_b
	\right)}{\left(\displaystyle\sum_bn_bX_b\right)}
\end{equation}

We now observe that the island diagram mechanism by which the Wilson area
law, and massive glueball saturation, arise in the group-variation 
equations, works in exactly the same way in leading order, i.e. if we
consider just the one-loop island diagrams, as it does in all orders.  The
higher-loop islands will correct the details, but they are not necessary
for quark confinement.

Furthermore, as we have just noted, if all the expansion coefficients in
$2g\beta(g)$ are negative in a natural renormalization scheme, then we may
expect that the expansions of all physical quantities in the explicit
powers of $g^2$ that multiply the right-hand side group-variation equation
diagrams, converge geometrically for all $g^2$ less than or equal to the
critical value.  Indeed, as we noted in the discussion after (\ref{E282}),
if the two terms displayed in (\ref{E282}) are the first two terms in a
geometric series, then the radius of convergence is given by $\frac{g^2}
{4\pi}=\frac{11\pi}{17}=2.03$, while, as we noted in the discussion after
(\ref{E261}), the largest value of our $\frac{g^2}{4\pi}$ for which there
is concrete experimental evidence, (based of $\alpha_s=0.3$ for
charmonium), is $\frac{g^2}{4\pi}=0.35$.  Thus there is ample room for the
expansions of all physical quantities in the explicit powers of $g^2$ that
multiply the right-hand side group-variation equation diagrams, to have a
convergence factor of $\frac{1}{2}$, or even better.

Thus we may obtain a good first approximation to the numerical value of
$\frac{m}{\mu}$, if the pre-exponential factor $\mathbf{F}(s)$ is 
non-marginal, by restricting the sums over the vacuum bubbles $b$ in the
numerator and denominator of (\ref{E356}), to the one-loop vacuum bubbles.
We then find immediately, since $n_b=2$ for each one-loop vacuum bubble,
(corresponding to the two fundamental representation Wilson loop ``sides''
of the closed-loop 45-path), that if the pre-exponential factor
$\mathbf{F}(s)$ in (\ref{E348})
is non-marginal, then the leading approximation, in
large-$N_c$ QCD, to the mass $m$ of the lightest glueball, in terms of the
Wilson area law parameter $\mu$, is given by:
\begin{equation}
	\label{E357}
	m=\sqrt{4\sqrt{2}}\mu=2.38\mu
\end{equation}
(In checking (\ref{E357}), we should remember that $m^2$ gets equal
contributions from corresponding Type-1 and Type-2 one-loop island
diagrams, while $\mu^2$ only gets contributions from Type-1 island
diagrams.)

Now as we noted between (\ref{E200}) and (\ref{E201}), the experimental
value of $\mu$ is 0.41 GeV, hence we find $m=0.98\textrm{ GeV}$.

We note that, if we estimate the mass $M$ of the lowest mass state of the
cylinder with two 45-paths along it as twice the effective mass $1.3\mu$
of a 45-path, then our assumption that $m$ is strictly less than $M$ is
satisfied, but only just.  Could this indicate that the mass of the second
lightest glueball is close above $m$?  Will ``hyperfine'' corrections,
(i.e. $F_{\mu\nu}$ insertions), destabilize this estimate of $M$?

Equation (\ref{E357}) suggests that our additional assumptions, made for
convenience, that $m$ is sufficiently small compared to $M$, that glueball
propagators do not significantly ``pull islands out of shape'', will
require further investigation.  This raises a quesion about the accuracy
of (\ref{E357}), but does not affect the verifications of the Wilson area
law and massive glueball saturation, which do not depend on those
assumptions.  (Note that, as discussed after (\ref{E200}), $1.3\mu$ is
essentially the \emph{smallest} reasonable estimate of the effective mass
of a 45-path.)

The best lattice value of $\frac{m}{\mu}$ is 3.56 \cite{Teper}, so the
zeroth-order estimate (\ref{E357}), which applies if the pre-exponential
factor $\mathbf{F}(s)$ in (\ref{E348}) is non-marginal, is about 33
percent smaller than the best lattice value.

Now as discussed after (\ref{E282}), there is evidence, from 't Hooft's
studies of planar diagrams \cite{'t Hooft a}, \cite{'t Hooft b}, that the
behaviour of the sums of diagrams in the right-hand sides of the 
group-variation equations, considered as an expansion in the 
\emph{explicit} powers of $g^2$ that multiply those diagrams, will be
geometric, at worst.  Since the zeroth-order estimate (\ref{E357}), which
applies if the pre-exponential factor $\mathbf{F}(s)$ in (\ref{E348}) is
non-marginal, corresponds to dropping all terms in the numerator and 
denominator of (\ref{E356}) with $n_b>2$, we can use the lattice result to
estimate what the convergence factor will have to be, if the 
pre-exponential factor $\mathbf{F}(s)$ is non-marginal.

As defined before (\ref{E326}), $n_b$ is the number of windows of the
vacuum bubble $b$, and the number of \emph{explicit} powers of $g^2$
multiplying the corresponding island diagram, is $(n_b-2)$.  Let us
suppose, for purposes of determining what the convergence factor would be,
that
\begin{equation}
	\label{E358}
	\sum_{b\vert n_b=n}X_b=X\alpha^{n-2}
\end{equation}
where $\alpha$ is a convergence factor to be determined.  Then 
(\ref{E356}), which applies if the pre-exponential factor $\mathbf{F}(s)$
in (\ref{E348}) is non-marginal, becomes:
\begin{equation}
	\label{E359}
	\frac{m^2}{\mu^2}=\frac{2\sqrt{2}\displaystyle\sum_{n=2}^\infty n^2
	\alpha^{n-2}}{\displaystyle\sum_{n=2}^\infty n\alpha^{n-2}}=2\sqrt{2}
	\frac{4-3\alpha+\alpha^2}{(1-\alpha)(2-\alpha)}
\end{equation}
which, according to the best lattice result, is to equal $(3.56)^2$.  This
implies that $\alpha$ is equal to 0.5919, which is not much larger than the
figure of $\frac{1}{2}$ suggested above.  With this value of $\alpha$, we
can study the convergence of $\frac{m}{\mu}$, as the sums in the numerator
and denominator of (\ref{E359}) are truncated at various values of $n$, and
we find that going to $n$, (or $n_b$), $=3$ gives a 26 percent error in
$\frac{m}{\mu}$, $n=4$ gives a 20 percent error, $n=5$ a 15 percent error,
$n=7$ an 8 percent error, $n=13$ a 1 percent error, and $n=18$ a 0.1
percent error.  After that, each extra decimal place of accuracy requires
increasing $n$, or $n_b$, by 5.

Looking at examples such as (\ref{E135}), (\ref{E140}), (\ref{E148}), and
(\ref{E149}), suggests that the chromatic polynomial factor
$\left.\frac{\mathrm{d}}{\mathrm{d}M}\mathbf{C}(M)\right|_{M=1}$ included
in $X_b$, as defined before (\ref{E326}), tends to alternate in sign as
$n_b$ increases, hence, bearing in mind the other points discussed in
connection with the signs of the island diagram contributions, such as in
arriving at (\ref{E317}), for example, and the fact that, as just noted, if
the pre-exponential factor $\mathbf{F}(s)$ in (\ref{E348}) is non-marginal,
then the convergence factor $\alpha$ must be strictly positive, it is clear
that the signs of the island diagram contributions will require very
careful study.

\chapter{Concluding Remarks}

\section{Dimensional Regularization}
\label{Dimensional Regularization}

It would be nice to be able to study the Group-Variation Equations
within dimensional regularization \cite{'t Hooft Veltman},
\cite{Bollini Giambiagi}, \cite{Ashmore}, in view of the large body of 
existing work within renormalization-group-improved perturbation theory, 
including such results as the $\beta$-function to four loops in 
$\overline{\mathrm{MS}}$ \cite{beta in MS bar 2}, which appears to be 
both compatible with, and very useful for, the Group-Variation Equations.

If so, it will be necessary to face the fact that not only the internal 
variables of diagrams, but also the observable physical quantities, 
namely the vacuum expectation values and correlation functions of Wilson 
loops, must be defined in $d$ dimensions, with $d$ complex.  Indeed, 
these vacuum expectation values and correlation functions provide the 
window weights for the diagrams in the right-hand sides of the 
Group-Variation Equations, and not only the vertices in these diagrams, 
but also the strings of vertices along the paths, in the sums over 
paths, must be freely moveable in $d$ dimensions, in order to obtain the 
correct $d$-dimensional propagators, for the leading terms, when the 
window weights are expanded in powers of $g^2$, for example.

Since we will divide the vacuum expectation values and correlation 
functions by short-distance factors, that remove the linear divergences, 
and soften the dependence on the fine details of the path, and re-write 
the Group-Variation Equations as equations for the ratios (\ref{E262}), 
restoring the short-distance factors perturbatively in the windows of 
the Group-Variation Equation right-hand side diagrams, by means of 
effective fields, it will be sufficient to study the ratios 
(\ref{E262}), the long-distance factors, for Wilson loops that consist 
of finite numbers of straight segments.  The long-distance factor of a 
vacuum expectation value, or correlation function, of Wilson loops 
defined on a total of $n$ straight segments, will be, in $d$ dimensions, 
$d$ complex, a function of the $\frac{1}{2}n(n-1)$ independent distances 
between the $n$ vertices at the ends of the segments.  It will have 
symmetries generated by independent cyclic permutations of the vertices 
around the separate Wilson loops involved, and a two-fold symmetry 
generated by the simultaneous reversal of the orientation of all the 
Wilson loops involved.  Let us denote such a long-distance factor by 
$F(r_{12},\ldots,r_{(n-1)n},g^2,d)$, where $r_{ij}$, $1\leq i<j\leq n$, 
are the distances between the $n$ vertices.

\subsection[Inequalities that might be satisfied by the distances among finite sets of points, for arbitrary complex $d$]{Inequalities that might be satisfied by the distances \\among finite sets of points, for arbitrary complex $d$}

We do not have a model for a $d$-dimensional Euclidean space, for 
general finite, complex, $d$, although an interesting construction has 
been proposed recently in reference \cite{Weinzierl}.  Therefore it is 
not really obvious that, when $d$ is complex, we can even assume that 
the distances $r_{ij}$ are real, in the ``physical'' domain, over which 
we integrate, and where we consequently have to calculate 
$F(r_{12},\ldots,r_{(n-1)n},g^2,d)$.  However while, in the process of 
solving the Group-Variation Equations, by iterative substitution of the 
left-hand sides into the right-hand sides, for example, we will 
inevitably have to calculate $F(r_{12},\ldots,r_{(n-1)n},g^2,d)$ for 
complex $d$, the target of the calculation is to determine 
$F(r_{12},\ldots,r_{(n-1)n},g^2,4)$, with all $r_{ij}$ real.  Therefore 
it might be reasonable to assume that, even at intermediate stages in 
the calculation, where we have to allow $d$ to be complex, we can choose 
the ``integration contours'', in the $d$-dimensional complex Euclidean 
space, such that the $r_{ij}$ are real.

Determining the domain of the $\{r_{ij}\}$, for which we actually have 
to calculate $F(r_{12},\ldots,r_{(n-1)n},g^2,d)$, at intermediate stages 
in the calculation, is very important, because we will not, in general, 
be able to find explicit, exact formulae for \\ 
$F(r_{12},\ldots,r_{(n-1)n},g^2,d)$.  What will, in practice, go into 
the right-hand sides of the Group-Variation Equations, and come out of 
the left-hand sides, will be various upper and lower bounds on the real 
and complex parts of $F(r_{12},\ldots,r_{(n-1)n},g^2,d)$, in various 
limited domains of the $r_{ij}$, $g^2$, and $d$.  The process of solving 
the Group-Variation Equations will involve finding successively tighter 
sets of upper and lower bounds, on the real and complex parts of 
$F(r_{12},\ldots,r_{(n-1)n},g^2,d)$, in the domain of the $\{r_{ij}\}$ 
that actually occurs in the integrals in the right-hand sides of the 
Group-Variation Equations, such that when we substitute these bounds 
into the right-hand sides of the Group-Variation Equations, we can prove 
that the left-hand side ``output'' of these bounds, satisfies the 
``input'' bounds.

Therefore it is important to know whether we can restrict the domain of 
the $\{r_{ij}\}$, at which we have to determine bounds on 
$F(r_{12},\ldots,r_{(n-1)n},g^2,d)$ at intermediate stages of the 
calculation, to domains where inequalities such as the triangle 
inequalities, $r_{ij}+r_{jk}\geq r_{ik}$, are valid.  These inequalities 
are valid for all known Euclidean spaces, namely those for positive 
integer $d$.  There are further inequalities, valid for all known 
Euclidean spaces, and independent of the triangle inequalities, which 
state, for example, that the sum of the areas of three faces of a 
tetrahedron, is greater than or equal to the area of the fourth face.  
That these inequalities are independent of the triangle inequality, can 
be seen by considering an example such as
\[ r_{12}=r_{13}=r_{23}=r
\]
\begin{equation}
\label{CR1}
r_{14}=r_{24}=r_{34}=s
\end{equation}
where
\begin{equation}
\label{CR2}
\frac{r}{2}<s<\frac{r}{\sqrt{3}}
\end{equation}
These inequalities can be expressed in terms of the $\{r_{ij}\}$ by 
means of Heron's formula \cite{Heron}, that the area of a triangle whose 
sides have lengths $a$, $b$, and $c$, is 
\begin{equation}
\label{CR2a}
\frac{1}{4}\sqrt{(a+b+c)(a+b-c)(a+c-b)(b+c-a)}
\end{equation}

\noindent$\ \ \ \ $Heron's formula, 
and the higher-dimensional volumes, required to express 
the higher-dimensional analogues of the triangle and tetrahedron 
inequalities, all of which are valid for all known Euclidean spaces, can 
be expressed in terms of Gramm determinants: in $r$-dimensional 
Euclidean space, the square of the volume of an $r$-dimensional simplex, 
(generalized triangle or tetrahedron), with vertices at 
$x_1,\ldots,x_{r+1}$, is given by $\left(\frac{1}{r!}\right)^2$ times
	\[
\left\vert\begin{array}{cccc}
(x_1-x_{r+1})_1 & (x_1-x_{r+1})_2 & \ldots & (x_1-x_{r+1})_r \\
(x_2-x_{r+1})_1 & (x_2-x_{r+1})_2 & \ldots & (x_2-x_{r+1})_r \\
     \vdots     &      \vdots     & \ddots &      \vdots     \\
(x_r-x_{r+1})_1 & (x_r-x_{r+1})_2 & \ldots & (x_r-x_{r+1})_r
\end{array}\right\vert\qquad\times\qquad\qquad\qquad
\]
	\[
\times\qquad\left\vert\begin{array}{cccc}
(x_1-x_{r+1})_1 & (x_2-x_{r+1})_1 & \ldots & (x_r-x_{r+1})_1 \\
(x_1-x_{r+1})_2 & (x_2-x_{r+1})_2 & \ldots & (x_r-x_{r+1})_2 \\
     \vdots     &      \vdots     & \ddots &      \vdots     \\
(x_1-x_{r+1})_r & (x_2-x_{r+1})_r & \ldots & (x_r-x_{r+1})_r
\end{array}\right\vert\qquad=\qquad
\]
\begin{equation}
\label{CR3}
=\left\vert\begin{array}{cccc}
(x_1-x_{r+1})^2             & (x_1-x_{r+1}).(x_2-x_{r+1}) & \ldots &
                                         (x_1-x_{r+1}).(x_r-x_{r+1}) \\
(x_2-x_{r+1}).(x_1-x_{r+1}) & (x_2-x_{r+1})^2 & \ldots    &
                                         (x_2-x_{r+1}).(x_r-x_{r+1}) \\
          \vdots            &           \vdots    & \ddots & \vdots  \\
(x_r-x_{r+1}).(x_1-x_{r+1}) & (x_r-x_{r+1}).(x_2-x_{r+1}) & \ldots &
                                         (x_r-x_{r+1})^2
\end{array}\right\vert
\end{equation}
The dot products can then be expressed in terms of the $\{r_{ij}\}$ by 
means of
	\[
(x_i-x_{r+1}).(x_j-x_{r+1})=\frac{1}{2}\left((x_i-x_{r+1})^2+
(x_j-x_{r+1})^2-(x_i-x_j)^2\right)\quad=\qquad
\]
\begin{equation}
\label{CR4}
\qquad\qquad=\quad\frac{1}{2}\left(r_{i(r+1)}^2+
r_{j(r+1)}^2-r_{ij}^2\right)
\end{equation}
The borderline case, where any of these inequalities becomes an 
equality, corresponds to the vanishing of the volume of the 
$(r+1)$-dimensional simplex, whose faces are the $r$-dimensional 
simplexes involved.

\subsection{The need for a Gaussian representation of the area law}

To perform the $d$-dimensional integrals, we can attempt to represent \\
$F(r_{12},\ldots,r_{(n-1)n},g^2,d)$ by a Laplace transform in the 
variables $\{r_{ij}^2\}$:
	\[
F(r_{12},\ldots,r_{(n-1)n},g^2,d)\quad=\qquad\qquad\qquad\qquad\qquad
\qquad\qquad\qquad\qquad\qquad\qquad\qquad
\]
\begin{equation}
\label{CR5}
=\int_0^\infty\ldots\int_0^\infty
\mathrm{d}s_{12}\ldots\mathrm{d}s_{(n-1)n}
\tilde{F}(s_{12},\ldots,s_{(n-1)n},g^2,d)e^{-(s_{12}r_{12}^2+\ldots+
s_{(n-1)n}r_{(n-1)n}^2)}
\end{equation}
The function $\tilde{F}(s_{12},\ldots,s_{(n-1)n},g^2,d)$, if it exists, 
is formally given by Bromwich's integral over the $\{r_{ij}^2\}$, which 
is an integral over lines parallel to the imaginary axis, in the complex 
$r_{ij}^2$ planes, where we do not expect to have good information on 
$F(r_{12},\ldots,r_{(n-1)n},g^2,d)$.  Nevertheless, we can apply 
Bromwich's integral to seek a representation of 
$F(r_{12},\ldots,r_{(n-1)n},g^2,d)$ in the form (\ref{CR5}), provided 
the integrals converge.  In practice, it might be important to know 
whether we can assume that the $\{r_{ij}\}$, in the $d$-dimensional 
Gaussian integrals, can be assumed to satisfy the triangle inequality, 
and its higher-dimensional generalizations, as discussed above, because 
if so, we might be able to choose a form of 
$F(r_{12},\ldots,r_{(n-1)n},g^2,d)$, in the regions of real 
$\{r_{ij}^2\}$ where these inequalities are violated, which leads to a 
more convenient form of $\tilde{F}(s_{12},\ldots,s_{(n-1)n},g^2,d)$.

In practice, it is a non-trivial problem to find an explicit 
representation of the form (\ref{CR5}) even for $e^{-\sigma A}$, where 
$A$ is the area of the triangle with edge lengths $r_{12}$, $r_{13}$, 
and $r_{23}$, as given by Heron's formula, and we can see that it might 
be important to have the freedom to choose 
$F(r_{12},r_{13},r_{23},g^2,d)$ to have a convenient dependence on the 
$r_{ij}$, for real $\{r_{ij}^2\}$ such that the triangle inequalities 
are violated.  To find an explicit representation of the form 
(\ref{CR5}) for $e^{-\sigma A}$, where $A$ is the area of the 
minimal-area spanning surface of a loop formed of four straight 
segments, we might expect it to be important to have the freedom to 
choose a convenient form of $F(r_{12},\ldots,r_{34},g^2,d)$, for real 
$\{r_{ij}^2\}$ such that any of the triangle inequalities or tetrahedron 
inequalities are violated.

\subsection{Application of Douglas's functional}

Conceivably, representations in the form (\ref{CR5}) of $e^{-\sigma 
A}$, where $A$ is the area of the minimal-area spanning surface of a 
loop formed of $n$ straight segments, might not actually be needed.  As 
we discussed after equation (\ref{E188}), the area of the minimal-area 
orientable spanning surface of \emph{any} simple closed path may be 
expressed in the quadratic form (\ref{E187}), summed over \emph{all} $d$ 
dimensions in which the path exists, provided the \emph{parametrization} 
of the path is the parametrization for which (\ref{E187}) takes its 
minimum value \cite{Courant}, \cite{Douglas}.  If we define 
$s=\tan\left(\frac{\theta}{2}\right)$, 
$t=\tan\left(\frac{\phi}{2}\right)$, (\ref{E187}), summed over all $d$ 
dimensions in which the path exists, can be expressed in the form:
\begin{equation}
\label{CR6}
A\leq\frac{1}{4\pi}\int_0^{2\pi}\int_0^{2\pi}\frac{(x(\theta)-x(\phi))^2}
{4\sin^2\left(\frac{\theta-\phi}{2}\right)}\mathrm{d}\theta\mathrm{d}\phi
\end{equation}
where $x$ is now a $d$-vector, and the inequality becomes an equality, 
for the parametrization of the path that minimizes the right-hand side.

We can explicitly introduce a reparametrization of the path into 
(\ref{CR6}), by introducing a continuous real function $f$, such that 
$\theta=f(u)$, and $\phi=f(v)$, and $f(0)=0$ and $f(2\pi)=2\pi$.  We may 
assume that $f$ is monotonic, so that the reparametrization is 
invertible, since a parametrization, that involved doubling backwards 
and forwards along the path, would certainly not minimize the right-hand 
side of (\ref{CR6}).  The reparametrized form of (\ref{CR6}) is then 
given by:
\begin{equation}
\label{CR7}
A\leq\frac{1}{4\pi}\int_0^{2\pi}\int_0^{2\pi}\frac{(x(u)-x(v)^2)}{4\sin^2
\left(\frac{f(u)-f(v)}{2}\right)}\frac{\mathrm{d}f(u)}{\mathrm{d}u}\frac{
\mathrm{d}f(v)}{\mathrm{d}v}\mathrm{d}u\mathrm{d}v
\end{equation}

Let us now suppose that our simple closed path consists of $n$ straight 
line segments, whose vertices are at $x_1,\ldots,x_n$.  Then since, at 
any finite stage of our limiting procedure, discussed in connection with 
equation (\ref{E278}), the closed path $x_1,\ldots,x_n$, that defines 
our long-distance factor (\ref{E262}), only approximately follows the 
path around the edge of the relevant simply-connected window of a 
Group-Variation Equation diagram, it is natural to consider, in 
conjunction with the discretization of the path, a ``discretization'' of 
Douglas's variational problem.  (The $n$ in (\ref{E262}) has no 
connection with our present $n$, and is in fact equal to 1, since we are 
considering, at present, the vacuum expectation value of a single Wilson 
loop.)  This means that, instead of seeking the minimum of the 
right-hand side of (\ref{CR7}) among all continuous monotonic real 
functions $f$, such that $f(0)=0$ and $f(2\pi)=2\pi$, we will only seek 
its minimum among piecewise linear such $f$, which are completely 
defined, for all $0\leq u\leq2\pi$, by the $(n-1)$ values
\begin{equation}
\label{CR8}
f_r=f\left(\frac{2\pi r}{n}\right)\qquad\qquad1\leq r\leq (n-1)
\end{equation}
such that
\begin{equation}
\label{CR9}
0\equiv f_0\leq f_1\leq f_2\leq\ldots\leq f_{n-1}\leq f_n\equiv2\pi
\end{equation}

If we suppose that our $x_s$, $1\leq s\leq n$, are $n$ samples of a 
$d$-vector function $x(u)$, it is clear, from the structure of 
(\ref{CR7}), that we should consider the samples to be taken at 
$u=\frac{2\pi\left(s-\frac{1}{2}\right)}{n}$, $1\leq s\leq n$, so that
\begin{equation}
\label{CR9a}
x_s=x\left(\frac{2\pi\left(s-\frac{1}{2}\right)}{n}\right)\qquad\qquad
1\leq s\leq n
\end{equation}

Thus we define, as the discrete version of (\ref{CR7}):
\begin{equation}
\label{CR10}
A_D\left(x_1,\ldots,x_n,f_1,\ldots,f_{n-1}\right)\equiv\frac{1}{4\pi}\sum_
{\begin{array}{c}1\leq s\leq n \\ 1\leq t\leq n \\ s\neq t\end{array}}
\frac{\left(x_s-x_t\right)^2\left(f_s-f_{s-1}\right)\left(f_t-f_{t-1}
\right)}{4\sin^2\left(\frac{f_{s-1}+f_s-f_{t-1}-f_t}{4}\right)}
\end{equation}
where the subscript $D$ stands for Douglas, and we expect that 
minimizing \\
$A_D\left(x_1,\ldots,x_n,f_1,\ldots,f_{n-1}\right)$ with respect to 
the $f_s$, $1\leq s\leq (n-1)$, subject to (\ref{CR9}), will give an 
approximation to the area of the minimal-area spanning surface of the 
closed loop of straight line segments, with vertices $x_1,\ldots,x_n$.

We therefore might consider, as an approximation to $e^{-\sigma A}$, 
where $A$ is the area of the actual minimal-area spanning surface of 
this closed loop of straight line segments, a constant, or slowly 
varying, multiple of the expression
\begin{equation}
\label{CR11}
\int_0^{2\pi}\mathrm{d}f_1\ldots\int_0^{2\pi}\mathrm{d}f_{n-1}\theta(f_2-
f_1)\ldots\theta(f_{n-1}-f_{n-2})e^{-\sigma 
A_D\left(x_1,\ldots,x_n,f_1,\ldots,f_{n-1}\right)}
\end{equation}
where $\theta(f)$ is the step function, as in (\ref{E1}), on the basis 
that a steepest descents approximation, to the integral (\ref{CR11}), 
will give an approximation to $e^{-\sigma A}$, times some 
pre-exponential factor.  What this pre-exponential factor will be, as 
well as the question of whether (\ref{CR11}) can be generalized to 
higher-topology minimal-area spanning surface terms, such as the 
cylinder-topology term, in our ansatz for the long-distance behaviour of 
the correlation function of two Wilson loops, I leave to further work.  
If the pre-exponential factor turns out to have a non-trivial dependence 
on the $x_s$, $1\leq s\leq n$, we could try to cancel it, at least in 
part, by including an additional function of the $f_s$, $1\leq 
s\leq(n-1)$, in the integrand of (\ref{CR11}), and then cancel any 
remaining dependence on the $x_s$ by another factor represented in the 
form (\ref{CR5}), provided a suitable $\tilde{F}$ exists.

The minimization of (\ref{CR10}) with respect to the $f_r$, $1\leq r\leq 
(n-1)$, subject to (\ref{CR9}), is not expected to give a good 
representation of the areas of the minimal-area spanning surfaces, of 
general paths formed of $n$ straight-line segments, with vertices at 
$x_1,\ldots,x_n$, because the piecewise-linear reparametrization of the 
path, defined by the $f_r$, does not have enough freedom.  Rather, we 
would attempt to determine whether the representation (\ref{CR11}) gives 
an adequate realization of the Wilson area law, when substituted into a 
path integral such as (\ref{E27}), where the kinematic factor, in 
(\ref{E27}), will suppress the contributions of paths where the 
successive $x_s$ are distant from one another, rather than following, in 
some approximation, the straight line between the endpoints $x$ and $y$ 
of the path.  For such paths, where the successive $x_s$ follow, in some 
approximation, some smooth curve, and just add jagged ``detail'' to the 
curve, we might expect that the piecewise-linear reparametrization 
function, defined by the $f_r$, will now begin to have enough freedom to 
approach the parametrization, of the corresponding smooth curve, that 
minimizes Douglas's functional (\ref{CR7}).

Indeed, if the curve $x(\theta)$, in (\ref{CR6}), takes the form of a 
simple planar polygon, then the parametrization of the curve, for which 
(\ref{CR6}) takes its minimum value, is the parametrization where 
$\theta$ represents the position on the unit circle in the complex 
plane, while the perimeter of the polygon is swept out, as $\theta$ goes 
around the unit circle, by the Schwartz-Christoffel transformation.

Let the complex variable $w$, be defined as an analytic function of the 
complex variable $z$, by the equation:
\begin{equation}
\label{SC1}
\frac{\mathrm{d}w}{\mathrm{d}z}=\kappa\left(e^{i\phi_1}-z\right)^
{-\alpha_1}\left(e^{i\phi_2}-z\right)^{-\alpha_2}\ldots\left(e^{i\phi_
n}-z\right)^{-\alpha_n}
\end{equation}
where $\phi_1,\phi_2,\ldots,\phi_n$ and 
$\alpha_1,\alpha_2,\ldots,\alpha_n$ are real angles, $\kappa$ is a fixed 
complex number, and for definiteness, we assume that 
$0<\phi_1<\phi_2<\ldots<\phi_n<2\pi$.  Then since the right-hand side of 
(\ref{SC1}) is analytic inside, and on, the unit circle, in the $z$ 
plane, except at the points on the unit circle where $z=e^{i\phi_s}$, 
$1\leq s\leq n$, it follows that, if $z$ traces out any simple closed 
curve, either in, or on, the unit circle, in the $z$-plane, avoiding the 
points $z=e^{i\phi_s}$, $1\leq s\leq n$, on the unit circle, $w$ will 
also trace out a closed curve in the $w$ plane.  In particular, on the 
unit circle, with $z=e^{i\theta}$, we see that
	\[
\frac{\mathrm{d}w}{\mathrm{d}\theta}=iz\frac{\mathrm{d}w}{\mathrm{d}z}
=i\kappa(2i)^{-(\alpha_1+\alpha_2+\ldots+\alpha_n)}e^{\frac{-i}{2}
(\phi_1\alpha_1+\phi_2\alpha_2+\ldots+\phi_n\alpha_n)}e^{\frac{i\theta}
{2}(2-(\alpha_1+\alpha_2+\ldots+\alpha_n))}\quad\times
\]
\begin{equation}
\label{SC2}
\times\quad\sin^{-\alpha_1}\left(
\frac{\phi_1-\theta}{2}\right)\sin^{-\alpha_2}\left(\frac{\phi_2-
\theta}{2}\right)\ldots\sin^{-\alpha_n}\left(\frac{\phi_n-\theta}{2}
\right)
\end{equation}
Thus if
\begin{equation}
\label{SC3}
\alpha_1+\alpha_2+\ldots+\alpha_n=2
\end{equation}
the phase of $\frac{\mathrm{d}w}{\mathrm{d}\theta}$ is constant, while 
$\theta$ is strictly between any two consecutive $\phi_s$.  Let us now 
assume (\ref{SC3}), and also that all the $\alpha_s$ are strictly less 
than 1, so that (\ref{SC1}) is integrable along the unit circle.  We see 
that if $z$ follows a contour that proceeds anticlockwise around the 
unit circle, except that, to avoid the points $e^{i\phi_s}$, $1\leq 
s\leq n$, it deviates inside the unit circle, along semicircles of very 
small radius, $\epsilon$, centred at these points, the path followed in 
the $w$ plane, in the limit where $\epsilon\to0$, has the form of a 
polygon, where the phase of $\frac{\mathrm{d}w}{\mathrm{d}\theta}$ 
increases by $\pi\alpha_s$, as $\theta$ passes $\alpha_s$.

Furthermore, since
\begin{equation}
\label{SC4}
\left\vert\frac{\mathrm{d}w}{\mathrm{d}\theta}\right\vert=\frac{\left
\vert\kappa\right\vert}{4}\left\vert\sin\left(\frac{\phi_1-\theta}{2}
\right)\right\vert^{-\alpha_1}\left\vert\sin\left(\frac{\phi_2-\theta}
{2}\right)\right\vert^{-\alpha_2}\ldots\left\vert\sin\left(\frac{\phi_n-
\theta}{2}\right)\right\vert^{-\alpha_n}
\end{equation}
we see that, in the region of a convex angle of the polygon, or in other 
words, an angle $\pi\alpha_s$ such that $\alpha_s>0$, $w$ sweeps very 
rapidly through the region of the corner, and infinitely rapidly through 
the corner itself, as $\theta$ passes steadily through $\theta_s$.

This means that the parametrization of a polygonal curve, which 
minimizes Douglas's functional (\ref{CR6}), must be such that, near any 
convex angle of the polygon, an increasingly large length of the 
perimeter of the polygon must be swept out, for a fixed increase of the 
parameter along the path, as the corner is approached.  Or in other 
words, the derivative of the parameter, with respect to distance along 
the perimeter of the polygon, must tend to zero, as a convex corner of 
the polygon is approached.

In particular, if the path $x(u)$ in (\ref{CR7}), is a simple planar 
polygon, and the parameter $u$ is the distance along the perimeter of 
the polygon, then the function $f(u)$, in (\ref{CR7}), which minimizes 
(\ref{CR7}), will be the angle $\theta$, in the Schwartz-Christoffel 
transformation, and $\frac{\mathrm{d}f(u)}{\mathrm{d}u}$ will be a 
constant multiple of 
$\frac{1}{\left\vert\frac{\mathrm{d}w}{\mathrm{d}\theta}\right\vert}$.

Thus the monotonic function $f(u)$, such that $f(0)=0$, and 
$f(2\pi)=2\pi$, which minimizes (\ref{CR7}), will be such that 
$\frac{\mathrm{d}f(u)}{\mathrm{d}u}$ tends to zero as $\vert 
u-u_0\vert^{\left(\frac{\alpha}{1-\alpha}\right)}$, as $u$ approaches a 
point, $u_0$, where the curve turns through a convex angle $\pi\alpha$, 
if the curve $x(u)$ is a simple planar polygon, and $u$ is proportional 
to distance along the curve.  In particular, if the curve is a 
rectangle, then $\frac{\mathrm{d}f(u)}{\mathrm{d}u}$ tends to zero, 
linearly, at each corner.  The approximate representation of the 
minimization of (\ref{CR7}), as the minimum of (\ref{CR10}), with 
respect to the $f_r$, $1\leq r\leq(n-1)$, subject to (\ref{CR9}), simply 
doesn't have enough flexibility to cope with this, if $n$ is small, and 
the points $x_s$ don't approximately follow some smooth path.

To see what this means in an explicit example, let us suppose that 
$n=4$, and the points $x_s$, $1\leq s\leq4$, form a rectangle, with 
sides of lengths $A$ and $B$.  Since the $f_s-f_{s-1}$ are associated 
with the corners of the rectangle, we assume they are all equal to 
$\frac{\pi}{2}$.  Then (\ref{CR10}) gives the value 
$\frac{3\pi}{16}(A^2+B^2)\geq\frac{3\pi}{8}AB=1.178\;AB$ for the area of 
the rectangle.

We might suppose that the situation improves if we allow more vertices 
along the perimeter of the rectangle, but instead, a new problem arises. 
Let us now suppose that $n=8$, and we have the same rectangle as 
before, with a vertex at each corner, and a vertex at the middle of 
each side.  Then, by the symmetries of the rectangle, we assume there 
can be three possible values of $f_s-f_{s-1}$, namely a value $a$, for 
$s$ in the middle of a side of length $A$, a value $b$, for $s$ in the 
middle of a side of length $B$, and a value $c$, for $s$ at a corner, 
such that $2a+2b+4c=2\pi$.  We find that the 
$\sin^2\left(\frac{f_{s-1}+f_s-f_{t-1}-f_t}{4}\right)$, in the 
denominators in (\ref{CR10}), depend on $a$, $b$, and $c$, only through 
the combination $a-b$, and some of them are independent of $a$, $b$, and 
$c$.  If $A=B$, we assume $a=b$, by symmetry, and we then find that 
(\ref{CR10}) takes its minimum value at $c=0$, which is on the boundary 
of the region allowed by (\ref{CR9}).  Since $c$ is the value of 
$f_s-f_{s-1}$, for $s$ at a corner, this is not unexpected, in view of 
the discussion of the Schwartz-Christoffel transformation.  For general 
$A$ and $B$, and general $a$ and $b$, subject to $c=0$, so that 
$a+b=\pi$, we find that (\ref{CR10}) gives:
\begin{equation}
\label{SC5}
\frac{1}{8\pi}\left(2(A^2+B^2)ab+A^2b^2+B^2a^2\right)
\end{equation}
This attains its minimum value, in the region allowed by (\ref{CR9}), 
either at $a=\pi$, $b=0$, or $a=0$, $b=\pi$, and the minimum value it 
takes is
\begin{equation}
\label{SC6}
\frac{\pi}{8}\mathrm{min}(A^2,B^2)
\end{equation}
which is \emph{less} than the correct value, $AB$.  However, we see 
that, when $c=0$, and either $a$ or $b$ is also equal to 0, some of the 
$\sin^2\left(\frac{f_{s-1}+f_s-f_{t-1}-f_t}{4}\right)$, in the 
denominators in (\ref{CR10}), vanish, although none of them vanish for 
$c=0$, if both $a$ and $b$ are non-zero.  In particular, suppose that 
$A>B$, so that the minimum of (\ref{SC5}) is attained at $a=\pi$, $b=0$. 
The terms in (\ref{CR10}), which come from $s$ and $t$ at the ends of an 
edge of the square, of length $B$, are:
\begin{equation}
\label{SC7}
\frac{B^2c^2}{4\pi\sin^2\left(\frac{b+c}{2}\right)}
\end{equation}
for general $a$, $b$, and $c$.  We see that, although this term vanishes 
when the point $a=\pi$, $b=c=0$ is approached by the route $c\to0$ for 
nonzero $b$, followed by $b\to0$, it can attain any value from 0 to 
$\frac{B^2}{\pi}$, by approaching the point $a=\pi$, $b=c=0$, keeping 
different fixed values of the ratio $b/c$.  Thus (\ref{CR10}) is not 
well-behaved, near the boundary of the allowed domain, (\ref{CR9}).

In fact, even if we take the value $a=b=\frac{\pi}{2}$ in (\ref{SC5}), 
we get the value $\frac{3\pi}{32}(A^2+B^2)$, which for $A=B$ is 
\emph{less} than the correct value, $AB$.  Thus it appears that the 
minimum value of (\ref{CR10}) can be less than the correct value of the 
area of the minimal-area spanning surface, of the path of straight line 
segments, with ends at the $x_s$, if the $f_r$ are allowed to approach 
the boundaries of the allowed region, (\ref{CR9}).  This problem must be 
due to the approximate way in which (\ref{CR10}) represents the problem 
of minimizing (\ref{CR7}), in the restricted space of piecewise-linear 
$f(u)$, with the points at which $\frac{\mathrm{d}f(u)}{\mathrm{d}u}$ is 
allowed to change, being halfway between the values of $u$ at successive 
corners of the path of straight segments, since (\ref{CR7}), correctly 
calculated, is strictly bounded below, by the true area of the 
minimal-area spanning surface of the path of straight segments, and 
attains this value, only for $f(u)$ that solve the variational problem.  
(The $f(u)$ that solve the variational problem are not unique, due to 
the freedom to re-parametrize the unit circle, by conformal 
transformations.)  Thus we might attempt to solve this problem by 
replacing (\ref{CR10}) by a more accurate representation of (\ref{CR7}), 
calculated with piecewise-linear $f(u)$, or by finding an appropriate 
subdomain of the allowed domain (\ref{CR9}), and a corresponding 
subdomain of the integration domain in (\ref{CR11}).

We observe that the requirement, noted above, that for an $f(u)$ that 
minimizes (\ref{CR7}), $\frac{\mathrm{d}f(u)}{\mathrm{d}u}$ must tend to 
zero, as $u$ approaches a ``convex'' corner of the path, if $u$ 
represents the distance along the path, can be re-expressed, for the 
approximate discrete form (\ref{CR10}), by saying that if, in 
(\ref{CR10}), we want the minimum to be attained for approximately 
constant values of the $f_r$, then the ``smooth'' path, which the 
vertices approximately follow, must be more sparsely populated by 
vertices, where it has a ``convex'' bend, or in other words, where it 
deviates maximally from its ``main'' route, and more densely populated 
by vertices, where it is approximately following its ``main'' route.  
This will automatically be the case, for typical paths, in path 
integrals such as (\ref{E27}).

In the search for a generalization of the representation (\ref{CR11}), 
for the correlation function of two or more Wilson loops, we might 
consider the way in which (\ref{CR11}), which is based on a 
discretization of the minimisation problem of Douglas's functional, 
nevertheless has some resemblance, in the integration domain of the 
$f_r$, $1\leq r\leq(n-1)$, to a dual resonance amplitude.

\subsection{The non-island diagrams spoil the Gaussian representation}

When we calculate a non-island diagram in the right-hand sides of the 
Group-Variation Equations, the Gaussian representation (\ref{CR5}) will 
be spoilt, when we integrate one end of a 45-path along a straight-line 
segment of one of the left-hand side Wilson loops.  This will give an 
incomplete Gaussian integral, which we then have to re-represent in the 
form (\ref{CR5}), with a suitable new $\tilde{F}$, if further 
$d$-dimensional integrals have to be performed.

\subsection{The short-distance factors}

The short-distance factors, which we divide the vacuum expectation 
values and correlation functions by, to obtain the long-distance factors 
(\ref{E262}), can be chosen in any convenient form, provided they cancel 
the linear divergences along the Wilson loops, and can be calculated in 
perturbation theory.  They should also soften the dependence of the 
long-distance factors on the fine details of the paths, which would be 
expected to occur, for example, for the form analysed in connection with 
equations (\ref{E262}) to (\ref{E277}).

Since the short-distance factors are going to be restored in the windows 
of the right-hand side Group-Variation equation diagrams, by means of 
effective fields, as we did in the simplest case, in the calculation 
from equation (\ref{E292}) to (\ref{E317}), they can in principle be 
defined by effective field theories, that depend on the shapes of the 
individual Wilson loops.  This means that there might be a possibility 
of defining gauge-invariant short distance factors, by the use of 
gauge-field actions of the form:
\begin{equation}
\label{CR12}
\frac{1}{4}\int\mathrm{d}^dx\frac{1}{g^2(x)}F_{\mu\nu a}F_{\mu\nu a}
\end{equation}
where $g^2(x)$ is a function of the shortest distance from $x$ to any 
point on the loop, and is chosen to equal the same value of $g^2$ as 
used in the Wilson loop vacuum expectation values and correlation 
functions, when this distance is zero, and to decrease sufficiently 
rapidly, as this distance increases, that these short-distance factors 
can be calculated in perturbation theory.  The use of a gauge-field 
action such as (\ref{CR12}) means that the free propagator used in the 
short-distance factor will not be translation-invariant, and I leave it 
to future work to determine whether such a technique is practical.

Given any two distinct choices of short-distance factor, their ratio 
should be calculable in perturbation theory, and we should then find, 
that if we multiply a long-distance factor, calculated with one 
particular choice of short-distance factor, by the appropriate ratio of 
short-distance factors, we should obtain the long-distance factor, as 
calculated with the other short-distance factor, which might provide a 
useful check on the calculations.

\subsection{Residual logarithmic divergences of subdiagrams with two ``path legs'' and one gluon leg}

A final point to be resolved, in the definition of the long-distance 
factors, (\ref{E262}), is the treatment of the residual logarithmic 
divergences of subdiagrams that have two ``legs'' that form part of the 
Wilson loop, and one gluon leg, as discussed after equation 
(\ref{E277}).  Within dimensional regularization, these divergences 
might be removed by the renormalization of the gluon field 
$A\raisebox{-4pt}{$\stackrel{\displaystyle{(x)}}{\scriptstyle{\mu a}}$}$ 
at the vertex, as was discussed by Dotsenko and Vergeles \cite{Polyakov}.

\subsection[The need to test the prescriptions by going backwards and 
forwards between the Group-Variation Equations and renormaliza-tion-group-improved perturbation theory at short distances]{The need to test the prescriptions by going backwards and 
forwards between the Group-Variation Equations and renormalization-group-improved perturbation theory at short distances}

All the prescriptions will then need to be tested by going backwards and 
forwards between the Group-Variation Equations and 
renormalization-group-improved perturbation theory, at short distances, 
or in other words, for a small value of the input coupling constant 
$g^2$, using the limiting procedure for calculating the path integrals, 
as described after equation (\ref{E278}).

\subsection{$F_{\mu\nu}$ insertions in the sums over paths}

One final prescription needed, is for dealing with the $\bar{F}_{\mu\nu}$ 
insertions in the path-ordered phase factors, that occur in the sums 
over paths for the gluon propagator, as in equation (\ref{E18}).

The solution to equation (\ref{E15}) can be expressed slightly more 
neatly if we define $H_{\sigma\nu}$ by:
\begin{equation}
\label{CR13}
\left(-\bar{D}^2\delta_{\mu\sigma}-2\bar{F}_{\mu\sigma}\right)
H_{\sigma\nu}=\delta_{\mu\nu}
\end{equation}
We then define
\begin{equation}
\label{CR14}
\tilde{E}\equiv-\bar{D}_\mu H_{\mu\nu}\bar{D}_\nu
\end{equation}
The solution to (\ref{E15}) can then be written as:
\begin{equation}
\label{CR15}
g^2\left(\begin{array}{c|c}H_{\sigma\nu}+H_{\sigma\tau}
\bar{D}_\tau\left(\frac{1-\beta}{(1-\beta)\tilde{E}+\beta}\right)\bar{D}_
\alpha
H_{\alpha\nu} & H_{\sigma\tau}\bar{D}_\tau\left(\frac{i}
{(1-\beta)\tilde{E}+\beta}\right) \\
\left(\frac{-i}{(1-\beta)\tilde{E}+\beta}\right)\bar{D}_\alpha
H_{\alpha\nu} & 
\left(\frac{1-\tilde{E}}{(1-\beta)\tilde{E}+\beta}\right)\end{array}
\right)
\end{equation}

All the $\bar{F}_{\sigma\tau}$ insertions are now contained within 
$H_{\mu\nu}$.  Furthermore, $\tilde{E}_{Ax,By}$ is equal to 
$\delta_{AB}\delta^4(x-y)$, plus terms of degree two and higher in 
$A_{\mu a}$, so the expansion
\begin{equation}
\label{CR16}
\frac{1}{\tilde{E}}=\frac{1}{1-(1-\tilde{E})}=1+(1-\tilde{E})+
(1-\tilde{E})^2+\ldots
\end{equation}
is slightly improved, in comparison with equation (\ref{E30}).

We can then express the $\bar{F}_{\mu\nu}$ insertions, in 
$H_{\alpha\beta}$, by using that, by equation (\ref{E29}):
	\[
\left(\bar{D}_\mu\bar{D}_\nu\left(\frac{-1}{\bar{D}^2}\right)\right)_
{Ax,By}\quad\simeq\qquad\qquad\qquad\qquad\qquad\qquad\qquad\qquad\qquad
\qquad\qquad\qquad
\]
\begin{equation}
\label{CR17}
\simeq\int\mathrm{d}^4w\frac{(w-x)_\mu}{2\sigma}\frac{e^{-\frac
{(w-x)^2}{4\sigma}}}{(4\pi\sigma)^2}W_{Ax,Ew}
\int\mathrm{d}^4z\frac{(z-w)_\nu}{2\sigma}\frac{e^{-\frac
{(z-w)^2}{4\sigma}}}{(4\pi\sigma)^2}W_{Ew,Cz}\left(\frac{-1}{\bar{D}^2}
\right)_{Cz,By}
\end{equation}
We thus see that an $F_{\mu\nu}$ insertion between two expressions 
(\ref{E27}), corresponds to inserting two extra straight segments, with 
the usual segment weights as in (\ref{E27}), and an additional 
pre-exponential factor:
\begin{equation}
\label{CR18}
\frac{(w-x)_\mu}{2\sigma}\frac{(z-w)_\nu}{2\sigma}-
\frac{(w-x)_\nu}{2\sigma}\frac{(z-w)_\mu}{2\sigma}
\end{equation}

As noted in connection with equations (\ref{E292}) to (\ref{E317}), it 
is these $\bar{F}_{\mu\nu}$ insertions that are responsible for 
reversing the sign of the island diagram contributions.

\subsection{The question of whether there might be any further independent relations involving the distances among a finite set of points, for general complex $d$}

It is perhaps worth asking, whether there are any further restrictions, 
in addition to the triangle inequalities, and their higher-dimensional 
generalizations, as discussed above, that might be imposed on real 
$\{r_{ij}\}$, all $\geq0$, in a $d$-dimensional Euclidean space, for 
general complex $d$.  For example, for integer $n$, $n\geq3$, if an 
arbitrary set of real $\{r_{ij}\}$, $1\leq i<j\leq n$, all $\geq0$, is 
given, which satisfy all the triangle inequalities, and their 
higher-dimensional generalizations, up to, and including, the 
$(n-1)$-dimensional simplex inequalities, can a set of $n$ points always 
be found, in some $d$-dimensional Euclidean space, with integer $d\geq 
0$, that realizes this set of $\{r_{ij}\}$?

\section{BPHZ Renormalization, Pauli-Villars Regulators and Higher Derivative Terms, and Lattice Regularization}
\label{BPHZ PVRHDT Lattice}

I have discussed BPHZ renormalization \cite{BP}, \cite{BS}, \cite{Hepp}, 
\cite{Zimmerman}, \cite{BPHZ}, in 
connection with the renormalization group, (equations (\ref{E253}) to 
(\ref{E261}), and with the reversal of the sign of the island diagram 
contributions, (equations (\ref{E292}) to (\ref{E317})).

In \cite{BPHZ}, in an effort to understand better how renormalization 
works in position space, I have given a BPHZ convergence proof directly 
in position space, without any parametrizations of the propagators, by 
finding a method of cutting up the Cartesian product of the 
configuration space of the vertices, and the set of all the forests, 
into a finite number of sectors, each of which is the Cartesian product 
of a subset of the configuration space of the vertices, and a set of 
forests, called a ``good set of forests,'' such that the position space 
integrals can be adequately bounded, and proved to converge, in each 
such sector.  I have given a brief discussion of this method in 
connection with equations (\ref{E283}) to (\ref{E287}).

The outstanding problem is to find the additional finite counterterms 
that restore the Ward-Takahashi identities and the Slavnov-Taylor 
identities.  A simple example of the restoration of a Ward-Takahashi 
identity is given in equation (\ref{E238}).  Recent work by Grassi, 
Hurth, and Steinhauser, \cite{Grassi Hurth Steinhauser}, might be 
helpful in the search for further finite counterterms.

The method of higher-derivative terms in the action, plus Pauli-Villars 
fields to regulate the one-loop divergences, \cite{Slavnov}, \cite{Lee 
Zinn-Justin}, \cite{Bakayev}, is natural for proving the absence of 
anomalies in higher-loop orders, for example, the Adler-Bardeen theorem 
\cite{Adler Bardeen}, because the higher-derivative terms covariantly 
regularize the higher-loop divergences, but not the one-loop 
divergences.  Lee and Zinn-Justin, \cite{Lee Zinn-Justin}, used scalar 
and spinor Pauli-Villars fields to regularize the one-loop divergences, 
while Slavnov, \cite{Slavnov}, used vector Pauli-Villars fields.  Some 
difficulties, which arose in the method with vector Pauli-Villars 
fields, are discussed, with a possible resolution, in reference 
\cite{Bakayev}.

In connection with the Group-Variation Equations, it is interesting to 
note that, due to the use of Landau gauge, it is sufficient to include 
higher-derivative terms in the gauge-invariant part of the action 
density, $\frac{1}{4g^2}F_{\mu\nu a}F_{\mu\nu a}$.  The gauge-fixing and 
Fadeev-Popov terms can be used exactly in the forms (\ref{E6}), and 
(\ref{E7}), at least, at the level of ordinary perturbation theory.  
(The analogue of equations (\ref{E16}) to (\ref{E18}), or of 
(\ref{CR13}) to (\ref{CR15}), in the presence of higher-derivative terms 
in the $\frac{1}{4g^2}F_{\mu\nu a}F_{\mu\nu a}$ part of the action 
density, has not yet been worked out.)  For non-zero $\alpha$ and 
$\beta$ in equation (\ref{E6}), the free gluon propagator has a term 
that is not properly regularized, if we only include higher-derivative 
terms in the $\frac{1}{4g^2}F_{\mu\nu a}F_{\mu\nu a}$ part of the action 
density, but for $\alpha=\beta=0$, or in other words, for Landau gauge, 
this term vanishes.  Power-counting arguments also indicate that, for 
Landau gauge, there is no need for any higher-derivative terms in the 
gauge-fixing, Fadeev-Popov, or Pauli-Villars actions, in order to obtain 
a negative degree of divergence, for all subdiagrams with two or more 
loops.

It would seem possible, furthermore, that the gauge-invariant 
higher-derivative terms in $\frac{1}{4g^2}F_{\mu\nu a}F_{\mu\nu a}$ will 
lead, in Landau gauge, to regularization of the linear divergences along 
Wilson loops, and softening their dependence on the fine details of the 
path, so that it would not be necessary to divide by short-distance 
factors.  There will be higher-dimension gauge-invariant insertions 
along Wilson loops, in the analogues of the propagators (\ref{E16}) to 
(\ref{E18}), or (\ref{CR13}) to (\ref{CR15}), in the presence of 
background fields in the subgroup, but these would be expected to be 
expressible in terms of extra segments along the path, with suitable 
extra pre-exponential weight factors, analogously to (\ref{E29}) and 
(\ref{CR17}).  If this possibility is utilized, it will be necessary to 
ensure that the correct dependence, on the fine details of the paths, 
which would normally be taken care of by the short-distance factors, is 
included in the ansatz for the vacuum expectation values and correlation 
functions.

An extra complication which we might face in this approach, in addition 
to the extra vertices coming from the higher covariant derivative terms
in the gauge-invariant part of the gauge-field action, would be due to
the possibility that we might have to allow the Pauli-Villars regulator
fields to wander in and out of the subgroup.  This would mean that the
analogues of the propagators (\ref{E16}) to (\ref{E18}), or (\ref{CR13})
to (\ref{CR15}), would have to be generalized to higher-dimensional
propagator matrices, with additional rows and columns for the
Pauli-Villars regulator fields, which would have to be calculated in the
presence of both background gauge fields in the subgroup, and background
Pauli-Villars regulator fields in the subgroup.

A compromise might involve using higher covariant derivative terms, in the
Yang-Mills action, to regularize the overall divergences of subdiagrams
with two or more loops, and BPHZ counterterms, with appropriate finite
counterterms to restore the Slavnov-Taylor and Ward-Takahashi identities,
for the one-loop divergences.

With regard to lattice regularization, I do not know, at present, 
whether a gauge-fixing such as the Landau-gauge case of (\ref{E6}) and 
(\ref{E7}) is possible, which ensures that Fadeev-Popov loops stay 
either in the subgroup, or out of it.

In this connection, it is natural to ask, both in the continuum, and on 
the lattice, whether the Group-Variation Equations might be generalized 
to gauges other than Landau gauge, by defining suitable BRS-invariant 
generalizations of Wilson loops, that have the form of Wilson loops, 
with Fadeev-Popov field insertions.  One would attempt to solve a 
generalization of equation (\ref{E14}), or (\ref{E15}), for 4 by 4 block 
matrices, that include rows and columns for the Fadeev-Popov fields not 
in the subgroup, and where the background fields now include the 
Fadeev-Popov fields in the subgroup, in addition to the gauge fields in 
the subgroup.  Then one would attempt to express the propagators, in the 
presence of the background fields in the subgroup, in terms of ``core'' 
non-local parts, that would be generalizations of 
$\left(\frac{-1}{\bar{D}^2}\right)$, and could be represented as path 
integrals, that would be generalizations of the expression (\ref{E27}).  
Then one would attempt to extract the appropriate BRS-invariant 
generalizations of Wilson loops, which would become the physical 
observables, in gauges other than Landau gauge, analogous to the Wilson 
loops, in Landau gauge, by examination of the expressions for the 
generalized path integrals, analogous to (\ref{E27}).  This, also, 
remains a topic for future work.

The renormalization of Wilson loops has been studied by Brandt, Neri,
and Sato \cite{Brandt:1981kf}.

\section{Derivation of the perturbative expansion of the propagator from the path integral form}
\label{Final Section}

Finally, I shall briefly sketch the derivation of equation (\ref{E28}), 
from the expression (\ref{E27}), as an additional check that 
$\left(\frac{-1}{\bar{D}^2}\right)_{Ax,By}$ really can be represented as 
the $\sigma\to0$ limit of the expression (\ref{E27}).

We first check that (\ref{E27}), with the path-ordered phase factor 
removed, is equal to the free, massless, scalar propagator, and indeed,
using the four-dimensional version of (\ref{E189}), we find that
(\ref{E27}) then becomes equal to:
\begin{equation}
\label{FS1}
\sigma\sum_{n=0}^\infty\frac{e^{-\frac{(x-y)^2}{4\sigma(n+1)}}}
{(4\pi\sigma(n+1))^2}
\end{equation}
This, in turn, becomes equal, in the limit $\sigma\to0$, to:
\begin{equation}
\label{FS2}
\int_0^\infty\mathrm{d}s\frac{e^{-\frac{(x-y)^2}{4s}}}{(4\pi s)^2}=
\frac{1}{4\pi^2(x-y)^2}=\left(\frac{-1}{\partial^2}\right)_{xy}
\end{equation}
since the integrand in the left-hand side of (\ref{FS2}) is
Riemann-integrable from 0 to $\infty$, and (\ref{FS1}) is a permitted
approximation to the integral, within the definition of Riemann
integrability, resulting from dividing the integration domain into
intervals of size $\sigma$.  For small, non-zero $\sigma$, the 
difference, between (\ref{FS1}) and (\ref{FS2}), is of order $\sigma$.

We now substitute equation (\ref{E1}) into equation (\ref{E27}), and
seek the coefficient of
\begin{equation}
\label{FS3}
A\raisebox{-4pt}{$\stackrel{\displaystyle{(x_1)}}
 	{\scriptstyle{\mu_1a_1}}$}
A\raisebox{-4pt}{$\stackrel{\displaystyle{(x_2)}}
 	{\scriptstyle{\mu_2a_2}}$}\ldots
A\raisebox{-4pt}{$\stackrel{\displaystyle{(x_m)}}
 	{\scriptstyle{\mu_ma_m}}$}
 	\left(t_{a_1}t_{a_2}\ldots t_{a_m}\right)_{AB}
\end{equation}

In consequence of the path ordering, we may split the sequence of
$A\raisebox{-4pt}{$\stackrel{\displaystyle{(x)}}
{\scriptstyle{\mu a}}$}$ into consecutive groups, which correspond to
$A\raisebox{-4pt}{$\stackrel{\displaystyle{(x)}}
{\scriptstyle{\mu a}}$}$'s that are on the same straight segment in the
path-ordered phase factor $\left(W_{xz_1z_2\ldots z_ny}\right)_{AB}$.
Suppose that among the subindices, 1 to $m$, those that represent the
first $A\raisebox{-4pt}{$\stackrel{\displaystyle{(x)}}
{\scriptstyle{\mu a}}$}$ in a group, or in other words, the first
$A\raisebox{-4pt}{$\stackrel{\displaystyle{(x)}}
{\scriptstyle{\mu a}}$}$ on a new straight segment, are
\begin{equation}
\label{FS4}
1=r_1<r_2<\ldots<r_p\leq m
\end{equation}

Let us consider a fixed $n$ in (\ref{E27}), (noting that $(n+1)\geq p$,
since we need at least $p$ distinct straight segments, for the term
under consideration), and consider the contribution where, for 
$1\leq q\leq p$, the $q^{\mathrm{th}}$ group of
$A\raisebox{-4pt}{$\stackrel{\displaystyle{(x)}}
{\scriptstyle{\mu a}}$}$'s is on the segment, in (\ref{E27}), from
$z_{t_q}$ to $z_{t_q+1}$, such that
\begin{equation}
\label{FS5}
0\equiv t_0\leq t_1<t_2<\ldots<t_p<t_{p+1}\equiv(n+1)
\end{equation}
and I have defined $z_0\equiv x$, and $z_{n+1}\equiv y$, in the notation
of (\ref{E27}).

We Taylor-expand all the $A\raisebox{-4pt}{$\stackrel{\displaystyle{(x)}}
{\scriptstyle{\mu a}}$}$ on a common segment, about the first end of that
segment, in the sequence $z_0,z_1,z_2,\ldots,z_n,z_{n+1}$.  To extract
the coefficient of the contribution with one, or more, 
$A\raisebox{-4pt}{$\stackrel{\displaystyle{(x)}}
{\scriptstyle{\mu a}}$}$, or its derivatives, at each of these points, we
set
\begin{equation}
\label{FS6}
z_{t_q}=w_q=\textrm{fixed}\qquad\qquad1\leq q\leq p
\end{equation}
and integrate only over the other $z_u$'s in (\ref{E27}), $1\leq u\leq 
n$.

For the integral with fixed endpoints $z_0=x$ and $z_{t_1}=w_1$, we get,
if $0<t_1$, the four-dimensional version of (\ref{E189}), with the $n+1$
in (\ref{E189}) set equal to $(t_1-t_0)=t_1$, and the $y$ in (\ref{E189})
set equal to $w_1$, while if $0=t_1$, we simply get $\delta^4(x-w_1)$.

For the integral with fixed endpoints $z_{t_q}=w_q$ and $z_{t_{(q+1)}}=
w_{q+1}$, $1\leq q\leq p$, (with $w_{p+1}\equiv z_{t_{(p+1)}}=z_{n+1}=
y$), we consider specific terms, in the Taylor expansions about $w_q$,
of the $A\raisebox{-4pt}{$\stackrel{\displaystyle{(x)}}
{\scriptstyle{\mu a}}$}$'s on the straight segment from $w_q$ to
$z_{t_q+1}$, and do the path-ordered integrals along the segment, as in
(\ref{E1}).  The result is a numerical coefficient, times an integral,
which is the four-dimensional version of the left-hand side of
(\ref{E189}), with the $n+1$ in (\ref{E189}) set equal to $t_{q+1}-t_q$,
the $x$ in (\ref{E189}) set equal to $w_q$, the $y$ in (\ref{E189}) set
equal to $w_{q+1}$, and the integrand multiplied by a factor
$(z_1-w_q)_\mu$ for each $A\raisebox{-4pt}{$\stackrel{\displaystyle{(x)}}
{\scriptstyle{\mu a}}$}$ on the segment from $w_q$ to
$z_{t_q+1}$, and a factor $(z_1-w_q)_\alpha$, for each derivative
$(\hat{w}_q)_\alpha\equiv\frac{\partial}{\partial(w_q)_\alpha}$ that acts
on a $A\raisebox{-4pt}{$\stackrel{\displaystyle{(w_q)}}
{\scriptstyle{\mu a}}$}$, in the Taylor expansion of a
$A\raisebox{-4pt}{$\stackrel{\displaystyle{(x)}}
{\scriptstyle{\mu a}}$}$ on that segment, about
$A\raisebox{-4pt}{$\stackrel{\displaystyle{(w_q)}}
{\scriptstyle{\mu a}}$}$.

The required integrals can all be obtained by suitable linear
combinations of derivatives, with respect to $x$, of the four-dimensional
version of (\ref{E189}).  In particular, (denoting the derivative, with
respect to a quantity, by that quantity with a hat over it):
\begin{equation}
\label{FS7}
\int\mathrm{d}^4z_1\ldots\int\mathrm{d}^4z_n(z_1-x)_\alpha\frac{e^
{-\frac{(x-z_1)^2}{4\sigma}}}{(4\pi\sigma)^2}\frac{e^
{-\frac{(z_1-z_2)^2}{4\sigma}}}{(4\pi\sigma)^2}\ldots\frac{e^
{-\frac{(z_n-y)^2}{4\sigma}}}{(4\pi\sigma)^2}=2\sigma\hat{x}_\alpha
\frac{e^{-\frac{(x-y)^2}{4\sigma(n+1)}}}{(4\pi\sigma(n+1))^2}
\end{equation}
	\[
\int\mathrm{d}^4z_1\ldots\int\mathrm{d}^4z_n(z_1-x)_\alpha(z_1-x)_\beta
\frac{e^{-\frac{(x-z_1)^2}{4\sigma}}}{(4\pi\sigma)^2}\frac{e^
{-\frac{(z_1-z_2)^2}{4\sigma}}}{(4\pi\sigma)^2}\ldots\frac{e^
{-\frac{(z_n-y)^2}{4\sigma}}}{(4\pi\sigma)^2}\quad=\qquad\qquad\qquad
\quad
\]
\begin{equation}
\label{FS8}
\qquad\qquad\qquad\qquad\qquad\qquad\qquad\qquad=\quad
\left(2\sigma\delta_{\alpha\beta}+(2\sigma)^2\hat{x}_\alpha\hat{x}_\beta
\right)\frac{e^{-\frac{(x-y)^2}{4\sigma(n+1)}}}{(4\pi\sigma(n+1))^2}
\end{equation}
	\[
\int\mathrm{d}^4z_1\ldots\int\mathrm{d}^4z_n(z_1-x)_\alpha(z_1-x)_\beta
(z_1-x)_\gamma
\frac{e^{-\frac{(x-z_1)^2}{4\sigma}}}{(4\pi\sigma)^2}\frac{e^
{-\frac{(z_1-z_2)^2}{4\sigma}}}{(4\pi\sigma)^2}\ldots\frac{e^
{-\frac{(z_n-y)^2}{4\sigma}}}{(4\pi\sigma)^2}\quad=\qquad
\]
\begin{equation}
\label{FS9}
\qquad\qquad=\quad
\left((2\sigma)^2\left(\delta_{\alpha\beta}\hat{x}_\gamma+\delta_{\alpha
\gamma}\hat{x}_\beta+\delta_{\beta\gamma}\hat{x}_\alpha\right)+(2\sigma)^
3\hat{x}_\alpha\hat{x}_\beta\hat{x}_\gamma
\right)\frac{e^{-\frac{(x-y)^2}{4\sigma(n+1)}}}{(4\pi\sigma(n+1))^2}
\end{equation}
We can show by induction that, for integer $a\geq1$, if there are 
$(2a-1)$ or $(2a)$ factors, like $(z_1-x)_\alpha$, in the left-hand side,
of one of these equations, then the leading term, in the right-hand side,
is of order $\sigma^a$.

We substitute these formulae into our contribution, specified by
(\ref{FS4}), (\ref{FS5}), (\ref{FS6}), and specific terms in the Taylor
expansions of the $A\raisebox{-4pt}{$\stackrel{\displaystyle{(x)}}
{\scriptstyle{\mu a}}$}$, about the initial points of the straight 
segments they are on, or in other words, about the appropriate points
$w_q$, $1\leq q\leq p$, and we see that the result is that, firstly, for
the integrals over the $z_u$, $1\leq u<t_1$, we get a factor
\begin{equation}
\label{FS10}
\frac{e^{-\frac{(x-w_1)^2}{4\sigma t_1}}}{(4\pi\sigma t_1)^2}
\end{equation}
if $0<t_1$, and a factor $\delta^4(x-w_1)$ if $0=t_1$.

And secondly, for each $q$, $1\leq q\leq p$, and the integrals over the 
$z_u$, $t_q<u<t_{q+1}$, we get the appropriate derivatives acting on the
$A\raisebox{-4pt}{$\stackrel{\displaystyle{(w_q)}}
{\scriptstyle{\mu_v a_v}}$}$, $r_q\leq v<r_{q+1}$, corresponding to the
specific Taylor terms we are considering, times a numerical coefficient,
that comes from the Taylor expansion coefficients, and the path-ordered
integrals of the appropriate Taylor expansion factors, such as $((w_q)_
\alpha+s_v(z_{t_q+1}-w_q)_\alpha)$, along the straight segment from 
$w_q$ to $z_{t_q+1}$, times a factor that is the 
right-hand side of one of the series of equations that begins with
(\ref{FS7}) to (\ref{FS9}), with $x$ replaced by $w_q$, $y$ replaced by
$w_{q+1}$, and $(n+1)$ replaced by $(t_{q+1}-t_q)$.

We now sum over $n$ in (\ref{E27}), (noting that the term we are
considering, only arises for $(n+1)\geq p$), and over the $t_q$, $1\leq
q\leq p$, subject to (\ref{FS5}).  We define $b_q\equiv(t_{q+1}-t_q)$,
$0\leq q\leq p$, and note that these sums are equivalent to summing over
$b_0$ from $0$ to $\infty$, and over the $b_q$ from $1$ to $\infty$, for
$1\leq q\leq p$.  Then since the only dependence on $b_q$, for each
$0\leq q\leq p$, is through the final factor, in the right-hand side, of
the appropriate equation, in the series of equations that begins with
the four-dimensional version of
(\ref{E189}), (\ref{FS7}), (\ref{FS8}), and (\ref{FS9}), (except for the
case $b_0=0$, which gives a factor $\delta^4(x-w_1)$),
we see that this factor gets transformed, for
each $q$, $0\leq q\leq p$, into the expression (\ref{FS1}), with $x$
replaced by $w_q$, $y$ replaced by $w_{q+1}$, and $(n+1)$ replaced by
$b_q$, and the overall factor of $\sigma$ removed, except that for $q=0$,
we get an additional term $\delta^4(w_0-w_1)$, where I define $w_0$ to be
equal to the $x$ in (\ref{E27}).

We next note that for each $q$, $1\leq q\leq p$, the right-hand side of 
the appropriate equation, in the series of equations that begins with
(\ref{FS7}) to (\ref{FS9}), includes \emph{at least} one factor of
$\sigma$.  Hence, with the overall factor of $\sigma$ in (\ref{E27}), we
have \emph{at least} $(p+1)$ powers of $\sigma$, and we can assign one
of these powers of $\sigma$ to each of the $(p+1)$ versions of the
expression (\ref{FS1}), with the overall factor of $\sigma$ removed, that
we have obtained, (one for each $q$, $0\leq q\leq p$).  We see that we
may therefore take the limit $\sigma\to0$, and that the extra term
$\delta^4(w_0-w_1)$ in the sum, in the version of (\ref{FS1}) we obtain
for $q=0$, is proportional to $\sigma$, and may be dropped, and that all
terms of order $\sigma^2$, or higher, in the series of equations that
begins with (\ref{FS7}) to (\ref{FS9}), may be dropped.

Thus for each $q$, $0\leq q\leq p$, the final factor, in the right-hand
side of the appropriate equation, in the series of equations that begins
with the four-dimensional version of
(\ref{E189}), (\ref{FS7}), (\ref{FS8}), and (\ref{FS9}), has now
been transformed to $\left(\frac{-1}{\partial^2}\right)_{w_qw_{q+1}}$.
And furthermore, we get no contribution, unless, for all $1\leq q\leq p$,
$(r_{q+1}-r_q)\leq 2$ holds, (where I define $r_{p+1}\equiv(m+1)$), and
we have only the zeroth order Taylor terms, if $(r_{q+1}-r_q)=2$, and
only the zeroth order, or the first order, Taylor term, if $(r_{q+1}-r_q)
=1$.  (More carefully, we would have to replace the Taylor expansions, by
the sum of the terms we will keep, plus Taylor remainder terms, and bound
and drop the contribution of every term that contains a Taylor remainder
term.)

If $(r_{q+1}-r_q)=1$, the zeroth-order Taylor term gives
\begin{equation}
\label{FS11}
2\;A\raisebox{-4pt}{$\stackrel{\displaystyle{(w_q)}}
{\scriptstyle{\mu_{r_q} a_{r_q}}}$}(\hat{w}_q)_{\mu_{r_q}}\left(\frac{-1}
{\partial^2}\right)_{w_qw_{q+1}}
\end{equation}
where the factor of 2 comes from the factor of 2 in the right-hand side
of (\ref{FS8}).

The first-order Taylor term gives
\begin{equation}
\label{FS12}
\frac{1}{2}\;2\;\delta_{\mu_{r_q}\alpha}\left((\hat{w}_q)_\alpha
A\raisebox{-4pt}{$\stackrel{\displaystyle{(w_q)}}
{\scriptstyle{\mu_{r_q} a_{r_q}}}$}\right)\left(\frac{-1}
{\partial^2}\right)_{w_qw_{q+1}}
\end{equation}
where the factor of $\frac{1}{2}$ comes from the integral of
$((w_q)_\alpha+s_{r_q}(z_{t_q+1}-w_q)_\alpha)$ with respect to $s_{r_q}$,
and the factor of 2 comes from the factor of 2, in the first term, in the
right-hand side of (\ref{FS9}).

If $(r_{q+1}-r_q)=2$, the zeroth-order Taylor terms give
\begin{equation}
\label{FS13}
\frac{1}{2}\;2\;\delta_{\mu_{r_q}\mu_{r_q+1}}
A\raisebox{-4pt}{$\stackrel{\displaystyle{(w_q)}}
{\scriptstyle{\mu_{r_q} a_{r_q}}}$}
A\raisebox{-4pt}{$\stackrel{\displaystyle{(w_q)\ \;\quad}}
{\scriptstyle{\mu_{r_q+1} a_{r_q+1}}}$}\left(\frac{-1}
{\partial^2}\right)_{w_qw_{q+1}}
\end{equation}
where the factor of $\frac{1}{2}$ comes from the path-ordering, on the 
straight segment, from $w_q$ to $z_{t_q+1}$, and the factor of 2 comes
from the factor of 2, in the first term, in the right-hand side of
(\ref{FS9}).

The sum of (\ref{FS11}), (\ref{FS12}), and (\ref{FS13}) is
\begin{equation}
\label{FS14}
\left((\partial_\mu A_\mu+A_\mu\partial_\mu+A_\mu A_\mu)
\left(\frac{-1}{\partial^2}\right)\right)_{w_qw_{q+1}}
\end{equation}
in the notation of (\ref{E28}).

\vspace{0.5cm}

We appreciate discussions with Dr. D. Fairlie, Prof. W.J. Stirling,
Dr. Malcom Perry, Prof. J.C. Taylor, Prof. Graham Ross, 
Prof. C. Sachrajda, Dr. Roger Phillips, Prof. Aubrey Truman, and
Dr. Graham Shore, and telephone discussions with Prof. D.J. Gross and
Dr. J. Ellis.

\end{document}